\documentclass[12pt]{article}
\usepackage{latexsym}
\usepackage{amsfonts}

\newcommand{\mysection}[1]{\section{#1}\setcounter{equation}{0}}

\newtheorem{theorem}{Theorem}[section]

\newcommand{\qed}{\hbox{${\vcenter{\vbox{
            \hrule height 0.4pt\hbox{\vrule width 0.4pt height 6pt
            \kern5pt\vrule width 0.4pt}\hrule height 0.4pt}}}$}}

\newtheorem{lemma}{Lemma}[section]

\newtheorem{corollary}{Corollary}[section]

\newcommand{\bea}{\begin{eqnarray}} 
\newcommand{\eea}{\end{eqnarray}}
\newcommand{\beann}{\begin{eqnarray*}} 
\newcommand{\eeann}{\end{eqnarray*}}
\newcommand{\beq}{\begin{equation}} 
\newcommand{\eeq}{\end{equation}}
\newcommand{\ba}{\begin{array}} 
\newcommand{\ea}{\end{array}}
\newcommand{\ben}{\begin{enumerate}} 
\newcommand{\een}{\end{enumerate}}

\newcommand{\5}{\bar }  
\newcommand{\6}{\partial } 
\newcommand{\7}{\hat } 
\newcommand{\4}{\tilde }

\newcommand{\gh}{\mathit{gh}} 
\newcommand{\agh}{\mathit{antifd}}

\newcommand{\cA}{{\cal A}}
\newcommand{\cB}{{\cal B}}
\newcommand{\cC}{{\cal C}}

\newcommand{\cF}{{\cal F}}
\newcommand{\cG}{{\cal G}}

\newcommand{\cI}{{\cal I}}

\newcommand{\cL}{{\cal L}}
\newcommand{\cM}{{\cal M}}
\newcommand{\cN}{{\cal N}}
\newcommand{\cO}{{\cal O}}
\newcommand{\cP}{{\cal P}}

\newcommand{\sfrac}[2]{\mbox{$\frac{{#1}}{{#2}}$}}
\newcommand{\f}[3]{{f_{{#1}{#2}}}^{#3}}

\newcommand{\um}{{\underline{m}}}
\newcommand{\uns}{{\underline{s}}}
\newcommand{\uk}{{\underline{k}}}
\newcommand{\tin}{\alpha}
\newcommand{\LRA}{\Leftrightarrow}
\newcommand{\then}{\Rightarrow}
\newcommand{\cov}{[F,\psi]_D}
\newcommand{\Ii}{\mathrm{i}}

\begin{document}

\thispagestyle{empty}

\begin{flushright}
hep-th/0002245
\\
ITP-UH-03/00
\\
ULB-TH-00/05
\end{flushright}
\begin{center}
Journal reference:\\
\ \\
{\sl Physics Reports, Volume 338, Number 5, 
November 2000, pp. 439-569}
\end{center}
\vspace{.5cm}

\begin{center}
{\LARGE
Local BRST cohomology in gauge theories
}
\end{center}
\vspace{.2cm}

\begin{center}
{\large
Glenn Barnich $^{a,b}$, Friedemann Brandt $^{c}$, 
Marc Henneaux $^{a,d}$
}
\end{center}
\vspace{.2cm}

{\sl 
\begin{center}
$^a$  
Physique Th\'eorique et Math\'ematique, Universit\'e
Libre de
Bruxelles,\\ 
Boulevard du Triomphe, Campus Plaine C.P. 231, B-1050
Bruxelles, Belgium
\end{center}
\begin{center}
$^b$ 
Departament de F\'{\i}sica Te\`orica, Universitat de 
Val\`encia, C. Dr. Moliner 50, E-46100 Burjassot (Val\`encia), Spain
\end{center}
\begin{center}
$^c$ 
Institut f\"ur Theoretische Physik, Universit\"at Hannover,
Appelstr. 2,\\ 
D-30167 Hannover, Germany
\end{center}
\begin{center}
$^d$
Centro de Estudios Cient\'{\i}ficos, Casilla 1469, Valdivia, Chile 
\end{center}
}
\begin{center}
E-mail: 
gbarnich@ulb.ac.be, 
brandt@itp.uni-hannover.de,
henneaux@ulb.ac.be
\end{center}

\begin{abstract}
The general solution of the anomaly consistency condition
(Wess-Zumino equation) has been found recently for Yang-Mills gauge
theory.
The general form of the counterterms arising in the
renormalization of gauge invariant operators (Kluberg-Stern and Zuber
conjecture) and in gauge theories of the Yang-Mills type
with non power counting
renormalizable couplings has also been worked out in any number
of spacetime dimensions. This Physics Report
is devoted to reviewing in a self-contained manner these results and
their proofs. This involves computing cohomology groups of the
differential introduced by Becchi, Rouet, Stora and Tyutin,
with the sources of the BRST variations of the fields (``antifields")
included in the problem.
        Applications of this computation to other physical questions 
(classical deformations of the action, conservation laws) are also
considered.
        The general algebraic techniques developed in the Report can be 
applied to other gauge theories, for which relevant references are given.
\end{abstract}
\newpage
\setcounter{page}{1}

   \tableofcontents

\newpage

\mysection{Conventions and notation}
\label{Conventions}

\paragraph{Spacetime.}

Greek indices from the middle of the alphabet generally
are spacetime indices, $\mu=0,1,\dots,n-1$.
We work in $n$-dimensional Minkowskian space
with metric $\eta_{\mu\nu}=\mathrm{diag}(-1,+1,...,+1)$.
The spacetime coordinates are denoted by $x^\mu$. 
The Levi-Civita tensor $\epsilon^{\mu_1\dots\mu_n}$ is 
completely antisymmetric in all its indices with 
$\epsilon^{01 \dots n-1}=1$. Its indices are lowered with the metric 
$\eta_{\mu\nu}$. The differentials $dx^\mu$ anticommute, 
the volume element is denoted by $d^nx$, 
\[
dx^\mu dx^\nu = dx^\mu \wedge dx^\nu = - dx^\nu dx^\mu,\quad
d^nx =dx^0\wedge\dots\wedge dx^{n-1}.
\]

\paragraph{Gauge theories of the Yang-Mills type.}

The gauge group is denoted by $G$, its Lie algebra by $\cG$. 
Capital Latin indices from the middle of the alphabet $I,J,\dots$ 
generally refer to a basis for $\cG$.
The structure constants of the Lie algebra in that basis are denoted by
$\f IJK$. 
\\
The gauge coupling constant(s) are denoted by $e$ and are
explicitly displayed in the formulae. However, in most formulae
we use a collective notation which does not distinguish the 
different gauge coupling constants when the gauge group is the 
direct product of several (abelian or simple) factors.
\\
The covariant derivative is defined by $D_\mu=\partial_\mu 
-eA^I_\mu\,\rho(e_I)$ where $\{e_I\}$ is a basis of $\cG$ and
$\rho$ a representation of $\cG$. The field strength is defined by 
$F_{\mu\nu}^I=\6^{}_\mu A_\nu^I-\6^{}_\nu A_\mu^I
+e\, \f JKI A_\mu^J A_\nu^K$, the corresponding 2-form 
by $F^I=\sfrac 12\, F_{\mu\nu}^Idx^\mu dx^\nu$.

\paragraph{General.}

The Einstein summation convention over repeated upper or lower 
indices generally applies.
\\
Complete symmetrization is denoted by
ordinary brackets $(\cdots)$, complete antisymmetrization by
square brackets $[\cdots]$
including the normalization factor: 
\[
M_{(\mu_1\dots\mu_k)}=\frac 1{k!}\sum_{\sigma\in {\cal S}^k}
M_{\mu_{\sigma(1)}\dots\mu_{\sigma{(k)}}}\ ,
\quad
M_{[\mu_1\dots\mu_k]}=\frac 1{k!}\sum_{\sigma\in {\cal S}^k}
(-)^{\sigma}M_{\mu_{\sigma(1)}\dots\mu_{\sigma{(k)}}}\ ,
\]
where the sums run over all 
elements 
$\sigma$ of the permutation group ${\cal S}^k$ of $k$ objects
and $(-)^{\sigma}$ is $1$ for an even and $-1$ for an odd permutation. 
\\
Dependence of a function $f$ on a set of fields $\phi^i$ and a 
finite number of their derivatives is collectively denoted by 
$f([\phi])$,
\[
f([\phi])\equiv f(\phi^i,\partial_\mu\phi^i,\dots,
\partial_{(\mu_1\dots\mu_r)}\phi^i).
\]
All our derivatives are left derivatives.
\\
The antifield of a field $\phi^i$ is denoted by $\phi^*_i$; 
for instance the antifield corresponding to the Yang-Mills gauge
potential $A^I_\mu$ is $A_I^{*\mu}$.
\\
The Hodge dual of a $p$-form $\omega$ is denoted by $\star\omega$,
\[  
\omega=\frac 1{p!}\,dx^{\mu_1}\dots dx^{\mu_p}\omega_{\mu_1\dots\mu_p}\ ,\
\star\omega=\frac 1{p!(n-p)!}\,dx^{\mu_1}\dots dx^{\mu_{n-p}}
\epsilon_{\mu_1\dots\mu_n}\omega^{\mu_{n-p+1}\dots\mu_n}.
\]

\newpage

\mysection{Introduction}
\label{Intro}

\subsection{Purpose of report}

Gauge symmetries underlie all known fundamental interactions.  While the 
existence of the gravitational force can be viewed as a consequence
of the invariance of the laws of physics under arbitrary spacetime
diffeomorphisms, the non-gravitational interactions are dictated by 
the invariance under an internal non-Abelian gauge symmetry.

It has been appreciated in the last twenty years or so
that many physical questions concerning local gauge
theories can be powerfully reformulated in terms of local BRST
cohomology.  The BRST differential was initially introduced in the
context of perturbative quantum Yang-Mills theory in four dimensions
\cite{Becchi:1974xu,Becchi:1974md,Becchi:1975nq,BecchiTira,Tyutin:1975qk}.
One of the aims was to relate  
the Slavnov-Taylor identities 
\cite{Slavnov:1972fg,Taylor:1971ff}  underlying the
proof of power-counting renormalizability 
\cite{'tHooft:1971fh,'tHooft:1971rn,'tHooft:1972fi,'tHooft:1972ue,%
Lee:1972fj,Lee:1972fk,Lee:1972fm,Lee:1973fn} 
to an invariance of the gauge-fixed action
(for a recent historical account, see \cite{Zinn-Justin:1999ze}).  
However, it was quickly realized that the scope
of BRST theory is much wider.  It can not only  
be formulated for any theory with a gauge freedom,
but also it is quite useful at a purely
classical level.

The purpose of this report is to 
discuss in detail the local BRST cohomology for gauge theories
of the Yang-Mills type.
We do so by emphasizing whenever possible
the general properties of the BRST cohomology
that remain valid in other contexts.
At the end
of the report, we give some references to papers where the
local BRST cohomology is computed for other gauge theories
by means of similar techniques.

\subsection{Gauge theories of the Yang-Mills type}

We first define the BRST differential in the Yang-Mills context.
The Yang-Mills gauge potential is a one-form, which we denote by
$A^I = dx^\mu A^I_\mu$.  The gauge group can be any finite dimensional
group of the form $G = G_0 \times G_1$, where $G_0$ is Abelian 
and $G_1$ is
semi-simple.  In practice, $G$ is compact so that 
$G_0$  is  a product of $U(1)$ factors while $G_1$ is compact and 
semi-simple.   
This makes the standard Yang-Mills
kinetic term definite positive.
However, it will not be necessary to make this assumption
for the cohomological calculation.  This is important as
non-compact gauge groups arise in gravity or
supergravity.
The Lie algebra of the gauge group is denoted by $\cG$.

The matter fields are denoted by $\psi^i$ and can be
bosonic or fermionic.  They are assumed to 
transform linearly under the gauge group according
to a completely reducible representation. The
corresponding representation matrices of $\cG$
are denoted by $T_I$ and the
structure constants of $\cG$ in that basis  are
written $\f IJK$,
\beq
[T_I,T_J]=\f IJK T_K\ .
\label{ym3}\eeq

The field strengths and the
covariant derivatives of the matter fields are denoted by
$F_{\mu\nu}^I$ and $D_\mu\psi^i$ respectively,
\bea
F_{\mu\nu}^I&=&\6^{}_\mu A_\nu^I-\6^{}_\nu A_\mu^I
+e\, \f JKI A_\mu^J A_\nu^K\ ,
\label{ym1}\\
D_\mu\psi^i&=&\6_\mu\psi^i+e\, A_\mu^I T^i_{Ij}\psi^j.
\label{ym2}\eea
Here $e$ denotes the gauge
coupling constant(s) (one for each simple or abelian factor).

The (infinitesimal) gauge transformations read
\begin{equation}
\delta_\epsilon A_\mu^I = D_\mu \epsilon^I,
\; \; \delta_\epsilon \psi^i =  - e \, \epsilon^I T^i_{Ij}\psi^j,
\label{gauge007}
\end{equation}
where the $\epsilon^I$ are the "gauge parameters" and
\beq
D_\mu \epsilon^I = \6_\mu \epsilon^I + 
e\, \f JKI A_\mu^J \epsilon^K.
\label{gauge008}
\eeq
In the BRST formalism, the gauge parameters are replaced by
anticommuting fields $C^I$; these are the ``ghost fields"
of \cite{Feynman:1963ax,DeWitt,Faddeev:1967fc}.

The Lagrangian $L$ is a function of the fields and their
derivatives up to a finite order (``local function"),
\beq
L = L([A^I_\mu], [\psi^i])
\eeq
(see  section \ref{Conventions}
for our notation and conventions).  It is invariant 
under the gauge transformations (\ref{gauge007}) up to a
total derivative, $\delta_\epsilon L = \partial_\mu
k^\mu$ for some $k^\mu$ that may be zero.
The detailed form of the Lagrangian is left
open at this stage except that we assume that the matter sector does
not carry a gauge-invariance of its own, so that (\ref{gauge007})
are the only gauge symmetries.  This requirement is made for
definiteness.  The matter fields
could carry further gauge symmetries (e.g., $p$-form
gauge symmetries) which would bring in
further ghosts; these could be discussed along the same lines but
for definiteness and simplicity, we exclude this possibility.

A theory with the above field content and gauge symmetries is said to be
of the ``Yang-Mills type". Specific forms of the Lagrangian are
$L = (-1/4) \delta_{IJ} F^I_{\mu \nu} F^{J \mu \nu}$,
for which $\delta_\epsilon L = 0$ (Yang-Mills
original theory \cite{Yang:1954ek}) or $d^3x\,L = \mathrm{Tr}
(A dA + (e/3) A^3)$ 
which is invariant only up to a non-vanishing surface term, 
$\delta_\epsilon L =\partial_\mu k^\mu$ with
$k^\mu \not= 0$ (Chern-Simons theory
in $3$ dimensions \cite{Deser:1982vy}).

We shall also consider ``effective Yang-Mills theories" for which
the Lagrangian contains all possible terms compatible with gauge invariance
\cite{Gomis:1996jp,Weinberg:1996kr} and thus involves derivatives of
arbitrarily high order.

In physical applications one usually assumes, of course, 
that $L$ is in addition Lorentz or Poincar\'e invariant. The
cohomological considerations actually go through without
this assumption.

The BRST differential acts in an enlarged space that contains
not only the original fields and the
ghosts, but also sources for the BRST variations of the
fields and the ghosts.  These sources are denoted by $A_I^{*\mu}$,
$\psi^*_i$ and $C_I^*$ respectively, and have Grassmann parity
opposite to the one of the corresponding fields.  They have been introduced
in order to control how the
BRST symmetry gets renormalized
\cite{Becchi:1974xu,Becchi:1974md,Becchi:1975nq,%
Zinn-Justin:1974mc,Zinn-Justin:1989mi}.  They play a crucial role in the
BV construction \cite{Batalin:1981jr,%
Batalin:1983wj,Batalin:1983jr} where they are known as the antifields;
for this reason, they will be indifferently called antifields or
BRST sources here. 

The BRST-differential decomposes into the sum of two differentials
\beq
s=\delta+\gamma
\label{ym6}
\eeq
with $\delta$ and $\gamma$ acting as 
\beq
\ba{c|c|c}
Z & \delta Z & \gamma Z
\\
\hline\rule{0em}{3ex}
A_\mu^I & 0 & D_\mu C^I
\\
\rule{0em}{3ex}
\psi^i & 0 & -e\, C^IT^i_{Ij}\psi^j
\\
\rule{0em}{3ex}
C^I & 0 & \sfrac 12\, e\, \f KJI\, C^J C^K
\\
\rule{0em}{3ex}
C_I^* & -D_\mu A_I^{*\mu} - e\,\psi^*_i T^i_{Ij} \psi^j &
   e\, \f JIK\, C^JC_K^*
\\
\rule{0em}{3ex}
A_I^{*\mu} & L^\mu_I &
   e\, \f JIK\, C^JA_K^{*\mu}
\\
\rule{0em}{3ex}
\psi^*_i & L_i & e\, C^I \psi^*_j T^j_{Ii}
\label{ym7}
\ea
\eeq
where 
\bea
D_\mu C^I &=& \6_\mu C^I+ e\, \f JKI A_\mu^J C^K
\label{ym8}\\
D_\mu A_I^{*\mu}&=&\6_\mu A_I^{*\mu}-e\, \f JIK A_\mu^J A_K^{*\mu}
\label{ym9}\\
L^\mu_I&=&\frac{\delta L}{\delta A_\mu^I}\ ,\quad
L_i=(-)^{\epsilon_i}\frac{\delta L}{\delta \psi^i}\ .
\label{ym10}
\eea
Here $\delta L/\delta \phi$ denotes the Euler-Lagrange derivative
of $L$ with respect to $\phi$ (taken from the left),
and $\epsilon_i$ is the Grassmann
parity of $\psi^i$ ($\epsilon_i=0$ for a boson, $\epsilon_i=1$ for
a fermion). The parity dependent sign in $L_i$ originates from the
convention that $\delta$ is defined through Euler-Lagrange 
right-derivatives of $L$.

$\delta$, $\gamma$ (and thus $s$) are extended to 
the derivatives of the variables 
by the rules $\delta \partial_\mu = \partial_\mu \delta$, 
$\gamma \partial_\mu = \partial_\mu \gamma$.  These rules imply
$\delta d + d \delta = \gamma d + d \gamma = s d + d s = 0$, where
$d$ is the exterior spacetime derivative $d = dx^\mu \partial_\mu$, because
$dx^\mu$ is odd.  Furthermore, $\delta$, $\gamma$, $s$ and $d$ act as
left (anti-)derivations, e.g., $\delta (ab) = (\delta a) b + (-1)^{\epsilon_a}
a \delta b$, where $\epsilon_a$ is the Grassmann parity of $a$.

The BRST differential is usually not presented in the above manner
in the literature; 
we shall make contact with the more familiar formulation in the 
appendix \ref{Intro}.A below.
However, we point out already that
on the fields $A_\mu^I$, $\psi^i$ and the ghosts $C^I$, $s$ reduces
to $\gamma$ and the BRST transformation takes the familiar form of a
``gauge transformation in which the gauge parameters are 
replaced by the ghosts" (for the fields $A_\mu^I$ and $\psi^i$),
combined with $s C^I = \frac{1}{2} e  \f KJI C^J C^K$ in order
to achieve nilpotency.  The differential $\delta$, known as
the Koszul-Tate differential, acts non-trivially only on the antifields.  It
turns out to play an equally important r\^ole in the
formalism.

The decomposition (\ref{ym6}) is related to the various gradings that
one introduces in the algebra generated by the fields, ghosts and antifields.
The gradings are the pure ghost number $\mathit{puregh}$, the antifield number
$\agh$ and the (total) ghost number $\gh$.  They are not independent
but related through
\begin{equation}
\gh = \mathit{puregh} - \agh.
\end{equation}
The antifield number is also known as ``antighost number".
The gradings of the basic variables are given by
\beq
\ba{c|c|c|c}
Z & \mathit{puregh}(Z) & \agh(Z) & \gh(Z)
\\
\hline\rule{0em}{3ex}
A_\mu^I & 0 & 0 & 0 
\\
\rule{0em}{3ex}
\psi^i & 0 & 0 & 0 
\\
\rule{0em}{3ex}
C^I & 1 & 0 & 1 
\\
\rule{0em}{3ex}
C_I^* & 0 & 2 & -2
\\
\rule{0em}{3ex}
A_I^{*\mu} & 0 & 1 & -1
\\
\rule{0em}{3ex}
\psi^*_i & 0 & 1 & -1 
\ea
\eeq                   

The BRST differential  and the differentials $\delta$
and $\gamma$ all increase the ghost number by one unit.
The differential $\delta$ does so by decreasing the antifield number
by one unit
while leaving the pure ghost number unchanged, while $\gamma$ does
so by increasing the pure ghost number by one unit while leaving
the antifield number unchanged.
Thus, one has $\agh(\delta) = -1$ and $\agh(\gamma) = 0$.
The decomposition (\ref{ym6}) corresponds to an expansion of 
$s$ according to the antifield number, $s = \sum_{k\geq -1} s_k$,
with $\agh(s_k) = k$, $s_{-1} = \delta$ and 
$s_0 = \gamma$.  For Yang-Mills gauge models, the expansion stops
at $\gamma$.  For generic gauge theories, in particular theories 
with open algebras, there are higher-order
terms $s_k$, $k \geq 1$ (when the gauge algebra closes off-shell,
$s_1,s_2,\dots$ vanish on the fields, but not necessarily
on the antifields).

\subsection{Relevance of BRST cohomology}

As stated above, the BRST symmetry provides an extremely efficient tool
for investigating many aspects of a gauge theory.
We review in this subsection the contexts in which it is 
useful.  
We shall only list the physical questions connected
with BRST cohomological groups without making explicitly the
connection here.  This connection can be found in the references listed
in the course of the discussion. [Some comments on the
link between the BRST symmetry and perturbative
renormalization  are surveyed -- very
briefly -- in appendix \ref{Intro}.A below to indicate
the connection with the gauge-fixed formulation in which
these questions arose first.] 

A crucial feature of the BRST differential $s$ is that it
is a differential, i.e., it squares to zero,
\begin{equation}
s^2 = 0.
\end{equation}
In Yang-Mills type theories (and in other gauge theories for which
$s=\delta+\gamma$) this is equivalent
to the fact that $\delta$ and $\gamma$ are differentials that
anticommute,
\begin{equation}
\delta^2 = 0, \; \delta \gamma + \gamma \delta = 0, \; \gamma^2 = 0.
\end{equation}

The BRST cohomology is defined as follows.  First, the BRST cocycles
$A$ are objects that are ``BRST -closed", i.e., in the kernel of $s$,
\begin{equation}
s A = 0. 
\end{equation}
Because $s$ squares to zero, the BRST-coboundaries, i.e., the
objects that are BRST-exact (in the image of $s$) are automatically closed,
\begin{equation}
A = s B \; \; \; \Rightarrow  \; \; \; sA = 0.
\end{equation}
The BRST-cohomology is defined as the quotient space Ker$\, s/$Im$\,s$,
\begin{equation}
H(s) = \frac{\hbox{Ker}\, s}{\hbox{Im}\, s}.
\end{equation}
An element in $H(s)$ is an equivalence class of BRST-cocycles, where
two BRST-cocycles are identified if they differ by a BRST-coboundary.
One defines similarly $H(\delta)$ and $H(\gamma)$. 

One can consider the BRST cohomology in the space of local functions or in the
space of local functionals.  In the first instance, the ``cochains" are local
functions, i.e., are functions of the fields, the ghosts, the
antifields and a finite number of their
derivatives.  In the second instance, the cochains are local functionals
i.e., integrals of local volume-forms, $A = \int a$, $a=f\, d^nx$, where
$f$ is a local function in the above sense ($n$ denotes the spacetime
dimension).  When
rewritten in terms of the integrands
(see subsection \ref{relationlocal}), the BRST cocycle and coboundary
conditions $s \int a = 0$ and $\int a = s \int b$ become
respectively
\begin{eqnarray}
s a + dm = 0 &\;& \hbox{(cocycle condition)}, \label{AAA}\\
a = sb + dn  &\;& \hbox{(coboundary condition)},
\label{AAAA}
\end{eqnarray}
for some local $(n-1)$-forms $m$ and $n$.  For this reason, we shall
reserve the notation $H(s)$ for the BRST cohomology in the
space of local functions and denote by $H(s\vert d)$ the BRST
cohomology in the space of local functionals. So $H(s|d)$ is defined
by (\ref{AAA}) and (\ref{AAAA}), while $H(s)$ is defined by 
\begin{eqnarray}
s a = 0 &\;& \hbox{(cocycle condition)}, \label{BBB}\\
a = sb &\;& \hbox{(coboundary condition)}.
\label{BBBB}
\end{eqnarray}
In both cases, $a$ and $b$ (and $m,n$) are local forms.

One may of course fix
the ghost number and consider cochains with definite ghost number.
At ghost number $j$,
the corresponding groups are denoted $H^j(s)$ and $H^{j,n}(s \vert d)$
respectively, where in the latter instance
the cochains have form-degree
equal to the spacetime dimension $n$.  
The calculation of $H^{j,n}(s \vert d)$ is more complicated than
that of $H^j(s)$ because one must allow for the possibility to
integrate by parts.  It is useful
to consider the problem defined by (\ref{AAA}), (\ref{AAAA})
for other values $p$ of
the form-degree.  The corresponding cohomology groups are denoted
by $H^{j,p}(s |d)$. 

Both cohomologies $H(s)$ and $H(s \vert d)$ capture important physical
information about the system.  The reason that the BRST symmetry is important
at the quantum level is that the standard method for quantizing
a gauge theory begins with fixing the gauge.  The BRST symmetry and
its cohomology become then substitutes for gauge invariance, which
would be otherwise obscure.  The
realization that BRST theory is also useful at the classical level is more 
recent.   We start by listing the quantum questions for which the
local BRST cohomology is relevant (points 1 though 4 below).
We then list the classical ones (points 5 and 6).

\begin{enumerate}
\item Gauge anomalies: $H^{1,n}(s \vert d)$.

Before even considering whether a quantum gauge
field theory is renormalizable, one must check whether
the gauge symmetry is anomaly-free.
Indeed,
quantum violations of the classical gauge symmetry
presumably spoil unitarity and 
probably render the quantum theory inconsistent.
As shown in \cite{Becchi:1974xu,Becchi:1974md,Becchi:1975nq},
gauge anomalies ${\cal A} = \int a$ are ghost number one local functionals
constrained by the cocycle condition $s {\cal A} = 0$, i.e.,
\begin{equation}
s a + dm = 0 \label{WZ},
\end{equation}
which is the BRST generalization 
of the Wess-Zumino consistency condition \cite{Wess:1971cm}.
Furthermore, trivial solutions can be removed by adding local
counterterms to the action.  Thus, the cohomology group
$H^{1,n}(s \vert d)$ characterizes completely the form of the non trivial
gauge anomalies.  Once $H^{1,n}(s \vert d)$ is known,
only the coefficients of the candidate anomalies need to be determined.

The computation of $H^{1,n}(s \vert d)$ was started
in \cite{Dixon:1979bs,Stora:1976kd,Bonora:1983ve,Stora:1983ct,%
Zumino:1983ew,Zumino:1984rz,%
Baulieu:1984iw,%
Thierry-Mieg:1984vx,Baulieu:1985tg,%
Manes:1985df,Dubois-Violette:1985hc,Dubois-Violette:1985jb,%
Bandelloni:1986wz,Bandelloni:1987kg} and completed in the
antifield-independent case in
\cite{Brandt:1989rd,Brandt:1990gv,Brandt:1990gy,Dixon:1991wi,%
Dubois-Violette:1992ye}.  Some aspects of
solutions with antifields included have been
discussed in
\cite{%Becchi:1974md,
Becchi:1975nq,Dixon:1976aa,Bandelloni:1978ke,%
Bandelloni:1978kf,Brandt:1994sn}~ ;
the general solution
was worked out more recently in
\cite{Barnich:1994ve,Barnich:1995db,Barnich:1995mt}.

\item Renormalization of Yang-Mills gauge models: $H^{0,n}(s \vert d)$.

Yang-Mills theory in four dimensions is power-counting renormalizable
\cite{'tHooft:1971fh,'tHooft:1971rn,'tHooft:1972fi,'tHooft:1972ue}.
However, if one includes interactions of higher mass dimensions, or
considers the Yang-Mills Lagrangian in higher spacetime
dimensions, one loses power-counting renormalizability.  To
consistently deal with these theories, the effective theory viewpoint is
necessary \cite{Weinberg:1979kz} (for recent reviews, see
\cite{Polchinski:1992ed,Georgi:1992xg,Kaplan:1995uv,Manohar:1996cq,%
Pich:1998xt}).
The question arises then as to whether these
theories are renormalizable in the ``modern sense", i.e., whether
the gauge symmetry constrains the divergences sufficiently, so that these
can be absorbed by gauge-invariant counterterms at each order of
perturbation theory \cite{Gomis:1996jp,Weinberg:1995mt,%
Weinberg:1996kr,Weinberg:1996kw}.
This is necessary for dealing meaningfully with loops -- and
not just tree diagrams.
This question can again be formulated in terms of BRST cohomology.
Indeed, the divergences and counterterms are constrained by
the cocycle condition (\ref{WZ}) but this time at ghost number zero;
and trivial solutions can be absorbed through field redefinitions
\cite{Zinn-Justin:1974mc,Becchi:1974xu,Becchi:1974md,Becchi:1975nq,%
Dixon:1975si,Tyutin:1981wa,Voronov:1982cp,Voronov:1982ur,Voronov:1982ph,%
Lavrov:1985hr,Anselmi:1994ry,Anselmi:1995zx,Weinberg:1996kr}. 

Thus it is $H^{0,n}(s \vert d)$ that controls the counterterms.  The
question raised above can be translated, in cohomological
terms, as to whether the most general solution
of the consistency condition $sa + dm =0$ at
ghost number zero can be written as $a = h\,d^nx + sb
+ dn$ where $h$ is a local function
which is off-shell gauge invariant up to a total derivative.
The complete answer to this question, which is affirmative
when the gauge-group is semi-simple,
has been worked out
recently 
\cite{Barnich:1994ve,Barnich:1995db,Barnich:1995mt}.

\item Renormalization of composite, gauge-invariant operators:
$H^0(s)$ and $H^{0,n}(s \vert d)$.

The renormalization of composite gauge-invariant operators arises in
the analysis of the operator product expansion, relevant, in particular,
to deep inelastic scattering.  In the mid-seventies, it was conjectured that
gauge-invariant operators can only mix, upon
renormalization,  with gauge-invariant operators
or with gauge-variant operators which vanish in physical matrix
elements \cite{Kluberg-Stern:1975rs,Kluberg-Stern:1975xv,Kluberg-Stern:1975hc}
(see also \cite{DixonTaylor,Deans:1978wn}).  

As established also there, the
conjecture is equivalent  to proving that in each BRST
cohomological class at ghost number zero, one can choose a representative
that is strictly gauge-invariant.  For local operators involving
the variables and their derivatives (at a given, unspecified, point)
up to some finite order,  the  relevant
cohomology is $H^0(s)$.  The
problem is to show that the most general solution of $sa = 0$ at
ghost number zero is of the form $a = I + sb$, where $I$ is
an invariant function of the curvatures $F^I_{\mu \nu}$,
the fields $\psi^i$ and a finite number of their covariant derivatives.
For operators of low mass dimension,  the problem involves a small
number of possible composite fields and can be analyzed rather
directly.
However, 
power counting becomes
a less constraining tool for operators of high mass dimension,
since the number of composite fields with the appropriate
dimensions proliferates as the dimension increases.  One must use 
cohomological techniques that do not rely on power counting.  The
conjecture turns out to be correct and
was proved first (in four spacetime dimensions) in
\cite{Joglekar:1976nu}.   The proof was streamlined in 
\cite{Henneaux:1993jn} using the above crucial decomposition of
$s$ into $\delta + \gamma$.  Further information on this topic
may be found in \cite{Collins:1984xc}.  Recent applications are given in
\cite{Collins:1994ee,Harris:1995tp}.

In the same way, the cohomological group $H^{0,n}(s \vert d)$ controls the
renormalization of integrated gauge-invariant operators $\int
d^nx\ O(x)$, or, as one also says, operators at zero momentum.
As mentioned above, its complete resolution has only been
given recently
\cite{Barnich:1994ve,Barnich:1995db,Barnich:1995mt}.  

\item Anomalies for gauge invariant operators: 
$H^1(s)$ and $H^{1,n}(s \vert d)$.

The question of mixing of gauge invariant operators as discussed in the 
previous paragraph may be obstructed by anomalies 
if a non gauge invariant
regularization and renormalization scheme is or has to be used 
\cite{Dixon:1980zp}. 
To lowest order, these anomalies
are ghost number $1$ local functions $a$ and 
have to satisfy the consistency condition $s a =0$. BRST-exact anomalies
are trivial because they 
can be absorbed through the addition of non BRST invariant operators, 
which compensate for the non invariance of the scheme. 
This means that the cohomology constraining these obstructions is 
$H^1(s)$.      
Similarly,
the group $H^{1,n}(s \vert d)$
controls the anomalies in the renormalization of
integrated gauge invariant composite operators \cite{Dixon:1980zp}.

\item Generalized conservation laws - ``Characteristic cohomology"
and BRST cohomology in negative ghost number

The previous considerations were quantum.  The BRST cohomology 
captures also important classical information about
the system. For instance, it has been proved in
\cite{Barnich:1995db} that the
BRST cohomology at negative ghost number is isomorphic to the
so-called ``characteristic cohomology" 
\cite{vinogradov1,vinogradov2,Bryant}, which generalizes
the familiar notion of (non-trivial) conserved currents.
A conserved current can be defined (in Hodge dual terms) as
an $(n-1)$-form that is $d$-closed modulo the equations of motion,
\begin{equation}
d j \approx 0
\end{equation}
where $\approx$ means ``equal when the equations of motion hold".
One says that a conserved current is (mathematically) trivial when
it is on-shell equal to an exact form (see
e.g. \cite{olver}),
\begin{equation}
j \approx dm,
\end{equation}
where $m$ is a {\em local} form.
The characteristic cohomology in form degree $n-1$ is by definition the
quotient space of conserved currents by trivial ones.  The 
characteristic cohomology in arbitrary form degree is defined by
the same equations, taken at the relevant form degrees.
The characteristic cohomology in form-degree $n-2$ plays a r\^ole
in the concept of ``charge without charge" \cite{misner-wheeler,Unruh}.

The exact correspondence between the BRST cohomology
and the characteristic cohomology is as follows \cite{Barnich:1995db}:
the characteristic cohomology $H^{n-k}_{char}$ in form-degree $n-k$
is isomorphic to the BRST cohomology $H^{-k,n}(s \vert d)$
in ghost number $-k$ and maximum form-degree $n$.\footnote{The
isomorphism assumes the De Rham cohomology of the spacetime manifold to be
trivial.  Otherwise, the statement needs to be refined along the lines of
%there are subtleties that can be
%resolved out using the results of 
\cite{Dubois-Violette:1991is}.
A proof of the isomorphism is given in sections \ref{Koszul2Section}
and \ref{homological}.}
Cocycles of the cohomology $H^{-k,n}(s \vert d)$ involve necessarily
the antifields since these are the only variables with negative
ghost number.

The reformulation of the characteristic cohomology in terms of BRST cohomology
is particularly useful in form degree $<n-1$, where it has yielded 
new results leading to a complete calculation of $H^{n-k}_{char}$
for $k>1$.  It
is also useful in form-degree $n-1$, where it enables one
to work out the explicit
form of all the conserved currents that are not gauge-invariant
(and cannot be invariantized by adding trivial terms)
\cite{Barnich:1995cq}.  The calculation of the gauge-invariant currents
is more complicated and depends on the specific choice of the
Lagrangian, contrary to
the characteristic cohomology in lower form-degree.

\item Consistent interactions: $H^{0,n}(s_0 \vert d)$, $H^{1,n}(s_0 \vert d)$ - 
Uniqueness of Yang-Mills cubic vertex

Is is generally believed that the only way to make a set of
massless vector fields consistently interact is through
the Yang-Mills construction - apart from interactions that do not
deform the abelian gauge transformations and involve only the abelian 
curvatures or abelian Chern-Simons terms.
Partial proofs of this result
exist \cite{arno-deser,wald1,Barnich:1994pa} but these always make
implicitly some restrictive assumptions on the
number of derivatives involved in the coupling or the
polynomial degree of the interaction.  In fact, a counterexample
exists in three spacetime dimensions, which generalizes the
Freedman-Townsend model for two-forms in four dimensions
\cite{Freedman:1981us,Anco:1995wt,Anco:1997mw}.  

The problem of constructing consistent (local) interactions
for a gauge field theory has been formulated in general
terms in \cite{Berends:1985rq}.  It turns out that
this formulation has in fact a natural interpretation
in terms of deformation theory and
involves the computation of the free
BRST cohomologies $H^{0,n}(s_0 \vert d)$ and 
$H^{1,n}(s_0 \vert d)$\cite{Barnich:1993vg}
(see also \cite{Gomis:1993ub,Stasheff:1997fe,Henneaux:1997bm}),
where $s_0$ is the free BRST differential. The BRST
point of view systematizes the search for consistent interactions
and the demonstration of their uniqueness up to field
redefinitions.

\end{enumerate}

We present in this report complete results on $H^k(s)$. We
also provide, for a very general class of Lagrangians, a complete
description of the cohomological groups $H(s \vert d)$ in terms of
the non trivial conserved currents that the model may have.
So, all these cohomology
groups are known once all the global symmetries
of the theory have been determined -- a problem that depends on the
Lagrangian.   Furthermore, we show that the cocycles in the cohomological
groups $H^{0,n}(s \vert d)$ (counterterms)
and $H^{1,n}(s \vert d)$ (anomalies) may be chosen
not to  involve the conserved currents
when the Yang-Mills gauge
group has no abelian factor (in contrast to the groups
$H^{g,n}(s \vert d)$ for other values of $g$).  So, in this case, we can also
work out completely $H^{0,n}(s \vert d)$ and $H^{1,n}(s \vert d)$, without
specifying $L$. 
We also present complete results for 
$H^{-k,n} (s \vert d)$ (with $k>1$) as well as partial results
for $H^{-1,n}(s \vert d)$.  Finally, we establish the 
uniqueness of the Yang-Mills cubic vertex in four spacetime dimensions,
using a result of Torre \cite{Torre:1995kb}.

\subsection{Cohomology and antifields}

The cohomological investigation of the BRST symmetry
was initiated as early as in the seminal papers
\cite{Becchi:1974xu,Becchi:1974md,Becchi:1975nq}, 
which gave birth to the modern algebraic
approach to the renormalization of gauge theories
(for a recent monograph on the subject, see
\cite{Piguet:1995er}; see also 
\cite{Piguet:1981nr,Bonneau:1990xu,Ferrari:1998jy,Grassi:1999tp}).
Many results on the
antifield-independent cohomology were established in the following
fifteen years. However, the antifield-dependent case remained largely
unsolved and almost not treated at all although a complete answer to the 
physical questions listed above requires one to tackle the BRST 
cohomology without a priori restrictions on the antifield dependence. 

In order to deal efficiently with the antifields in the BRST
cohomology, a new qualitative
ingredient is necessary.  This new ingredient is the understanding that
the antifields are algebraically associated with the
equations of motion in a well-defined fashion, which
is in fact quite standard in cohomology theory.  With this
novel interpretation of the antifields, new progress could be made
and previous open conjectures could be proved.

Thus, while the original point of view on the antifields
(sources coupled to the BRST variations of the fields
\cite{Becchi:1974xu,Becchi:1974md,Becchi:1975nq,%
Zinn-Justin:1974mc,Zinn-Justin:1989mi}) is useful
for the purposes of renormalization theory, the
complementary interpretation in terms of equations of
motion is quite crucial for cohomological calculations.
Because this interpretation of the antifields,
related to the so-called ``Koszul-Tate complex",  plays a central r\^ole
in our approach,
we shall devote two entire sections to explaining it (sections 
\ref{KoszulSection} and \ref{Koszul2Section}).

The relevant interpretation of the antifields originates from
work on the Hamiltonian formulation of the BRST symmetry, 
developed by the Fradkin school 
\cite{Fradkin:1975cq,Fradkin:1977wv,Fradkin:1977hw,Batalin:1977pb,%
Fradkin:1978xi,Batalin:1983pz},
where a similar interpretation can be given
for the momenta conjugate to the ghosts
\cite{Henneaux:1985kr,Dubois-Violette:1987uu,Mcmullan:1987hb,%
Browning:1987hc,Henneaux:1988ej,%
Fisch:1989dq,Stasheff:1991eb}.  The reference \cite{Fisch:1989dq}
deals in particular with the case of reducible constraints, which is
the closest to the Lagrangian case from the algebraic
point of view.
The extension of the  work on the Hamiltonian
Koszul-Tate complex to the antifield formalism
was carried out in \cite{Fisch:1990rp,Henneaux:1989jq}. 
Locality was analyzed in \cite{Henneaux:1991rx}.

What enabled one to identify the Koszul-Tate differential as a key
building block was the attempt to generalize the BRST construction
to more general settings, in which the algebra of
the gauge transformations closes only
``on-shell".  In that case, it is crucial to introduce
the Koszul-Tate differential from the outset; the BRST differential
is then given by $s = \delta + \gamma + ``\mbox{more}"$, where ``more"
involves derivations of higher antifield number.  Such a generalization
was first uncovered in the case of supergravity 
\cite{Kallosh:1978de,Sterman:1978ds}.  The construction was then
systematized in \cite{deWit:1978cd,vanholten} and given its present
form in \cite{Batalin:1981jr,Batalin:1983jr,Batalin:1985qj}.  However, even
when the algebra closes off-shell, the Koszul-Tate differential
is a key ingredient for cohomological purposes.

\subsection{Further comments}

The antifield formalism is extremely rich and we shall exclusively 
be concerned here with its cohomological aspects in the context of
local gauge theories.  So, many of its properties (meaning
of antibracket and 
geometric interpretation of the master equation \cite{Witten:1990wb}, 
anti-BRST symmetry with antifields \cite{Batalin:1991qy}, quantum
master equation and its regularization 
\cite{Batalin:1984ss,Troost:1990,Howe:1990pz,Tonin:1992wf}) will not be
addressed.  A review with the emphasis on the algebraic interpretation
of the antifields used here is \cite{Henneaux:1992ig}.  Other reviews are
\cite{Troost:1994xw,Troost:1993mr,Gomis:1995he}.  

It would be impossible - and out of place - to list here
all references dealing with one aspect or the other of the antifield 
formalism; we shall thus quote only some papers from the last
years which appear to be representative of the general trends.
These are \cite{Batalin:1996mp,Khudaverdian:1993ji,Alexandrov:1997kv,%
Grigorev:1999qz,Grigorev:2000zg} (geometric aspects of the antifield 
formalism),
\cite{Tataru:1998pn,Geyer:1999qi,Batalin:1999fi} 
($Sp(2)$-formalism), 
\cite{Batalin:1999gf} (quantum antibrackets),
\cite{Bering:1996kw} (higher antibrackets).  We are fully aware
that this list is incomplete but we hope that the interested
reader can work her/his way through
the literature from these references.

Finally, a  monograph dealing with aspects of
anomalies complementary to those discussed here is 
\cite{Bertlmann:1996xk}.

\subsection{Appendix \ref{Intro}.A: 
Gauge-fixing and antighosts}
\label{historical}

The BRST differential as we have introduced it is manifestly gauge-independent
since nowhere in the definitions did we ever fix the gauge.
That this is the relevant differential for the classical questions
described above (classical deformations of the action, conservation
laws) has been established in \cite{Barnich:1993vg,Barnich:1995db}.  Perhaps
less obvious is the fact that this is also the relevant differential for
the quantum questions.  Indeed, the BRST differential is usually introduced
in the quantum context only after the gauge has been fixed and one may wonder
what is the connection between the above  definitions and the more usual ones.

A related aspect is that we have not included
the antighosts.  There is in fact a good reason
for this, because these do not enter the cohomology: they
only occur through ``trivial" terms because, as one
says, they are in the contractible part of the algebra
\cite{sullivan}.

To explain both issues, we first recall the usual derivation of the BRST
symmetry.  First, one fixes the gauge through gauge conditions
\beq
{\cal F}^I = 0,
\eeq
where the ${\cal F}^I$ involve the gauge potential, the matter fields and their
derivatives.  For instance, one may take ${\cal F}^I = \partial^\mu A^I_\mu$
(Lorentz gauge).  Next, one writes down the gauge-fixed action
\beq
S_F = S^{\mathit{inv}} + S^{\mathit{gf}} + S^{\mathit{ghost}}\label{1.24}
\eeq
where $S^{\mathit{inv}} = \int d^n x\ L$ is
the original gauge-invariant action, $S^{gf}$
is the gauge-fixing term
\beq
S^{\mathit{gf}} = \int d^nx\ b_I ({\cal F}^I + \frac{1}{2} b^I)
\eeq
and $S^{\mathit{ghost}}$ is the ghost action,
\beq
S^{\mathit{ghost}} = -\Ii \int d^n x
\left[
D_\mu C^J\,\frac {\delta (\bar{C}_I {\cal F}^I)}{\delta A^J_\mu}
-e\, C^J T_{Jj}^i\psi^j\, \frac {\delta (\bar{C}_I {\cal F}^I)} 
{\delta \psi^i}
\right].
\eeq
The $b^I$'s are known as the auxiliary fields, while the $\bar{C}_I$ are
the antighosts.  We take both the ghosts and the antighosts to be real.
 
The gauge-fixed action is invariant under the BRST symmetry
$\sigma A^I_\mu = D_\mu C^I$, $\sigma \psi^i = -e C^I T^i_{Ij} \psi^j$,
$\sigma C^I = (1/2) e \f KJI C^J C^K$,
$\sigma \bar{C}_I = \Ii b_I$ and
$\sigma b_I = 0$.  This follows from the gauge-invariance of the
original action as well as from $\sigma^2 = 0$.  $\sigma$ coincides
with $s$ on $A^I_\mu$, $\psi^i$ and $C^I$ but we use a
different letter to avoid confusion.                                             
One can view the auxiliary fields $b^I$ as the ghosts
for the (abelian) gauge shift symmetry $\bar{C}_I
\rightarrow \bar{C}_I + \epsilon_I$ under which
the original Lagrangian is of course invariant since it
does not depend on the antighosts.
 
To derive the Ward-Slavnov-Taylor identities associated with the original
gauge symmetry and this additional shift symmetry,
one introduces
sources for the BRST variations of all the variables,
including the antighosts and the auxiliary
$b^I$-fields.
This yields
\beq
S^{\mathit{total}} = S^{\mathit{inv}} + S^{\mathit{gf}} + S^{\mathit{ghost}}
 + S^{\mathit{sources}}
\label{1.27}
\eeq
with 
\beq
S^{\mathit{sources}} = -\int d^n x\ (\sigma A^I_\mu K^\mu_I +\sigma \psi^i K_i
+\sigma C^I L_I + \Ii b_I M^I)
\eeq
where $K^\mu_I$, $K_i$,
$L_I$ and $M_I$ are respectively the sources for
the BRST variations of $A^I_\mu$, $\psi_i$, $C^I$
and $\bar{C}_I$.  We shall also denote by $N^I$ the sources associated
with $b_I$.  
The antighosts $\bar{C}_I$, the auxiliary fields $b_I$ and
their sources define the ``non-minimal sector".
 
The action $S^{\mathit{total}}$ fulfills the identity
\beq
(S^{total}, S^{total}) = 0
\label{nilpo0}
\eeq          
where the ``antibracket" $(\ ,\ )$ is defined by declaring the sources
to be conjugate to the corresponding fields, i.e.,
\beq
(A^I_\mu (x), K^\nu_J(y)) = \delta^I_J \delta^\nu_\mu \delta^n(x-y),
\; \; (C^I(x), L_J(y)) = \delta^I_J \delta^n(x-y)
\quad\mbox{etc.}
\label{antibracket}
\eeq  
The generating functional $\Gamma$ of the one-particle irreducible proper
vertices coincides with $S^{total}$ to zeroth order in $\hbar$,
\beq
\Gamma = S^{\mathit{total}} + \hbar \Gamma^{(1)} + O(\hbar^2).
\eeq
 
The perturbative quantum problem is to prove that the renormalized finite
$\Gamma$, obtained through the addition of
counterterms of higher orders in $\hbar$ to $S^{\mathit{total}}$,
obeys the same identity,
\beq
(\Gamma,\Gamma)=0\label{WSTI} 
\eeq
These are the Ward-Slavnov-Taylor
identities (written in Zinn-Justin form) associated with the original gauge
symmetry and the antighost shift symmetry.
The problem involves two aspects: anomalies and stability.
 
General theorems \cite{Lowenstein:1971jk,Lam:1972mb,Lam:1973qa} guarantee
that to lowest order in $\hbar$, the breaking $\Delta_k$ of the Ward identity,
\beq
(\Gamma,\Gamma)
=\hbar^k\Delta_k+O(\hbar^{k+1}),
\eeq
is a local functional. The identity $(\Gamma
,(\Gamma,\Gamma))=0$, then gives the lowest order consistency condition
\beq
{\cal S}\Delta_k=0,
\eeq
where ${\cal S}$ is the so-called (linearized) Slavnov operator,
${\cal S} \; \cdot = (S^{\mathit{total}}, \cdot)$
which fulfills ${\cal S}^2 = 0$ because of (\ref{nilpo0}). 
Trivial anomalies of the form ${\cal S}\Sigma_k$ can be absorbed
through the addition of finite counterterms, so that, in the absence of
non trivial anomalies,
(\ref{WSTI}) can be fulfilled to that order. Hence, non trivial anomalies
are constrained by the cohomology of ${\cal S}$ in ghost number $1$.
 
The remaining 
counterterms $S_k$ of that order must satisfy
\beq
{\cal S} S_k = 0,
\eeq
in order to preserve the Ward identity to that order.
Solutions of the form ${\cal S}$(something) can be removed through
field redefinitions or a change of the gauge conditions.
The question of stability in the minimal, physical, sector is
the question whether any non trivial solution of this equation
can be brought back to
the form $S^{\mathit{inv}}$ by redefinitions of the coupling
constants and field redefinitions.
Thus, it is the cohomology of ${\cal S}$ in ghost number $0$
which is relevant for the analysis of stability.
 
As it has been defined, ${\cal S}$ acts also on the
sources and depends on the gauge-fixing.  However,
the gauge conditions can be completely absorbed through a
redefinition of the antifields
\begin{eqnarray}
A^{* \mu}_I &=& K^\mu_I + \frac{\delta \Psi}{\delta A^I_\mu}, \\
\psi^{*}_i &=& K_i + \frac{\delta \Psi}{\delta \psi^i}, \\
\bar{C}^{* I} &=& M^I + \frac{\delta \Psi}{\delta \bar{C}_I}, \\
C^{*}_I &=& L_I + \frac{\delta \Psi}{\delta C^I}, \\
b^{*I} &=& N^I + \frac{\delta \Psi}{\delta b_I},
\end{eqnarray}
where, in our case, $\Psi$ is given by
\beq
\Psi = \Ii\, \bar{C}_I ({\cal F}^I + \frac{1}{2} b^I).
\eeq
This change of variables does not affect 
the antibracket ("canonical transformation
generated by $\Psi$").
In terms of the new variables, $S^{\mathit{total}}$ becomes
\beq
S^{\mathit{total}} = S^{\mathit{inv}} - \int d^nx \Big[ (sA^I_\mu)
 A^{* \mu}_I +
( s \psi^i)\psi^{*}_i +  (s C^I)C^{*}_I\Big] + S^{\mathit{NonMin}}
\eeq
where the non-minimal part is just
\beq
S^{\mathit{NonMin}} = -\Ii \int\ d^nx\ b_I\, \bar{C}^{* I}
\eeq
The Slavnov operator becomes then
\beq
{\cal S} = s + \Ii \int d^n x\ \Big[b_I(x) \frac{\delta}{\delta \bar{C}_I(x)}
-\bar{C}^{* I}(x) \frac{\delta}{\delta b^{* I}(x)} \Big]
\eeq
where here, $s$ acts on the variables $A^I_\mu$, $\psi^i$, $C^I$
of the ``minimal sector" and on their antifields $A^{* \mu}_I$,
$\psi^{*}_i$, $C^*_I$ and takes exactly the form given
in (\ref{ym7}), while the remaining piece is contractible and does not
contribute to the cohomology (see appendix B).  Thus, $H({\cal S})$
and $H(s)$ are isomorphic, as are $H({\cal S} \vert d)$ and
$H(s \vert d)$: any cocycle of ${\cal S}$ may be assumed not
to depend on the variables $\bar{C}_I$, $\bar{C}^{* I}$,
$b^{* I}$ and $b_I$ of the ``non-minimal sector" and is then a
cocycle of $s$.  Furthermore, for chains depending on the variables of
the minimal sector only, $s$-coboundaries and ${\cal S}$-coboundaries
coincide. Hence, the cohomological problems indeed reduce to computing
$H(s)$ and $H(s \vert d)$.
\medskip

\noindent{\bf Remarks:} 
 
(i) Whereas the choice made above to introduce sources also for the
antighosts
$\bar C_I$ and the auxiliary $b^I$ fields is motivated by the desire to
have
a symmetrical description of fields and sources with respect to
the antibracket (\ref{antibracket}) and a BRST transformation that is
canonically generated on all the variables, other authors prefer to
introduce
sources for the BRST variations of the variables
$A^I_\mu$, $\psi_i$ and $C^I$ only. In that approach,
the final BRST differential in the physical sector is the same as
above, but the non minimal sector is smaller and consists only of
$\bar C_I$ and $b^I$.  

(ii) In the case of standard
Yang-Mills theories in four dimensions, one may wonder
whether one has stability of the complete action 
(\ref{1.27}) not only in the relevant, physical sector but also in the
gauge-fixing sector, i.e., whether linear gauges are stable.
Stability of linear gauges can be established by imposing
legitimate auxiliary conditions.  In the formalism where
the antighosts and auxiliary fields
have no antifields, these auxiliary conditions are
the gauge condition and the ghost equation, fixing the dependence
of $\Gamma$ on $b^I$ and $\bar C_I$ respectively.
The same result can be recovered in the approach with antifields for
the antighosts and the auxiliary fields. The dependence of
$\Gamma$
on the variables of the non minimal sector is now fixed by the same ghost
equation 
as before, while the gauge condition is modified through the
additional term
$-\Ii\bar C^*_Ib^I$.  In addition, one imposes:
$\delta\Gamma/\delta \bar C^*_I=- \Ii b^I$ and
$\delta\Gamma/\delta b^*_I=0$.

(iii) One considers sometimes a different cohomology, the so-called
gauge-fixed BRST cohomology, in which there is no
antifield and the equations of motion of the gauge-fixed theory
are freely used.  This cohomology is particularly
relevant to the "quantum Noether method" for
gauge-theories \cite{Hurth:1998nq,Hurth:1998ib}.
The connection between the BRST cohomology discussed here 
and the gauge-fixed cohomology is studied
in \cite{Henneaux:1996ex,Barnich:1999cy}, where it is shown that
they are isomorphic under appropriate conditions which are
explicitly stated.

\subsection{Appendix \ref{Intro}.B: Contractible pairs}
\label{IIBContractible}

We show here that the antighosts and the auxiliary fields
do not contribute to the cohomology of ${\cal S}$.  This
is because $\bar{C}_I$ and $b_I$ form ``contractible pairs",
\beq
{\cal S}\bar  {C}_I = \Ii \, b_I, \; {\cal S} b_I = 0,
\eeq
and furthermore, the ${\cal S}$-transformations of the other
variables do not involve $\bar  {C}_I$ or $b_I$.
We shall repeatedly meet the concept of ``contractible pairs"
in this work.

Let $N$ be the operator counting $\bar C_I$ and $b_I$
and their derivatives,
\beq
N =\bar C_I \frac{\partial}{\partial \bar C_I}+ b_I
\frac{\partial}{\partial b_I} +
\sum_{t >0} \Big[\partial_{\mu_1 \dots \mu_t} \bar C_I
\frac{\partial}{\partial (\partial_{\mu_1 \dots \mu_t}
\bar C_I)}
+\partial_{\mu_1 \dots \mu_t} b_I
\frac{\partial}{\partial (\partial_{\mu_1 \dots \mu_t} 
b_I)}\Big] .
\eeq
One has $[N, {\cal S}] = 0$ and in fact 
$N= {\cal S} \varrho  + \varrho  {\cal S}$
with 
\beq
\varrho  = -\Ii\, \bar C_I\frac{\partial}{\partial b_I} - \Ii
\sum_{t >0} \partial_{\mu_1 \dots \mu_t} \bar C_I
\frac{\partial}{\partial (\partial_{\mu_1 \dots \mu_t}
b_I)}  .
\eeq
$\varrho$ is called a contracting homotopy for $N$ with respect to
${\cal S}$. Now,
let $a$ be ${\cal S}$-closed.
One can expand $a$ according to the $N$-degree,
$a = \sum_{k \geq 0} a_k$ with $N a_k = k a_k$.
One has ${\cal S} a_k = 0$ since $[N, {\cal S}] = 0$.  It
is easy to show that the components of $a$ with $k>0$ are
all ${\cal S}$-exact.  Indeed, for $k>0$ one has
$a_k = (1/k)Na_k=(1/k)({\cal S} \varrho + \varrho  {\cal S}) a_k$ 
and thus $a_k= {\cal S}b_k$
with $b_k=(1/k) \varrho a_k$.  Accordingly, 
$a = a_0 + {\cal S} (\sum_{k>0} b_k)$.
Hence, the "non-minimal part" of an $s$-cocycle
is always trivial. Furthermore, by analogous arguments, 
an $s$-cocycle $a$ is trivial if and only if
its "minimal part" $a_0$ is trivial
in the minimal sector. 

Similarly, if $a$ is a solution of the
consistency condition, ${\cal S} a + dm = 0$,
one has ${\cal S} a_k + dm_k = 0$ and one gets,
for $k\not=0$, $a_k = {\cal S} b_k + dn_k$ with
$b_k=(1/k) \varrho a_k$ and  $n_k = (1/k)\varrho  m_k$ (thanks to
$[N,d] = 0$,
$d \varrho  + \varrho  d = 0$).
Thus, the cohomology of ${\cal S}$ can be non-trivial only in
the space of function(al)s not involving the antighosts and
the auxiliary fields $b_I$.

The same reasoning
applies to the antifields
$\bar{C}^{*I}$ and $b^{*I}$, which form also contractible
pairs since
\beq
{\cal S} \bar{C}^{*I} = 0, \; {\cal S} b^{*I} =  - \Ii \, \bar{C}^{*I}.  
\eeq

The argument is actually quite general and constitutes
one of our primary tools for computing cohomologies.
We shall make a frequent use
of it in the report, whenever we have a pair of 
independent variables $(x,y)$ and
a differential $\Delta$ such that $\Delta x = y$ and 
$\Delta y = 0$, and the action of $\Delta$ on the remaining variables 
does not involve $x$ or $y$.

\newpage

\mysection{Outline of report}

The calculation of the BRST cohomology is based on the decomposition
of $s$ into $\delta + \gamma$, on the computation of the
individual cohomologies of $\delta$ and $\gamma$ and
on the descent equations. Our approach
follows \cite{Barnich:1994ve,Barnich:1995db,Barnich:1995mt},
but we somewhat streamline and systematize the developments of these papers
by starting the calculation {\em ab initio}.  This enables one
to make some shortcuts in the derivation of the results.

We start by recalling some useful properties of the exterior
derivative $d$ in the algebra of local forms (section \ref{Loc}).
In particular, we establish the important ``algebraic Poincar\'e lemma"
(theorem \ref{Loct2}), which is also
a tool repeatedly used in the whole report.

We  compute then $H(\delta)$ and $H(\delta \vert d)$
(sections \ref{KoszulSection} and \ref{Koszul2Section},
respectively).
0ne can establish general properties of
$H(\delta)$ and $H(\delta \vert d)$, independently of the
model and valid for other gauge theories such as gravity
or supergravity.  
In particular, the relationship between $H(\delta \vert d)$
and the (generalized) conservation laws is quite general,
although the detailed form of the conservation laws does depend
of course on the model.
We have written
section \ref{Koszul2Section} with the desire to make  
these general features explicit.
We specialize then the analysis to gauge theories of the
Yang-Mills type.  Within this set of theories (and on natural
regularity and normality conditions on the Lagrangian), 
the groups $H(\delta \vert d)$ are completely
calculated, except $H^{n}_1(\delta \vert d)$, which is related
to the global symmetries of the model and can be fully determined
only when the Lagrangian is specified.

In section \ref{homological}, we establish the general link
between $H(s)$, $H(\delta)$ and $H(\gamma)$ (respectively,
$H(s \vert d)$, $H(\delta \vert d)$ and $H(\gamma \vert d))$.
The connection follows the line of ``homological perturbation
theory" and applies also to generic field theories with a gauge
freedom. 

We compute next $H(\gamma)$ (section \ref{LieAlgebraCoho}). 
The calculation is tied to
theories of the Yang-Mills type but within this class
of models, it does not depend on the form of the Lagrangian since
$\gamma$ involves only the gauge transformations and not
the detailed dynamics.
%We first show that the 
%derivatives of the ghosts and the naked gauge potentials
%and their symmetrized derivatives 
%$\6_{(\mu_1}\dots \6_{\mu_k} A^I_{\mu_{k+1})}$ 
%disappear from $H(\gamma)$
%so that
%only the undifferentiated ghosts and gauge covariant
%tensor fields survive in cohomology.
%The calculation of $H(\gamma)$
%reduces then 
%to the problem of computing the Lie
%algebra cohomology of the gauge group in
%the representation space of the polynomials in the curvature
%components, the matter fields, the antifields, and their covariant
%derivatives.  
We first show that the
calculation of $H(\gamma)$
reduces to the problem of computing the Lie
algebra cohomology of the gauge group formulated in terms of
the curvature
components, the matter fields, the antifields, and their covariant
derivatives, and the undifferentiated ghosts.  
This is a well-known mathematical problem
whose general solution has been worked out
long ago (for reductive Lie algebras).
Knowing the connection between
$H(\delta)$, $H(\gamma)$ and $H(s)$ makes it easy to compute
$H(s)$ from $H(\delta)$ and $H(\gamma)$.

We  then turn to the computation of
$H(s \vert d)$.  The relevant mathematical
tool is that of the descent equations, which we first review
(section \ref{Des}).  Equipped with this tool, we 
calculate $H(s \vert d)$ in all form
and ghost degrees in a smaller algebra
involving only the forms $C^I$, $dC^I$, $A^I$, $dA^I$
and their exterior products (section \ref{small}).
Although this problem is a sub-problem of the general
calculation of $H(s \vert d)$, it turns out to be crucial
for investigating solutions of the consistency
condition $sa + dm =0$ that ``descend non trivially".  The
general case (in the algebra of all local forms not necessarily
expressible as exterior products of $C^I$, $dC^I$, $A^I$
and $dA^I$) is treated next (section \ref{solution}), paying due attention
to the antifield dependence.

Because the developments in section \ref{solution} are rather involved,
we discuss their physical implications in a separate section 
(section \ref{DisDisDis}).  The reader who is not interested in the
proofs but only in the results may skip section \ref{solution}
and go directly to section \ref{DisDisDis}.  We explain in particular
there why the
antifields can be removed from the general solution of the
consistency condition at ghost
numbers zero (counterterms) and one (anomalies)
when the gauge group is semisimple.  We also
explain why the anomalies can be expressed solely in terms of
$C^I$, $dC^I$, $A^I$
and $dA^I$ in the semisimple case
\cite{Brandt:1989rd,Brandt:1990gy,Dubois-Violette:1992ye}.  
These features do not hold,
however, when there are abelian factors or for different values
of the ghost number.

The case of a system of free abelian gauge fields, relevant to
the construction of consistent couplings among 
massless vector particles,
has special features and is therefore treated separately
in section \ref{Free}.  We also illustrate the general results
in the case of pure Chern-Simons theory in three dimensions,
where many solutions disappear because the Yang-Mills curvature
vanishes when the equations of motion hold (section \ref{CS}).

Finally, the last section reviews the literature on the calculation of
the local BRST cohomology $H(s \vert d)$
for other local field theories with a gauge freedom. 

\newpage

\mysection{Locality - Algebraic Poincar\'e lemma: $H(d)$}
\label{Loc}

Since locality is a  fundamental ingredient in
our approach, we introduce in this subsection
the basic
algebraic tools that allow one to deal with locality.
The central idea is to consider the
``fields" $\phi^i$ and their partial derivatives of
first and higher order
as independent coordinates of so-called
jet-spaces.  The approach is familiar from the variational
calculus where the fields and their partial
derivatives are indeed treated as independent variables when 
computing the derivatives $\partial L/ \partial \phi^i$ or
$\partial L / \partial(\partial_\mu \phi^i)$ of the Lagrangian
$L$.  
%The jet-spaces are equipped with
%useful differentials which we study.  
Fields, in the usual
sense of ``functions of the space-time
coordinates'', emerge in this
approach as sections of the corresponding
jet-bundles.

What are the ``fields" will depend on the context.
In the case of gauge theories of the Yang-Mills type, the ``fields"
$\phi^i$ may be the original classical fields
$A^I_\mu$ and $\psi^i$, or may be these fields plus the ghosts.  
In some other instances, they could be the original fields plus the
antifields, or the complete set of all variables introduced in
the introduction.  The
considerations of this section are quite general and do not depend on  
any specific model or field content.  So, we shall develop the argument
without committing ourselves to a definite set of variabes.

\subsection{Local functions and jet-spaces}

A local function $f$ is a smooth function of the
spacetime coordinates, the field variables $\phi^i$ and
a finite number of their derivatives, 
$f=f(x,[\phi])$,
where the notation $[\phi]$ means dependence on
$\phi^i,\phi^i_{\mu},\dots, \phi^i_{(\mu_1\dots
\mu_k)}$
for some finite $k$ (with $\phi^i_{\mu}\equiv \6_\mu\phi^i$,
$\phi^i_{(\mu_1\dots\mu_k)}\equiv \6_{(\mu_1}\dots \6_{\mu_k)}\phi^i$).
A local function is thus a function on
the ``jet space of order $k$" 
$J^k(E)=M\times V^k$ (for some $k$), where $M$ is Minkowski (or Euclidean)
spacetime and where $V^k$ is the space with coordinates
given by 
$\phi^i,\phi^i_{\mu},\dots,
\phi^i_{(\mu_1\dots\mu_k)}$ -- some of which may be
Grassmannian.
The fields and their various derivatives are
considered as
completely independent in $J^k(E)$
except that the various
derivatives commute,
so that only the completely symmetric combinations
are independent coordinates.  

In particular, the jet space of order zero 
$J^0(E) \equiv E$ is 
coordinatized by $x^\mu$ and $\phi^i$.
A field history is a section $s: M\longrightarrow E$,
given in coordinates by $x\longmapsto (x,\phi(x))$.
A section of $E$ induces a section of $J^k(E)$, 
with $\phi^i_{(\mu_1\dots\mu_k)}|_s =
\partial^k\phi^i(x)/\partial x^{\mu_1}\dots\partial
x^{\mu_k}$.   Evaluation of a local function at a section
yields a spacetime function.  The independence of the
derivatives reflects the
fact that the only local function
$f(x, [\phi])$ which is zero on all sections  is the
function $f \equiv 0$.               

Because the order in the derivatives 
of the relevant functions
is not always known a priori, it is
useful to introduce the
infinite jet-bundle $\pi^\infty:
J^\infty(E)=M\times
V^\infty\rightarrow M$,
where coordinates on $V^\infty$ are given by
$\phi^i,\phi^i_{\mu},\phi^i_{(\mu_1\mu_2)},
\dots$.

In our case where spacetime is Minkowskian (or Euclidean)
and the field manifold is homeomorphic to $\mathbb{R}^m$
(with $m$ the number of independent real fields $\phi^i$, $i=1,\dots,m$),
the jet-bundles are trivial and the use of bundle
terminology may appear a bit pedantic.  However, this approach
is crucial for dealing with more complicated situations
in which the spacetime manifold and/or the field manifold
is topologically non-trivial.   Global aspects related
to these features will not be discussed here.  They
come over and above the local cohomologies analyzed in
this report, which must in any case be understood\footnote{Our
considerations are furthermore sufficient for the purposes of
perturbative quantum field theory.}.  
We refer to \cite{Manes:1985df,Dubois-Violette:1992ye,%
Barnich:1995ap} for an analysis of BRST cohomology
taking into account some of these extra global features.

In the case of field theory, the local functions are usually
polynomial in the derivatives.  This restriction will turn
out to be important in
some of the next sections.  However, for the present
purposes, it is not necessary. The theorems of
this section and the next are valid both in the space of
polynomial local functions and in the space of arbitrary
smooth local functions.  For this reason, we shall not restrict
the functional space of local functions at this stage.

\subsection{Local functionals - Local $p$-forms}

An important class of objects are local
functionals. They are given by
the integrals over space-time of local $n$-forms
evaluated at a section.
An example is of course
the classical action.

Local $p$-forms are by definition exterior forms
with coefficients that are local functions,
\begin{equation}
\omega = \frac{1}{p!}\,
dx^{\mu_1} \wedge \dots \wedge dx^{\mu_p}
\omega_{\mu_1 \dots \mu_p}(x, [\phi]).
\end{equation}
We shall drop the exterior product symbol $\wedge$
in the sequel since no confusion can arise.

Thus, local functionals evaluated at the section $s$
take the form
${\cal F}(f,s) = \int_M \omega|_s$ with
$\omega$ the $n$-form $\omega = f d^nx$, where
$d^nx = dx^0 \cdots dx^{n-1}$.
In the case of the action, $f$ is the Lagrangian density.

It is customary to identify local functionals with the
formal expression $\int_M \omega $ prior to evaluation.

\subsection{Total and Euler-Lagrange
derivatives}

The total derivative $\partial_\mu$ is the vector
field
defined on local functions by
\bea
\partial_\mu=\frac{\partial}{\partial x^\mu}
+\phi^i_{\mu}\frac{\partial}{\partial \phi}+
\phi^i_{(\mu\nu)}\frac{\partial}{\partial\phi^i_{\nu}}
+\dots
%\nonumber\\
=\frac{\partial}{\partial x^\mu}+\sum_{k\geq 0}
\phi^i_{(\mu\nu_1\dots\nu_k)}
\frac{\partial}{\partial\phi^i_{(\nu_1\dots\nu_k)}},
\label{totalderivative}\eea
where for convenience, we define the index
$(\nu_1\dots\nu_k)$ to be
absent for $k=0$.
These vector fields commute,
$[\partial_\mu,\partial_\nu]=0$. 
Note that $\partial_\mu \phi = \phi_\mu$,
$\partial_\mu \partial_\nu \phi = \phi_{\mu \nu}$ etc...
Furthermore,
evaluation at a section and differentiation
commute as well:
\bea
(\partial_\mu
f)|_s=\frac{\partial}{\partial x^\mu}(
f|_s).\label{Loce2}
\eea

The Euler-Lagrange derivative
$\frac{\delta}{\delta \phi^i}$
is defined on a local function $f$ by
\bea
\frac{\delta f}{\delta \phi^i}=\frac{\partial
f}{\partial \phi^i}
-\partial_{\mu}\frac{\partial
f}{\partial\phi^i_{\mu} }
+\dots
=\sum_{k\geq 0}(-)^k\partial_{(\mu_1\dots\mu_k)}\frac{\partial
f}{\partial\phi^i_{(\mu_1\dots\mu_k)}},
\eea
where
$\partial_{(\mu_1\dots\mu_k)}=\partial_{\mu_1}\dots
\partial_{\mu_k}$.

\subsection{Relation between local functionals
and local functions}
\label{relationlocal}

A familiar property of total derivatives 
$\partial_\mu j^\mu$ is that they have vanishing
Euler-Lagrange derivatives.  The converse is also
true.  In fact, one has

\begin{theorem}\label{Loct1}

(i) A local function is a total derivative iff it has
vanishing Euler-Lagrange derivatives with respect to all
fields,
\beq
f=\partial_\mu j^\mu\quad\LRA\quad
\frac{\delta f}{\delta \phi^i}=0\quad\forall\,\phi^i\ .
\eeq

(ii) Two local functionals $\cF,\cG$ agree for all
sections
$s$, $\cF(f,s)=\cG(g,s)$  iff their integrands
differ by a total
derivative, $f=g+\partial_\mu j^\mu$, for some
local functions
$j^\mu$, whose boundary integral
vanishes, $\oint_{\partial M}j=0$.
\end{theorem}

\paragraph{Proof:}

(i) Let 
\[
N=\sum_{k\geq 0}\phi^i_{(\mu_1\dots\mu_k)}\,
\frac{\partial }{\partial
\phi^i_{(\mu_1\dots\mu_k)}}\ .
\] 
We start from
the identity
\bea
f(x,[\phi])-f(x,0)=\int_0^1d\lambda
\frac{d}{d\lambda}
f(x,[\lambda\phi])=
\int_0^1\frac{d\lambda}{\lambda}[Nf](x,[\lambda\phi]).\label{A1}
\eea
Using 
integrations by parts
and
$f(x,0)=\partial_\mu k^\mu(x)$ 
(which holds as a consequence of the standard Poincar\'e       
lemma for $\mathbb{R}^n$),
one gets
\beq
f(x,[\phi])=\partial_\mu j^\mu+ \int_0^1\
\frac{d\lambda}{\lambda}
[\phi^i \frac{\delta
f}{\delta \phi^i}](x,[\lambda\phi]).\label{A1.1}
\eeq
for some local functions $j^\mu$.
Thus, $\delta f/\delta \phi^i=0$ $\forall i$ implies $f=\partial_\mu j^\mu$.
Evidently, $j^\mu$ is polynomial whenever $f$ is.
                                                   
Conversely,
we have
\beq
[\frac{\partial}{\partial\phi^i_{(\mu_1\dots\mu_k)}},\partial_\nu]
=\delta^{\mu_1}_{(\nu} \cdots \delta^{\mu_k}_{\lambda_{k-1})}
\frac{\partial}{\partial\phi^i_{(\lambda_1\dots\lambda_{k-1})}}.
\label{prooe1}
\eeq
[For $k=0$: $[{\partial}/\partial\phi^i,\partial_\nu]=0$.]
This gives
\bea
\sum_{k\geq 0}(-)^k\partial_{(\mu_1\dots
\mu_k)}\frac{\partial(\partial_\nu j^\nu)}{\partial\phi^i_{(\mu_1\dots\mu_k)}}
=\sum_{k\geq 0}(-)^k\partial_{(\mu_1\dots\mu_k\nu)}
\frac{\partial
j^\nu}{\partial\phi^i_{(\mu_1\dots\mu_k)}}\nonumber\\
+\sum_{k\geq 1}(-)^{k}\partial_{(\nu\lambda_1\dots
\lambda_{k-1})}\frac{\partial
j^\nu}{\partial\phi^i_{(\lambda_1\dots\lambda_{k-1})}}=0.\label{prooe2}
\eea

(ii) That two local functionals whose integrands
differ by a
total derivative with vanishing boundary integral agree
for all sections $s$
follows from (\ref{Loce2}) and Stokes
theorem. Conversely, $\cF(f,s)=\cG(g,s)$ for all
$s$, implies
$I=\int_M d^nx\ (f-g)|_{\phi^i(x)+\varepsilon
\eta^i(x)}=0$ for all
$\varepsilon$. Thus, for sections $\eta^i(x)$ which
vanish with a sufficient
number of their derivatives at $\partial M$, using
integrations by parts and
(\ref{Loce2}), one gets
$0=\frac{d}{d\varepsilon}|_{\epsilon=0}I=
\int_M d^nx\ 
\eta^i(x)\frac{\delta(f-g)}{\delta\phi^i}|_{\phi^i(x)}$,
which 
implies $\frac{\delta(f-g)}{\delta\phi^i}=0$ for all $\phi^i$ and
concludes the proof by
using (i).~\qed

{\bf Remarks:}

(i) The first part of this theorem 
can be reformulated as
the statement that ``terms
in a classical Lagrangian give no contributions to
the classical equations of motion iff they are total derivatives''.
It is crucial to work in jet space for this
statement to be true since the Poincar\'e lemma for the standard
De Rham cohomology in ${\mathbb R}^n$ implies that any function
of $x$ can be written as a total derivative.  Thus,
if we were to  consider na\"{\i}vely the Lagrangian
$L$ as
a function of the spacetime coordinates
($L=L(x)$), we would get an apparent contradiction, since on the
one hand the Euler-Lagrange equations are in general
not empty while on the other hand $L$ can be written as
a total derivative in $x$-space.
The point is of course that $L$  would not be given in general
by the total derivative of a {\em local} vector density.

(ii) The theorem leads to the following view on local functionals,
put forward in explicit terms in \cite{gelfand}:
local functionals can be identified with
equivalence classes of local functions modulo
total derivatives.  This is legitimate whenever the surface terms can be
neglected or are not under focus. This is
the case in the physical situations described in
the introduction (e.g. in renormalization theory, the classical
fields in the effective action $\Gamma$ are in
fact sources yielding 1-particle irreducible Green functions.
They can be assumed to be of compact support in that context).
In the remainder of the report, we always use this
identification.

\subsection{Algebraic Poincar\'e lemma - $H(d)$}\label{APL}

Let $\Omega$ be the algebra of local forms.
The exterior (also called horizontal)
differential in $\Omega$ is defined by $d=dx^\mu\partial_\mu$
with $\partial_\mu$ given by formula (\ref{totalderivative}).

The derivative $d$ satisfies $d^2=0$ because the
$dx^\mu$ anticommute while the $\partial_\mu$
commute. The complex
$(\Omega,d)$ is called the horizontal complex.
Its elements, the local forms, are also called
the horizontal forms,
or just the ``forms". 
The $p$-forms $\omega^p$ satisfying $d\omega^p=0$
are called closed,
or cocycles (of $d$), the ones given by
$\omega^p=d\omega^{p-1}$, which are
necessarily closed, are called exact or
coboundaries (of $d$). The cohomology
group $H(d,\Omega)$ is defined to be the space of
equivalence classes
of cocycles modulo coboundaries, $[\omega^p]\in
H^p(d,\Omega)$ if
$d\omega^p=0$ with
$\omega^p\sim\omega^p+d\omega^{p-1}$.

Integrands of local functionals are local
$n$-forms.
A local $n$-form $\omega^n = d^nx\, f$ is exact, 
$\omega^n = d\omega^{n-1}$ (with $\omega^{n-1}$ a local $(n-1)$-form),
if and only if
$f$ is a total derivative; one has 
\beq
d^nx f=d\omega^{n-1}\quad \LRA\quad 
f=\6_\mu j^\mu\quad \LRA\quad
\frac{\delta f}{\delta \phi^i}=0\quad \forall\,\phi^i
\eeq
where the first equivalence holds with
$\omega^{n-1}=-\frac{1}{(n-1)!}
\epsilon_{\mu\nu_1\dots\nu_{n-1}}dx^{\nu_1}\dots
dx^{\nu_{n-1}} j^\mu$ and the second one holds by theorem
\ref{Loct1}.
Due to $d\omega^n=0$, theorem \ref{Loct1} can thus be
reformulated
as the statement that local functionals are
described by
$H^n(d,\Omega)$, 
and that $H^n(d,\Omega)$ is given by the
equivalence
classes $[\omega^n]$ having identical
Euler-Lagrange derivatives,
$\omega^n = f d^nx \sim{\omega^\prime}^n= f^\prime d^nx$ if
$\frac{\delta}{\delta\phi^i}(f-f^\prime)=0$.
An important
result is the following on the cohomology of $d$
in lower form
degrees.

\begin{theorem}\label{Loct2}
The cohomology of $d$ in form degree strictly smaller
than $n$ is
exhausted by the constants in form degree $0$,
\bea
0<p<n:&& d\omega^p=0\quad\LRA\quad \omega^p=d\omega^{p-1}\ ;
\nonumber\\
p=0: && d\omega^0=0\quad\LRA\quad \omega^0=\mathit{constant}\ .
\eea
\end{theorem}

Theorem \ref{Loct1} (part (i)) and theorem \ref{Loct2},
which give the cohomology of $d$ in the algebra of local forms,
are
usually collectively referred to as the ``algebraic Poincar\'e lemma". 
The proof below applies to the case of a polynomial dependence on 
derivatives. Proofs covering the smooth case can be found in the literature 
cited at the end of this section. 

\paragraph{Proof:}

As in (\ref{A1}), we have the decomposition
$\omega(x,dx,[\phi])=\omega_0+\tilde\omega$,
where $\omega_0=\omega(x,dx,0)$ and
$\tilde\omega(x,dx,[\phi])=
\int^1_0\frac{d\lambda}{\lambda}[N\omega](x,dx,[\lambda\phi])$.

The condition $d\omega=0$ then implies separately
$dx^\mu\frac{\partial}{\partial
x^\mu}\omega_0=0$ and
$d\tilde\omega(x,dx,[\phi])=0$ because
$d$ is homogeneous of degree zero in $\lambda$ and
commutes with $N$.        

Defining
$\rho^\prime=x^\nu\frac{\partial}{\partial
dx^\nu}$, we have
$\{dx^\mu\frac{\partial}{\partial
x^\mu},\rho^\prime\}=x^\mu\frac{\partial}{\partial
x^\mu}+dx^\nu\frac{\partial}{\partial
dx^\nu}$. Using a homotopy formula analogous to
(\ref{A1})
for the variables $x^\mu,dx^\mu$, we get
$\omega_0(x,dx,0)=\omega(0,0,0)
+d\int_0^1\frac{d\lambda}{\lambda}[\rho^\prime\omega_0](\lambda
x,\lambda dx,0)$, which is the standard Poincar\'e
lemma.        

Let $t^{\nu}=\sum_{k\geq 0}k
\delta_{(\mu_1}^{\nu}\delta_{\mu_2}^{\lambda_1}\dots
\delta_{\mu_k)}^{\lambda_{k-1}}
\phi^i_{(\lambda_1\dots\lambda_{k-1})}\frac{\partial}
{\partial\phi^i_{(\mu_1\dots\mu_k)}}$. Then
\beq
[t^{\nu},\partial_\mu]=\delta^\nu_\mu
N.\label{A1ecom}
\eeq
If one defines $D^{+\nu}\tilde\omega=
\int^1_0 \frac{d\lambda
}{\lambda}[t^\nu\tilde\omega](x,dx,[\lambda\phi])$, one gets
\beq
[D^{+\nu},\partial_\mu]\tilde\omega=\delta^\nu_\mu\tilde\omega,\label{A4}
\eeq                                     
because $\6_\mu$ is homogeneous of degree $0$ in
$\lambda$.
With $\rho=D^{+\nu}\frac{\partial}{\partial
dx^\nu}$, one has 
\beq
\{d,\rho\}\tilde\omega=
[D^{+\nu}\partial_\nu-dx^\mu\frac{\partial}{\partial
dx^\mu}]\tilde\omega=[\partial_\nu
D^{+\nu}+(n-dx^\mu\frac{\partial}{\partial
dx^\mu})]\tilde\omega.\label{A5}
\eeq
Let $\alpha=n-p$, for $p<n$.
Apply the previous relation to a $d$-closed $p$-form
$\tilde\omega^p$ to get
\beq
d\tilde\omega^p=0\quad\then \quad
\tilde\omega^p=d\frac{\rho}{\alpha}\tilde\omega^p
-\frac{1}{\alpha}\partial_\nu
D^{+\nu}\tilde\omega^p.\label{A6}
\eeq 
We want to use this formula recursively.
In order to do so, we need some relations for the
operators 
$P_m=\partial_{\nu_1}\dots\partial_{\nu_m}D^{+\nu_1}\dots
D^{+\nu_m}$ where, by definition $P_0=1$.
(\ref{A4}) implies
$[P_1,d]\tilde\omega=d\tilde\omega$ and
$P_1P_m\tilde\omega=[P_{m+1}+mP_m]\tilde\omega$.
The latter allows one to express
$P_m$ in terms of $P_1$:
$P_m\tilde\omega=\Pi_{l=0}^{m-1}(P_1-l)\tilde\omega$.
If
$\tilde\omega$ is closed,
it follows that $dP_m\tilde\omega=0$. Together
with
(\ref{A5}) (applied to a $p$-form $P_m\tilde\omega^p$), this yields
\beq
d\tilde\omega^p=0\quad\then \quad
(\alpha+m)P_m\tilde\omega^p=d\rho
P_m\tilde\omega^p-P_{m+1}\tilde\omega^p.
\eeq
Injecting this relation for $m=1$ into (\ref{A6}),
we find
\beann
\tilde\omega^p=d[\frac{\rho}{\alpha}
-\frac{\rho}{\alpha(\alpha+1)}P_1]\tilde\omega^p+\frac{1}{\alpha(\alpha+1)}
P_2\tilde\omega^p.
\eeann
Going on recursively, the procedure will stop
because $\tilde\omega^p$ is polynomial in derivatives: the total number 
of derivatives contained $\tilde\omega^p$ is bounded by some $\bar m$,
so that
$P_{\bar m+1}\tilde\omega^p=0$.
Hence, the final result is $d\tilde\omega^p=0$ $\then$ $\tilde\omega^p=
d\eta^{p-1}$, with
\beann
\eta^{p-1}=\sum_{m=0}^{\bar m}\ \frac{(-)^{m}\rho\,
P_m\,\tilde\omega^p}{\alpha(\alpha+1)\cdots(\alpha+m)}\ ,\quad
\alpha=n-p.
\eeann 
                                                                                                                                        
This proves
the theorem, by noting that $\omega(0,0,0)$ can
never be $d$
exact.~\qed

Note that the above construction preserves polynomiality:
$\eta^{p-1}$ is polynomial in the fields and their derivatives if
$\omega^p$ is.

\subsection{Cohomology of $d$ in the complex of $x$-independent local forms}

The previous theorem holds in the space of forms that are allowed
to have an explicit $x$-dependence.  It is sometimes necessary to
restrict the analysis to translation-invariant local forms, which have
no explicit $x$-dependence.  In that case, the cohomology is
bigger, because the constant forms (polynomials in the
$dx^\mu$'s with constant coefficients) are closed but not exact
in the algebra of forms without explicit $x$-dependence ($dx^\mu =
d(x^\mu)$, but $x^\mu$ is not in the algebra). 
In fact, as an adaptation of the proof of the previous
theorem easily shows, this is the only additional cohomology.

\begin{theorem}\label{Loct''2} (algebraic
Poincar\'e lemma in the $x$-independent case):
In the algebra of local forms without 
explicit $x$-dependence, the 
cohomology of $d$ in form degree strictly less
than $n$ is
exhausted by the constant forms:
\beq
p<n:\quad d\omega^p=0\quad\LRA\quad 
\omega^p=d\omega^{p-1}+c_{\mu_1 \dots \mu_p}
dx^{\mu_1} \dots dx^{\mu_p}
\eeq
where the
$c_{\mu_1 \dots \mu_p}$ are constants.
\end{theorem}           

If one imposes Lorentz invariance,
the cohomology in form-degree $0<p<n$ disappears since
there is no Lorentz-invariant constant form except
for $p=0$ or $p=n$ (see Sections \ref{AssUMPtions}
and \ref{Strategy} for a discussion of Lorentz invariance).

\subsection{Effective field theories}

The Lagrangian of effective Yang-Mills theory
contains all terms compatible with gauge invariance 
\cite{Gomis:1996jp,Weinberg:1996kr}.
Consequently, it is not
a local function since it contains an infinite number of
derivatives.  

However, the above considerations are relevant to the
study of effective theories.  Indeed, in that case
the Lagrangian $L$ is in fact a formal power series
in some free parameters (including the gauge coupling constants).  
The coefficients of the
independent powers of the parameters in this
expansion are local functions in the above sense
and are thus defined in a jet-space of finite order.
Equality of two formal power series in the parameters
always means equality of all the coefficients.  For this
reason, formal power series in parameters with
coefficients that are local function(al)s -- in fact
polynomials in derivatives by dimensional analysis -- can be investigated 
by means of the tools introduced for local function(al)s.
These are the objects that we shall manipulate in the context of effective
field theories.

\subsection{A guide to the literature}

Useful references on jet-spaces 
are \cite{olver,Saunders,Anderson}.
The algebraic Poincar\'e lemma has been proved by many
authors and repeatedly rediscovered.  Besides
the references just quoted, one can list
(without pretendence of being exhaustive!)
\cite{vinogradov1,takens,tulczyjew,anderson1,DeWilde,tsujishita,%
Brandt:1990gy,Dubois-Violette:1991is,Dickey,Wald}.
We have followed the proof given
in \cite{Dragon:1996md}.

The very suggestive
terminology ``algebraic Poincar\'e lemma" appears to
be due to Stora \cite{Stora:1976kd}.

\newpage

\mysection{Equations of motion and
Koszul-Tate differential: $H(\delta)$}
\label{KoszulSection}

The purpose of this section is to compute the
homology\footnote{One speaks of ``homology", rather than
cohomology, because $\delta$ acts like a boundary
(rather than coboundary) operator: it decreases the degree
(antifield number) of the
objects on which it acts.} of the Koszul-Tate differential
$\delta$ in the algebra of local forms $\Omega_F$ on the jet-space 
$J^{\infty}(F)$ of 
the original fields $A^I_\mu$ and $\psi$, the ghosts $C^I$, the antifields,
and their derivatives.  To that end, it is necessary to 
make precise a few properties of the equations of motion. 

The action of the Koszul-Tate
differential $\delta$ for gauge theories of the Yang-Mills type
has been defined in the introduction on 
the basic variables $A^I_\mu$, $\psi$, $C^I$,
$A^{*I}_\mu$, $\psi^{*}_i$ and $C^*_I$, through 
formula (\ref{ym7}).  
The differential $\delta$ is then
extended to the jet space $J^{\infty}(F)$
by requiring that $\delta$ be a derivation (of odd
degree) that commutes with $\partial_\mu$.
This yields explicitly
\beann
\delta &=& \sum_{l\geq 0}\partial_{(\mu_1\dots\mu_l)} L^\mu_I
\frac{\partial}{\partial
A^{* \mu}_{I|(\mu_1\dots\mu_l)}}
+ \sum_{l\geq 0}\partial_{(\mu_1\dots\mu_l)} L_i
\frac{\partial}{\partial \psi^*_{i |(\mu_1\dots\mu_l)}}
\\
&+&\sum_{m\geq 0}\partial_{(\nu_1\dots\nu_m)} \big[
- D_\mu A^{* \mu}_I-e \psi_i^* T^i_{Ij} \psi^j \big]
\frac{\partial}{\partial
C^*_{I|(\nu_1\dots\nu_m)}}\ ,
\eeann
an expression that makes it obvious that $\delta$ is a derivation.
By setting $\delta(dx^\mu) = 0$, one extends $\delta$
trivially to the algebra
$\Omega_F$ of local forms.

%One has $[A, \delta] = -\delta$, where $A$ is the antifield
%number operator, 
%\[
%A =\sum_{l\geq 0} \big[\phi^*_{(\mu_1\dots\mu_l)}
%\frac{\partial}{\partial
%\phi^*_{(\mu_1\dots\mu_l)}}+2C^*_{I|(\mu_1\dots\mu_l)}
%\frac{\partial}{\partial
%C^*_{I|(\mu_1\dots\mu_l)}}\big]
%\]
%where $\phi^*$ denotes collectively $A^{* \mu}_{I}$ and
%$\psi^*_{i}$.
%$A$ is also a derivation (of even degree).

\subsection{Regularity conditions}

\subsubsection{Stationary surface}

The Euler-Lagrange equations of motion and
their derivatives define  surfaces in the jet-spaces
$J^r(E)$, $J^\infty (E)$ of the original fields $A^I_\mu$,
$\psi^i$, the ghosts $C^I$ and their derivatives.

Consider the collection $R^\infty$ of equations
$L_I^\mu=0$, $L_i = 0$, $\partial_{\nu}L_I^\mu =0$
$\partial_{\nu}L_i=0$, $\partial_{(\nu_1\nu_2)}L_I^\mu=0$,
$\partial_{(\nu_1\nu_2)}L_i=0,
\dots$ defined on $J^\infty(E)$. They define the                
so-called
stationary surface $\Sigma^\infty$ on
$J^\infty(E)$.
In a given jet-space $J^{r}(E)$ of finite order $r$, the stationary
surface $\Sigma^r$ is the
surface
defined by the subset $R^r\subset R^\infty$ of the
above collection of equations which is relevant in $J^{r}(E)$. 

Note that the equations of motion involve only the original
classical fields $A^I_\mu$ and $\psi^i$ and their derivatives.  They
do not constrain the ghosts because one is dealing
with the original gauge-covariant equations and not
those of the gauge-fixed theory.  This fact will turn out to be
quite important later on.

\subsubsection{Noether identities}

Because of gauge invariance, the left hand sides of the
equations of motion  are
not all independent functions on $J^{\infty}(E)$,
but they satisfy
some relations, called 
Noether identities.
These read
\begin{equation}
\partial_{(\nu_1\dots\nu_k)}[D_\mu
L^\mu_I+e L_i T^i_{Ij} \psi^j]=0,
\label{Noether007}
\end{equation}
for all $k=0,1,\dots$ and $I=1,\dots,\mathit{dim}(\cG)$.

\subsubsection{Statement of regularity conditions}
\label{regcond}

The Yang-Mills Lagrangian $L = (-1/4) \delta_{IJ}F^I_{\mu \nu}
F^{J \mu \nu}$ fulfills important 
regularity conditions which we now spell out in detail.
For all $r$, the collection $R^r$ of equations of motion
can be split
into two groups, the
``independent equations" $L_a$ and the ``dependent
equations"
$L_\Delta$.\footnote{By an abuse of terminology,
we use ``equations of motions'' both for the actual equations and 
for their left hand sides.} 
The independent equations are by definition such
that they can be taken
to be some of the  coordinates of a new coordinate
system on $V^r$,
while the dependent
equations
hold as consequences of the independent ones:
$L_a=0$ implies
$L_\Delta=0$.  Furthermore, there is one and only one
dependent equation for each Noether identity: 
these identities are the only
relations among the equations and they are not redundant. 
To be precise, when viewed as equations for the
$L_a$ and $L_\Delta$, the Noether identities 
$\partial_{(\nu_1\dots\nu_k)}D_\mu
L^\mu_I=0$ are strictly
equivalent to 
\beq
L_\Delta - L_a\, k_\Delta^a = 0,
\label{equivNoether}
\eeq
for some local functions $k_\Delta^a$ of the fields and their derivatives
(a complete set $\{L_\Delta,L_a\}$ is given explicitly below).
Thus, the left hand sides of the Noether identities
$\partial_{(\nu_1\dots\nu_k)}D_\mu
L^\mu_I=0$, viewed as equations for the $L_a$ and $L_\Delta$, 
are of the form 
\beq
\partial_{(\nu_1\dots\nu_k)}D_\mu L^\mu_I
=  (L_\Delta - L_a\, k_\Delta^a) \cM_{(\nu_1\dots\nu_k)I}^\Delta,
\label{equivNoetherbis}
\eeq
for some invertible matrix $\cM_{(\nu_1\dots\nu_k)I}^\Delta$
that may depend on the dynamical variables (the range of
$(\nu_1\dots\nu_k)I$ is equal to the range of $\Delta$).

Similarly,
one can split the gauge fields and their
derivatives into
``independent
coordinates" $y_A$, which are not constrained
by the equations of motion in the jet-spaces,
and ``dependent coordinates" $z_a$, which can be expressed in
terms of the $y_A$ on the stationary surface(s).
For fixed $y_A$, the change of variables from $L_a$
to $z_a$ is smooth and invertible
(a complete set $\{y_A,z_a\}$ is given explicitly below).

The same regularity properties hold for the Chern-Simons
action, or if one minimally couples scalar or spinor
fields to the Yang-Mills potential as in the standard model. 

The importance of the regularity conditions is that they will enable us to
compute completely the homology of $\delta$
by identifying appropriate contractible pairs.
We shall thus verify them explicitly.

We successively  list the
$L_a,L_\Delta,y_A$ and $z_a$ for the massless free Dirac field,
for the massless free
Klein-Gordon field, for
pure Yang-Mills theory and for
the Chern-Simons theory in three dimensions.

\paragraph{Dirac field:}
We start with the simplest case, that of the free (real) Dirac field,
with equations of motion $\cL \equiv \gamma^\mu \partial_\mu \Psi = 0$.
These equations are clearly independent (no Noether identity)
and imply no restriction on the undifferentiated field components.
So, the stationary surface in $J^0(E)$ is empty.  The equations
of motion start ``being felt" in $J^1(E)$ since they are of
the first order.  One may rewrite them as $\partial_0 \Psi=
\gamma^0 \gamma^k \partial_k \Psi$, so the derivatives
$\partial_k \Psi$ may be regarded as independent, while $\partial_0 \Psi$
is dependent.  Similarly, the successive derivatives of $\partial_0 \Psi$
may be expressed in terms of the spatial derivatives of $\Psi$ by
differentiating the equations of motion, so one gets the following
decompositions:
\begin{eqnarray}                      
\{L_a\}&\equiv&\{ {\cal L},\partial_\mu {\cal
L},\partial_{(\mu_1\mu_2)}{\cal L},\dots\},\quad
\{L_\Delta\}\ {\rm is\ empty},\label{LoceracD}
\\
\{y_A\}&\equiv&\{\Psi,\Psi_s,
\Psi_{(s_1 s_2)},\dots,\Psi_{(s_1\dots
s_m)},\dots\},\\
\{z_a\}&\equiv&\{\Psi_{0},\Psi_{(\rho_10)},
\dots,\Psi_{(\rho_1\dots\rho_m 0)},\dots\}.
\end{eqnarray}
The subset of equations $R^r$ relevant in $J^r(E)$ is given
by the Dirac equations and their derivatives up to order $r-1$.

\paragraph{Klein-Gordon field:}
 Again, the equation of motion 
$\cL \equiv \partial_\mu \partial^\mu \phi = 0$ and
all its differential consequences
are
independent.  Furthermore, they can clearly be used to express
any derivative of $\phi$ involving at least two temporal derivatives
in terms of the other derivatives.
\begin{eqnarray}
\{L_a\}&\equiv&\{ {\cal L},\partial_\mu {\cal
L},\partial_{(\mu_1\mu_2)}{\cal L},\dots\},\quad
\{L_\Delta\}\ {\rm is\ empty},%\label{Loce7}
\\
\{y_A\}&\equiv&\{\phi,\phi_\rho,
\phi_{(s_1\rho)},\dots,\phi_{(s_1\dots
s_m\rho)},\dots\},\\
\{z_a\}&\equiv&\{\phi_{(00)},\phi_{(\rho_100)},
\dots,\phi_{(\rho_1\dots\rho_m 00)},\dots\}.
\end{eqnarray}

\paragraph{Pure Yang-Mills field:}
The equations of motion $L_I^\mu\equiv D_\nu F^{\nu \mu}_I = 0$
are defined in $J^2(E)$, where they are independent.  The
Noether identities involve the spacetime derivatives of the Euler-Lagrange
derivatives of the pure Yang-Mills Lagrangian and so start
playing a r\^ ole in $J^3(E)$.  They can be used to
express $\partial_0$ of the field equation for $A^I_0$ in terms of
the other equations.  Thus, in $J^3(E)$, $ L^\mu_I = 0$,
$\partial_{\rho} L_I^m =0$
and $\partial_{k} L_I^0 = 0$ are independent equations, while
$\partial_{0}
L_I^0 = 0$ are dependent equations following from the others.
One can solve the equations $ L^m_I= 0$ and
$ L^0_I= 0$ for $A^I_{m|(00)}$ and $A^I_{0|(11)}$
in terms of the other second order derivatives of the
fields.  Similarly, the derivatives of the equations
$ L^m_I= 0$ can be solved for
$A^I_{m|(\rho_1\dots\rho_s00)}$ while the independent derivatives
of $ L^0_I= 0$ can be solved for $A^I_{0|(s_1\dots
s_n11)}$.  A possible split of the equations and the variables
fulfilling the above requirements is  therefore given by
\begin{eqnarray}
\{L_a\}&\equiv&\{  L^\mu_I,\partial_\rho  L_I^m,
\dots,\partial_{(\rho_1\dots\rho_s)}L_I^m,\dots,\nonumber\\
&&\phantom{\{  L^\mu_I,}
\partial_{k} L_I^0,\dots,\partial_{(k_1\dots k_2)}
L_I^0,\dots\},
\label{YMLa}
\\
\{L_\Delta\}&\equiv&\{\partial_{0}
L_I^0,\partial_{\rho}\partial_{0}L_I^0,\dots,
\partial_{(\rho_1\dots \rho_s)}\partial_{0}L_I^0,\dots\},
\label{YMLDelta}
\\
\{y_A\}&\equiv&\{A^I_\mu,
A^I_{\mu|\rho},A^I_{m|(s_1\rho)},
\dots,
A^I_{m|(s_1\dots s_k\rho)},\dots,
\nonumber\\
&&\phantom{\{A^I_\mu,A^I_{\mu|\rho},}
A^I_{0|(\lambda 0)},\dots,
A^I_{0|(\lambda_1\dots\lambda_k 0)},\dots,
\nonumber\\
&&\phantom{\{A^I_\mu,A^I_{\mu|\rho},}
A^I_{0|(\bar l m )},
\dots,
A^I_{0|(\bar l_1\dots \bar l_k  m )}
,\dots\}\quad (\bar
l,\bar l_i > 1),\\
\{z_a\}&\equiv&\{
A^I_{m|(00)},
A^I_{m|(\rho00)}
,\dots,
A^I_{m|(\rho_1\dots\rho_s00)}
,\dots,\nonumber\\
&&\phantom{\{}
A^I_{0|(11)},
\dots,
A^I_{0|(s11)},\dots,A^I_{0|(s_1\dots
s_n11)},
\dots\}.
%\label{Loce13}
\end{eqnarray}
Finally, the matrix $\cM_{(\nu_1\dots\nu_k)I}^\Delta$ 
in Eq.\ (\ref{equivNoetherbis}) associated
with this split of the equations of motion is easily constructed: it
is a triangular matrix with entries 1 on the diagonal and thus it is
manifestly invertible.  Indeed, one has, for the undifferentiated
Noether identities, $ D_\mu  L^\mu_I = \delta_I^J \partial_0
L^0_J + \hbox{ ``more"}$, where ``more" denotes 
terms involving only the independent equuations.  By differentiating
these relations, one gets $\partial_{(\nu_1\dots \nu_k)}
D_\mu  L^\mu_I =  \delta_I^J
\partial_{(\nu_1\dots \nu_k)}  \partial_0  L^0_J
+$ ``lower" $+$ ``more", where ``lower" denotes terms involving
the previous dependent equations.

\paragraph{Chern-Simons theory in three dimensions:}
The equations 
are this time $L^\mu_I \equiv \varepsilon^{\mu \rho \sigma}
F_{I \rho \sigma} = 0$.  They can be split as above
since the Noether identities take the same form.
But there are less independent field components since the
equations of motion are of the first order and thus start
being relevant already in $J^1$.
From the equations $L^i_I = 0$ and their derivatives,
one can express the derivatives of the spatial components $A^I_i$
of the vector potential with at least one $\partial_0$ in terms 
of the derivatives of $A^I_0$, which are unconstrained.
Similarly, from the equations $L^0_I = 0$ and their spatial
derivatives, which are independent, one can express all spatial
derivatives of $A^I_2$ with at least one $\partial_1$
in terms of the spatial derivatives of $A^I_1$.
Thus, we have $L_a$ and $L_\Delta$ as in (\ref{YMLa}) and
(\ref{YMLDelta}), but the
$y_A$ and $z_a$ are now given by
\begin{eqnarray}
\{y_A\} &\equiv& \{A^I_\mu, A^I_{0 |\rho}, \dots ,
A^I_{0 |(\rho_1 \dots \rho_k)}, \dots,\nonumber\\
&&\phantom{\{\{A^I_\mu,}
 A^I_{1 |s},
\dots,A^I_{1|(s_1 \dots s_k)}, \dots,\nonumber\\
&&\phantom{\{\{A^I_\mu,}
A^I_{2 |2},
\dots, A^I_{2 |2 \dots 2}, \dots \}, \\
\{z_a\} &\equiv& \{A^I_{m |0}, A^I_{m |(0 \rho)}, \dots,
A^I_{m |(0 \rho_1 \dots \rho_k)}, 
\dots, \nonumber \\
&&\phantom{\{}
A^I_{2|1}, A^I_{2 |(1s)}, \dots, A^I_{2 |(1 s_1 \dots s_k)},
\dots\}.
\label{CScoordsplit} 
\eea

\vspace{.2cm}

We have systematically used $\Psi_{(s\rho)} =
\partial_{(s\rho)} \Psi$ etc and $A^I_{0|(s11)}
= \partial_{(s11)} A^I_{0}$ etc (with $s=1,\dots,n-1$ and
$\rho=0,\dots,n-1$).
Note that the above splits are not unique. Furthermore, they are
not covariant.  We will in practice not use any of these splits.
The only thing that is needed is the fact that such splits exist.

It is clear that the regularity conditions continue to hold if one
minimally couples the Klein-Gordon or Dirac fields to the
Yang-Mills potential since the coupling terms involve terms
with fewer derivatives.  Therefore, the regularity
conditions hold in particular
for the Lagrangian of the standard model.  

For more general local Lagrangians of the
Yang-Mills type, the regularity conditions are not
automatic. For instance, they are not fulfilled in pure Chern-Simons theory
in five dimensions because the equations of motion of that theory
have no part linear in the fields and can therefore not be used as new
admissible jet coordinates. 
The results on $H(\delta)$ derived in this section are valid 
only for Lagrangians fulfilling the regularity conditions. 

For non-local Lagrangians of the type appearing in the
discussion of effective field theories, 
the question of whether the regularity conditions
are fulfilled does not arise since
the equations of motion imply no restriction in the jet-spaces
$J^r(E)$ of finite order. In some definite sense to be made
precise below, one can say, however, that these theories also
fulfill the regularity conditions.

\subsubsection{Weakly vanishing forms}
An antifield independent 
local form vanishing when the equations of
motion hold is
said to be weakly vanishing. This is 
denoted by $\omega\approx 0$.
An immediate consequence of the regularity
conditions is
\begin{lemma}\label{Locl1}
If an antifield independent local form 
$\omega\in\Omega_E$ is weakly vanishing, $\omega\approx 0$,
it can be written as a
linear combination of equations of motion with coefficients which
are local forms, and is thus $\delta$-exact in the space
$\Omega_F$ of local forms,
\[
\omega\approx 0,\quad \agh(\omega)=0 \quad
\Leftrightarrow\quad \omega=\delta \eta,\quad \agh(\eta)=1.
\]
\end{lemma}

\paragraph{Proof}

In the coordinate system $(x,dx,L_a,y_A)$,
$\omega$ satisfies
$\omega(x,dx,0,y_A)=0$. Using
a homotopy formula like in (\ref{A1}), one 
gets
$\omega=L_a\int_0^1
d\lambda[\frac{\partial}{\partial L_a}\omega]
(x,dx,\lambda L_b,y_A)$. Since the $L_a$ are equations of motion,
there are antifield variables $\phi^*_a$ such that
$\delta \phi^*_a=L_a$. This gives
$\omega=\delta \eta$ where
$\eta=\phi^*_a\int_0^1
d\lambda[\frac{\partial}{\partial L_a}\omega]
(x,dx,\lambda L_b,y_A)$. Going back to the
original coordinate system
proves the lemma. \qed

If both the equations of motion and the form $\omega$ 
are polynomial, $\eta$ is also polynomial.
                                                                                
\subsection{Koszul-Tate
resolution}\label{koszul}

Forms defined on the stationary
surface can be viewed
as equivalence classes of forms defined on the whole of jet-space
modulo forms that vanish when the equations of motion hold.
It turns out that the homology of $\delta$ is precisely
given, in degree zero, by this quotient space.  Furthermore,
its homology in all other degrees is
trivial.
This is why one says that the Koszul-Tate differential
implements the equations of motion.

More precisely, one has
\begin{theorem}\label{Loct4}
(Homology of $\delta$ in the algebra $\Omega_F$ of local
forms involving the original fields, the ghosts and the
antifields)

The homology of $\delta$ 
in antifield number $0$
is given by the
equivalence classes of local forms ($\in \Omega_E$)
modulo weakly
vanishing ones,
$H_0(\delta,\Omega_F)=\{[\omega_0]\}$, with
$\omega_0\sim\omega_0^\prime$
if $\omega_0-\omega_0^\prime\approx 0$. 

The
homology of $\delta$ in
strictly positive antifield number is trivial,
$H_m(\delta,\Omega_F)=0$
for $m>0$.
\end{theorem}

In mathematical terminology, one says that the Koszul-Tate
complex provides a ``resolution" of the algebra of local 
forms defined on the stationary surface.

\paragraph{Proof}

The idea is to exhibit appropriate contractible pairs
using the regularity conditions. 
First, one can replace the jet-space coordinates
$A^I_\mu, \psi^i$ and their derivatives by
$y_A$ and $L_a$.  As we have seen, this change of variables is 
smooth and invertible.  

In the notation $(a,\Delta)$, the
antifields $A^{*I}_\mu, \psi^*_i$ and their derivatives
are $(\phi^*_a, \phi^*_\Delta)$ with $\delta \phi^*_a
= L_a$ and $\delta \phi^*_\Delta = L_\Delta$.   The
second step is to redefine the antifields $\phi^*_\Delta$
using the matrix $\cM_{(\nu_1\dots\nu_k)I}^\Delta$ for the 
Noether identities (\ref{Noether007}) analogous to the matrix
in (\ref{equivNoetherbis}) for the pure Yang-Mills case. One defines
$\4\phi^*_{(\nu_1\dots\nu_k)I} := (\phi^*_\Delta -
\phi^*_a k^a_\Delta)\cM_{(\nu_1\dots\nu_k)I}^\Delta$.  
This definition makes the
action of the Koszul-Tate differential particularly
simple since one has $\delta \4\phi^*_{(\nu_1\dots\nu_k)I} = 
0$.  Indeed $(L_\Delta - L_a k_\Delta^a)
\cM_{(\nu_1\dots\nu_k)I}^\Delta$
identically vanishes by the Noether identity.
In fact, one has $\4\phi^*_{(\nu_1\dots\nu_k)I}
= \delta C^*_{I \vert (\nu_1\dots\nu_k)}$.

In terms of the new variables, the
Koszul-Tate
differential reads
$$\delta=L_a\,\frac{\partial}{\partial\phi^*_a}+
\sum_{k\geq 0}\4\phi^*_{(\nu_1\dots\nu_k)I}\,
\frac{\partial}{\partial  C^*_{I |(\nu_1\dots\nu_k)}},$$
which makes it clear that $L_a, \phi^*_a, \4\phi^*_{(\nu_1\dots\nu_k)I}$
and $C^*_{I |(\nu_1\dots\nu_k)}$ form contractible
pairs dropping from the homology.  This leaves only
the variables $y_A$, as well as the ghosts $C^I$ and their
derivatives, as generators of the homology of $\delta$.
In particular, the antifields disappear from the homology and
there is thus
no homology in strictly positive antifield number. \qed                              

A crucial ingredient of the proof is the fact that the
Noether identities are independent and exhaust all the 
independent Noether
identities.  This is what guaranteed the change of variables
used in the proof of the theorem to be invertible.
It allows one to generalize the theorem to theories fulfilling 
regularity conditions analogous to those of the Yang-Mills case.
For theories with ``dependent" Noether identities
(``reducible case"),
one must add further antifields at higher antifield
number.  With these additional variables, the theorem  still
holds.  The homological rationale for the antifield spectrum
is explained in \cite{Fisch:1989dq,Fisch:1990rp}.

\paragraph{Remarks:}

(i) We stress that the equations of motion that appear 
in the theorem are
the gauge-covariant equations of motion derived from 
the gauge-invariant Lagrangian $L$
(and not any gauge-fixed form of these equations).

(ii) When the Lagrangian is Lorentz-invariant, 
it is natural
to regard the antifields $A^{*\mu}_I$, $\psi^*_i$ and $C^*_I$
as transforming in the representation of the Lorentz group
contragredient to the representation of $A^I_\mu$,
$\psi^i$ and $C^I$, respectively.  Thus, the $A^{*\mu}_I$
are Lorentz vectors while the $C^*_I$ are Lorentz scalars.
Because $\delta A^{*\mu}_I$, $\delta \psi^*_i$
and $\delta C^*_I$ have the same transformation properties
as $A^{*\mu}_I$, $\psi^*_i$ and $C^*_I$, $\delta$ commutes
with the action of the Lorentz group.  One can consider
the  homology of $\delta$ in the algebra of Lorentz-invariant 
local forms.  Using that the Lorentz group is semi-simple,
one  checks that  this homology is trivial in strictly
positive antifield number, and given by the equivalence classes
of Lorentz-invariant local forms modulo weakly vanishing ones in
antifield number zero (alternatively one may verify this
directly be means of the properties of the contracting homotopy 
used in the proof).
Similar considerations apply to other 
linearly realized global symmetries of the Lagrangian.

(iii) Again, if the equations of motion are polynomial,
theorem \ref{Loct4} holds in the algebra of local, polynomial
forms.

\subsection{Effective field theories}
\label{Hdeltaeff}

The results for the homology of $\delta$ in the Yang-Mills
case extends to the analysis of effective
field theories.  The problem is to compute the homology
of $\delta$ in the space of formal power series in the
free parameters, generically denoted by $g_\alpha$, with
coefficients that are local forms.  We normalize the fields
so that they have canonical dimensions.  This means,
in particular, that the Lagrangian takes the form
\beq
L = L_0 + O(g_\alpha)
\eeq
where the zeroth order Lagrangian $L_0$ is the free Lagrangian and is
the sum of the standard kinetic term for free massless vector
fields and of the free Klein-Gordon or Dirac Lagrangians.

Corresponding to this decomposition of $L$, there is a decomposition
of $\delta$,
\beq
\delta = \delta_0 + O(g_\alpha),
\eeq
where $\delta_0$ is the Koszul-Tate differential of the
free theory.
Now, $\delta_0$ is acyclic (no homology)
in positive antifield number.  The point is that
this property passes on to the
complete $\delta$.  Indeed, let $a$ be a form which is $\delta$-closed,
$\delta a = 0$, and has positive antifield number.  Expand $a$ 
according to the degree in the
parameters, $a = a_i + a_{i+1} + \dots$. Since there are many parameters, 
$a_j$ is in fact the 
sum of independent monomials of
degree $j$ in the $g_\alpha$'s.
The terms $a_i$ of lowest order in
the parameters must be $\delta_0$-closed, $\delta_0
a_i = 0$.  But then, they are  
$\delta_0$-exact, $a_i = \delta_0 b_i$, where $b_i$
is a local form (since $a_i$ has positive antifield number by assumption).  
This implies that
$a - \delta b_i$ starts at some higher order $i'$ ($i'>i$)
if it does not
vanish.  By repeating
the reasoning at order $i'$, and then successively at the higher orders, 
one sees that $a$ is indeed equal to the $\delta$
of a (in general infinite, formal) power series in the parameters, 
where each coefficient is a local form.

Similarly, at antifield number zero, a formal power
series is $\delta$-exact
if and only if it is a combination of the Euler-Lagrange derivatives
of $L$ (such forms may be called ``weakly vanishing formal
power series").
Thus Theorem
\ref{Loct4}  holds also in
the algebra of formal power series in the parameters with
coefficients that are local forms, relevant to
effective field theories. In that sense, effective theories fulfill
the regularity conditions because the leading term $L_0$ does. 

\newpage

\mysection{Conservation laws and symmetries: 
$H(\delta \vert d)$}
\label{Koszul2Section}

In this chapter, we relate $H(\delta \vert d)$ to 
the characteristic cohomology of the theory. 
The argument is quite general and not
restricted to gauge theories of the Yang-Mills type.
It only relies on the fact that the Koszul-Tate complex
provides a resolution of the algebra of local $p$-forms
defined on the stationary surface, so that lemma \ref{Locl1}
and theorem \ref{Loct4} hold for the gauge theory under study.

We then compute this cohomology for irreducible gauge theories
in antifield number higher than $2$ on various assumptions
and specialize the results to the Yang-Mills case.

\subsection{Cohomological version of Noether's first theorem}
%{Conservation laws and global symmetries}

In this section, the fields $\phi^i$ are the original
classical fields and $\cL_i$ the Euler-Lagrange derivatives
of $L$ with respect to $\phi^i$. The corresponding jet-spaces are
denoted by $J^r(D)$, $J^\infty(D)$. 
In order to avoid cluttered formulas, we shall 
assume for simplicity that the fields are all bosonic. 
The inclusion of fermionic fields leads only to extra sign 
factors in the formulas below. 

An infinitesimal field transformation is
characterized by local
functions $\delta_Q\phi^i=Q^i(x,[\phi])$,
to which one associates
the vector 
field 
\[
\vec Q= \sum_{l\geq 0}\partial_{(\mu_1\dots\mu_l)}
Q^i\,\frac{\partial}{\partial
\phi^i_{(\mu_1\dots\mu_l)}}
\]
on the jet-space $J^\infty(D)$. It commutes with the
total derivative,
\beq
[\partial_\mu, \vec Q]=0. \label{Loctbd}
\eeq
The $Q^i$ are called the ``characteristics" of the field
transformation.

A symmetry of the theory is an infinitesimal
field transformation leaving the Lagrangian
invariant up to a total
derivative:
\beq
\vec Q L \equiv \delta_Q L =\partial_\mu k^\mu,\label{Locgs}
\eeq
for some local functions $k^\mu$.
One can rewrite this equation
using integrations by parts as
\beq
Q^i\cL_i+\partial_\mu j^\mu=0,\label{Loces}
\eeq
for some local vector density $j^\mu$. This
equation can also be read
as 
\beq
\partial_\mu j^\mu\approx 0,\label{curr}
\eeq
which means that $j^\mu$
is a conserved current.
This is just Noether's result that to every 
symmetry there corresponds
a conserved current. Note that
this current could be zero in the case where
$Q^i=M^{[ij]}\cL_j$ (such $Q^i$ are examples of trivial symmetries,
see below), which means that the
correspondence is not
one to one. On the other hand, one can associate
to a given symmetry the family of
currents $j^\mu+ \partial_\nu
k^{[\nu\mu]}$, which means that the correspondence
is not onto
either. As we now show,
one obtains bijectivity by passing to appropriate
quotient spaces. 

Defining
$\omega^{n-1}_0=\frac{1}{(n-1)!}\,
dx^{\mu_1}\dots dx^{\mu_{n-1}}\epsilon_{\mu_1\dots\mu_{n}}j^{\mu_n}$,
we can
rewrite Eq.\ (\ref{curr}) in terms of the antifields as
\beq
d\omega^{n-1}_0+\delta\omega^{n}_1=0.\label{Locecoc}
\eeq
This follows from lemma \ref{Locl1} and theorem
\ref{Loct4}, the superscript
denoting the form degree and the subscript the
antifield
number (we do not write the pure ghost number
because the ghosts do not enter at this stage; the
pure ghost number is always zero in this and the next 
subsection). Because $\delta$ and $d$ anticommute, a
whole class of
solutions to this equation is provided by
\beq
\omega^{n-1}_0=d\eta^{n-2}_0+\delta\eta^{n-1}_1,\label{Locecob}
\eeq
with
$\omega^{n}_1=d\eta^{n-1}_1$. This suggests to
define equivalent
conserved currents
$\omega^{n-1}_0\sim\omega^{\prime n-1}_0$ as conserved
currents that differ 
by terms of the form given in the right-hand side of
(\ref{Locecob}).  In other words,
equivalence classes of conserved
currents are just  the elements of the cohomology
group $H^{n-1}_0(d|\delta)$ (defined through the
cocycle condition
(\ref{Locecoc}) and the coboundary condition
(\ref{Locecob})). Expliciting the coboundary
condition in dual notation,
one thus identifies conserved currents which differ
by
identically conserved currents of the form
$\partial_\nu
k^{[\nu\mu]}$ modulo weakly vanishing currents,
\beq
j^\mu\sim j^\mu+\partial_\nu
k^{[\nu\mu]}+t^\mu,\ t^\mu\approx 0.\label{trcurr}
\eeq
Equations (\ref{curr}) and (\ref{trcurr}) define the characteristic 
cohomology $H^{n-1}_\mathrm{char}$ in form degree $n-1$, 
which can thus be identified with $H^{n-1}_0(d|\delta)$. 

Let us now turn to symmetries of the theory.  In a gauge
theory, gauge transformations do not change the
physics.  It is therefore natural to identify
two symmetries that differ by a gauge transformation.
A general gauge transformation involves not only standard
gauge transformations, but also, ``trivial gauge transformations"
that vanish on-shell \cite{Kallosh:1977ik,deWit:1978cd,vanholten} 
(for a recent discussion, see e.g. \cite{Henneaux:1992ig}).  

A trivial, local, gauge symmetry reads
\beq
\delta_M \phi^i = \sum_{m,k\geq 0}(-)^k
\partial_{(\mu_1\dots\mu_k)}[
M^{j(\nu_1\dots\nu_m)i(\mu_1\dots\mu_k)}
\partial_{(\nu_1\dots\nu_m)}\cL_j],
\label{trivial}
\eeq
where the functions $M^{j(\nu_1\dots\nu_m)i(\mu_1\dots\mu_k)}$
are arbitrary local functions antisymmetric for the
exchange of the indices
\beq
M^{j(\nu_1\dots\nu_m)i(\mu_1\dots\mu_k)}
= - M^{i(\mu_1\dots\mu_k)j(\nu_1\dots\nu_m)}.\label{asym}
\eeq
It is direct to verify that trivial gauge symmetries leave
the Lagrangian invariant up to a total derivative.

If 
\beq
\delta_f \phi^i = 
\sum^{\bar l_\alpha}_{l=0} R^{i(\mu_1\dots\mu_l)}_\alpha
\partial_{(\mu_1\dots\mu_l)} f^\alpha, 
\label{complete}
\eeq
where the $f^\alpha$ are arbitrary local functions,
provides a complete set of nontrivial gauge symmetries in
the sense of \cite{Henneaux:1992ig}\footnote{For
more information on complete sets of gauge
transformations, see appendix A.}, 
then, the most general
gauge symmetry is given by the sum of a transformation
(\ref{complete}) and a trivial gauge transformation (\ref{trivial})
\beq
\delta_{f,M} \phi^i = \delta_f \phi^i + \delta_M \phi^i.
\label{generalgauge}
\eeq
We thus define equivalent global symmetries
as symmetries of the theory that differ by a 
gauge transformation of the form (\ref{generalgauge}) 
with definite choices of the local functions $f^\alpha$ and 
$M^{j(\nu_1\dots\nu_m)i(\mu_1\dots\mu_k)}$.
The resulting quotient space is called the
space of non trivial global symmetries.

The Koszul-Tate differential is defined through
\bea
\delta
\phi^*_i &=& \cL_i, \\
\delta C^*_\alpha &=& \sum^{\bar l_\alpha}_{l= 0} 
R^{+i(\mu_1\dots\mu_l)}_\alpha \partial_{(\mu_1\dots\mu_l)} \phi^*_i,
\label{Def}
\eea
where the $R^{+i(\mu_1\dots\mu_l)}_\alpha$ are defined in terms
of the $R^{i(\mu_1\dots\mu_l)}_\alpha$ through
\beq
\sum^{\bar l_\alpha}
_{l=0}(-)^l
\partial_{(\mu_1\dots\mu_l)}[R^{+i(\mu_1\dots\mu_l)}_\alpha f^\alpha ]
= \sum^{\bar l_\alpha}_{l=0} R^{i(\mu_1\dots\mu_l)}_\alpha
\partial_{(\mu_1\dots\mu_l)} f^\alpha
\eeq
and where the $ C^*_\alpha$ are the antifields conjugate to
the ghosts.  For instance, for pure
Yang-Mills theory, one has $R^I_{\mu J} = e f_{KJ}^{\;\;\;\; \; I}
A_\mu^K = R^{+I}_{\mu J}$, $R^{I \nu}_{\mu J} =
\delta^\nu_\mu \delta^I_J = - R^{+ I \nu}_{\mu J}$  
and one recovers from (\ref{Def})
the formula (\ref{ym7}) for $\delta C^*_I$.

To any infinitesimal field transformation $Q^i$, one can associate
a local $n$-form linear in the antifields $\phi^*_i$ through
the formula $\omega^n_1= d^nx\, a_1 = d^nx\, 
Q^i \phi^*_i$.  Conversely, given an arbitrary local $n$-form
of antifield number one,
$\omega^n_1= d^nx\, a_1 =
d^nx\,
\sum_{l\geq 0}a^{i(\mu_1\dots\mu_l)}\phi^*_{i|(\mu_1\dots\mu_l)}$,
one can add to it a $d$-exact term in order to remove the
derivatives of $\phi^*_i$.  The coefficient of $\phi^*_i$ in
the resulting expression defines an infinitesimal transformation.
Explicitly,
$Q^i=\frac{\delta
a_1}{\delta\phi^*_i}
=\sum_{l\geq 0}(-)^l\partial_{(\mu_1\dots\mu_l)}a^{i(\mu_1\dots\mu_l)}$.
There is thus a bijective correspondence between infinitesimal
field transformations and equivalence classes of local $n$-forms
of antifield number one (not involving the ghosts), where one
identifies two such local $n$-forms that differ by a $d$-exact
term.

Now, it is clear that $Q^i$ defines a symmetry of the theory
if and only if the corresponding $\omega^n_1$'s are 
$\delta$-cocycles modulo $d$. In fact, one has

\begin{lemma}\label{Locl3}
Equivalence classes of global symmetries 
are in bijective correspondence with
the elements of $H^n_1(\delta|d)$.
\end{lemma}

\paragraph{Proof:}
The proof simply follows by expanding the most
general local $n$-form $a_2$ of antifield number $2$,
computing $\delta a_2$ and making
integrations by parts.  It is left to the reader. We only remark that the 
antisymmetry in (\ref{asym}) follows from the fact that the antifields 
are Grassmann odd.

\vspace{.2cm}

This cohomological set-up allows to prove
Noether's first theorem
in the case of (irreducible) gauge theories in a
straightforward way, using theorems \ref{Loct2}
and \ref{Loct4}.

\begin{theorem}\label{Loct5}
The cohomology groups $H^n_1(\delta|d)$ and
$H^{n-1}_0(d|\delta)$ are
isomorphic in spacetime dimensions $n>1$. In classical mechanics
(n=1), they are isomorphic up to the constants,
$H^1_1(\delta|d)\simeq
H^{0}_0(d|\delta)/\mathbb{R}$.\footnote{We derive and
write cohomological results
systematically for the case that the cohomology under study is
computed over $\mathbb{R}$. They hold analogously over $\mathbb{C}$.}
In other words, there is an isomorphism between
equivalence classes of global symmetries and
equivalence classes of conserved currents (modulo
constant currents when
$n=1$).
\end{theorem}

\paragraph{Proof:}
The proof relies on the triviality of the (co)homologies
of $\delta$ and $d$ in appropriate degrees and follows a standard pattern.
We define a mapping from $H^n_1(\delta|d)$ to
$H^{n-1}_0(d|\delta)/\delta^{n-1}_0{\mathbb R}$ as follows.  
Let $\omega^n_1$ be a $\delta$-cocycle
modulo $d$ in form-degree $n$ and antifield number $1$,
\beq
\delta \omega^n_1 + d \omega^{n-1}_0 = 0
\label{mapmap}
\eeq
for some $\omega^{n-1}_0$ of form-degree $(n-1)$ and antifield
number $0$. Note that $\omega^{n-1}_0$ is a $d$-cocycle
modulo $\delta$. Furthermore, given $\omega^n_1$, $\omega^{n-1}_0$
is defined up to a $d$-closed term, i.e., up to a $d$-exact
term ($n>1$) or a constant ($n=1$) (algebraic Poincar\'e
lemma).  If one changes $\omega^n_1$
by a term which is $\delta$-exact modulo $d$, $\omega^{n-1}_0$ is
changed by a term which is $d$-closed modulo $\delta$.  Formula (\ref{mapmap})
defines therefore a mapping from $H^n_1(\delta|d)$ to
$H^{n-1}_0(d|\delta)/\delta^{n-1}_0{\mathbb R}$.  This mapping is 
surjective because
(\ref{mapmap}) is the cocycle condition
both for $H^n_1(\delta|d)$ and for $H^{n-1}_0(d|\delta)$.
It is also injective because
$H^{n}_1(\delta)=0$.  \qed 

\subsection{Characteristic cohomology and $H(\delta|d)$}

We now consider the cohomology groups
$H^{n-p}_0(d|\delta)$ for all values 
$p=1,\dots,n$, and not just for $p=1$. 
Using again lemma \ref{Locl1} and theorem
\ref{Loct4}, these groups can be described independently of 
the antifields. 
In that context, they are
known as the ``characteristic cohomology groups'' $H^{n-p}_\mathrm{char}$
of the
stationary surface
$\Sigma^\infty$ and define the ``higher order
conservation laws''.
For instance, for $p=2$, they are given, in dual
notation, by ``supercurrents'' $k^{[\mu\nu]}$ such
that
$\partial_{\mu}k^{[\mu\nu]}\approx 0$, where two
such supercurrents are 
identified if they differ on shell by an
identically conserved
supercurrent: $k^{[\mu\nu]}\sim k^{[\mu\nu]}
+\partial_\lambda
l^{[\lambda\mu\nu]}+t^{[\mu\nu]}$, where
$t^{[\mu\nu]}\approx 0$.

In the same way, one generalizes non-trivial global
symmetries by considering
the cohomology groups $H^n_k(\delta|d)$, for
$k=1,2,\dots$. These
groups are referred to as ``higher order (non-trivial) global
symmetries''.

The definitions are such that the isomorphism
between
higher order conserved currents and higher order
symmetries still
holds:
\begin{theorem}\label{Loct6}
One has the following isomorphisms
\bea
k<n:&&
H^n_k(\delta|d)\simeq
H^{n-1}_{k-1}(\delta|d)\simeq\dots \simeq
H^{n-k+1}_1(\delta|d)\simeq  H_0^{n-k}(d|\delta)\, ;
\label{Loct6e2}
\\[4pt]
&&
H^n_n(\delta|d)\simeq
H^{n-1}_{n-1}(\delta|d)\simeq\dots \simeq
H^{1}_1(\delta|d)\simeq  H_0^{0}(d|\delta)/\mathbb{R}\, ;\label{Loct6e3}
\\[4pt]
k>n:&&
H^n_k(\delta|d)\simeq
H^{n-1}_{k-1}(\delta|d)\simeq\dots \simeq H^{1}_{k-n+1}(\delta|d)\simeq
H^{0}_{k-n}(\delta)=0\, .\quad
\label{Loct6e4}
\eea
and
\beq
H^p_k(\delta|d)\simeq H^{p-1}_{k-1}(d|\delta)
\quad {\rm for}\ k>1\ {\rm
and}\ 0<p\leq n.\label{Loct6e1}
\eeq
In particular, the cohomology groups
$H^n_k(\delta|d)$ $(1 \leq k\leq n)$
are isomorphic to the characteristic cohomology
groups $H_0^{n-k}(d|\delta)$ (modulo the
constants for $k=n$).  
\end{theorem}

\paragraph{Proof:}
One proves equation (\ref{Loct6e1}) and the last isomorphisms
in (\ref{Loct6e2}),
(\ref{Loct6e3}) as Theorem \ref{Loct5}.
The last equality in (\ref{Loct6e4}) holds because of the acyclicity
of $\delta$ in all positive antifield numbers 
(see Section \ref{KoszulSection}).
The proof of the remaining isomorphisms 
illustrates the general technique of the
descent equations of which we shall make ample use in
the sequel.
Let $a^i_j$ be a $\delta$-cocycle modulo $d$, $\delta
a^i_j + da^{i-1}_{j-1} = 0$, $i>1$, $j>1$.  Then,
$d \delta a^{i-1}_{j-1} = 0$, from which one infers, using the
triviality of $d$ in form-degree $i-1$ ($0<i-1<n$)
that $\delta a^{i-1}_{j-1} + d a^{i-2}_{j-2} = 0$ for
some $a^{i-2}_{j-2}$.  Thus, $a^{i-1}_{j-1}$ is also a
$\delta$-cocycle modulo $d$.  If $a^i_j$ is modified by trivial
terms ($a^i_j \rightarrow a^i_j + \delta  b^{i}_{j+1} +
d b^{i-1}_j$), then, $a^{i-1}_{j-1}$ is also modified
by trivial terms.  This follows again from $H^{i-1}(d) = 0$.
Thus the ``descent" $[a^i_j] \rightarrow [a^{i-1}_{j-1}]$
from the class of $a^i_j$ in $H^i_j(\delta \vert d)$ to
the class of $a^{i-1}_{j-1}$ in $H^{i-1}_{j-1}(\delta \vert d)$
defines a well-defined application from $H^i_j(\delta \vert d)$
to $H^{i-1}_{j-1}(\delta \vert d)$.  This application is both
injective (because $H_j(\delta) = 0$) and surjective
(because $H_{j-1}(\delta) =0$).  Hence, the groups $H^i_j(\delta \vert d)$
and $H^{i-1}_{j-1}(\delta \vert d)$ are isomorphic.
\qed 

\paragraph{Remark:}
The isomorphism $H^{n-k+1}_1(\delta|d)\simeq  H_0^{n-k}(d|\delta)$
($n>k$) uses $H^{n-k}(d) =0$, 
which is true only in the space of
forms with an explicit $x$-dependence.  If one does not allow
for an explicit $x$-dependence, $H^{n-k}(d)$ 
is isomorphic
to the space $\wedge^{n-k}{\mathbb
R}$ of constant forms.  The last equality in
(\ref{Loct6e2}) reads then $H^{n-k+1}_1(\delta|d)\simeq
H_0^{n-k}(d|\delta) / \wedge^{n-k}{\mathbb R}$. 

\vspace{.2cm}

The results on the groups $H^p_k(\delta|d)$
are summarized in the table below. 
The first row contains the
characteristic cohomology groups
$H_\mathrm{char}^{n-p}$ while ${\cal F}(\Sigma)$
corresponds to the local
functionals
defined on the stationary surface, i.e., the
equivalence classes of
$n$-forms depending on the original fields
alone, where two such
forms are identified if they differ, on the
stationary surface,
by a $d$-exact $n$-form, $\omega^n_0\sim
\omega^n_0+
d\eta^{n-1}_0+\delta \eta^n_1$.  The characteristic
cohomology group $H_\mathrm{char}^0$, in particular, 
contains the functions that
are constant when the equations of motion hold.
All the cohomology groups $H^i_i(\delta|d)$
along the principal 
diagonal
are isomorphic to $H_\mathrm{char}^0/{\mathbb R}$; those along the
parallel diagonals are isomorphic among themselves.
The unwritten
groups $H^p_k(\delta \vert d)$ with $k>n$ all
vanish.  

\begin{eqnarray*}
\begin{array}{|c||c|c|c|c|c|c|c||c|}
\hline k\backslash p & 0 & 1 & 2 & \dots & \dots
& n-2 & n-1
& n \\
\hline 0 & H_\mathrm{char}^0/{\mathbb R} & H_\mathrm{char}^1 & 
H_\mathrm{char}^2 & & &
H_\mathrm{char}^{n-2} &
H_\mathrm{char}^{n-1}
&
{\cal F}(\Sigma)
 \\ \hline 1 & 0 & H_\mathrm{char}^0/{\mathbb R} & 
H^2_1 & & & H^{n-2}_1
& H^{n-1}_1 &
H^n_1\\
\hline 2 & 0 & 0 & H_\mathrm{char}^0/{\mathbb R} & & & H^{n-2}_2  &
H^{n-1}_2  & H^n_2\\
\hline \vdots & & & & & & & & \\ \hline \vdots & &
& & & & &
&\\
\hline n-2 & 0 & 0 & 0 & & & H_\mathrm{char}^0/{\mathbb R} & 
H^{n-1}_{n-2} &
H^n_{n-2}\\
\hline n-1 & 0 & 0 & 0 & & & 0 & H_\mathrm{char}^0/{\mathbb R} &
H^n_{n-1}\\
\hline n & 0 & 0 & 0 & & & 0 & 0 & H_\mathrm{char}^0/{\mathbb R} \\
\hline
\end{array}
\end{eqnarray*}

\subsection{Ghosts and
$H(\delta|d)$\label{Csdeltamodd}}

So far, we have not taken the ghosts into account
in the calculation of the homology of $\delta$ modulo $d$
(the ghosts are denoted by $C^\alpha$; in Yang-Mills theories one has 
$C^\alpha \equiv C^I$).
These are easy to treat since they are not constrained
by the equations of motion. As we have seen, $\delta$ acts trivially
on them, $\delta C^\alpha = 0$, and they do not occur in the 
$\delta$-transformation of any antifield.
Let
$N_C$ be the counting operator for
the ghosts and their derivatives, $N_C=
\sum_{n\geq 0}
C^\alpha_{(\mu_1\dots\mu_n)}{\partial/\partial
C^\alpha_{(\mu_1\dots\mu_n)}}$. This counting operator
is of course the pure ghost number.  The vector space of
local forms is the direct
sum of the vector space of forms with zero pure
ghost number ($=$ forms which
do not depend on the $C^\alpha_{(\mu_1\dots\mu_n)}$) and the vector
space of
forms that
vanish when one puts the ghosts and their derivatives equal to
zero,
$\Omega^*=\Omega^*_{N_C=0}\oplus
\Omega^*_{N_C>0}$.  In fact, $\Omega^*_{N_C>0}$ is itself the
direct sum of the vector spaces of forms with pure ghost number
one, two, etc.
Since $[N_C,\delta]=0$,  $\delta (\Omega^*_{N_C=0})$
is included in $\Omega^*_{N_C=0}$ and $\delta (\Omega^*_{N_C>0})$
is included in $\Omega^*_{N_C>0}$.

\begin{theorem}\label{Loct7}
The cohomology of $\delta$ modulo $d$ in form
degree $n$ and positive antifield
number vanishes for forms in 
$\Omega^*_{N_C>0}$,
$H^n_k(\delta|d,\Omega^*_{N_C>0})=0$ for  $k\geq 1$.
\end{theorem}

\paragraph{Proof:}

Let $\omega_k=d^nx\ a_k$, with $k\geq 1$, be
a cycle of $H^n_k(\delta|d,\Omega^*_{N_C>0})$,
$\delta a_k + \partial_\mu k^\mu = 0$.
Because of theorem \ref{Loct1},
$\frac{\delta}{\delta C^\alpha}\delta
a_k$ = 0.  Since $\delta$ does not involve the
ghosts, this implies $\delta(\frac{\delta}{\delta
C^\alpha} a_k)=0$.
Theorem \ref{Loct4} then yields 
\beq
\frac{\delta a_k}{\delta C^\alpha}
=\delta b_{\alpha\,k+1}.
\label{63.1}
\eeq
Now, by an argument similar to the one that leads
to the homotopy formula (\ref{A1.1}), one finds
that $a_k$  satisfies 
\beq
a_k(\xi,[C])-a_k(\xi,0) = \partial_\mu k^\mu
+ \int^1_0 \frac{d\lambda}{\lambda}
[C^\alpha \frac{\delta a_k}{\delta C^\alpha}](\xi, 
[\lambda C^\beta])
\label{63.2}
\eeq
where $\xi$ stands for $x$, the antifields, and all the
fields but the ghosts.  The second term in the left-hand 
side of (\ref{63.2}) is zero when $a_k$ belongs to
$\Omega^*_{N_C>0}$.  Using this information
and (\ref{63.1}) in (\ref{63.2}),
together with
$[\delta,C^\alpha]=0$, one finally gets that $a_k=\delta
b_{k+1}+\partial_\mu 
k^\mu$ for some $b_{k+1},k^\mu$.
\qed

\vspace{.2cm}

Using that the isomorphisms of theorem \ref{Loct6}
remain valid when the ghosts are included
in $\Omega^*_{N_C>0}$ (because they are only based on the
algebraic Poincar\'e lemma and on
the vanishing homology of $\delta$ in all positive antifield
numbers), we have
\begin{corollary}\label{Locc3}
The cohomology groups 
$H^p_k(\delta|d,\Omega^*_{N_C>0})$
and $H^{n-k}_0(d|\delta,\Omega^*_{N_C>0})$
vanish for all $k\geq 1$.
\end{corollary}

Note that the constant forms do not appear in 
$H^{n-k}_0(d|\delta,\Omega^*_{N_C>0})$, even if one
considers forms with no explicit $x$-dependence.
This is because the constant forms do
not belong to $\Omega^*_{N_C>0}$.

To summarize, any mod-$d$ $\delta$-closed form can be decomposed as a
sum of terms of definite pure ghost number,
$\omega = \sum_l \omega^l$,  where $\omega^l$ has pure
ghost number $l$.  Each component $\omega^l$ is $\delta$-closed
modulo $d$.  According to the above discussion, it is then necessarily
$\delta$-exact modulo $d$, unless $l=0$.

\subsection{General results on $H(\delta \vert d)$
\label{sdeltamodd}}

\subsubsection{Cauchy order}

In order to get additional vanishing theorems on
$H^n_k(\delta|d)$, we
need more information on the detailed structure of
the theory.

An inspection of the split of the field variables
in Eqs.\ (\ref{LoceracD}) through (\ref{CScoordsplit})
shows that for the Dirac and Klein-Gordon field, 
the set
of independent variables $\{y_A\}$ is
closed under spatial
differentiation:
$\partial_\mu y_A\subset \{y_B\}$ for $\mu=1,\dots n-1$, 
while there are $y_A$ such that $\partial_0
y_A$ involves $z_a$.
For the standard Yang-Mills and Chern-Simons theories, we
have
$\partial_\mu y_A\subset \{y_B\}$ for $\mu=2,\dots n-1$,
while there are $y_A$ such that $\partial_0 y_A$ or 
$\partial_1 y_A$ involves $z_a$.

We define the Cauchy order of a theory to be the
minimum value of
$q$ such that the space of local functions $f(y)$ 
is stable under $\partial_\mu$ for
$\mu=q,q+1,\dots,n-1$ (or, equivalently, $\6_\mu y_A=f_{\mu A}(y)$
for all $A$ and all $\mu= q,\dots,n-1$ where $f_{\mu A}(y)$
are local functions which can be expressed solely in terms of the
$y_A$). The minimum is taken
over all sets of space-time
coordinates and all choices of $\{y_A\}$.
The Dirac and Klein-Gordon theories are of Cauchy
order one, while Chern-Simons and Yang-Mills theories
are of Cauchy order two.  The Lagrangian of the standard
model defines therefore a theory of Cauchy order two.

The usefulness of the concept of Cauchy order lies in the
following theorem.
\begin{theorem}\label{Loct8}
For theories of Cauchy order $q$, the
characteristic
cohomology is trivial for all form-degrees $p=1,\dots,n-q-1$:
\[
H^{p}_0(d|\delta)=\delta^{p}_0\,{\mathbb R}\quad\mbox{for}\quad
p<n-q.
\]
Equivalently, among all
cohomological groups $H^n_k(\delta|d)$
only those with $k\leq q$ may possibly be
nontrivial.
\end{theorem}

The proof of the theorem is given in the appendix 
\ref{Koszul2Section}.B.

In particular, for Klein-Gordon or Dirac theory, only
$H^{n-1}_0(d|\delta)\simeq
H^n_1(\delta|d)$ may be
nonvanishing (standard conserved currents), while for Yang-Mills or
Chern-Simons theory, there can
be in addition a nonvanishing
$H^{n-2}_0(d|\delta)\simeq H^n_2(\delta|d)$.
We shall strengthen this result by showing that this
latter group is in fact zero unless there are free 
abelian factors.

\paragraph{Remark:}
The results on $H^n_k(\delta|d)$ in theorem \ref{Loct8} 
hold in the space of 
forms with or without an explicit coordinate dependence.
By contrast, the results on the characteristic cohomology
hold only in the space of $x$-dependent forms.  If one
restricts the forms to have no explicit $x$-dependence,
there is additional cohomology: the constant forms encountered
above are nontrivial even if one uses the field equations.

\subsubsection{Linearizable theories}
Let $N=N_\phi+N_{\phi^*}+N_{C^*}$, where
$N_{\phi}=\sum_{l\geq 0}
\phi^i_{(\lambda_1\dots\lambda_l)}\frac{\partial}{\partial
\phi^i_{(\lambda_1\dots\lambda_l)}}$ (and
similarly for the other
fields), i.e., $N$ is the counting operator for the
fields, the antifields and their derivatives. 
Decompose the Lagrangian $L$ and the reducibility
functions $R^{+i(\mu_1\dots\mu_l)}_\alpha$ according to
the $N$-degree,
$L=\sum_{n\geq 2}L^{(n)}$,
$R^{+i(\mu_1\dots\mu_l)}_\alpha=\sum_{n\geq 0} 
({R^{+i(\mu_1\dots\mu_l)}_\alpha})^{(n)}$,
$\delta=\sum_{n\geq 0}\delta^{(n)}$.  So,
$L^{(2)}$ is quadratic in the fields
and their derivatives, $L^{(3)}$ is cubic etc, while
$\delta^{(0)}$ preserves the polynomial degree,
$\delta^{(1)}$ increases it by one unit, etc.
We say that a gauge
theory can be linearized 
if the cohomology of $\delta^{(0)}$ (i) is trivial
for all positive antifield numbers and (ii) is in
antifield number $0$ given by the equivalence classes of local
forms modulo forms
vanishing on the surface in the jet space
defined by the linearized equations of motion
$\partial_{\mu_1\dots\mu_k}
\frac{\delta L^{(2)}}{\delta\phi^j}=0$.
This just means  
that the field independent
$({R^{+i(\mu_1\dots\mu_l)}_\alpha})^{(0)}$ provide an
irreducible generating set of
Noether identities for the linearized theory.

The Lagrangian of the standard model is clearly 
linearizable since its quadratic piece is the sum
of the Lagrangians for free Klein-Gordon, Dirac and
$U(1)$ gauge fields.  Pure Chern-Simons theory in three dimensions
is linearizable too, and so are
effective field theories sketched in Section \ref{Hdeltaeff}.

One may view the condition of linearizability as a regularity
condition on the Lagrangian, which is not necessarily
fulfilled by all conceivable Lagrangians of the Yang-Mills type, although
it is fulfilled in the cases met in practice in the usual physical models.
An example of a non-linearizable theory is pure Chern-Simons
theory in $(2k+1)$ dimensions with $k>1$.  The lowest order
piece of the Lagrangian is of order $(k+1)$ and so $L^{(2)}
=0$ when $k>1$.  The zero Lagrangian has a much bigger gauge
symmetry than the Yang-Mills gauge symmetry.  The non-linearizability
of pure Chern-Simons theory in $(2k+1)$ dimensions ($k>1$)
explains some of its pathologies. [By changing the ``background" from
zero to a non-vanishing one, one
may try to improve on this, but the issue will not be addressed here].  

\begin{theorem}\label{Loct9}
For irreducible linear gauge theories, (i)
$H^n_k(\delta|d)=0$
for $k\geq 3$, (ii) if
$N_{C^*}\omega^n_2=0$, then $\delta \omega^n_2
+d\omega^{n-1}_1=0$
implies $\omega^n_2=\delta\eta^n_3+d\eta^{n-1}_2$
and (iii) if
$\omega^n_1 \approx 0$, then
$\delta\omega^n_1+d\omega^{n-1}_0=0$
implies $\omega^n_1=\delta\eta^n_2+d\eta^{n-1}_1$.

For irreducible linearizable gauge theories, the
above results hold
in the space of forms with coefficients that are
formal power series in the
fields, the antifields and their derivatives.
\end{theorem}

The proof is given in the appendix
\ref{Koszul2Section}.B. 

The theorem settles the case of effective field theories
where the natural setting is the space of formal power series.
In order to go beyond this and to make sure
that the power series
stop and are thus in fact local forms in the
case of theories with a local Lagrangian, an additional condition
is needed.

\subsubsection{Control of locality. Normal theories}

For theorem \ref{Loct9} to be valid in the space of local
forms, we need more
information on how the derivatives appear in the Lagrangian.
Let $N_\partial$ be the counting operator of the
derivatives of the
fields and antifields,
$N_\partial=N_{\partial\phi}+N_{\partial\phi^*}
+N_{\partial C^*}$, where
$N_{\partial\phi}=\sum_{k}k\phi^i_{(\lambda_1\dots\lambda_k)}
\frac{\partial}{\partial\phi^i_{(\lambda_1\dots\lambda_k)}}$
and
similarly for the antifields.
The equations of motions $\cL_i=0$ are partial
differential equations
of order $r_i$ and gauge transformations
involve a maximum of $\bar l_\alpha$
derivatives ($\bar l_\alpha = 1$ for theories of the Yang-Mills
type). We define
\[
A=\sum_{k\geq 0}\Big[r_{i}\,\phi^*_{i(\lambda_1\dots\lambda_k)}\ 
\frac{\partial}{\partial\phi^*_{i(\lambda_1\dots\lambda_k)}}+
m_{\alpha}C^*_{\alpha(\lambda_1\dots\lambda_k)}\ 
\frac{\partial}{\partial
C^*_{\alpha(\lambda_1\dots\lambda_k)}}\Big]
\]
where $m_{\alpha}=\bar l_{\alpha}+{\rm
max}_{i,l,(\nu_1\dots\nu_l)}\{
r_i+n_{\partial\phi}(R^{+i(\nu_1\dots\nu_l)}_{\alpha})\}$
with $n_{\partial\phi}(R^{+i(\nu_1\dots\nu_l)}_{\alpha})$
the largest eigenvalue of $N_{\partial\phi}$ contained in
$R^{+i(\nu_1\dots\nu_l)}_{\alpha}$.
In standard, pure Yang-Mills theory, $A$ reads explicitly
\beq
A = \sum_{k\geq 0}\Big[ 2 A^{*\mu}_{I(\lambda_1\dots\lambda_k)}
\frac{\partial}{\partial A^{*\mu}_{I(\lambda_1\dots\lambda_k)}}
+ 3 C^*_{I(\lambda_1\dots\lambda_k)}
\frac{\partial}{\partial C^*_{I(\lambda_1\dots\lambda_k)}}\Big],
\eeq
since the equations of motion are of second order.  For
pure Chern-Simons theory in three dimensions, $A$ is
\beq
A = \sum_{k\geq 0}\Big[  A^{*\mu}_{I(\lambda_1\dots\lambda_k)}
\frac{\partial}{\partial A^{*\mu}_{I(\lambda_1\dots\lambda_k)}}
+ 2 C^*_{I(\lambda_1\dots\lambda_k)}
\frac{\partial}{\partial C^*_{I(\lambda_1\dots\lambda_k)}}\Big] ,
\eeq
since the equations of motion are now of first order.

The degree $K=N_{\partial}+A$ is such that
$[K,\partial_{\mu}]=[N_\partial,\partial_{\mu}]=\partial_{\mu}$
and
\bea
[K,\delta]=\sum_{k\geq 0}\Big[\partial_{(\mu_1\dots\mu_k)}
[(N_{\partial\phi}-r_{i})
\cL_i]\frac{\partial}{\partial\phi^*_{i(\mu_1\dots\mu_k)}}
\nonumber\\+\partial_{(\mu_1\dots\mu_k)}[\sum_{l\geq 0}^{\bar
l_\alpha}(N_{\partial\phi}+r_{i}+l
-m_{\alpha})
R^{+i(\nu_1\dots\nu_l)}_{\alpha}\phi^*_{i(\nu_1\dots\nu_l)}]
\frac{\partial}{\partial
C^*_{\alpha(\mu_1\dots\mu_k)}}\Big].
\eea
It follows that $\delta=\sum_{t}\delta^t$, 
$[K,\delta^t]=t\delta^t$ with
$t\leq 0$. For a linearizable theory, we have now
two degrees:
the degree of homogeneity in the
fields, antifields and their derivatives, for
which $\delta$ has only
nonnegative eigenvalues and the $K$ degree, for which
$\delta$ has only
nonpositive eigenvalues.

A linearizable theory is called a normal theory if
the homology of $\delta^{(0),0}$ is  trivial
in positive antifield number. 
Let us define furthermore
$\delta^{{\rm int},t}:=\sum_{n\geq
1}\delta^{(n),t}$.  Examples of normal theories are (i)
pure Chern-Simons theory in three dimensions, (ii) pure
Yang-Mills theory; (iii) standard model.
For instance, in the first case, $\delta^{(0),0}$ reduces 
to the Koszul-Tate differential of the $U(1)^{\mathit{dim}(G)}$ 
Chern-Simons theory,
while in the second case, it reduces 
to the Koszul-Tate differential for a set of free Maxwell fields.
For these free theories, we have seen that theorem
\ref{Loct4} holds, and thus, indeed, we have ``normality" of
the full theory.

\begin{theorem}\label{Loct10}
For normal theories, the results of theorem
\ref{Loct9} extend to the
space of forms with coefficients that are
polynomials in the differentiated fields, the
antifields and
their derivatives and
power series in the
undifferentiated fields. Furthermore, if
$\delta^{\mathrm{int},0}=0$, they
extend to the space of polynomials in the fields,
the antifields and
their derivatives.
\end{theorem}

The proof is given in the appendix
\ref{Koszul2Section}.B.    

The condition in the last part of this theorem is
fulfilled by the Lagrangian of pure Chern-Simons theory,
pure Yang-Mills theory or the standard model, because
the interaction terms in the Lagrangian of those theories 
contain less derivatives than the quadratic terms.
Thus theorem \ref{Loct10} holds in full in these cases.
The condition would not be fulfilled if the theory contained for instance
the local function $\exp (\partial_{\mu}\partial^\mu\phi/k)$.

We thus see that normal, local theories and effective theories 
have the same properties from the point of view of the cohomology 
groups $H(\delta|d)$. For this reason, the terminology 
``normal theories'' will cover both cases in the sequel. 
\medskip

{\bf Remark.} Part (iii) of theorems \ref{Loct9},
\ref{Loct10} means
that global symmetries with on-shell vanishing
characteristics are
necessarily trivial global symmetries in the sense
of lemma
\ref{Locl3}. In particular, in the absence of non
trivial Noether
identities, weakly vanishing global symmetries are
necessarily related to
antisymmetric combinations of the equations of
motions through
integrations by parts.

\subsubsection{Global reducibility identities and $H^n_2(\delta \vert d)$}

We define a ``global reducibility identity'' by a collection
of local functions $f^\alpha$ such that they give a 
gauge transformation
$\delta_f\phi^i$ as in Eq.\ (\ref{complete}) which
is at the same time an on-shell trivial gauge symmetry
$\delta_M\phi^i$ as in Eqs.\ (\ref{trivial}).
Explicitly a global reducibility identity requires thus
\beq
\sum^{\bar l_\alpha}_{l=0} R^{i(\mu_1\dots\mu_l)}_\alpha
\partial_{(\mu_1\dots\mu_l)} f^\alpha
=
-\sum_{k,m\geq 0}(-)^k\partial_{(\mu_1\dots
\mu_k)}[M^{j(\nu_1\dots\nu_m)i(\mu_1\dots
\mu_k)}
\partial_{(\nu_1\dots\nu_m)}\cL_j]\label{Loce232}
\eeq
for some local functions $M^{j(\nu_1\dots\nu_m)i(\mu_1\dots \mu_k)}$ 
with the antisymmetry
property (\ref{asym}).
Note that this is a stronger condition than just requiring
that the transformations $\delta_f\phi^i$ vanish on-shell.

A global reducibility identity is defined to
be trivial if all $f^\alpha$ vanish on-shell,
$f^\alpha \approx 0$, because $f^\alpha \approx 0$ implies
$\delta_f\phi^i=-\delta_M\phi^i$ for some $\delta_M\phi^i$.
This is seen as follows: $f^\alpha \approx 0$ means $f^\alpha=\delta
g^\alpha_1$ for some $g^\alpha_1$ and implies 
$\phi^*_i\delta_f\phi^i =\delta(C^*_\alpha\delta
g^\alpha_1)+\partial_\mu(\ )^\mu=
-\delta[(\delta C^*_\alpha)
g^\alpha_1]+\partial_\mu(\ )^\mu$; 
taking now the Euler-Lagrange derivative with respect to $\phi^*_i$
yields indeed
$\delta_f\phi^i=-\delta_M\phi^i$ because
$(\delta C^*_\alpha) g^\alpha_1$ is quadratic
in the $\phi^*_{i(\mu_1\dots\mu_m)}$.

The
space of nontrivial global reducibility
identities is defined to be
the space of equivalence classes of global
reducibility identities
modulo trivial ones.

\begin{theorem}\label{Loct11}
In normal theories, $H_2^n(\delta|d)$ is
isomorphic to the space of
non trivial global reducibility identities.
\end{theorem}

\paragraph{Proof: }
Every
cycle of $H_2^n(\delta|d)$ 
can be assumed to be of the form $d^nx\, a_2$ with
$a_2=C^*_\alpha f^\alpha+M_2+\partial_{\mu}(\
)^\mu$, where $M_2=
\frac{1}{2}\sum_{n,m\geq 0}
\phi^*_{j(\nu_1\dots\nu_n)}\phi^*_{i(\mu_1\dots\mu_m)}
M^{j(\nu_1\dots\nu_n)i(\mu_1\dots\mu_m)}$ such
that (\ref{asym}) holds
(indeed, all derivatives can be removed from $C^*_\alpha$
by subtracting a total derivative from $a_2$; the antisymmetry
of the $M$'s follows from the odd grading of the $\phi^*_i$).
Taking the
Euler-Lagrange derivative of the cycle condition
with respect to
$\phi^*_i$ gives (\ref{Loce232}). Conversely,
multiplying
(\ref{Loce232}) by $\phi^*_i$ and integrating
by parts  an appropriate
number of times yield 
$\delta a_2
+\partial_{\mu}(\ )^\mu=0$. Hence, cycles of $H_2^n(\delta|d)$
correspond to global reducibility
identities and vice versa.

We still have to show that an element of $H_2^n(\delta|d)$
is trivial iff the corresponding global reducibility
identity is trivial.
The term $b_3$ in the coboundary condition
$a_2=\delta
b_3+\partial_{\mu}(\ )^\mu$ contains terms
with one $C^*$ and one $\phi^*$
and terms trilinear in $\phi^*$'s.
Taking the Euler-Lagrange derivative
with respect to $C^*_{\alpha}$ of the coboundary
condition implies
$f^{\alpha}\approx 0$, or $f^{\alpha}=\delta
g^\alpha_1$. Conversely, $f^{\alpha}=\delta g^\alpha_1$ implies that
$a_2-\delta(C^*_{\alpha} g^\alpha_1)$ is a
$\delta$-cycle modulo a total derivative in
antifield number $2$,
which does not depend on $C^*_{\alpha}$. Part (ii)
of theorems
\ref{Loct9} or  \ref{Loct10} then implies that 
$a_2-\delta(C^*_{\alpha} g^\alpha_1)$ is
a $\delta$-boundary modulo a total derivative, and thus
that $d^nx\, a_2$ is trivial in $H^n_2(\delta|d)$. \qed

The same result applies to effective field theories
since one can then use theorem \ref{Loct9}.

\subsubsection{Results for Yang-Mills gauge models}

For irreducible normal gauge theories, we have  
entirely reduced the
computation of the higher order characteristic
cohomology groups
to properties of the 
gauge transformations.

We now perform explicitly the calculation of
the global reducibility identities 
in the case of gauge theories of the Yang-Mills
type, which are irreducible.
We start with free
electromagnetism, which has a non vanishing
$H^n_2(\delta|d)$.
The Koszul-Tate differential is defined on the
generators by $\delta A_\mu=0$,
$\delta A^{*\mu}=\partial_{\nu}F^{\nu\mu}$,
$\delta
C^*=-\partial_{\mu}A^{*\mu}$.
\begin{theorem}\label{Loct12}
For a free abelian gauge field with Lagrangian
$L=-\frac 14 F_{\mu\nu} F^{\mu\nu}$ in  dimensions $n>
2$,
$H_2^n(\delta|d)$ is represented by $d^nx\, C^*$. A
corresponding
representative of the characteristic cohomology 
$H^{n-2}_\mathrm{char}$ is
$\star F=\frac{1}{(n-2)!2}
dx^{\mu_1}\dots dx^{\mu_{n-2}}\epsilon_{\mu_1\dots\mu_n}
F^{\mu_{n-1}\mu_n}$.
\end{theorem}

\paragraph{Proof: }

According to theorem \ref{Loct11}, $H_2^n(\delta|d)$ is determined
by the nontrivial global reducibility conditions.
A necessary condition for the existence
of a global reducibility condition is that a gauge transformation
$\delta_f A_\mu=\partial_{\mu}f$ vanishes weakly, i.e.
$df\approx 0$. Hence, $f$ is a cocycle of $H^0_0(d|\delta)$.
By the isomorphisms (\ref{Loct6e3}) we have $H^0_0(d|\delta)\simeq
H^n_n(\delta|d)+\mathbb{R}$. $H^n_n(\delta|d)$
vanishes for $n>2$ according to part (i) of theorem \ref{Loct9}.
Hence, if $n>2$, we conclude $f\approx \mathit{constant}$, i.e.
the nontrivial global reducibility conditions are exhausted
by constant $f$. The nontrivial representatives of
$H_2^n(\delta|d)$ can thus be taken proportional to $d^nx\, C^*$
if $n>2$ which
proves the first part of the theorem.
The second part of the theorem follows from
the chain of equations
$\delta C^*+\partial_\mu A^{*\mu}=0$,
$\delta A^{*\mu}-\partial_\nu F^{\nu\mu}=0$ by
the isomorphisms
(\ref{Loct6e2}) (see also proof of these isomorphisms).
\qed 

The reason that there is a non-trivial group $H^n_2(\delta
\vert d)$ for free electromagnetism is that there is in that case a
global reducibility identity associated with gauge transformations
with constant gauge parameter. As the proof of the theorem shows,
this property remains true if one
includes gauge-invariant self-couplings of the Born-Infeld or
Euler-Heisenberg type. 
The corresponding representatives of 
$H^{n-2}_\mathrm{char}$ are obtained through the descent equations.
Furthermore the result extends straightforwardly to models
with a set of abelian gauge fields $A_\mu^I$, $I=1,2,\dots$: 
then $H^n_2(\delta \vert d)$ is represented by $d^nx\, C^*_I$, 
$I=1,2,\dots$

However, if one turns on self-couplings of the
Yang-Mills type, which are not invariant under the abelian
gauge symmetries, or if one includes minimal couplings to charged
matter fields, the situation changes: there is no
non-trivial reducibility identity any more.   Indeed, gauge
transformations leaving the Yang-Mills field $A^I_\mu$ and the
matter fields $\psi^i$ invariant on-shell fulfill
\beq
D_\mu f^I \approx 0, \; \; f^I T^i_{Ij} \psi^j \approx 0
\eeq
whose only solution $f^I([A],[\psi])$ is $f^I \approx 0$.  
By theorem \ref{Loct11}, $H^n_2(
\delta \vert d)$ vanishes in those cases. To summarize, we get
the following result.

\begin{corollary}
For normal theories of the Yang-Mills type in dimensions $n>2$, 
the cohomology
groups $H^n_k(\delta \vert d)$ vanish for $k>2$.  The group
$H^n_2(\delta \vert d)$ also vanishes, unless there are abelian
gauge symmetries under which all matter fields are uncharged, in which
case $H^n_2(\delta \vert d)$ is represented by those $d^nx\, C^*_I$
which correspond to these abelian
gauge symmetries.
\end{corollary} 
 
This theorem covers pure 
Chern-Simons theory in three dimensions,
pure Yang-Mills theory, the standard model as well as
effective theories of the Yang-Mills type (this
list is not exhaustive).

Finally,
the group $H^n_1(\delta \vert d)$ is related to the 
standard conserved currents through theorem \ref{Loct5}.  Its
dimension depends on the specific form of the Lagrangian, which
may or may not have non trivial global symmetries.  The 
complete calculation
of $H^n_1(\delta \vert d)$ is a  question that must be
investigated on a case by case basis. 
For free theories, there is an infinite number
of conserved currents. At the other extreme,
for effective theories, which include all possible
terms compatible with gauge symmetry and a definite set of global
symmetries (such as Lorentz invariance),   the only global
symmetries and conservation laws should be the prescribed ones.

\subsection{Comments}

The characteristic cohomology associated with a system of 
partial differential equations has been investigated in
the mathematical literature for some time 
\cite{vinogradov2,tsujishita,vinogradov3,Bryant}.
The connection with the Koszul-Tate differential is more
recent \cite{Barnich:1995db}.  This new point of view has even enabled one
to strengthen and generalize some results on the characteristic cohomology,
such as the result on $H^{n-2}_0 (d \vert \delta)$
(isomorphic to $H^n_2(\delta \vert d)$). The connection
with the reducibility properties of the gauge transformations
was also worked out in this more recent work and turns out
to be quite important for $p$-form gauge theories,
where higher order homology groups $H^n_k(\delta \vert d)$ are
non-zero \cite{Henneaux:1997ws,Verbovetsky:1997fi}.

The relation to the characteristic cohomology provides a
physical interpretation of the nontrivial homology 
groups $H(\delta \vert d)$
in terms of conservation laws. In particular it establishes
a useful cohomological formulation of Noether's first
theorem and a direct interpretation of the (nontrivial) homology 
groups $H^n_1(\delta \vert d)$ in terms of the (nontrivial) 
global symmetries. Technically,
the use of the antifields allows one, among other things, 
to deal with trivial symmetries 
in a very efficient way. For instance the rather cumbersome
antisymmetry property (\ref{asym}) of
on-shell trivial symmetries is automatically reproduced
through the coboundary condition in $H^n_1(\delta \vert d)$ thanks
to the odd Grassmann parity of the antifields.

\subsection{Appendix \ref{Koszul2Section}.A: Noether's second theorem}

We discuss in this appendix the general relationship between Noether
identities, gauge symmetries and ``dependent" field equations.

In order to do so, it is convenient to extend the jet-spaces by 
introducing a new field $\epsilon$. A gauge symmetry on the enlarged 
jet-space is defined
to be an infinitesimal field transformation
$\vec Q(\epsilon)$ leaving the Lagrangian invariant up to a 
total derivative,
\beq
\vec Q(\epsilon)L_0+ \partial_\mu
j^{\prime\mu}(\epsilon)=0.\label{Loceg}
\eeq
The 
characteristic $Q^i(\epsilon)=\sum_{l\geq 0}
Q^{i(\mu_1\dots\mu_l)}\epsilon_{(\mu_1\dots\mu_l)}$
depends linearly and homogeneously on $\epsilon$ and 
its derivatives $\epsilon_{(\mu_1\dots\mu_l)}$ up to some 
finite order.

A ``Noether operator" is a differential operator 
$N^{i(\mu_1\dots\mu_l)}
\partial_{(\mu_1\dots\mu_l)}$ that yields an
identity between the equations of motion,
\beq
\sum_{l\geq 0}
N^{i(\mu_1\dots\mu_l)}
\partial_{(\mu_1\dots\mu_l)} \cL_i = 0.
\eeq

We consider theories described by a Lagrangian 
that fulfills regularity conditions as described in
section \ref{regcond} (``irreducible gauge theories''). Namely, 
the original equations of motion are equivalent to a set of  
independent equations 
$\{L_a\}$ (which can be taken as coordinates in a new coordinate system on 
the jet-space) and to a set of dependent 
equations $\{L_\Delta\}$ (which hold as a consequence of the independent
ones). Explicitly, $\partial_{(\mu_1\dots\mu_l)} \cL_i
=L_\Delta \cN^\Delta_{(\mu_1\dots\mu_l)i}+L_a \cN^a_{(\mu_1\dots\mu_l)i}$, 
where the matrix $\cN^M_{(\mu_1\dots\mu_l)i}$, with $M=\{a,\Delta\}$ is
invertible. 

Furthermore, we assume that the dependent equations are
generated by a finite set $\{L_\alpha\}$ of equations
(living on finite dimensional jet-spaces) through repeated 
differentiations, $\{L_\Delta\}\equiv
\{L_\alpha,\partial_{\rho}L_\alpha,
\partial_{(\rho_1\rho_2)}L_\alpha,\dots\}$, and that    
these successive derivatives are independent among themselves.
For instance,
the split $\{L_\Delta,L_a\}$ made in section \ref{regcond} 
in the pure Yang-Mills case corresponds to 
$\{L_\alpha\}\equiv\{\6_0 L^0_I\}$.

\begin{lemma}\label{Locl2}
Associated to the dependent equations of motions,
there exists a set of Noether operators
$\{\sum_{l\geq 0}^{\bar
l_\alpha}R^{+i(\mu_1\dots\mu_l)}_\alpha
\partial_{(\mu_1\dots\mu_l)}\}$,
which are non trivial, in the sense that they
do not vanish weakly, and which are
irreducible, in the sense that if
$\sum_{m\geq 0}
Z^{+\alpha(\nu_1\dots\nu_m)}
\partial_{(\nu_1\dots\nu_m)}\circ[\sum_{l\geq 0}
^{\bar l_\alpha}
R^{+i(\mu_1\dots\mu_l)}_\alpha\partial_{(\mu_1\dots\mu_l)}]\approx
0$ (as an operator identity) then
$\sum_{m\geq 0}Z^{+\alpha(\nu_1\dots\nu_m)}\partial_{(\nu_1\dots\nu_m)}
\approx 0$ (i.e.\ all $Z$'s vanish weakly).
\end{lemma}

\paragraph{Proof:}                                   

Applying the equivalent of lemma \ref{Locl1} to
the equations $L_\alpha$,
we get $L_\alpha= L_{\bar a}k^{\bar a}_\alpha$ 
where $\{L_{\bar a}\}$ is a finite subset of 
$\{L_a\}$. 
These are Noether
identities 
whose left hand sides can be written 
in terms of the original equations of motion,
\bea
\sum_{l\geq 0}^{\bar l_\alpha}
R^{+i(\mu_1\dots\mu_l)}_\alpha\partial_{(\mu_1\dots\mu_l)}\cL_i=
L_\alpha -L_{\bar a}k^{\bar a}_\alpha,\label{schun}
\eea
for some $R^{+i(\mu_1\dots\mu_l)}_\alpha$.
Note that the expression on the right hand side takes the form
\bea
L_\alpha -L_{\bar a}k^{\bar a}_\alpha=
R^{+ \beta}_\alpha L_\beta + R^{+\bar a}_\alpha L_{\bar
 a},
\eea
where $R^{+ \beta}_\alpha=\delta^{\beta}_{\alpha}$ and $
R^{+\bar a}_\alpha=-k^{\bar a}_\alpha$, i.e., 
$R^{+i(\mu_1\dots\mu_l)}_\alpha\equiv(R^{+ \beta}_\alpha,
R^{+\bar a}_\alpha)$.
The presence of $\delta^{\beta}_{\alpha}$ then implies the first
part of the lemma. 

Taking derivatives 
$\partial_{(\rho_1\dots\rho_m)}$, $m=0,1,\dots$, of (\ref{schun})
we get
the identities 
\beq
\partial_{(\rho_1\dots\rho_m)}
[R^{+i(\mu_1\dots\mu_l)}_\alpha\partial_{(\mu_1\dots\mu_l)}\cL_i]
=[L_\Delta- L_ak^{a}_\Delta]\cM^\Delta_{(\rho_1\dots\rho_m)\alpha},
\label{schlud}
\eeq
for some functions
$k^{a}_\Delta$ and for an invertible matrix 
$\cM^\Delta_{(\rho_1\dots\rho_m)\alpha}$
analogous to the one in (\ref{equivNoetherbis}). 
The Noether identities $L_\Delta- L_ak^{a}_\Delta=0$
are equivalent to $R^{+ \Gamma}_\Delta L_\Gamma + R^{+ a}_\Delta
L_{a}=0$, where $R^{+ \Gamma}_\Delta=\delta^{\Gamma}_{\Delta}$ and $
R^{+ a}_\Delta=-k^{ a}_\Delta$.

Thus, because of $\delta^\Gamma_\Delta $, if
$Z^{+\Delta} (R^{+\Gamma}_\Delta,R^{+a}_\Delta) \approx 0$ then
$Z^{+\Delta}\approx
0$. The lemma follows from the (\ref{schlud}), the fact that 
$Z^{+\Delta}$ is related to $Z^{+\alpha(\rho_1\dots\rho_m)}$,
$m=0,1,\dots$ through the invertible matrix
$\cM^\Delta_{(\rho_1\dots\rho_m)\alpha}$ and the fact that the 
$(L_\Delta,L_a)$
are related to $\partial_{(\mu_1\dots\mu_l)}\cL_i$ through the
invertible matrix $\cN^M_{(\mu_1\dots\mu_l)i}$. \qed

In terms of the equations $L_\Delta=0$ and $L_a=0$, the acyclicity of
the Koszul-Tate operator $\delta C^*_\Delta=\phi^*_\Delta-\phi^*_a
k^a_\Delta$, $\delta \phi^*_\Delta=L_\Delta$, $\delta \phi^*_a= L_a$
follows directly by introducing new generators $\tilde
\phi^*_\Delta=\phi^*_\Delta-\phi^*_a k^a_\Delta$. 
Using both matrices $\cN^M_{(\mu_1\dots\mu_l)i}$ and 
$\cM^\Delta_{(\rho_1\dots\rho_m)\alpha}$, one can verify that the
Koszul-Tate operator is given in terms of the original equations
by (\ref{Def}).

\begin{lemma}\label{Locc1}
The irreducible set of Noether operators
associated to the dependent
equations of motion is a
generating set of
non trivial Noether identities in the following sense:
every Noether operator
$\sum_{m\geq 0}N^{i(\mu_1\dots\mu_m)}\partial_{(\mu_1\dots\mu_m)}$
can be
decomposed into the direct sum of
\beq
\sum_{n\geq 0}Z^{+\alpha(\rho_1\dots\rho_n)}\partial_{(\rho_1\dots\rho_n)}
\circ [\sum_{l\geq 0}^{\bar
l}R^{+i(\lambda_1\dots\lambda_l)}_\alpha
\partial_{(\lambda_1\dots\lambda_l)}]
\label{Loce17}
\eeq
with $Z^{+\alpha(\rho_1\dots\rho_n)}\not\approx
0$, $n=0,1,\dots$
and the weakly vanishing piece
\beq
\sum_{m,n\geq 0}M^{j(\nu_1\dots\nu_n)i(\mu_1\dots\mu_m)}
[\partial_{(\nu_1\dots\nu_n)}
\cL_j]\partial_{(\mu_1\dots\mu_m)},
\label{Loce18}
\eeq
where
$M^{j(\nu_1\dots\nu_n)i(\mu_1\dots\mu_m)}$
is
antisymmetric in the exchange of
$j(\nu_1\dots\nu_n)$ and
$i(\mu_1\dots\mu_m)$.
\end{lemma}

\paragraph{Proof:}

Every Noether identity can
be written as a $\delta$ cycle in antifield number
$1$,
$\delta(\sum_{m}N^{i(\mu_1\dots\mu_m)}
\phi^*_{i|(\mu_1\dots\mu_m)})=0$, which implies
because of theorem \ref{Loct4}
that $\sum_{m}N^{i(\mu_1\dots\mu_m)}
\phi^*_{i|(\mu_1\dots\mu_m)}=\delta b_2$ with
\beq
b_2=\sum_{n\geq 0} 
{C^*_{\alpha|(\rho_1\dots\rho_n)}}
Z^{+\alpha(\rho_1\dots\rho_n)}+\frac{1}{2}\sum_{n,m\geq 0}
\phi^*_{j(\nu_1\dots\nu_n)}\phi^*_{i(\mu_1\dots\mu_m)}
M^{j(\nu_1\dots\nu_n)i(\mu_1\dots\mu_m)},\label{A9bis}
\eeq
or explicitly
\bea
\sum_{m\geq 0}N^{i(\mu_1\dots\mu_m)}
\phi^*_{i|(\mu_1\dots\mu_m)}
=\sum_{n\geq 0}Z^{+\alpha(\rho_1\dots\rho_n)}\partial_{(\rho_1\dots\rho_n)}
[\sum_{l\geq 0}^{\bar
l_\alpha}R^{+i(\lambda_1\dots\lambda_l)}_\alpha
\phi^*_{i(\lambda_1\dots\lambda_l)}]\nonumber
\\+
\sum_{m,n\geq 0}M^{j(\nu_1\dots\nu_n)i(\mu_1\dots\mu_m)}
[\partial_{(\nu_1\dots\nu_n)}
\cL_j]\phi^*_{i(\mu_1\dots\mu_m)}\label{A9}.
\eea

Identification of
the coefficients of the independent
$\phi^*_{i|(\mu_1\dots\mu_m)}$
gives the result that every Noether operator can
be written as the sum
of (\ref{Loce17}) and (\ref{Loce18}). In order to
prove that the
decomposition is direct for non weakly vanishing
$Z^{+\alpha(\rho_1\dots\rho_n)}$, we have to show
that
every weakly vanishing Noether identity can be
written as in (\ref{Loce18})
(and in particular we have to show this
for a Noether identity of the form (\ref{Loce17}),
without sum over $n$ and
$Z^{+\alpha(\rho_1\dots\rho_n)}\approx 0$ for this
$n$).
Using the set of
indices $(a,\Delta)$, a weakly vanishing Noether
identity is defined by
$N^{a}L_a+N^{\Delta}L_\Delta=0$ with $N^{a}\approx
0$ and
$N^{\Delta}\approx 0$. The last equation implies
as in the proof
of lemma \ref{Locl1}, that $N^{\Delta}=l^{\Delta a}L_a$ so
that the Noether
identity becomes $(N^{a}+L_\Delta l^{\Delta
a})L_a=0$.
In terms of the new generators
$\tilde
\phi^*_\Delta=\phi^*_\Delta-k^{a}_\Delta\phi^*_a$, 
$\tilde\phi^*_a=\phi^*_a$  
the Koszul-Tate differential
$\delta=L_a\frac{\partial}{\partial\tilde\phi^*_a}+\tilde\phi^*_\Delta
\frac{\partial}{\partial C^*_\Delta}$ involves only
the contractible pairs. The Noether identity
$\delta[(N^{a}+L_\Delta l^{\Delta a})\tilde\phi^*_a]=0$ then implies
$(N^{a}+L_\Delta l^{\Delta a})\tilde\phi^*_a
=\delta[\frac{1}{2}\tilde\phi^*_b\tilde\phi^*_a
\mu^{[ba]}]$. This
proves the corollary, because we get
$N^{a}=\mu^{[ba]}L_b-L_\Delta
l^{\Delta a}$ and $N^{\Delta}=l^{\Delta a}L_a$.
\qed

\begin{theorem}\label{Loct3} (Noether's second theorem)
To every Noether operator $\sum_{l\geq 0}
N^{i(\mu_1\dots\mu_l)}\partial_{(\mu_1\dots
\mu_l)}$
there corresponds a gauge symmetry
$\vec Q(\epsilon)$ given by
$Q^i(\epsilon)=\sum_{l\geq 0}
(-)^l\partial_{(\mu_1\dots\mu_l)}[N^{i(\mu_1\dots\mu_l)}\epsilon]$
and, vice versa, to every gauge symmetry
$Q^i(\epsilon)$, there
corresponds the Noether operator defined by
$\sum_{l\geq 0}Q^{+i(\nu_1\dots\nu_l)}
\partial_{(\nu_1\dots\nu_l)}\equiv
\sum_{l\geq 0}
(-)^l\partial_{(\mu_1\dots\mu_l)}
[Q^{i(\mu_1\dots\mu_l)}\ \cdot\ ]$.
\end{theorem}

\paragraph{\bf Proof:}

The first part follows by multiplying the Noether
identity
\beq
\sum_{l\geq 0}
N^{i(\mu_1\dots\mu_l)}\partial_{(\mu_1\dots
\mu_l)} \cL_i
\eeq
by $\epsilon$ and then removing the
derivatives from the
equations of motion by integrations by parts to
get
$\sum_{l\geq 0}(-)^l\partial_{(\mu_1\dots
\mu_l)}[\epsilon
N^{i(\mu_1\dots\mu_l)}]\cL_i+\partial_\mu
j^\mu=0$, which
can be transformed to (\ref{Loceg}).

The second part
follows by starting from (\ref{Loceg}) and doing
the reverse integrations
by parts  to get
$\epsilon(\sum_{l\geq 0}(-)^l\partial_{(\mu_1\dots
\mu_l)}[Q^{i(\mu_1\dots\mu_l)}\cL_i])+\partial_\mu
j^{\prime\prime\mu}=0$. Taking
the Euler-Lagrange derivatives with respect to
$\epsilon$, which
annihilates the total derivative according to
theorem \ref{Loct1},
proves the theorem. \qed 

The gauge transformations associated with a generating
(or ``complete") set of Noether identities are said to
form a generating (or complete) set of gauge symmetries.

Trivial gauge symmetries are defined as those that
correspond to weakly vanishing Noether operators:
\bea
Q_T^i(\epsilon)=\sum_{m,k\geq 0}(-)^k
\partial_{(\mu_1\dots\mu_k)}[\partial_{(\nu_1\dots\nu_m)}\cL_j
M^{j(\nu_1\dots\nu_m)i(\mu_1\dots\mu_k)}\epsilon]\nonumber\\
=\sum_{m,n\geq 0}
M^{+j(\nu_1\dots\nu_m)i(\lambda_1\dots\lambda_n)}
\partial_{(\nu_1\dots\nu_m)}\cL_j\epsilon_{(\lambda_1\dots\lambda_n)},
\eea
where the last equation serves as the definition
of the functions
$M^{+j(\nu_1\dots\nu_m)i(\lambda_1\dots\lambda_n)}$.

Note that trivial gauge symmetries do not only
vanish weakly, they are
moreover related to antisymmetric combinations of
equations of
motions through integrations by parts.

Non trivial gauge transformations are defined as
gauge transformations corresponding to non weakly
vanishing
Noether identities. In particular, the gauge
transformations corresponding to the generating
set constructed above
are given by
\beq
R^i_\alpha(\epsilon)=\sum^{\bar l_\alpha}
_{l\geq 0}(-)^l
\partial_{(\mu_1\dots\mu_l)}[R^{+i(\mu_1\dots\mu_l)}_\alpha\epsilon]
=
\sum^{\bar l_\alpha}_{l\geq 0}
R^{i(\mu_1\dots\mu_l)}_\alpha
\partial_{(\mu_1\dots\mu_l)}\epsilon\ .
\label{RDEF}
\eeq

The operator
$Z^{+\alpha}\equiv\sum_{m\geq 0}Z^{+\alpha(\rho_1\dots\rho_m)}
\partial_{(\rho_1\dots\rho_m)}\approx 0$ iff
the operator $Z^\alpha\equiv\sum_{m\geq 0}(-)^m
\partial_{(\rho_1\dots\rho_m)}\circ Z^{+\alpha(\rho_1\dots\rho_m)}
\approx 0$. A direct consequence of
theorem \ref{Loct3} and lemma \ref{Locc1} is
then
\begin{corollary}\label{Locc2}
Every gauge symmetry $Q^i(\epsilon)$ can be
decomposed into the direct
sum
$Q^i(\epsilon)=R^i_{\alpha}(Z^{\alpha}(\epsilon))+Q_T^i(\epsilon)$,
where the operator $Z^{\alpha}$ is not weakly
vanishing, while
$Q_T^i(\epsilon)$ is weakly vanishing.
Furthermore, $Q_T^i(\epsilon)$
is related to an antisymmetric combination of
equations of motion
through integrations by parts.
\end{corollary}

It is in that sense that a complete set of gauge transformations
generates all gauge symmetries.

\subsection{Appendix \ref{Koszul2Section}.B: Proofs of theorems
\ref{Loct8}, \ref{Loct9} and \ref{Loct10}}

\paragraph{Proof of theorem \ref{Loct8}:}
We decompose the spacetime indices into two subsets,
$\{\mu\}=\{a,\ell\}$ where $a=0,\dots,q-1$ and $\ell=q,\dots,n-1$.
The cocycle condition $da+\delta b=0$ decomposes into
\beq
d^1a^M+\delta b^{M+1}=0,\quad
d^0a^M+d^1a^{M-1}+\delta b^M=0,\quad\dots
\label{cau1}
\eeq
where the superscript is the degree in the $dx^\ell$ and $M$ is
the highest degree in the decomposition of $a$,
\[
a=\sum_{m\leq M}a^m,\quad
d^1=dx^\ell\6_\ell\ ,\quad d^0=dx^a\6_a\ .
\]
Note that $M$
cannot exceed $n-q$ because the $dx^\ell$
anticommute.
Without loss of generality we can assume that $a$ depends
only on the $y_A$, $dx^\mu$, $x^\mu$ because this
can be always achieved by adding a $\delta$-exact 
piece to $a$ if necessary. 
In particular, $a^M$ can thus be assumed to be of the form
\[
a^M=dx^{\ell_1}\dots dx^{\ell_M}
f_{\ell_1\dots \ell_M}(dx^a,x^\mu,y_A).
\]
Since we assume that the theory has Cauchy order $q$,
$d^1a^M$ depends also
only on the $y_A$, $dx^\mu$, $x^\mu$ and therefore vanishes 
on-shell only if it vanishes even off-shell.
The first equation in (\ref{cau1}) implies thus
\beq
d^1 a^M=0.
\label{cau3}
\eeq
To exploit this equation, we need
the cohomology of $d^1$. It is given by a variant of the algebraic 
Poincar\'e lemma in section \ref{APL} and can be
derived by adapting the derivation of that lemma 
as follows. Since $d^1$ contains only
the subset $\{\6_\ell\}$ of $\{\6_\mu\}$, the 
jet coordinates $\6_{(a_1\dots a_k)}\phi^i$ play now the
same r\^ole as the $\phi^i$ in the derivation of the
algebraic Poincar\'e lemma (it does not matter that the
set of all $\6_{(a_1\dots a_k)}\phi^i$ is infinite because
a local form contains only finitely many
jet coordinates). The $dx^a$ and $x^a$ are
inert to $d^1$ and play the r\^ole of constants.
Forms of degree $n-q$ in the $dx^\ell$ 
take the r\^ole of the volume forms.
One concludes that the $d^1$-cohomology is trivial
in all $dx^\ell$-degrees $1,\dots,n-q-1$ and
in degree 0 represented by functions $f(dx^a,x^a)$.
Eq.\ (\ref{cau3}) implies thus:
\bea
0<M<n-q:&& a^M=d^1\eta^{M-1}\ ;
\nonumber
\\
M=0:&& a^0=f(dx^a,x^a).
\label{cau4}
\eea
In the case $0<M<n-q$ we introduce
$a':=a-d\eta^{M-1}$ which is equivalent to $a$. Since $a'$ contains
only pieces with $dx^\ell$-degrees strictly smaller than $M$, one
can repeat the arguments until the $dx^\ell$-degree 
drops to zero. 
In the case $M=0$, one has $a=a^0=f(dx^a,x^a)$ and the cocycle
condition imposes
$d^0f(dx^a,x^a)\approx 0$. This requires $d^0f(dx^a,x^a)=0$ 
which implies $f(dx^a,x^a)=\mathit{constant}+d g(dx^a,x^a)$
by the ordinary Poincar\'e lemma in $\mathbb{R}^q$.
Hence, up to a constant, $a$ is trivial in
$H(d|\delta)$ whenever $M<n-q$.
This proves the theorem because one has $M<n-q$ whenever the
form-degree of $a$ is smaller than $n-q$
since $M$ cannot exceed the form-degree. \qed

\paragraph{Proof of theorem \ref{Loct9}:}

Let us first establish two additional properties
satisfied by
Euler-Lagrange derivatives.  These are
\beq
\sum_{k\geq 0}(-)^k\partial_{(\lambda_1\dots\lambda_k)}
\left[\frac{\partial(\partial_\mu
f)}{\partial\phi_{(\lambda_1\dots\lambda_k)}}\ g\right]=
-\sum_{k\geq 0}(-)^k\partial_{(\lambda_1\dots\lambda_k)}
\left[\frac{\partial
f}{\partial\phi_{(\lambda_1\dots\lambda_k})}\ \partial_\mu
g\right],\label{A1e18}
\eeq
for any local functions $f,g$ and 
\bea
\vec Q\,\frac{\delta f}{\delta\phi^j}
=(-)^{\epsilon_j\epsilon_Q}\frac{\delta}{\delta\phi^j}
\left[Q^i\frac{\delta f}{\delta\phi^i}\right]-
(-)^{\epsilon_j\epsilon_Q}
\sum_{k\geq 0}(-)^k\partial_{(\lambda_1\dots\lambda_k)}
\left[\frac{\partial
Q^i}{\partial\phi^j_{(\lambda_1\dots\lambda_k)}}\, 
\frac{\delta f}{\delta\phi^i}\right]\label{A1e19}
\eea
for local functions $f$ and $Q^i$ ($\epsilon_i$ and $\epsilon_Q$
denote the Grassmann
parities of $\phi^i$ and $\vec Q$ respectively).     
Indeed, using (\ref{prooe1}), the left hand side
of (\ref{A1e18}) is
\[
\sum_{k\geq 0}(-)^k\partial_{(\lambda_1\dots\lambda_k)}
\left[\partial_\mu(\frac{\partial
f}{\partial\phi_{(\lambda_1\dots\lambda_k)}})\ g\right]+
\sum_{k\geq 0}(-)^k\partial_{(\nu_1\dots\nu_{k-1}\mu)}\left[\frac{\partial
f}{\partial\phi_{(\nu_1\dots\nu_{k-1})}}\ g\right].
\]
Integrating by parts
the $\partial_\mu$ in the first term and using the
same cancellation
as before in (\ref{prooe2}) gives (\ref{A1e18}).
Similarly, because of (\ref{Loctbd}), the left hand side of (\ref{A1e19})
is
$\sum_{k\geq 0}(-)^k\partial_{(\lambda_1\dots\lambda_k)}[\vec
Q
(\frac{\partial
f}{\partial\phi^j_{(\lambda_1\dots\lambda_k)}})]$.
Commuting $\vec Q$
with $\frac{\partial
}{\partial\phi^j_{(\lambda_1\dots\lambda_k)}}$ gives
(\ref{A1e19}): the terms with $\vec Q$ and 
$\frac{\partial
}{\partial\phi^j_{(\lambda_1\dots\lambda_k)}}$
in reverse order
reproduce the first term on the right hand side of 
(\ref{A1e19}) due to theorem \ref{Loct1},
while the commutator terms yield the second term
upon repreated use of
Eq.\ (\ref{A1e18}).

Let us now turn to the proof of the theorem.
Using $\omega_k=d^nx\,a_k$,
the cocycle condition reads $\delta a_k+\partial_\mu
j^\mu=0$. Using theorem \ref{Loct1}, the Euler-Lagrange derivatives
of this condition 
with respect to a field and or antifield $Z$ gives
\beq
\frac{\delta}{\delta Z}\sum_{\Phi^*=C^*_\alpha,\phi^*_i}
(\delta \Phi^*)\, 
\frac{\delta a_k}{\delta \Phi^*}=0.
\label{forreader}
\eeq
Using Eq.\ (\ref{A1e19}) (for $\vec Q\equiv \delta$),
the previous formula is now exploited for $Z\equiv C^*_\alpha$,
$Z\equiv\phi^*_i$ and $Z\equiv \phi^i$.
For $Z\equiv C^*_\alpha$ it gives
\[
\delta\ \frac{\delta a_k}{\delta C^*_\alpha}=0.
\]
When $k\geq 3$, $\delta a_k/\delta C^*_\alpha$
has positive antifield number. Due to
the acyclicity of
$\delta$ in positive antifield number, the previous equation gives
\beq
\frac{\delta a_k}{\delta
C^*_\alpha}=\delta\sigma^\alpha_{k-1}.\label{A1e20}
\eeq
For the proof of part (ii) of the theorem, we note
that this
relation holds trivially with
$\sigma^\alpha_{k-1}=0$.
For $Z\equiv \phi^*_i$, Eq.\ (\ref{forreader}) gives
\[
\delta\ \frac{\delta a_k}{\delta\phi^*_i}
=
R^i_\alpha(\frac{\delta a_k}{\delta
C^*_\alpha})
\]
where we used the same notation as in Eq.\ (\ref{RDEF}).
Using (\ref{A1e20}) in the previous equation, and
once again the
acyclicity of
$\delta$ in positive antifield number gives
\beq
\frac{\delta a_k}{\delta\phi^*_i}=R^{i}_\alpha(\sigma^\alpha_{k-1})+\delta
\sigma^i_k.\label{A1e21}
\eeq
For the proof of part (iii) of the theorem, we
note that this relation
holds with $\sigma^\alpha_{k-1}=0$ because by
assumption
$\delta a_1/\delta\phi^*_i$ is weakly
vanishing, and so it is
$\delta$-exact. Finally Eq.\ (\ref{forreader})
gives for $Z\equiv \phi^i$, using (\ref{A1e19}) and (\ref{A1e18})
repeatedly,
\beann
\delta\ \frac{\delta a_k}{\delta\phi^i}
&=&-\sum_{k\geq 0}(-)^k\partial_{(\lambda_1\dots\lambda_k)}\left[
\frac{\partial 
\cL_j}{\partial\phi^i_{(\lambda_1\dots\lambda_k)}}
\frac{\delta a_k}{\delta\phi^*_j}\right.\\
&&+\left.
\phi^*_{j}\sum_{l\geq 0}\frac{\partial 
R^{j(\mu_1\dots\mu_l)}_\alpha }
{\partial\phi^i_{(\lambda_1\dots\lambda_k)}}
\partial_{(\mu_1\dots\mu_l)}\ \frac{\delta a_k}{\delta
C^*_\alpha}\right].
\eeann
One now inserts (\ref{A1e20}), (\ref{A1e21}) in the previous
equation and uses
\[
\sum_{k\geq 0}\sum_{l\geq 0}(-)^k
\partial_{(\lambda_1\dots\lambda_k)}\left[
\frac{\partial (R^{j(\mu_1\dots\mu_l)}_\alpha \cL_j)}
{\partial\phi^i_{(\lambda_1\dots\lambda_k)}}\
\partial_{(\mu_1\dots\mu_l)}
\sigma^\alpha_{k-1}\right]=0,
\] 
which follows from
repeated application of
(\ref{A1e18}) and the fact that
$R^{+i(\mu_1\dots\mu_l)}_\alpha$ defines a Noether
identity. This gives an expression $\delta(\dots)=0$.
Using then acyclicity of $\delta$ in positive antifield number,
one gets $(\dots)=\delta \sigma_{i k+1}$:
\bea
\frac{\delta a_k}{\delta\phi^i}&=&
-\sum_{k\geq 0}(-)^k\partial_{(\lambda_1\dots\lambda_k)}\left[
\frac{\partial 
\cL_j}{\partial\phi^i_{(\lambda_1\dots\lambda_k)}}\ \sigma^j_k
\right.\nonumber\\
&&-\left.
\sum_{l\geq 0}\frac{\partial (R^{+j(\mu_1\dots\mu_l)}_\alpha
\phi^*_{j(\mu_1\dots\mu_l)})}
{\partial\phi^i_{(\lambda_1\dots\lambda_k)}}\ 
\sigma^\alpha_{k-1}\right]
+\delta \sigma_{i k+1}\ .
\label{A1e22}
\eea
On the other
hand, we have
\beq
Na_k=\phi^i\frac{\delta a_k}{\delta\phi^i}
+\phi^*_i\frac{\delta a_k}{\delta\phi^*_i}
+C^*_\alpha\frac{\delta a_k}{\delta
C^*_\alpha}+\partial_\mu j^\mu.
\eeq
Using (\ref{A1e20})-(\ref{A1e22}),
integrations by parts and
(\ref{A1e19}), we get
\bea
Na_k&=&\delta(\phi^i\sigma_{i
k+1}-\phi^*_i\sigma^i_k+C^*_\alpha
\sigma^\alpha_{k-1})+\partial_{\mu}{j^\prime}^\mu\nonumber\\
&&+\left[2\cL_j-\frac{\delta (N_{\phi}L)}{\delta\phi^j}\right]\sigma^j_k
+\sum_{l\geq 0}
(N_{\phi }R^{+i(\mu_1\dots\mu_l)}_\alpha)\phi^*_{i(\mu_1\dots\mu_l)}
\sigma^\alpha_{k-1}.\label{A1e24}
\eea
If the theory is linear, the two terms in the last
line vanish. We can
then use this result in the homotopy formula
$a_k=\int^1_0\frac{d\lambda}{\lambda}[Na_k]
(x,\lambda\phi,\lambda\phi^*,\lambda C^*)$ and the fact that
$\delta=\delta^{(0)}$ and $\partial_\mu$ are
homogeneous of degree
$0$ in $\lambda$ to conclude that $a_k=\delta(\ )
+\partial_\mu (\ )^\mu$. This ends the proof in the
case of irreducible linear gauge theories.

If a theory is linearizable, we decompose $a_k$,
with
$k\geq 1$ into pieces of definite homogeneity $n$
in all the fields, antifields
and their derivatives, $a_k=\sum_{n\geq l}a_k^{(n)}$
where
$l\geq 2$ due to the assumptions of the theorem.
We then use the acyclicity of $\delta^{(0)}$ to
show that if
$c_k=\delta e_{k+1}$ with the
expansion of $c$
starting at homogeneity $l\geq 1$,
then the expansion of $e$ can be taken to start
also
at homogeneity $l$. Indeed,
we have $\delta^{(0)}e^{(1)}_{k+1}=0$,
$\delta^{(0)}e^{(2)}_{k+1}+\delta^{(1)}e^{(1)}_{k+1}
=0,\dots,$
$\delta^{(0)}e^{(l-1)}_{k+1}+\dots+\delta^{(l-2)}e^{(1)}_{k+1}=0$.
The first equation
implies $e^{(1)}_{k+1}=\delta^{(0)}f^{(1)}_{k+2}$,
so that the
redefinition $e_{k+1}-\delta f^{(1)}_{k+2}$, which
does not modify
$c_k$ allows to absorb $e^{(1)}_{k+1}$. This
process can be continued
until $e^{(l-1)}_{k+1}$ has been absorbed.

Hence, we can choose $\sigma_{i
k+1},\sigma^i_k,\sigma^\alpha_{k-1}$
to start at homogeneity $l-1$. This implies that
the two last terms in
(\ref{A1e24}) are of homogeneity $\geq l+1$.
Due to $a_k=\frac{1}{l}Na_k+\sum_{n>l}a^{\prime (n)}$,
Eq.\ (\ref{A1e24}) yields
$a_k=\delta(\ )+\partial_\mu (\
)^\mu+a^{\prime\prime}_k$, where
$a^{\prime\prime}_k$ starts at homogeneity $l+1$ (unless it
vanishes). Going on
recursively proves the theorem. \qed

\paragraph{Proof of theorem \ref{Loct10}:}

In the space of forms which are polynomials in the
derivatives of the
fields, the antifields and their derivatives with
coefficients that
are power series in the fields, the $K$ degree is
bounded. It is of
course also bounded in the space of forms that are
polynomials in the
undifferentiated fields as well.
We can use the acyclicity of
$\delta^{(0),0}$ to prove the acyclicity of
$\delta^0$ in the
respective spaces. Indeed,
suppose that $c$ is of strictly positive antifield
number, its
polynomial expansion starts with $l$ and its $K$
bound is $M$. From
$\delta^0 c_M=0$ it follows that $\delta^{(0),0}
c_{l,M}=0$ and then
that $c_{l,M}=\delta^{(0),0}e_{l,M}$. This means
that
$c-\delta^{0}e_{l,M}$ starts at homogeneity $l+1$.
Going on in this
way allows to absorb all of $c_M$. Note that if
$c$ is a polynomial in
the undifferentiated fields and $\delta^{{\rm
int},0}=0$, the
procedure stops after a finite number of steps
because the terms
modifying the terms in $c_M$ of homogeneity higher
than $l$ in the
absorption of $c_{l,M}$ have a $K$ degree which is
strictly smaller
than $M$.
One can then go on recursively to remove
$c_{M-1},\dots$.
Because $\delta^0$ is acyclic, we can assume that
in the equation
$c_{k}=\delta e_{k+1}$, the $K$ degrees of
$e_{k+1}$ is bounded by the
same $M$ bounding the $K$ degree of $c_{K}$.
Indeed, if the $K$ degree
of $e_{k+1}$ were bounded by $N>M$, we have
$\delta^0
e_{k+1,N}=0$. Acyclicity of $\delta^0$ then
implies
$e_{k+1,N}=\delta^0 f_{k+1,N}$ and, the
redefinition $e-\delta
f_{k+1,N}$, which does not affect $c_{k}$ allows
to absorb
$e_{k+1,N}$.

If the $K$ degree of $a_k$ is bounded by $M$, the
$K$ degree of
$\frac{\delta a_k}{\delta \phi^i}$,
$\frac{\delta a_k}{\delta \phi^*_i}$,
$\frac{\delta a_k}{\delta C^*_\alpha}$ is bounded
respectively by
$M$, $M-r_i$, $M-m_\alpha$, because of
$[K,\partial_{\mu}]=\partial_\mu$. It follows
from  the definitions of
$r_i$ and $m_{\alpha}$ that the $K$ degree of
$\sigma_{i
k+1},\sigma^i_k,\sigma^\alpha_{k-1}$ is also
bounded respectively by
$M,M-r_i,M-m_{\alpha}$ and that the $K$ degree of
the second line of
equation (\ref{A1e24}), modifying the terms of
higher homegeneity in
the fields in the absorption of the term of order
$l$ in the proof of
theorem \ref{Loct9}, is also bounded by $M$. This
proves the theorem,
by noticing as before that in the space of
polynomials in all the
variables, with $\delta^{{\rm int},0}=0$, the
procedure stops after a
finite number of steps. \qed

\newpage

\mysection{Homological perturbation theory}
\label{homological}

\subsection{The longitudinal differential
$\gamma$}

In the introduction, we have defined
the $\gamma$-differential for Yang-Mills gauge models
in terms of generators.  Contrary to the
Koszul-Tate differential, $\gamma$ does not depend on the
Lagrangian but only on the gauge symmetries.  Thus, it
takes the same form for all gauge theories of the Yang-Mills type.
One has explicitly
\begin{eqnarray}
\gamma &=&\sum_m \big[\partial_{\mu_1 \dots \mu_m} (D_\mu C^I)
\frac{\partial}{\partial (\partial_{\mu_1 \dots \mu_m}
A_\mu^I)} + \partial_{\mu_1 \dots \mu_m} (-e\, C^IT^i_{Ij}\psi^j)
\frac{\partial}{\partial (\partial_{\mu_1 \dots \mu_m}
\psi^i)}\big]\nonumber \\
&&+ \sum_m \partial_{\mu_1 \dots \mu_m} (\sfrac 12\, e\, \f KJI\, 
C^J C^K )
\frac{\partial}{\partial (\partial_{\mu_1 \dots \mu_m}
C^I)} \nonumber \\ 
&&+ \sum_m \big[\partial_{\mu_1 \dots \mu_m} (e\, \f JIK\, C^JA_K^{*\mu})
\frac{\partial}{\partial (\partial_{\mu_1 \dots \mu_m}
A_I^{*\mu})}
+ \partial_{\mu_1 \dots \mu_m} (e\, C^I \psi^*_j T^j_{Ii})
\frac{\partial}{\partial (\partial_{\mu_1 \dots \mu_m}
\psi^*_i)}\big] \nonumber \\
&&+ \sum_m \partial_{\mu_1 \dots \mu_m} (e\, \f JIK\, C^JC_K^* )
\frac{\partial}{\partial (\partial_{\mu_1 \dots \mu_m}
C_I^*)} .\label{definitiongamma} 
\end{eqnarray}
Clearly,  $\gamma^2 = 0$.
The differential $\gamma$ increases the pure ghost number by
one unit, $[N_C, \gamma] = \gamma$.

One may consider the restriction $\gamma_R$ of $\gamma$ to the 
algebra generated
by the original fields and the ghosts, without the antifields,
\begin{eqnarray}
\gamma_R &=&\sum_m \big[\partial_{\mu_1 \dots \mu_m} (D_\mu C^I) 
\frac{\partial}{\partial (\partial_{\mu_1 \dots \mu_m}A_\mu^I)} 
+ \partial_{\mu_1 \dots \mu_m} (-e\, C^IT^i_{Ij}\psi^j)\frac{\partial}
{\partial (\partial_{\mu_1 \dots \mu_m}\psi^i)}\big]\nonumber \\
&&+ \sum_m \partial_{\mu_1 \dots \mu_m} (\sfrac 12\, e\, \f KJI\,
C^J C^K )
\frac{\partial}{\partial (\partial_{\mu_1 \dots \mu_m}
C^I)}. 
\end{eqnarray}
One has also $\gamma_R^2 = 0$, i.e., the antifields
are not necessary for nilpotency of $\gamma$. It is sometimes this
differential which is called the BRST differential. 

However, the fact that this restricted differential 
-- or even (\ref{definitiongamma}) --
is nilpotent
is an accident of gauge theories of the Yang-Mills type.
For more general gauge theories with so-called 
``open algebras", $\gamma_R$ (known as the ``longitudinal
exterior differential along
the gauge orbits") is nilpotent only on-shell, $\gamma_R^2 
\approx 0$.  Accordingly, it is a differential only on the stationary
surface.  Alternatively,
when the antifields are included,
$\gamma$ fulfills 
$\gamma^2 = - (\delta s_1 + s_1 \delta)$
and is a differential only in the homology of $\delta$.  Thus, one
can define, in general, only the cohomological groups 
$H(\gamma, H(\delta))$. [For Yang-Mills theories, however, $H(\gamma)$ makes
sense even in the full algebra since $\gamma$ is strictly nilpotent
on all fields and antifields.
The cohomology $H(\gamma)$
turns out to be important and will be computed below.]

In the general case
the BRST differential $s$ is not simply
given by $s = \delta + \gamma$, but contains higher order terms
\beq
s = \delta + \gamma + s_1 + \hbox{ ``higher order terms"},
\eeq
where the higher order terms have higher antifield number, and
$s_1$ and possibly higher order terms are necessary for $s$ to be nilpotent, 
$s^2 = 0$. This can even
happen for a ``closed algebra". Indeed, in the case of 
non constant structure functions, 
$\gamma^2$ does not necessarily vanish on the antifields
and a non vanishing $s_1$ may be needed.

The construction of $s$ from $\delta$ and $\gamma$ follows a recursive
pattern known as ``homological perturbation theory".  We shall not
explain it here since this machinery is not needed in the Yang-Mills
context where $s$ is simply given by $s = \delta + \gamma$.  However,
even though the ideas of homological perturbation theory are
not necessary for constructing $s$ in the Yang-Mills case, they are
crucial in elucidating some aspects of the BRST cohomology and in relating
it to the cohomologies $H(\delta)$ and $H(\gamma, H(\delta))$.
In particular, they show the importance of the
antifield number as  auxiliary degree useful to split the
BRST differential.  They also put into light the importance of the
Koszul-Tate differential in the BRST 
construction\footnote{In fact, the explicit decomposition $s = \delta
+ \gamma$ appeared in print relatively recently, even though it
is of course rather direct.}.  It is this step that has enabled one,
for instance, to solve long-standing conjectures regarding the 
BRST cohomology.

\subsection{Decomposition of BRST cohomology}

The BRST cohomology groups are entirely determined
by cohomology
groups involving the first two terms $\delta$ and $\gamma$
in the
decomposition $s = \delta + \gamma + s_1 +\dots$
This result is quite general, so we shall state and demonstrate 
it without sticking to theories of the Yang-Mills type.
In fact, it is based solely on the acyclicity of $\delta$ in positive
antifield number which is crucial for the whole BRST construction, 
\beq
H_k(\delta)=0\quad\mbox{for}\quad k>0.
\label{Hdelta}
\eeq
\begin{theorem}\label{Loctdec}
In the space of local forms, one has the following
isomorphisms:
\bea
H^g(s)&\simeq& H^g_0(\gamma,H(\delta)),
\label{74}
\\[4pt]
H^{g,p}(s|d)&\simeq&
\left\{
\ba{ccc}
H^{g,p}_0(\gamma,H(\delta|d)) & \mathrm{if} & g\geq 0,
\\[4pt]
H_{-g}^p(\delta|d) & \mathrm{if} & g<0,
\ea
\right.
\label{75}
\eea
where the superscripts $g$ and $p$ indicate the
(total) ghost number and the form-degree respectively 
and the subscript indicates the antifield number.
\end{theorem}

{\bf Explanation and proof.}
Both isomorphisms are based upon the expansion in the 
antifield number and state that solutions $a$ to
$sa=0$ or $sa+dm=0$ can be fully characterized in the
cohomological sense (i.e., up to respective trivial solutions)
through properties of the lowest term in their expansion.
Let $g$ denote the ghost number of $a$.
When $g$ is nonnegative, $a$ may contain a piece that
does not involve an antifield at all;
in contrast, when $g$ is negative,
the lowest possible term in the expansion of $a$ has
antifield number $-g$,
\beq
a = a_\uk + a_{\uk+1} + a_{\uk+2} + \dots\ , \ \agh(a_k) = k,\
\uk\geq\cases{\phantom{-}0\ \mathrm{if}\ g\geq 0,\cr
-g\ \mathrm{if}\ g< 0,}
\label{expansion000}
\eeq
because there are no fields of negative ghost number.

Now, (\ref{74}) expresses on the one hand
that every nontrivial solution $a$ to $sa=0$
has an antifield independent piece $a_0$ which fulfills
\beq 
\gamma a_0+\delta a_1=0,\quad a_0\neq \gamma b_0+\delta b_1
\label{exp1}
\eeq
and is thus a nontrivial element of $H^g_0(\gamma,H(\delta))$
since $\gamma a_0+\delta a_1=0$ and $a_0= \gamma b_0+\delta b_1$
are the cocycle and coboundary condition in that cohomology
respectively (a cocycle of $H^g_0(\gamma,H(\delta))$
is $\gamma$-closed up to a $\delta$-exact form
since $\delta$-exact forms vanish in $H(\delta)$). 
Note that (\ref{exp1}) means 
$\gamma a_0\approx 0$ and $a_0\not\approx \gamma b_0$.
In particular, $H(s)$ thus vanishes at all negative ghost numbers
because then $a$ has no antifield independent piece $a_0$.
Furthermore (\ref{74}) expresses that each solution $a_0$ to
(\ref{exp1}) can be
completed to a nontrivial $s$-cocycle $a=a_0+a_1+\dots$
and that this correspondence between $a$ and $a_0$
is unique up to terms which are trivial in $H^g(s)$ and
$H^g_0(\gamma,H(\delta))$ respectively.

To prove these statements,
we show first that $sa=0$ implies $a=s b$ whenever $\uk>0$,
for some $b$ whose expansion starts at antifield number
$\uk+1$,
\beq
sa=0,\quad \uk> 0\quad \then\quad a=s(b_{\uk+1}+\dots).
\label{exp2}\eeq
This is seen as follows.
$sa=0$ contains the equation $\delta a_\uk=0$ 
which implies $a_\uk=\delta b_{\uk+1}$ when $\uk>0$,
thanks to (\ref{Hdelta}). 
Consider now $a':=a-sb_{\uk+1}$. 
If $a'$ vanishes we get $a=sb_{\uk+1}$ and thus that $a$ is trivial.
If $a'$ does not vanish, its expansion
in the antifield number reads
$a' = a'_{\uk'} +\dots$ where $\uk'>\uk$ because of
$a'_\uk=a_\uk-\delta b_{\uk+1}=0$. 
Furthermore we have $sa'=sa-s^2b_{\uk+1}=0$. 
Applying the same arguments to $a'$ as before to $a$, we conclude
$a'_{\uk'}=\delta b'_{\uk'+1}$ for some
$b'_{\uk'+1}$. We now consider $a''=a'-sb'_{\uk'+1}=
a-s(b_{\uk+1}+b'_{\uk'+1})$. If $a''$ vanishes
we get $a=s(b_{\uk+1}+b'_{\uk'+1})$ and stop.
If $a''$ does not vanish we continue until we
finally get $a=sb$ for some
$b=b_{\uk+1}+b'_{\uk'+1}+b''_{\uk''+1}+\dots$
(possibly after infinitely many steps).

(\ref{exp2}) shows that every $s$-cocycle with
$\uk>0$ is trivial.
When $\uk=0$, $a_0$ satisfies automatically $\delta a_0=0$
since it contains no antifield. The
first nontrivial equation in the expansion of $sa=0$ 
is then $\gamma a_0+\delta a_1=0$,
while $a=sb$ contains $a_0= \gamma b_0+\delta b_1$.
Hence, every nontrivial $s$-cocycle contains indeed
a solution to (\ref{exp1}).

To show that every solution to (\ref{exp1}) can be completed
to a nontrivial $s$-cocycle, we consider
the cocycle condition in $H^g_0(\gamma,H(\delta))$, 
$\gamma a_0+\delta a_1=0$,
and define $X:=s(a_0+a_1)$. 
When $X$ vanishes we have $sa=0$ with $a=a_0+a_1$ and thus
that $a$ is an $s$-cocycle.
When $X$ does not vanish, its expansion
starts at some antifield number $\geq 1$ due to
$X_0=\gamma a_0+\delta a_1=0$.
Furthermore we have $sX=s^2(a_0+a_1)=0$.
Applying (\ref{exp2}) to $X$ yields thus
$X=-sY$ for some $Y=a_k+\dots$ where $k\geq 2$.
Hence we get $X=s(a_0+a_1)=-s(a_k+\dots)$ and thus $sa=0$ where
$a=a_0+a_1+a_k+\dots$ with $k\geq 2$. So, each solution of
$\gamma a_0+\delta a_1=0$ can indeed be completed to an $s$-cocycle
$a$. Furthermore $a$ is trivial if $a_0= \gamma b_0+\delta b_1$
and nontrivial otherwise. Indeed,
$a_0= \gamma b_0+\delta b_1$ implies that $Z:=a-s(b_0+b_1)$
fulfills $sZ=0$ and $Z_0=a_0-(\gamma b_0+\delta b_1)=0$, and thus,
by arguments used before,
either $Z=0$ or $Z=-s(b_k+\dots)$, $k\geq 2$
which both give $a=sb$.
$a_0\neq \gamma b_0+\delta b_1$ guarantees
$a\neq sb$ because $a=sb$ would imply $a_0= \gamma b_0+\delta b_1$.
We have thus seen that (non)trivial elements of $H(s)$ correspond
to (non)trivial elements of $H^g_0(\gamma,H(\delta))$ and vice versa
which establishes (\ref{74}).
\medskip

(\ref{75}) expresses
that every nontrivial solution $a$ to $sa+dm=0$
has a piece $a_0$ if $g\geq 0$, or $a_{-g}$ if $g<0$,
fulfilling
\bea
g\geq 0:&&
\gamma a_0+\delta a_1+dm_0=0,\quad a_0\neq \gamma b_0+\delta b_1+dn_0,
\label{exp3}
\\
g<0:&&
\delta a_{-g}+dm_{-g-1}=0,\quad a_{-g}\neq \delta b_{-g+1}+dn_{-g}\ .
\label{exp4}
\eea
(\ref{exp3}) states that $a_0$ is a nontrivial cocycle
of $H^{g,*}_0(\gamma,H(\delta|d))$ because
$\gamma a_0+\delta a_1+dm_0=0$ and $a_0= \gamma b_0+\delta b_1+dn_0$
are the cocycle and coboundary condition in that cohomology
respectively. (\ref{exp4}) states that
$a_{-g}$ is a nontrivial cocycle of $H_{-g}^p(\delta|d)$.
Furthermore (\ref{75}) expresses that every solution to
(\ref{exp3}) or (\ref{exp4}) can be completed to a nontrivial 
solution $a=a_0+a_1+\dots$ or $a=a_{-g}+a_{-g+1}+\dots$ of
$sa+db=0$.
Note that $a_0$ contains no antifield, while $a_{-g}$ contains
no ghost due to $\gh(a)=g$. Hence, (\ref{exp3}) means
$\gamma a_0+dm_0\approx 0$ and $a_0\not\approx \gamma b_0+dn_0$,
while (\ref{exp4}) means that $a_{-g}$ is related to
a nontrivial element of the characteristic cohomology as explained
in detail in Section \ref{Koszul2Section}.

These statements can be proved along lines whose logic 
is very similar to the derivation of (\ref{74}) given above.
Therefore we shall only sketch the proof, leaving the
details to the reader. 
The derivation is based on corollary \ref{Locc3} which itself is
a direct consequence of (\ref{Hdelta}) as the proof of that
theorem shows. The r\^ole of Eq.\ (\ref{exp2}) is now taken
by the following result:
\beq
sa+dm=0,\ \uk>\cases{\phantom{-}0\ \mathrm{if}\ g\geq 0\cr
-g\ \mathrm{if}\ g< 0}\ \then\
a=s(b_{\uk+1}+\dots)+d(n_\uk+\dots).
\label{exp5}
\eeq
This is proved as follows. $sa+dm=0$ contains the equation
$\delta a_\uk+dm_{\uk-1}=0$. When $\uk>0$ and $g\geq 0$,
or when $\uk>-g$ and $g<0$, $a_\uk$ has both positive
antifield number and positive pureghost number
(due to $\gh=\agh+\mathit{puregh}$).
Using theorem \ref{Loct7}, we then conclude $a_\uk=\delta b_{\uk+1}
+d n_\uk$ for some $b_{\uk+1}$ and $n_\uk$.
One now considers $a':=a-sb_{\uk+1}-dn_\uk$ which fulfills
$sa'+dm'=0$ ($m'=m-sn_\uk$) and derives (\ref{exp5}) using
recursive
arguments analogous to those in the derivation of (\ref{exp2}).
The only values of $\uk$ which are not covered in (\ref{exp5})
are $\uk=0$ if $g\geq 0$, and $\uk=-g$ if $g<0$.
In these cases, $sa+dm=0$ contains the equation
$\gamma a_0+\delta a_1+dm_0=0$ if $g\geq 0$, or 
$\delta a_{-g}+dm_{-g-1}=0$ if $g<0$. These
are just the first equations in (\ref{exp3}) and (\ref{exp4})
respectively. To finish the proof one 
finally shows that $a$ is trivial ($a=sb+dn$)
if and only if $a_0= \gamma b_0+\delta b_1+dn_0$ for
$g\geq 0$, or $a_{-g}= \delta b_{-g+1}+dn_{-g}$
for $g<0$ by arguments
which are again analogous to those used in the derivation of
(\ref{74}). \qed

\subsection{Bounded antifield number}

As follows from the proof, 
the isomorphisms in theorem \ref{Loctdec}
hold under the assumption that the local forms
in the theory may contain terms of arbitrarily high
antifield number.  That is, if one expands the
BRST cocycle $a$ associated with a given element $a_0$ of
$H^g_0(\gamma,H(\delta))$ or $H^{g,p}_0(\gamma,H(\delta|d))$
according to the antifield number as in Eq.\ (\ref{expansion000}),
there is no guarantee, in the general case,
that the expansion stops even if $a_0$ is a local form.  So,
although each term in the expansion would be a local form
in this case, $a$
may contain arbitrarily high derivatives if the
number of derivatives in $a_k$ grows with $k$.
This is not a problem for effective field theories,
but is in conflict with locality otherwise.

In the case of normal theories with a local Lagrangian, 
which include, as we have seen,
the original Yang-Mills theory,  the standard model
as well as pure Chern-Simons theory in $3$ dimensions (among
others), one can easily refine the theorems and show that the
expansion (\ref{expansion000}) stops, so that $a$ is a local form.
This is done by introducing a degree that appropriately controls
the antifield number as well as the number of derivatives.

To convey the idea, we illustrate the procedure in the simplest
case of pure electromagnetism.  We leave it to the reader to
extend the argument to the general case.  The degree in question 
-- call it $D$ -- may then be taken to be the sum of the degree 
counting the number
of derivatives plus the degree assigning weight one to the
antifields $A^{*\mu}$ and $C^*$.  Our assumption
of locality and polynomiality in the
derivatives for  $a_0$ implies that it has bounded degree $D$.
In fact, since the differentials $\delta$,
$\gamma$ and $d$ all increase this degree by one unit,  one
can assume that $a_0$ is homogeneous of definite (finite) $D$-degree $k$.
The recursive equations in $sa+dm=0$ determining $a_{i+1}$ from $a_i$ read
in this case
$\delta a_{i+1} + \gamma a_i + dm_i = 0$ (thanks to $s=\delta+\gamma$), 
and so, one
can assume that $a_{i+1}$ has also $D$-degree $k$.  Thus, all terms
in the expansion (\ref{expansion000}) have same $D$-degree equal
to $k$.  This means that
as one goes from one term $a_i$ to the next $a_{i+1}$ in
(\ref{expansion000}), the antifield
number increases (by definition) while the number of derivatives decreases
until one reaches $a_m = a_{m+1} = \cdots = 0$ after a finite number
of (at most $2k$) steps.

The fact that the expansion (\ref{expansion000}) stops is particularly
convenient because it enables one to analyse the BRST cohomology
starting from the last term in (\ref{expansion000}) (which exists).
Although this is not always necessary, this turns out to be often a
convenient procedure.

\subsection{Comments}

The ideas of homological perturbation theory appeared in the 
mathematical literature in 
\cite{hirsch,gugmay,gug,gugsta,lamsta,guglam,guglamsta}.
They have been applied in the context of the antifield formalism in
\cite{Fisch:1990rp} (with locality analyzed in 
\cite{Henneaux:1991rx,Barnich:1995db})
and are reviewed in \cite{Henneaux:1992ig}, 
chapters 8 and 17. 

\newpage

\mysection{Lie algebra cohomology: $H(s)$ and $H(\gamma)$ 
in Yang-Mills type theories}
\label{LieAlgebraCoho}

\subsection{Eliminating the derivatives of the ghosts}
\label{AdCelim}

Our first task in the computation of $H(s)$ for gauge theories of the
Yang-Mills type is to get rid of the derivatives of
the ghosts.  This can be achieved for every Lagrangian $L$ 
fulfilling the conditions of the introduction; it
is performed by making a change of
jet-space coordinates adapted to the
problem at hand (see \cite{Brandt:1990gy,Dubois-Violette:1992ye}). 

We consider subsets $W^k$ ($k=-1,0,1,\dots$) of jet coordinates 
where $W^{-1}$ contains
only the undifferentiated ghosts $C^I$, while $W^k$ for
$k\geq 0$ contains
\bea
A^I_{\mu|(\nu_1\dots\nu_l)},\ \psi^i_{(\nu_1\dots\nu_l)},
\ C^I_{(\nu_1\dots\nu_{l+1})},\nonumber\\
A^{*\mu}_{I(\nu_1\dots\nu_{l-2})},\ \psi^*_{i(\nu_1\dots\nu_{l-m})},\
C^*_{I(\nu_1\dots\nu_{l-3})}
\label{defofW}
\eea
for $l=0,\dots,k$ and $m=1,2$, for matter fields with first and second
order field equations respectively.  [These definitions are in fact taylored
to the standard model; if the gauge fields obey equations of
motion of order $k$, $\nu_{l-2}$ should be replaced by $\nu_{l-k}$
and $\nu_{l-3}$ by $\nu_{l-k-1}$
in (\ref{defofW}); similarly, $m$ is generally the derivative order
of the matter field equations.]

We can take as new coordinates on $W^k$ 
the following functions of the old ones:
\begin{eqnarray}
&\partial_{(\nu_1\dots\nu_l}A^I_{\mu)},\
\partial_{(\nu_1\dots\nu_l}D_{\mu)}C^I,&
\label{y1}\\
&C^I,&\label{z1}\\
&D_{(\nu_1}\dots D_{\nu_{l-1}}F^I_{\nu_l)\mu},\
D_{(\nu_1}\dots D_{\nu_{l})}\psi^i,&\label{x1}
\\
&D_{(\nu_1}\dots
D_{\nu_{l-2})}A^{*\mu}_I,\ D_{(\nu_1}\dots D_{\nu_{l-3})}C^*_I,\
D_{(\nu_1}\dots D_{\nu_{l-m})}\psi^*_i,&
\label{x2}
\end{eqnarray}
for $l=0,\dots,k$ and $m=1,2$. 
This change of coordinates is invertible because
$\partial_{\nu_1\dots\nu_l}
A^I_{\mu}=\partial_{(\nu_1\dots\nu_l}A^I_{\mu)} + {l\over 
l+1}D_{(\nu_1}\dots
D_{\nu_{l-1}}F^I_{\nu_l)\mu}+O(l-1)$
where $O(l-1)$ collects terms with less than $l$
derivatives. 
There are no algebraic 
relations
between the $\partial_{(\nu_1\dots\nu_l}A^a_{\mu)}$ and the
$D_{(\nu_1}\dots D_{\nu_{l-1}}F^a_{\nu_l)\mu}$, which 
correspond to the independent irreducible components of
$\partial_{\nu_1\dots\nu_l}
A^I_{\mu}$. Similarly,  one has
$D_{(\nu_1}\dots
D_{\nu_{l})}\psi^i=\partial_{\nu_1\dots\nu_l}\psi^i+O(l-1)$.

The new coordinates can be grouped 
into two sets : 
contractible pairs (\ref{y1}) on the one hand, 
and gauge covariant coordinates 
(\ref{x1}), (\ref{x2})
plus undifferentiated ghosts (\ref{z1}) on the
other hand. These sets transform
among themselves under $s$. Indeed we have $s
\partial_{(\nu_1\dots\nu_l}A^I_{\mu)}=
\partial_{(\nu_1\dots\nu_l}D_{\mu)}C^I$
and $s \partial_{(\nu_1\dots\nu_l}D_{\mu)}C^I = 0$. 
Similarly, if we
collectively denote by $\chi^u_\Delta$
 the coordinates
(\ref{x1}) and (\ref{x2}), we have 
$\gamma \chi^u_\Delta= -eC^IT^u_{Iv}\chi^v_\Delta$ where
the $T^u_{Iv}$ are the entries of representation matrices
of $\cG$ and $\Delta$ labels the various multiplets of $\cG$ formed
by the $\chi$'s. For instance, for every fixed
set of spacetime indices, the $D_{(\nu_1}\dots
D_{\nu_{l-1}}F^K_{\nu_l)\mu}$ ($K=1,\dots,\mathit{dim}(\cG)$)
form a multiplet $\chi^K_\Delta$ (with fixed $\Delta$)
of the coadjoint representation with
$T^u_{Iv}$ given by $\f IJK$.\footnote{
Our notation is slightly sloppy because the index $u$ (and in 
particular its range) really depends on the given multiplet and 
thus should carry a subindex $\Delta$. The substitution
 $u\longrightarrow u_\Delta$ should thus be understood in the formulas 
below.
}  

Finally, $sC^I$ is a function of the ghosts
alone, and $\delta$ on the antifields (\ref{x2}) only involves the 
coordinates (\ref{x2}) and (\ref{x1}). 
The latter statement about the $\delta$-transformations
is equivalent to the gauge covariance of the
equations of motion. It
can be inferred from $\gamma \delta+\delta \gamma =0$, without
referring to a particular Lagrangian.
Namely $(\gamma \delta+\delta \gamma)A^{*\mu}_I=0$
gives $\gamma L_I^\mu=eC^J\f JIK L_K^\mu$ while
$(\gamma \delta+\delta \gamma)\psi^*_i=0$ gives
$\gamma L_i=eC^I T_{Ii}^jL_j$, see Eq.\ (\ref{ym7}).
The absence of derivatives of the ghosts in $\gamma L_I^\mu$
and $\gamma L_i$ implies that $L_I^\mu$ and $L_i$
can be expressed solely in terms of the variables (\ref{x1}) (and
the $x^\mu$ when the Lagrangian involves $x^\mu$ explicitly)
because $\gamma$ is stable in the subspaces of local functions 
with definite degree in the variables
(\ref{y1}).

The coordinates (\ref{y1}) form thus indeed contractible pairs and do
not contribute to the cohomology of $s$ according to a reasoning 
analogous to the one followed in
section \ref{IIBContractible}. 

Note that the removal of the vector potential, its
symmetrized derivatives, 
and the derivatives of the ghosts, works
both for $H(s)$ and $H(\gamma)$ since $s$ and $\gamma$
coincide in this sector.

\subsection{Lie algebra cohomology with coefficients in a representation}

One of the interests of the elimination of the derivatives of the
ghosts is that the connection between
$H(\gamma)$,  $H(s)$ and ordinary Lie algebra cohomology becomes
now rather direct.

We start with $H(\gamma)$, for which matters are straightforward.
We have reduced the computation of the cohomology of $\gamma$
in the algebra of all local forms to the calculation of the
cohomology of $\gamma$ in the algebra ${\cal K}$ of local 
forms depending on the covariant objects $\chi^u_\Delta$ and
the undifferentiated ghosts $C^I$. More precisely, the relevant
algebra is now
\beq
{\cal K} = \Omega(\mathbb{R}^n) \otimes {\cal F} \otimes \Lambda(C)
\eeq
where $\Omega(\mathbb{R}^n)$ is the algebra of exterior forms on $\mathbb{R}^n$,
${\cal F}$ the algebra of functions of the covariant objects
$\chi^u_\Delta$, and $\Lambda(C)$ the algebra of polynomials in
the ghosts $C^I$ (which is just the antisymmetric algebra
with $\mathit{dim} (\cG)$ generators).

The subalgebra $\Omega(\mathbb{R}^n) \otimes {\cal F}$ 
provides a representation
of the Lie algebra ${\cal G}$, the factor $\Omega(\mathbb{R}^n)$ being trivial
since it does not transform under ${\cal G}$.  
We call this representation $\rho$.
The differential $\gamma$
can be written as 
\beq
\gamma=eC^I\rho(e_I)+\frac{e}{2}\,C^JC^I\f I J
K\frac{\partial}{\partial C^K}
\eeq
where the $e_I$ form a basis for
the Lie algebra ${\cal G}$ and the $\rho(e_I)$ are
the corresponding ``infinitesimal generators" in the
representation,
\beq
\rho(e_I)=-T^u_{Iv} \; \chi^v_\Delta\,
\frac{\partial}{\partial\chi^u_\Delta}\ .
\eeq
The identification of the polynomials in $C^I$ with the 
cochains on ${\cal G}$ then allows to identify 
the differential $\gamma$ with the standard
Chevalley-Eilenberg differential \cite{ChevEil}
for Lie algebra cochains with values in the representation space
$\Omega(\mathbb{R}^n)\otimes
C^\infty(\chi^u_\Delta)$.

Thus, we see that the cohomology of $\gamma$ is just 
standard Lie algebra cohomology with coefficients in the
representation $\rho$ of functions in the covariant objects
$\chi^u_\Delta$ (times the spacetime exterior forms).

The space of smooth functions in the variables $\chi^u_\Delta$
is evidently infinite-dimensional.  In order to be able to
apply theorems on Lie algebra cohomology, it is necessary to make
some restrictions on the allowed functions so as to effectively
deal with finite dimensional representations.  This condition
will be met, 
for instance, if one considers
polynomial local functions in the $\chi^u_\Delta$
with coefficients that can possibly
be smooth functions of invariants (e.g. $\exp \phi$ can
occur if
$\phi$ does not transform under ${\cal G}$).  This 
space is still infinite-dimensional,
but splits as the direct sum of finite-dimensional representation
spaces of ${\cal G}$.
Indeed, 
because $\rho(e_I)$ is homogeneous of degree 0 in the $\chi^u_\Delta$,
we can consider separately polynomials of a given
homogeneity in the $\chi^u_\Delta$, which form
finite dimensional representation spaces.
Thus, the problem of computing the Lie algebra cohomology
of ${\cal G}$ with coefficients in the representation $\rho$
is effectively reduced to the problem of computing the Lie algebra cohomology
of ${\cal G}$ with coefficients in a finite-dimensional representation.
The same argument applies, of course, to effective field theories.
From now on, it will be understood that such restrictions are made on
${\cal F}$.

\subsection{$H(s)$ versus $H(\gamma)$}
\label{Xvariables}

The previous section shows that the
computation of $H(\gamma)$ boils down to a standard problem of
Lie algebra cohomology with coefficients in a definite
representation.  This is also true for $H(s)$, but the
representation space is now different.

Indeed, we have seen that $H(s) \simeq H(\gamma, H(\delta))$.
This result was established in the algebra of all local forms
depending also on the differentiated ghosts and symmetrized
derivatives of the vector potential, but also holds in the algebra
${\cal K}$. Moreover, the cohomology of the Koszul-Tate differential
in ${\cal K}$ can be computed in the same manner as above.
The antifields drop out with the variables constrained to
vanish with the equations of motion.  More precisely, among
the field strength components, the matter field components, and
their covariant derivatives, some can be viewed as constrained
by the equations of motion and the others can be viewed as independent.
Let $X^u_A$ be the independent ones  
and ${\cal F}^R$ be the algebra of smooth functions in $X^u_A$ 
(with restrictions analogous to those made in the previous subsection). 
Because the equations of motion are gauge covariant
(see above), one
can take the $X^u_A$ to transform in a linear representation
of ${\cal G}$, which we denote by $\rho^R$.  
Again, since one can
work order by order in the derivatives, the representation
$\rho^R$ effectively splits as a direct sum of finite-dimensional
representations. [See below for an explicit construction of 
the $X^u_A$ in the standard model.]

By our general discussion of section \ref{homological},
it follows that
\begin{lemma}\label{lemsgam}
The cohomology of $s$ is isomorphic to the cohomology of
$\gamma$ in the space of local forms depending only on the
undifferentiated ghosts $C^I$ and the $X_A^u$.
\end{lemma}

In the case of the standard model, the $X^u_A$ may be
constructed as follows.  First,
the $D_{(\nu_1}\dots D_{\nu_{l-1}}F^I_{\nu_l)\mu}$ are split
into the algebraically
independent completely
traceless combinations $(D_{(\nu_1}\dots
D_{\nu_{l-1}}F^I_{\nu_l)\mu})_\mathrm{tracefree}$
(i.e., $\eta^{\nu_1\nu_2}(D_{(\nu_1}\dots
D_{\nu_{l-1}}F^I_{\nu_l)\mu})_\mathrm{tracefree}=0$) and the traces
$D_{(\nu_1}\dots D_{\nu_{l-2}}D_{\lambda)}F^{I\lambda}_{\mu}$
\cite{Torre:1995kb}. 
Second the covariant derivatives $D_{(s_1}\dots D_{s_{l})}\psi^i$
of the matter fields are replaced by 
$D_{(s_1}\dots D_{s_{l})}\hat\psi^i$
and $D_{(\nu_1}\dots D_{\nu_{l-1})}\hat {\cal L}_i$ for matter fields
$\hat \psi^i$
with first order equations and by
$D_{(s_1}\dots D_{s_{l})}\tilde\psi^i$,
$D_{(s_1}\dots D_{s_{l-1}}D_{0)}\tilde\psi^i$, $D_{(\nu_1}\dots
D_{\nu_{l-2})} \tilde{\cal L}_i$ for matter fields $\tilde \psi^i$
with second order
field equations. The 
$X_A^u$ are then the coordinates 
$(D_{(\nu_1}\dots D_{\nu_{l-1}}F^I_{\nu_l)\mu})_\mathrm{tracefree}$, 
$D_{(s_1}\dots
D_{s_{l})}\hat\psi^i$, $D_{(s_1}\dots D_{s_{l})}\tilde\psi^i$,
$D_{(s_1}\dots D_{s_{l-1}}D_{0)}\tilde\psi^i$.                                
Other splits are of course available.

In the algebra 
\beq
{\cal K}^R = \Omega(\mathbb{R}^n) \otimes {\cal F}^R \otimes \Lambda(C)
\eeq  
the differential $\gamma$ reads
\beq
\gamma=eC^I\rho^R(e_I)+\frac{e}{2}\,C^JC^I\f I JK
\frac{\partial}{\partial C^K}
\eeq
where 
the $\rho^R(e_I)$ are
the infinitesimal generators in the
representation $\rho^R$,
\beq
\rho^R(e_I)=-T^u_{Iv} \, X^v_A\,
\frac{\partial}{\partial X^u_A}\ .
\eeq
The representations of $\cG$ which occur in ${\cal F}$ and ${\cal F}^R$
are the same ones, but with
a smaller multiplicity in ${\cal F}^R$.
One can thus identify the cohomology of $s$ with
the Lie-algebra cohomology of ${\cal G}$, with coefficients
in the representation $\rho^R$.  The difference between
$H(s)$ and $H(\gamma)$ lies only in the space of coefficients.

\subsection{Whitehead's theorem}

In order to proceed, we shall now assume that the gauge group $G$
is the direct product of a compact abelian group
$G_0$ times a semi-simple group $G_1$.  Thus, $G= G_0 \times
G_1$, with $G_0 = (U(1))^q$.
This assumption on $G$ was not necessary for the previous
analysis, but is used for the subsequent developments since
in this case one has complete results on the Lie
algebra cohomology.
On these conditions, it can then be shown
that any finite-dimensional representation of
${\cal G}$ is completely reducible \cite{Bourbaki}.

We can now follow the standard literature (e.g. \cite{GHV}). 
For any representation space $V$ with representation $\rho$, let 
$V_{\rho=0}$ be the invariant subspace of $V$ carrying the trivial
representation, which
may occur several times 
($v \in V_{\rho=0}\ \then\ \rho(e_I)v=0\ \forall I$).
Note that the space $\Lambda(C)$ of ghost polynomials 
is a representation of 
${\cal G}$ for 
\[
\rho^C(e_I)=-C^J\f I JK\frac{\partial}{\partial C^K}\ .
\] 
(With the above interpretation of
the $C^I$, it is the extension of the
coadjoint representation to $\Lambda {\cal G}^*$.) The total
representation on $V\otimes \Lambda(C)$ is
$\rho^T(e_I)=\rho(e_I)+\rho^C(e_I)$. It satisfies 
\beq
\Big\{\gamma,\frac{\partial}{\partial C^I}\Big\}=e\rho^T(e_I),\quad
[\gamma,\rho^T(e_I)]=0,\label{commga}
\eeq
the first relation following by direct computation, the second one
from the first and $\gamma^2=0$. 
Hence the cohomology $H(\gamma,(V\otimes \Lambda(C))_{\rho^T=0})$ is
well defined. We can write
\[
\gamma=eC^I\rho(e_I)+\frac{e}{2}\,C^I \rho^C(e_I)=eC^I\rho^T(e_I)
+\hat\gamma\ ,\]
where $\hat\gamma$ is the
restriction of $\gamma$ to $\Lambda(C)$, up to the sign:
\[
\hat\gamma=\frac{e}{2}\,C^IC^J\f I J K \frac{\partial}{\partial C^K}\ .
\] 
It follows that 
\bea
H(\gamma,(V\otimes \Lambda(C))_{\rho^T=0})\simeq
H(\hat\gamma,(V\otimes \Lambda(C))_{\rho^T=0}).\label{redu}
\eea
Note that we also have,
\beq
\Big\{\hat \gamma,\frac{\partial}{\partial C^I}\Big\}=-e\rho^C(e_I),\quad
[\hat \gamma,\rho^C(e_I)]=0.
\label{commgaha}
\eeq

The first mathematical result that we shall need reduces the problem
of computing the Lie algebra cohomology of ${\cal G}$ with
coefficients in $V$ to that of finding the invariant subspace
$V_{\rho=0}$
and computing the 
Lie algebra cohomology of ${\cal G}$ with
coefficients in the trivial, one-dimensional, representation.
\begin{theorem}\label{thlico6}
(i)
$H(\gamma,V\otimes\Lambda(C))$ is
isomorphic to $H(\gamma,(V\otimes
\Lambda(C))_{\rho^T=0})$. In particular, $H(\hat \gamma,\Lambda(C))$
is isomorphic to 
$\Lambda(C)_{\rho^C=0}$.

(ii) $H(\gamma,(V\otimes
\Lambda(C))_{\rho^T=0})$ is isomorphic to
$V_{\rho=0}\otimes\Lambda(C)_{\rho^C=0}$.

\end{theorem}

The proof of this theorem is given in the appendix
\ref{LieAlgebraCoho}.A. 
The result $H(\gamma,(V\otimes \Lambda(C))\simeq 
V_{\rho=0}\otimes H(\hat\gamma,\Lambda(C))$ is known 
as Whitehead's theorem.
                                                                             
To determine the cohomology of $s$, we thus need to determine on the
one hand the
invariant monomials in the $X_A^u$, which depends on the precise
form of the matter field representations and which will not be discussed
here; and on the other
hand, $H(\hat\gamma,\Lambda(C))\simeq
\Lambda(C)_{\rho^C=0}$. This latter cohomology
is known as the Lie algebra cohomology
of ${\cal G}$ and is discussed in the next section. 

\subsection{Lie algebra cohomology - Primitive elements}
\label{primitive5}

We shall only give the results (in ghost notations),
without proof.  We refer to the mathematical literature for
the details \cite{Koszul,GHV}.

The cohomology $H^g(\hat\gamma,\Lambda(C))$ can be described in terms
of particular ghost polynomials 
$\theta_r(C)$ representing the so-called primitive elements. These
are in bijective correspondence with the 
independent Casimir operators $\cO_r$,
\beq
\cO_r=d^{I_1\dots I_{m(r)}}\delta_{I_1}\dots \delta_{I_{m(r)}},
\quad
r=1,\dots,rank(\cG).
\label{c4.4}
\eeq
The $d^{I_1\dots I_{m(r)}}$ are symmetric invariant tensors, 
\bea
\sum_{i=1}^{m(r)} \f JLK d^{I_1\dots I_{i-1}LI_{i+1}\dots I_{m(r)}}=0,
\eea 
while $\delta_I=\rho(e_I)$ for some representation $\rho(e_I)$ of
${\cal G}$. 
The ghost polynomial $\theta_r(C)$ corresponding to $\cO_r$
is homogeneous of degree $2m(r)-1$, and given by
\bea
& & \theta_r(C)=(-e)^{m(r)-1}\,
\frac {m(r)!(m(r)-1)!}{2^{m(r)-1}(2m(r)-1)!}\, 
f_{I_1\dots I_{2m(r)-1}} C^{I_1}\dots C^{I_{2m(r)-1}}\ ,
\nonumber\\
& & f_{I_1J_1\dots I_{m(r)-1}J_{m(r)-1}K_{m(r)}}=
d_{K_1\dots K_{m(r)-1}[K_{m(r)}}
\f {I_1}{J_1}{K_1}\dots \f {I_{m(r)-1}}{J_{m(r)-1}]}{K_{m(r)-1}}\ .
\label{thetas}
\eea
This definition of $\theta_r(C)$
involves, for later purpose, a normalization factor containing both the
gauge coupling constant and the
order $m(r)$ of $\cO_r$. The $d_{K_1\dots K_{m(r)}}$ arise from
the invariant symmetric tensors
in (\ref{c4.4}) by lowering the indices with the invertible
metric $g_{IJ}$ obtained by adding the Killing metrics for each simple
factor, trivially extended to the whole of $\cG$, and 
with the identity for an abelian $\delta_I$.  
Using an appropriate (possibly complex) 
matrix representation $\{T_I\}$ of $\cG$
(cf.\ example below), $\theta_r(C)$ can also be written as
\beq
\theta_r(C)=(-e)^{m(r)-1}\,\frac {m(r)!(m(r)-1)!}{(2m(r)-1)!}
         \, \mathrm{Tr}(C^{2m(r)-1}),\quad C=C^I T_I\ .
\label{thetas'}\eeq
Indeed, using that the $C^I$ anticommute and that $\{T_I\}$ 
represents $\cG$ ($[T_I,T_J]=\f IJK T_K$), 
one easily verifies that (\ref{thetas'}) agrees
with (\ref{thetas}) for
\[ d_{I_1\dots I_{m(r)}}=
\mathrm{Tr}[T_{(I_1}\dots T_{I_{m(r)})}]\ .
\]
Those $\theta_r(C)$ with degree 1 coincide with the
abelian ghost fields (if any), in accordance
with the above definitions:  the abelian
elements of $\cG$ count among the Casimir operators as they
commute with all the other elements of $\cG$. We thus set
\beq
\{\theta_r(C):m(r)=1\}=\{\mbox{abelian ghosts}\}.
\label{c4.5}
\eeq
Note that this is consistent with (\ref{thetas'}), as it
corresponds to the choice $\{T_I\}=\{0,\dots,0,1,0,\dots,0\}$,
where one of the abelian elements of $\cG$
is represented by the number 1, while all the other
elements of $\cG$ are represented by 0. 

Each $\theta_r(C)$
is $\hat\gamma$ closed, as is easily verified using
the matrix notation (\ref{thetas'}),
\[ \hat\gamma\,\mathrm{Tr}(C^{2m-1})= e\,\mathrm{Tr}(C^{2m})=0,\]
where the first equality holds due to $\hat\gamma C=eC^2$ 
and the second equality holds because the
trace of any even power of a Grassmann odd matrix vanishes.
The cohomology of $\hat\gamma$ is generated 
precisely by the $\theta_r(C)$, i.e.,
the corresponding cohomology classes are represented 
by polynomials in the $\theta_r(C)$, and
no nonvanishing polynomial of the $\theta_r(C)$ is
cohomologically trivial,
\bea
&\hat\gamma h(C)=0\quad \Leftrightarrow\quad  h(C)=P(\theta(C))+\hat \gamma
g(C)\ ;&
\nonumber\\
&P(\theta(C))=\hat\gamma g(C)\quad \Leftrightarrow\quad P=0.&
\label{c4.6a}\eea

Note that the $\theta_r(C)$ anticommute because they
are homogeneous polynomials of odd degree in the ghost fields.
Therefore the dimension of the cohomology of $\hat\gamma$ 
is $2^{\,rank(\cG)}$.
Note also that the highest nontrivial cohomology class
(i.e., the one with highest ghost number) is represented by
the product of all the $\theta_r(C)$. This product is always 
proportional
to the product of all the ghost fields (see, e.g., \cite{Brandt:1990gv}), 
and has thus ghost number equal to the dimension of $\cG$,
\beq
\prod_{r=1}^{rank(\cG)}\theta_r(C)\propto
\prod_{I=1}^{dim(\cG)}C^I.
\label{c4.7}
\eeq

\paragraph{Example 1.}

Let us spell out the result for the gauge group of the standard model,
$G=U(1)\times SU(2)\times SU(3)$.
The $U(1)$-ghost is a $\theta$ by itself, see (\ref{c4.5}). We set
\[ \theta_1(C)=\mbox{U(1)-ghost}.\]

$SU(2)$ has only one Casimir operator which has order 2. 
The corresponding $\theta$ has thus degree 3 and is given by
\[ 
\theta_2(C)=-\frac {e_\mathrm{su(2)}}3\, \mathrm{Tr}_\mathrm{su(2)}(C^3),
\]
with $C=C^IT_I$,
$\{T_I\}=\{0,\mbox{i}\sigma_\alpha,0,\dots,0\}$ where the
zeros represent $u(1)$ and $su(3)$, and $\sigma_\alpha$ 
are the Pauli matrices.

$SU(3)$ has two independent Casimir operators, with degree
2 and 3 respectively. This gives two additional $\theta$'s
of degree 3 and 5 respectively,
\[
\theta_3(C)=-\frac {e_\mathrm{su(3)}}3\, \mathrm{Tr}_\mathrm{su(3)}(C^3),\quad
\theta_4(C)=\frac {e_\mathrm{su(3)}^2}{10}\, \mathrm{Tr}_\mathrm{su(3)}(C^5),
\]
with $\{T_I\}=\{0,0,0,0,\mbox{i}\lambda_a\}$
where $\lambda_a$ are the Gell-Mann matrices.
 
A complete list of inequivalent 
representatives of $H(\hat\gamma,\Lambda(C))$ is:
\[ 
\ba{c|c}
\mbox{ghost number} & \mbox{representatives of $H(\hat\gamma,\Lambda(C))$}
\\
\hline\rule{0em}{3ex}
0 & 1
\\
\hline\rule{0em}{3ex}
1 & \theta_1
\\
\hline\rule{0em}{3ex}
3 & \theta_2\ ,\ \theta_3
\\
\hline\rule{0em}{3ex}
4 & \theta_1\theta_2\ ,\ \theta_1\theta_3
\\
\hline\rule{0em}{3ex}
5 & \theta_4
\\
\hline\rule{0em}{3ex}
6 & \theta_2\theta_3\ ,\ \theta_1\theta_4
\\
\hline\rule{0em}{3ex}
7 & \theta_1\theta_2\theta_3
\\
\hline\rule{0em}{3ex}
8 & \theta_2\theta_4\ ,\ \theta_3\theta_4
\\
\hline\rule{0em}{3ex}
9 & \theta_1\theta_2\theta_4\ ,\ \theta_1\theta_3\theta_4
\\
\hline\rule{0em}{3ex}
11 & \theta_2\theta_3\theta_4
\\
\hline\rule{0em}{3ex}
12 & \theta_1\theta_2\theta_3\theta_4
\ea
\]

\paragraph{Example 2.}  

Let us denote by $C^{I_a}$ the ghosts of the abelian factors,
${I_a}=1,\dots, l$.
A basis of the first cohomology groups $H^g(\hat\gamma, \Lambda(C))$
($g=0, 1, 2$) is given by  (i) $1$ for $g=0$; (ii) $C^{I_a}$,
with ${I_a}=1,\dots, l$, for $g=1$; and (iii) $C^{I_a} C^{J_a}$,
with ${I_a} < J_a$, for $g=2$.
In particular, $H^1(\hat\gamma, \Lambda(C))$ and
$H^2(\hat\gamma, \Lambda(C))$ are trivial if there is no 
abelian factor. For a compact group, 
a basis for $H^3(\hat\gamma, \Lambda(C))$
is given by $C^{I_a} C^{J_a} C^{K_a}$ ($I_a < J_a < K_a$)
and $f_{I_s J_s K_s} C^{I_s} C^{J_s} C^{K_s} \equiv
\mathrm{Tr}_\mathrm{\cG_s} C^3$, where $s$ runs over 
the simple factors $G_s$ in $G$.

\subsection{Implications for the renormalization of local 
gauge invariant operators}

Class I local operators are defined as 
local, non integrated, gauge invariant operators (built out of the
gauge potentials and the matter fields) that are linearly
independent, even when the gauge covariant equations of motions are
used \cite{Kluberg-Stern:1975hc,Joglekar:1976nu,Deans:1978wn}. 
In the absence of anomalies, it can be shown 
that these operators only mix, under renormalization, with BRST closed local
operators (built out of all the fields and antifields).
BRST exact operators are called class II operators. They can be shown
not to contribute to the physical S matrix and to renormalize only
among themselves. The question is then whether class I operators can
only mix with class I operators and class II operators under
renormalization. 

That the answer is affirmative follows from 
lemma \ref{lemsgam} in the case of ghost number $0$. Indeed, 
the $\gamma$ cohomology in the space of forms in the $X^u_A$, (i.e.,
combinations of the covariant derivatives of the field strength
components and the matter field components not constrained by the
equations of motions) reduces to the gauge invariants in these
variables. This is precisely the required statement 
that class I operators give a 
basis of the BRST cohomology $H^{0,*}(s,\Omega)$ in ghost number $0$.

The statement was first proved in \cite{Joglekar:1976nu}. A different
proof has been given in \cite{Henneaux:1993jn}.

\subsection{Appendix \ref{LieAlgebraCoho}.A: 
Proof of theorem \ref{thlico6}}

Our proof of theorem \ref{thlico6} uses ghost notations.

(i) Let us first of all prove general relations for a
completely reducible representation commuting with a differential,
i.e., take into account only the second relation of (\ref{commga}).
This relation implies that the
representation $(\rho^T)^{\#}(e_I)$ induced in cohomology by
$(\rho^T)^{\#}(e_I)[a]=[\rho^T(e_I)a]$ is well defined.
The induced representation is completely reducible: since the space
of $\gamma$ cocycles
$Z$ is stable under $\rho^T(e_I)$,
there exists a stable subspace $E\subset
V\otimes\Lambda(C)$, such that $V\otimes\Lambda(C)=Z
\oplus E$. Similarly, because
the space of $\gamma$ coboundaries $B$ is stable under
$\rho^T(e_I)$, there exists a stable subspace $F\subset
Z$ such that
$Z=F\oplus B$.
It follows that $H(\gamma,V\otimes\Lambda(C))$ is
isomorphic to $F$. Since $F$
is completely reducible for
$\rho^T(e_I)$, so is $H(\gamma,V\otimes\Lambda(C))$ for 
$(\rho^T)^{\#}(e_I)$. Complete reducibility also implies that
$Z(\rho^T (V\otimes\Lambda(C)))=Z\cap \rho^T (V\otimes\Lambda(C))=\rho^T
Z$. This means
that $H(\gamma,\rho^T (V\otimes \Lambda(C)))\subset (\rho^T)^{\#}
H(\gamma,V\otimes\Lambda(C))$. In the same way,
$H(\gamma,(V\otimes\Lambda(C))_{\rho^T=0})\subset
H(\gamma,V\otimes\Lambda(C))_{(\rho^T)^{\#}=0}.$
Complete reducibility of $\rho^T$
then implies $H(\gamma,V\otimes\Lambda(C))=
H(\gamma,(V\otimes\Lambda(C))_{\rho^T=0})\oplus
H(\gamma,\rho^T (V\otimes\Lambda(C)))$, while complete reducibility of
$(\rho^T)^{\#}$ implies $H(\gamma,V\otimes\Lambda(C))
=H(\gamma,V\otimes\Lambda(C))_{
(\rho^T)^{\#}=0}\oplus (\rho^T)^{\#}H(\gamma,V\otimes\Lambda(C))$. It
follows that $H(\gamma,(V\otimes\Lambda(C))_{\rho^T=0})
=H(\gamma,V\otimes\Lambda(C))_{
(\rho^T)^{\#}=0}$ and
$H(\gamma,\rho^T(V\otimes\Lambda(C)))=(\rho^T)^{\#}H(\gamma,V\otimes
\Lambda(C))$.
Using now the first relation of (\ref{commga}), it follows that
$(\rho^T)^{\#}=0$ so that $H(\gamma,\rho^T(V\otimes\Lambda(C)))=0$ and
$H(\gamma,V\otimes\Lambda(C))          
=H(\gamma,(V\otimes\Lambda(C))_{\rho^C=0})$.
This proves the first part of (i).

For the second part, we note that in $\Lambda(C)$,
the representation $\rho^C(e_I)$ reduces to the
representation of the semi-simple factor. Let us denote
 the generators of this factor by $e_a$ and its representation
on $\Lambda(C)$ by $\rho^C(e_a)$.
This representation is completely reducible, because defining
properties of semi-simple Lia algebras are (a) the Killing metric
$g_{ab}=\f a c d \f b d c$ is invertible; (b) it is the direct sum of
simple ideals, (a Lie algebra being simple if it is non abelian and
contains no proper non
trivial ideals); (c) all its representations
 in finite dimensional vector spaces are completely reducible
 \cite{Bourbaki}.
This proves then the second part by the same reasoning as above with
$V=\rho=0$, respectively (\ref{commgaha}) instead of (\ref{commga}).

(ii)
$\Lambda(C)=\Lambda(C)_{\rho^C=0}\oplus\rho^C\Lambda(C)$
implies $(V\otimes\Lambda(C))_{\rho^T=0}
=V_{\rho=0}\otimes\Lambda(C)_{\rho^C=0}\oplus
(V\otimes\rho^C\Lambda(C))_{\rho^T=0}$. Because it follows from
$[\hat\gamma,\rho^T]=0=[\hat\gamma,\rho^C] =[\hat\gamma,\rho]$ that
all the spaces are stable under $\hat \gamma$,
the K\"unneth formula gives $H(\hat \gamma,(V\otimes
\Lambda(C))_{\rho^T=0})=V_{\rho=0}\otimes
H(\hat\gamma,\Lambda(C)_{\rho^C=0})\oplus H(\hat\gamma,
(V\otimes\rho^C\Lambda(C))_{\rho^T=0})$. The result (ii) then
follows from (\ref{redu}) and $\hat\gamma=-\frac{e}{2}C^I\rho^C(e_I)$,
if we can show that $H(\hat\gamma,
(V\otimes\rho^C\Lambda(C))_{\rho^T=0})=0$.

The contracting homotopy that allows to prove
$H(\hat\gamma,\rho^C\Lambda(C))=0$ can be constructed explicitely as
follows.
Define the Casimir operator
$\Gamma=\frac{1}{2}g^{ab}\rho^C(e_a)\rho^C(e_b)$,
where $g^{ab}$ is the inverse of the Killing metric associated to the
semi-simple Lie subalgebra of ${\cal G}$. From the complete skew
symmetry of the structure constants lowered or raised through the
Killing metric or its inverse (this being a consequence of the Jacobi
identity), it follows that this operator commutes with
all the operators of the representation,
$[\Gamma,\rho^C(e_a)]=0$, while the first relation of (\ref{commgaha})
implies that $[\hat\gamma,\Gamma]=0$.

A property of semi-simple Lie algebras is that the
Casimir operator $\Gamma$ is invertible on
$\rho^C\Lambda(C)$. Obviously, in this case,
$[\Gamma^{-1},\rho^C(e_a)]=0$ and
$[\hat\gamma,\Gamma^{-1}]=0$. Take $a\in
\rho^C\Lambda(C)$, with $\hat\gamma a=0$. We have $a=\Gamma\Gamma^{-1}
a=-\frac{1}{2}g^{ab}\{\hat\gamma,\frac{\partial}{\partial
  C^a}\}\rho^C(e_b)\Gamma^{-1}a=-\hat\gamma\rho^C(e_b)
\frac{1}{2}g^{ab}
\frac{\partial}{\partial C^a}\Gamma^{-1}a$, where we have used in
addition $g^{ab}[\frac{\partial}{\partial C^a},\rho^C(e_b)]=0$, as
follows from the first relation of (\ref{commgaha}) and the graded
Jacobi identity for the graded commutator of operators. Hence,
$H(\hat\gamma,\rho^C\Lambda(C))=0$.

Similarly, let $a=v\otimes b$, where $b\in \rho^C\Lambda(C)$, with
$\gamma a=0=eC^I \rho(e_I)v\otimes b-e(-)^vv\otimes \hat \gamma b$
and
$\rho^T(e_I) a=0=\rho(e_I)v\otimes b+v\otimes\rho(e_I) b$. We have
$a=-v\otimes(\hat\gamma\rho^C(e_b)\frac{1}{2}g^{ab}
\frac{\partial}{\partial C^a}\Gamma^{-1} b+\rho^C(e_b)
\frac{1}{2}g^{ab}
\frac{\partial}{\partial C^a}\Gamma^{-1}\hat\gamma b)=
\gamma\rho^C(e_b)\frac{1}{2}g^{ab}
\frac{\partial}{\partial C^a}\Gamma^{-1}(v\otimes b )$.
Furthermore,
direct computation using skew symmetry of structure constants with
lifted indices shows that
$[\rho^T(e_I),\rho^C(e_b)\frac{1}{2}g^{ab}
\frac{\partial}{\partial C^a}\Gamma^{-1}]=0$, which proves that
$H(\hat\gamma,
(V\otimes\rho^C\Lambda(C))_{\rho^T=0}=0$ and thus (ii).
\qed  

\newpage

\mysection{Descent equations: $H(s \vert d)$}\label{Des}

\subsection{Introduction}

The descent equation technique is a powerful tool to
calculate $H(s|d)$ which we shall use
below. Its usefulness rests on the fact that it
relates $H(s|d)$ to $H(s)$ which is often much
simpler than $H(s|d)$ - and which we have determined. 

In subsections \ref{properties}, \ref{liftsobstr} and 
\ref{destheo}, we shall
review general properties of the descent equations and work
out the relation between $H(s|d)$ and $H(s)$ in detail.
Our only assumption for doing so will be
that in the space of local forms under study, 
the cohomology of $d$ is trivial at
all form-degrees $p=1,\dots,n-1$ and is represented at $p=0$
by  pure numbers,
\beq
H^p(d)=\delta^p_0\,\mathbb{R}\quad\mbox{for}\quad p<n.
\label{des2}
\eeq
Since this is
the only assumption being made at this stage,
the considerations in subsections \ref{properties} through
\ref{destheo} are not restricted to
gauge theories of the Yang-Mills type but apply whenever
(\ref{des2}) is fulfilled. 

In the case of theories of the Yang-Mills type (in
${\mathbb R}^n$), the
considerations apply in the space of local smooth
forms or in the space of local polynomial forms, 
provided one allows for an explicit $x$-dependence.
Indeed,
the algebraic Poincar\'e  lemma guarantees then that
\ref{des2} holds (theorem \ref{Loct2}).  Although our
ultimate goal is to cover the polynomial (or formal power
series) case, such a restriction is not necessary in this section.
The considerations are also valid in 
subalgebras of the algebra of local forms for which \ref{des2}
remains true.

However the considerations of this section do not 
immediately apply, for instance,
if no  explicit 
spacetime coordinate dependence is allowed.  In this case,
the cohomology of $d$ is non-trivial in degrees $\not= 0$ 
and contains the constant forms (see theorem \ref{Loct''2}).
It turns out, however, that the constant forms cannot come in the
way, so that the same descent equation techniques
in fact apply.  This
is explained in subsection 
\ref{weaker}.

Finally, we carry out the explicit derivation of the descent equations
in the case of the differential $s$, but a similar discussion applies to 
$\gamma$
or $\delta$ (or, for that matter, any differential $D$ such that
$Dd+dD=0$).  In fact, 
this tool has already been used in section \ref{Koszul2Section}
for the mod $d$ cohomology of $\delta$,
to prove the isomorphism $H^p_k(\delta|d)\simeq H^{p-1}_{k-1}(\delta|d)$ for
$p>1,k>1$.  The same techniques can be followed for $H(s|d)$ (or
$H(\gamma |d)$), with,
however, one complication.  While one had $H_{k}(\delta)=0$ ($k>0$)
for $\delta$, the cohomology of $s$ is non trivial.  As a result,
while it is easy to ``go down" the descent (because this
uses the triviality of $d$ -- see below), it is more intricate
to ``go up". 

\subsection{General properties of the descent equations}
\label{properties}

We shall now review the derivation and some basic properties of
the descent equations, assuming that (\ref{des2}) holds.

\paragraph{Derivation of the descent equations.}
Let $\omega^m$ be a cocycle of $H^{*,m}(s|d)$,
\beq
s\omega^m+d\omega^{m-1}=0.
\label{des6}
\eeq
Two cocycles are equivalent in $H^{*,m}(s|d)$ when they differ by a 
trivial solution of the consistency condition,
\beq
\omega^m\sim\omega'{}^m
\quad\LRA\quad
\omega^m-\omega'{}^m=s\eta^m+d\eta^{m-1}.
\label{des7}
\eeq
Applying $s$ to Eq.\ (\ref{des6})
yields $d(s\omega^{m-1})=0$ (due to $s^2=0$ and $sd+ds=0$), i.e.,
$s\omega^{m-1}$ is a $d$-closed $(m-1)$-form.
Let us assume that $m-1> 0$ (the case $m-1=0$ is treated below).
Using (\ref{des2}), we conclude that
$s\omega^{m-1}$ is $d$-exact, i.e., there is 
an $(m-2)$-form such that
$s\omega^{m-1}+d\omega^{m-2}=0$.
Hence, $\omega^{m-1}$ is a cocycle of $H^{*,m-1}(s|d)$.
Moreover, due to the ambiguity (\ref{des7}) in $\omega^m$,
$\omega^{m-1}$ is also determined only up to a coboundary
of $H^{*,m-1}(s|d)$. Indeed, when
$\omega^m$ solves Eq.\ (\ref{des6}), then
$\omega'{}^m=\omega^m+s\eta^m+d\eta^{m-1}$ fulfills
$s\omega'{}^m+d\omega'{}^{m-1}=0$ with
$\omega'{}^{m-1}=\omega^{m-1}+s\eta^{m-1}+d\eta^{m-2}$.
Now, two things can happen:

(a) either $\omega^{m-1}$ is trivial in $H^{*,m-1}(s|d)$,
$\omega^{m-1}=s\eta^{m-1}+d\eta^{m-2}$; then
we can substitute 
$\omega'{}^{m-1}=\omega^{m-1}-s\eta^{m-1}-d\eta^{m-2}=0$
and 
$\omega'{}^m=\omega^m-d\eta^{m-1}$ for
$\omega^{m-1}$ and $\omega^m$ respectively and obtain
$s\omega'{}^m=0$; we say that
$\omega^m$ has a trivial descent;

(b) or $\omega^{m-1}$ is nontrivial in $H^{*,m-1}(s|d)$; then
there is no way to make $\omega^m$ $s$-invariant by
adding a trivial solution to it; we say that $\omega^m$ has a nontrivial
descent.

In case (b), we treat $s\omega^{m-1}+d\omega^{m-2}=0$ 
as Eq.\ (\ref{des6}) before:
acting with $s$ on it gives
$d(s\omega^{m-2})=0$; if $m-2\neq 0$, this implies
$s\omega^{m-2}+d\omega^{m-3}=0$ for some
$(m-3)$-form thanks to (\ref{des2}).
Again there are two possibilities: either $\omega^{m-2}$ is trivial
and can be removed through suitable redefinitions such that
$s\omega'{}^{m-1}=0$; or it is nontrivial. In the latter case one 
continues the procedure until one arrives at $s\omega^{\um}=0$
at some nonvanishing form-degree $\um$ (possibly after
suitable redefinitions), or
until the form-degree drops to zero and one gets the equation
$d(s\omega^{0})=0$. 

From the equation $d(s\omega^{0})=0$, one derives,
using once again (\ref{des2}),
$s\omega^{0}=\alpha$ for some constant $\alpha\in\mathbb{R}$. If one
assumes that the equations of motion are consistent - which one
better does! - , then $\alpha$ must vanish and the
conclusion is the same as in the previous case.
This is 
seen by decomposing $s\omega^{0}=\alpha$ into pieces
with definite antifield number and pure ghost number. Since
$\alpha$ is a pure number and has thus vanishing antifield number 
and pure ghost number, the decomposition yields
in particular the equation $\delta a=\alpha$ 
where $a$ is the piece contained in $\omega^{0}$ which
has antifield number $1$ and pure ghost number 0.
Due to $\delta a\approx 0$, this makes only sense if
$\alpha=0$ because otherwise the equations of motion
would be inconsistent (as one could have, e.g., 
$0=1$ on-shell\footnote{Such an inconsistency would arise, for
instance, if one had a neutral scalar field $\Phi$ with Lagrangian
$L = \Phi$.  This Lagrangian yields the equation of motion
$1 = 0$ and must be excluded.  Having $1 = s \omega$ would of
course completely kill the cohomology of $s$.  We have not investigated
whether inconsistent Lagrangians (in the above sense) are eliminated
by the general conditions imposed on $L$ in the introduction, and so,
we make the assumption separately.  Note that a similar difficulty
does not arise in the descent associated with $\gamma$; an
equation like $1 = \gamma \omega$ is simply impossible,
independently of the Lagrangian, 
because $1$ has vanishing pure ghost number while $\gamma \omega$
has pure ghost number equal to or greater than $1$.  For
$s$, the relevant grading is the total ghost number and can be
negative.}).

We conclude:
when (\ref{des2}) holds, Eq.\ (\ref{des6})
implies in physically meaningful theories that there are
forms $\omega^p$, $p=\um,\dots,m$ fulfilling
\beq
s\omega^p+d\omega^{p-1}=0\quad \mbox{for}\quad p=\um+1,\dots,m\ ,
\quad s\omega^{\um}=0
\label{des8}
\eeq
with $\um\in\{0,\dots,m\}$. 
Eqs.\ (\ref{des8}) are called the descent equations.
We call the forms $\omega^p$, $p=\um,\dots,m-1$ descendants
of $\omega^m$, and $\omega^{\um}$ the bottom form of the
descent equations.

Furthermore we have seen that 
$\omega^m$ and its descendants are determined
only up to coboundaries of $H(s|d)$. 
In fact, for given cohomology class
$H^{*,m}(s|d)$ represented by $\omega^m$, this is the only ambiguity
in the solution of the descent equations, modulo constant forms
at form-degree 0. 
This is so because
a trivial solution of the consistency condition
can only have trivial descendants, except
that $\omega^{0}$ can contain a constant. 
Indeed, assume that
$\omega^p$ is trivial, $\omega^p=s\eta^{p}+d\eta^{p-1}$.
Inserting this in $s\omega^p+d\omega^{p-1}=0$ gives
$d(\omega^{p-1}-s\eta^{p-1})=0$ and thus, by
(\ref{des2}), $\omega^{p-1}=s\eta^{p-1}+d\eta^{p-2}+\delta^{p-1}_0\alpha$
where $\alpha\in\mathbb{R}$ can occur only when $p-1=0$.
Hence, when $\omega^p$ is trivial, its first descendant $\omega^{p-1}$
is necessarily trivial too, except for a possible pure number when $p=1$.
By induction this applies to all further descendants too.

\paragraph{Shortest descents.}
The ambiguity in the solution of the descent equations
implies in particular that all nonvanishing forms
which appear in the descent equations can be chosen such 
that none of them
is trivial in $H(s|d)$ because otherwise we can
``shorten'' the descent equations. In particular,
there is thus a 
``shortest descent'' (i.e., a maximal value of $\um$) for 
every nontrivial cohomology class $H^{*,m}(s|d)$. A
shortest descent
is realized precisely when 
all the forms in the descent equations are nontrivial.
An equivalent characterization of a shortest descent is 
that the bottom form $\omega^\um$ is nontrivial
in $H^{*,\um}(s|d)$ if $\um>0$, respectively that
it is nontrivial
in $H^{*,0}(s|d)$ even up to a constant
if $\um=0$ (i.e., that
$\omega^\um\neq s\eta^\um+d\eta^{\um-1}+\delta^{\um-1}_0\alpha$,
$\alpha\in\mathbb{R}$). 
The latter statement holds because the triviality of any
nonvanishing form in the descent equations implies necessarily that
all its descendants, and thus in particular $\omega^\um$, are
trivial too except for a number that can contribute to
$\omega^0$.
Of course, the shortest descent is not unique since
one may still make trivial redefinitions which do not change
the length of a descent.  

\subsection{Lifts and obstructions}
\label{liftsobstr}

We have seen that the bottoms $ \omega^\um$
of the descent equations associated with solutions
$\omega^m$ of the consistency conditions
$s\omega^m + d \omega^{m-1} = 0$
are cocycles of $s$, $s \omega^\um = 0$, which are non
trivial in $H^{*,0}(s|d)$ (even up to a constant
if $\um=0$).  In particular, they are non trivial in
$H(s)$.

One can conversely ask the following questions. Given a non
trivial cocycle of $H(s)$: (i) Is it trivial in $H(s\vert d)$?;
(ii) Can it be viewed as bottom of a non trivial descent?
These questions were raised for the first time in
\cite{Dubois-Violette:1985hc,Dubois-Violette:1985jb} and 
turn out to contain the key to the calculation of
$H(s \vert d)$ in theories of the Yang-Mills type.

We say that an $s$-cocycle $\omega^p$ can be ``lifted'' $k$ times
if there are forms $\omega^{p+1},\dots,\omega^{p+k}$ such that
$d\omega^p+s\omega^{p+1}=0$, \dots , $d\omega^{p+k-1}+s\omega^{p+k}=0$.
Contrary to the descent, which is never obstructed,
the lift of an element of $H(s)$ can be obstructed
because the cohomology of $s$ is non trivial.  Let
$a$ be an $s$-cocycle and let us try to construct an element
``above it".  To that end, one must compute $da$ and see
whether it is $s$-exact.  It is clear that $da$ is $s$-closed;
the obstructions to it being $s$-exact are thus in
$H(s)$.  Two things can happen. Either $da$ is not $s$-exact,
\beq
da = m
\eeq
with $m$ a non trivial cocycle of $H(s)$.
Or $da$ is $s$-exact, in which case
one has 
\beq
da+sb =0
\eeq
for some $b$.  Of course, $b$ is defined up to a cocycle
of $s$.

In the first case, it is clear that ``the obstruction"
$m$ to lifting $a$, although non trivial in $H(s)$
is trivial in $H(s \vert d)$.  Furthermore
$a$ itself cannot be trivial in
$H(s \vert d)$ since trivial elements $a = su + dv$ can always be
lifted ($da = s(-du)$).

In the second case, one may try 
to lift $a$ once more.  Thus one computes $db$.
Again, it is easy to verify that $db$ is 
an $s$-cocycle.  Therefore, either $db$ is not
$s$-exact,
\beq
db + sc = n \; \; \; \; \; \; \hbox{ ``Case A"}
\eeq
for some non trivial cocycle $n$ of $H(s)$ (we allow here for the presence of
the exact term $sc$ - which can be absorbed in $n$ -
because usually, one has natural representatives
of the classes of $H(s)$, and $db$ may differ from such a representative
$n$ by an $s$-exact term).  Or $db$ is $s$-exact,
\beq
db + sc = 0 \; \; \; \; \; \; \hbox{ ``Case B"}
\eeq
for some $c$.

Note that in case A, $b$ is defined up to the addition of
an $s$-cocycle, so the ``obstruction" $n$ to lifting $a$ a second time
is really present only if $n$ cannot be written as
$dt + sq$ where $t$ is an $s$-cocycle, i.e., if $n$ is not
in fact the obstruction to the first lift of some $s$-cocycle.
The obstructions to second lifts are therefore
in the space $H(s)/\mathrm{Im}\,d$ of the cohomology of $s$ quotientized by
the space of obstructions to first lifts.  If the obstruction
to lifting $a$ a second time is really present, then $a$ is clearly
non trivial in $H(s \vert d)$.  And in any case, $n$ is trivial in
$H(s \vert d)$.  

In case B, one can continue and try to lift $a$ a third
time. This means computing $dc$. The analysis proceeds as above
and is covered by the results of the next subsection. 

\subsection{Length of chains and structure of
$H(s|d)$}\label{destheo}

By following the above procedure, one can construct a basis
of $H(s)$ which displays explicitly the lift structure
and the obstructions.

\begin{theorem}\label{descth1} 
If $H^p(d,\Omega)=\delta^p_0\mathbb{R}$ for $p=0,\dots,n-1$ and the
equations of motions are consistent,    
there exists a basis
\begin{eqnarray}
\{[1],[h^0_{i_r}],[\hat h_{i_r}],[e^0_{\alpha_s}]\}\label{bass}
\end{eqnarray}
of $H(s)$ such that the representatives
fulfill 
\begin{eqnarray}
&s h^{r+1}_{i_r}+ d h^{r}_{i_r}=\hat h_{i_r},&\nonumber\\
&s h^r_{i_r}+ d
h^{r-1}_{i_r}=0,&\nonumber\\
&\vdots&\label{prop1}\\
&s h^1_{i_r}+ dh^0_{i_r}=0,& \nonumber \\
&s h^0_{i_r} = 0&
\nonumber
\end{eqnarray}             
and
\begin{eqnarray}
&{\rm form\ degree}\ e^s_{\alpha_s}=n,&\nonumber\\
&s e^{s}_{\alpha_s}+ d e^{s-1}_{\alpha_s} =0,&\nonumber\\
&\vdots&\label{prop2}\\
&s e^1_{\alpha_s}+ d e^0_{\alpha_s}=0,&
\nonumber \\
&s e^0_{\alpha_s}=0,&
\nonumber
\end{eqnarray}
for some forms $h^q_{i_r}$, $q=1,\dots, r+1$ and $e^p_{\alpha_s}$,
$p=1,\dots,s$.   
Here, $[a]$ denotes the class of the $s$-cocycle $a$ in
$H(s)$.  
\end{theorem}

We recall that a set $\{ f_A\}$ of $s$-cocycles is such that
the set $\{ [f_A]\}$ forms a basis of $H(s)$ if and only if the
following two properties hold: (i) any $s$-cocycle is a linear combination
of the $f_A$'s, up to an $s$-exact term; and (ii) if $\lambda^A f_A
= s g$, then the coefficients $\lambda^A$ all vanish. 

The elements of the basis (\ref{bass})
have the following properties:
The $h^0_{i_r}$ can be lifted $r$ times, until one hits an obstruction
given by $\hat h_{i_r}$.  By contrast, the $e^0_{\alpha_s}$ can
be lifted up to maximum degree without meeting any obstruction. 
We stress that the superscripts of $h^q_{i_r}$ and $e^p_{\alpha_s}$ 
in the above theorem
do {\em not} indicate the form-degree but the increase of the
form-degree relative to $h^0_{i_r}$ and $e^0_{\alpha_s}$ respectively.
The form-degree of $h^0_{i_r}$ is not determined by the above
formulae except that it is smaller than $n-r$.
$e^0_{\alpha_s}$ has form-degree $n-s$.

We shall directly 
construct  bases with such properties in the Yang-Mills
setting, for various (sub)algebras fulfilling
(\ref{des2}), thereby proving explicitly
their existence in the concrete cases relevant for
our purposes.
The proof of the theorem in the general case is given in appendix 
\ref{Des}.A following \cite{Henneaux:1999rp}.
We refer the interested reader to the pioneering work of 
\cite{Dubois-Violette:1985hc,Dubois-Violette:1985jb,Dubois-Violette:1986cj}
for a proof involving more powerful homological tools (``exact couples''). 

For a basis of $H(s)$ with the
above properties, the eqs (\ref{prop1}) provide optimum lifts
of the $h^0_{i_r}$.  The $\hat h_{i_r}$ 
represent true obstructions;  
by using the ambiguities in the successive lifts of $h^0_{i_r}$, one
cannot lift $h^0_{i_r}$ more than $r$ times.  
This is seen by using a recursive
argument.  It is
clear that the $h^0_{i_0}$ cannot be lifted at all since
the $\hat h_{i_0}$ are independent in $H(s)$.  Consider next
$h^0_{i_1}$ and the corresponding chain, $sh^0_{i_1}=0$,
$d h^0_{i_1} + s h^1_{i_1} = 0$, $dh^1_{i_1} +
s h^2_{i_1} = \hat h_{i_1}$.
Suppose that the
linear combination 
$\alpha ^{i_1} h^0_{i_1}$ could be lifted more than once, which would
occur if and only if $\alpha ^{i_1}
\hat h_{i_1}$ was the obstruction to the
single lift of an $s$-cocycle, i.e., $\alpha ^{i_1} \hat h_{i_1} = da + sb$,
$sa = 0$.  Since (\ref{bass}) provides a basis of $H(s)$, one could
expand $a$ in terms of $\{1,h^0_{i_r},\hat h_{i_r},e^0_{\alpha_s}\}$
(up to an $s$-exact term), $a = \alpha^{i_0} h^0_{i_0}
+ \cdots$.  Computing $da$ using
(\ref{prop1}) and (\ref{prop2}), and
inserting the resulting expression into
$\hat h_{i_1} = da + sb$,  one gets $\hat h_{i_1} = \alpha^{i_0}
\hat h_{i_0} + s ( \cdot)$, leading to a contradiction since
the $\hat h_{i_1}$ and $\hat h_{i_0}$ are independent in
cohomology. The argument can be  
repeated in the same way for bottoms leading to longer
lifts and is left to the reader.   

It follows from this analysis that the $h^0_{i_r}$
are $s$-cocycles that are non-trivial in $H(s\vert d)$ -
while, of course, the $\hat h_{i_r}$ are trivial.
In fact, the advantage of a basis of the type of 
(\ref{bass}) for $H(s)$ is that it gives immediately
the cohomology of $H(s\vert d)$.

\begin{theorem}\label{descth2}
If $\{[1],[h^0_{i_r}],[\hat h_{i_r}],[e^0_{\alpha_s}]\}$ is
a basis of $H(s)$ with the properties of theorem
\ref{descth1}, then an associated basis of $H(s|d)$ is given by
\begin{eqnarray}
\{[1],[h^q_{i_r}],[e^p_{\alpha_s}]:q=0,\dots, r,\ p=0,\dots,s\} \label{bassd}
\end{eqnarray}
where in this last list, $[ \ {} ]$ denotes the class in
$H(s|d)$.
\end{theorem}  

\paragraph{Proof:}
The proof is given in the appendix \ref{Des}.B.

\vspace{.2cm}

The theorem shows in particular that
the $e^0_{\alpha_s}$, just like the
$h^0_{i_r}$,  are non trivial $s$-cocycles that remain
non trivial in $H(s \vert d)$. This property
holds even though they can be lifted all the
way to maximum form-degree $n$ (while the lifts of the
$h^0_{i_r}$ are obstructed before).
The basis of $H(s \vert d)$ is given by the non trivial
bottoms $h^0_{i_r}$, $e^0_{\alpha_s}$ and all the
terms in the descent above them (up to the obstructions in 
the case of $h^0_{i_r}$).

\subsection{Descent equations with weaker assumptions on $H(d)$}
\label{weaker}

So far we have assumed that (\ref{des2}) holds.
We shall now briefly discuss the modifications
when (\ref{des2}) is replaced by an appropriate weaker prerequisite.
Applications of these modifications
are described below.

Let $\{\alpha^p_{i_p}\}$ be a set 
of $p$-forms representing
$H^p(d)$ (hence, the superscript of the $\alpha$'s indicates
the form-degree, the subscript $i_p$ labels the inequivalent 
$\alpha$'s for fixed $p$). That is, any $d$-closed $p$-form is a 
linear combination of the $\alpha^p_{i_p}$ with constant
coefficients $\lambda^{i_p}$, modulo a
$d$-exact form,
\beq
 d\omega^p=0 \quad\LRA\quad
\omega^p=\lambda^{i_p}\alpha^p_{i_p}+d \eta^{p-1}\ ,
\quad d\alpha^p_{i_p}=0,
\label{des2a}
\eeq
and no nonvanishing linear
combination of the $\alpha^p_{i_p}$ is $d$-exact.

Now, in order to derive the descent equations, it is quite crucial
that the equality
$d(s \omega ^p) = 0$ implies $s \omega ^p +
d \omega^{p-1} = 0$.  However, if $H^p(d)$ is non trivial, we must
allow for a combination of the $\alpha^p_{i_p}$ on the
right hand side, and this spoils the descent.
This phenomenon cannot occur if no 
$\alpha^p_{i_p}$ is $s$-exact modulo $d$.  
Thus, in order to be able to use the tools provided by the descent
equations, we shall  assume that the forms non trivial in $H^p(d)$
remain non trivial
in $H(s \vert d)$ for $p<n$.  More precisely,  
it is assumed that
the $\alpha^p_{i_p}$ with $p<n$ have the 
property that no nonvanishing linear
combination of them is trivial in $H(s|d)$, 
\beq
p<n:\quad
\lambda^{i_p}\alpha^p_{i_p}=s\eta^p+d\eta^{p-1} \quad\then\quad
\lambda^{i_p}=0\quad \forall\, i_p\ .
\label{des4}
\eeq
This clearly implies the central
property of the descent,
\beq
p<n:\quad d(s\omega^p)=0\quad\LRA\quad
\exists\,\omega^{p-1}:\quad s\omega^p+d\omega^{p-1}=0.
\label{des2b}
\eeq
Indeed, by (\ref{des2a}), $d(s\omega^p)=0$
implies
$s\omega^p+d\omega^{p-1}=\lambda^{i_p}\alpha^p_{i_p}$ for
some $\omega^{p-1}$ and $\lambda^{i_p}$.
(\ref{des4}) implies now $\lambda^{i_p}=0$
whenever $p<n$.

When (\ref{des4}) holds, the discussion
of the descent equations proceeds as before. The only new feature
is the fact that the
$\alpha^p_{i_p}$ yield additional nontrivial classes of $H(s|d)$
and $H(s)$. Indeed,
$d\alpha^p_{i_p}=0$ implies $d(s\alpha^p_{i_p})=0$ and thus,
due to (\ref{des2b}),
$s\alpha^p_{i_p}+d\alpha^{p-1}_{i_p}=0$ for some
$\alpha^{p-1}_{i_p}$ (which may vanish). Hence, the $\alpha^p_{i_p}$
are cocycles of $H(s|d)$ and they are nontrivial by
(\ref{des4}). In particular, some of the $\alpha^p_{i_p}$
may have a nontrivial descent.
Theorems (\ref{descth1}) and (\ref{descth2}) get modified because
the $\alpha^p_{i_p}$ and their nontrivial descendants (if any)
represent classes of $H(s|d)$ in addition to the 
$[h^q_{i_r}]$ and $[e^p_{\alpha_s}]$ ($q=0,\dots, r$, $p=0,\dots,s$),
while $H(s)$ receives additional classes represented by
nontrivial bottom forms corresponding to the $\alpha^p_{i_p}$.

\paragraph{Applications.}

1. The above discussion is important to cover
the space of local forms which are not allowed to depend explicitly on
the $x^\mu$. Indeed, in that space $H^p(d)$
is represented for $p<n$ by the constant
forms $c_{\mu_1 \dots \mu_p}
dx^{\mu_1} \dots dx^{\mu_p}$ (theorem \ref{Loct''2}). Now,
the equations of motion may be such that some of the constant forms
become trivial in $H(s|d)$.  We know that this cannot happen in 
form-degree zero, but nothing prevents it from
happening in higher form-degrees.  For instance,
for a single abelian gauge field with Lagrangian $L = (-1/4) F^{\mu \nu}
F_{\mu \nu} + k^\mu A_\mu$ with constant $k^\mu$, the
equations of motion read $\6_\nu F^{\nu\mu}+k^\mu = 0$ and imply
that the constant $(n-1)$-form $\star k=\frac 1{(n-1)!}
dx^{\mu_1}\dots dx^{\mu_{n-1}}\epsilon_{\mu_1\dots \mu_n}k^{\mu_n}$ 
is trivial in $H(s|d)$,
$s\star A^*+d\star F+(-)^n\star k=0$. 
Hence, for this Lagrangian (\ref{des4}) is not fulfilled.
Of course, the example is academic
and  the Lagrangian is not Lorentz-invariant. The triviality
of $\star k$ in $H(s|d)$ is a consequence of the linear 
term in the Lagrangian.

(\ref{des4}) is fulfilled in the space
of $x$-independent local forms for Lagrangians
having no linear part in the fields, for which
the equations of motion reduce identically to $0=0$
when the fields are set to zero.  It is also
fulfilled if one restricts one's attention to the space 
of Poincar\'e invariant local forms. [And it is also trivially
fulfilled in the space of all local forms with a possible
explicit $x$-dependence, as we have seen].  For this
reason, (\ref{des4}) does not appear to be a drastic restriction
in the space
of $x$-independent local forms.

Note that the classification
of the elements (and the number of these elements) in a basis
of $H(s)$
having the properties of theorem \ref{descth1} depends on the context.
For instance, for a single abelian gauge field with ghost $C$,
$dx^\mu C$ is non trivial in the algebra of local
forms with no explicit $x$-dependence, and so can be taken as
a $h^0_{i_1}$; but it becomes trivial if one allows for an 
explicit $x$-dependence, $dx^\mu C  = d (x^\mu  C) + s (x^\mu A)$,
and so can be regarded in that case as a $\hat{h}_{i_0}$.

2. Another instance where 
the descent equations with the above assumption on $H(d)$ play
a r\^ole is the cohomology $H(\delta|d,\cI_\chi)$, where
$\cI_\chi$ is the space of $\cG$-invariant local forms depending
only on the variables $\chi_\Delta^u$ defined in section \ref{AdCelim} and
on the $x^\mu$ and $dx^\mu$. 
$H^p(d,\cI_\chi)$ is represented for $p<n$ by the 
``characteristic classes'', i.e., by
$\cG$-invariant polynomials $P(F)$ in the
curvature 2-forms $F^I$ \cite{Brandt:1990gy,Dubois-Violette:1992ye}
(these polynomials are $d$-exact in the space 
of all local forms, but they are not $d$-exact in $\cI_\chi$).
Using this result on $H^p(d,\cI_\chi)$, one proves straightforwardly
by means of the descent equations that
$H(\delta|d,\cI_\chi)$ is isomorphic
to the characteristic cohomology in the space of gauge invariant
local forms (``equivariant
characteristic cohomology'') which
will play an important r\^ole in the
analysis of the consistency condition performed in
section \ref{solution}.

3. Finally we mention that the above discussion was
used within the computation of the local BRST cohomology
in Einstein-Yang-Mills theory \cite{Barnich:1995ap}. In that case
$H^p(d)$ is nontrivial in certain form-degrees $p<n$ due to
the nontrivial De Rham cohomology of the manifold in which the
vielbein fields take their values. The corresponding $\alpha^p_{i_p}$
fulfill (\ref{des4}) and have a nontrivial descent (in contrast,
the constant forms and the characteristic classes met in the
two instances discussed before do not descend).

\subsection{Cohomology of $s+d$}

The descent equations establish a useful relation between
$H^{*,n}(s|d)$, $H(d)$ and the cohomology of the differential 
$\tilde s$ that combines $s$ and $d$,
\beq
\tilde s=s+d.
\label{des12}
\eeq
Note that $\tilde s$ squares to zero thanks to $s^2=sd+ds=d^2=0$ and defines
thus a cohomology $H(\tilde s)$ 
in the space of formal sums
of forms with various degrees,
\[
\tilde \omega=\sum_p \omega^p\ .
\]
We call such sums total forms. Cocycles
of $H(\tilde s)$ are defined through
\beq
\tilde s\tilde \omega=0,
\label{des13}
\eeq
while coboundaries take the form $\tilde \omega
=\tilde s\tilde\eta$.
Consider now a cocycle $\tilde\omega=\sum_{p=\um}^m\omega^p$
of $H(\tilde s)$. The cocycle condition
(\ref{des13}) decomposes into
\[
d\omega^m=0,\quad s\omega^m+d\omega^{m-1}=0,\quad\dots\quad,\quad
s\omega^\um=0.
\]
These are the descent equations with top-form $\omega^m$ and
the supplementary condition $d\omega^m=0$. The
extra condition is of course automatically fulfilled if $m=n$.
Hence, assuming that Eq.\ (\ref{des4}) holds,
every cohomology class of $H^{*,n}(s|d)$
gives rise to a cohomology class of $H(\tilde s)$. Evidently,
this cohomology class of $H(\tilde s)$ is nontrivial if its
counterpart in $H^{*,n}(s|d)$ is nontrivial (since
$\tilde \omega=\tilde s\tilde\eta$ implies
$\omega^n=s\eta^n+d\eta^{n-1}$). All additional cohomology
classes of $H(\tilde s)$ (i.e.\ those without
counterpart in $H^{*,n}(s|d)$) 
correspond precisely to the cohomology classes of
$H(d)$ in form degrees $<n$ and thus to
the $\alpha_{i_p}^p$ with $p<n$ in the notation used above. 
Indeed, $d\omega^m=0$ implies
$\omega^m=\lambda^{i_m}\alpha_{i_m}^m+d\eta^{m-1}$ by (\ref{des2a}).
Furthermore, as shown above,
each $\alpha_{i_m}^m$ gives rise to a
solution of the descent equations and thus to a
representative of $H(\tilde s)$. These
representatives are nontrivial and inequivalent because of
(\ref{des4}).
This yields the following isomorphism whenever (\ref{des4}) holds:
\beq
H(\tilde s)\simeq  H^{*,n}(s|d)\oplus 
H^{n-1}(d)\oplus\cdots\oplus H^0(d).
\label{des17}
\eeq
This isomorphism can be used, in particular, to determine
$H^{*,n}(s|d)$ by computing $H(\tilde s)$.
The cohomology of $s$ modulo $d$ in lower form degrees is however not
given by $H(\tilde s)$.

\subsection{Appendix \ref{Des}.A: Proof of theorem \ref{descth1}}

In order to prove the existence of (\ref{bass}), we note first 
that the space $\Omega$ of local forms admits the decomposition
$\Omega=E_0\oplus G\oplus sG$ for some space $E_0\simeq H(s,\Omega)$
and some space $G$. 

Following \cite{Dubois-Violette:1985jb,Talon:1985dz,Henneaux:1999rp},
we define recursively differentials
$d_r:H(d_{r-1})\longrightarrow 
H(d_{r-1})$ for $r=0,\dots, n$, with
$d_{-1}\equiv s$ as follows: the space $H(d_{r-1})$ is given by
equivalence classes $[X]_{r-1}$ of elements $X\in \Omega$ such that 
there exist
$c_1,\dots, c_r$ satisfying $sX=0,dX+s
c_1=0,\dots,dc_{r-1}+sc_r=0$ (i.e., $X$ can be lifted at least
$r$ times). We define $c_0\equiv X$.
The equivalence relation $[\ ]_{r-1}$ is
defined by 
$X\sim_{r-1} Y$ iff $X-Y=sZ +d(v^0_0+\dots+v^{r-1}_{r-1})$ where
$sv_j^j+dv^{j-1}_j=0,\dots, sv^0_j=0$. $d_r$ is
defined by $d_r[X]_{r-1}=[dc_r]_{r-1}$. Let us check that this definition
makes sense.
Applying $d$ to $dc_{r-1}+sc_r=0$ gives 
$sdc_r=0$, so that $[dc_r]_{r-1}\in H(d_{r-1})$ (one can
simply choose the required $c'_1,\dots, c'_r$ to be zero because
of $d(dc_r)=0$). Furthermore,
$[dc_r]_{r-1}$ does not depend on the ambiguity in the definition of
$X,c_1,\dots,c_r$. Indeed, the ambiguity in the definition of $X$ is 
$sZ +d(v^0_0+\dots+v^{r-1}_{r-1})$, the ambiguity in the definition of
$c_1$ is $dZ+m^0_{r-1}$, where $sm^0_{r-1}=0,d
m^0_{r-1}+s m^1_{r-1}=0,\dots,dm^{r-2}_{r-1}+s
m^{r-1}_{r-1}=0$. Similarly, the ambiguity in $c_2$ is
$m^1_{r-1}+m^0_{r-2}$, where $s
m^0_{r-2}=0,dm^0_{r-2}+sm^1_{r-2}=0,\dots,
dm^{r-3}_{r-2}+sm^{r-2}_{r-2}=0$. Going on in the same way, one finds
that the ambiguity in $c_r$ is $m^{r-1}_{r-1}+\dots +m^0_0$, where $s m_j^j+d
m_j^{j-1}=0$. This means that the ambiguity in $d c_r$ is
$d(m^{r-1}_{r-1}+\dots +m^0_0)$, which is zero in $H(d_{r-1})$. (For $r=0$,
the ambiguity in $dc_0\equiv dX$ is $-sdZ$, which is zero in $H(s)$.) 
Finally, $d_r^2=0$ follows from $d^2=0$. 

The cocycle condition $d_r [X]_{r-1}=0$ reads explicitly
$d c_r=sW + d(w^0_0+\dots+w^{r-1}_{r-1})$, where 
$sw_j^j+dw^{j-1}_j=0,\dots, sw^0_j=0$. This means that $sX=0$, 
$dX+s(c_1-w^{0}_{r-1})=0$,
$d(c_1-w^{0}_{r-1})+s(c_2-w^{1}_{r-1}-w^0_{r-2})=0$, $\dots$, 
$d(c_r- w^{r-1}_{r-1}-\dots-w^0_0)+s(-W)=0$. The coboundary condition 
$[X]_{r-1}=d_r[Y]_{r-1}$ gives
$X=db_r+sZ+d(v^0_0+\dots+v^{r-1}_{r-1})$, where
$sY=0,dY+sb_1=0,\dots,db_{r-1}+sb_r=0$. The choices 
$c^\prime_1=c_1-w^{0}_{r-1},\dots, c^\prime_{r}=c_r-
w^{r-1}_{r-1}-\dots-w^0_0$, $c^\prime_{r+1}=-W$, respectively
${v^\prime}^j_j=v^j_j$ for $j=0,\dots,r-1$ and $v^r_r=b_r$,
$v^{r-1}_r=b_{r-1}$, $\dots$, $v^{1}_r=b_{1}$, $v^{0}_r=Y$, show that 
$H(d_r)$ is defined by the same equations as $H(d_{r-1})$ with $r-1$
replaced by $r$, as it should. 

Because the maximum form degree is $n$, one has $d_n\equiv 0$
and the construction stops. 

It is now possible to define spaces
$E_r\subset E_{r-1} \subset E_0\subset \Omega$ and spaces $F_{r-1}\subset
E_{r-1}\subset\Omega$ 
for $r=1,\dots,n$ such that $E_{r-1}=E_r\oplus d_{r-1}F_{r-1}\oplus
F_{r-1}$ with $E_r\simeq H(d_{r-1})$. This leads to the decomposition
\bea
E_0=E_n\oplus (d_{n-1}F_{n-1}\oplus F_{n-1})\oplus 
(d_{n-1}F_{n-1}\oplus F_{n-1})\oplus\dots\oplus(d_{0}F_{n-1}\oplus F_{0}).
\eea
Note that this decomposition involves choices of representatives $(E_i)$ and 
of supplementary subspaces $(F_i)$ and is not ``canonical''. 
But it does exist.
 
The $e^0_{\alpha_s}$ are elements of a basis of $E_n$. They 
can be lifted $s$ times to form degree $n$, i.e., 
they are of form degree $n-s$. The element $1$ also belongs to $E_n$.
The $\hat h_{i_r}$ and $h^0_{i_r}$ are 
elements of a basis
of $d_r F_r$ and $F_r$ respectively.
\qed

\subsection{Appendix \ref{Des}.B: Proof of theorem \ref{descth2}}

We verify here that if a basis (\ref{bass}) with properties
(\ref{prop1}) and (\ref{prop2}) exists, then a basis of $H(s|d)$ is given by
(\ref{bassd}). 

\paragraph{The set (\ref{bassd}) is complete.}

Suppose that $s\omega^l+d\omega^{l-1}=0,$
$s\omega^{l-1}+d\omega^{l-2}=0,\dots,$ $s\omega^0=0$ where the superscript
indicates the length of the descent (number of lifts)
rather than the form-degree.  We shall
prove that then
\beq
\omega^l=C +
\sum_{0\leq q\leq l}\sum_{r\geq l-q} \lambda^{i_r}_{q}h^{l-q}_{i_r}
+\sum_{0\leq p\leq l}\sum_{s\geq l-p} \mu^{\alpha_s}_{p}e^{l-p}_{\alpha_s}
+s\eta^l +d[\eta^{l-1}+\sum_{r\geq 0}\nu^{(l)i_r}h^r_{i_r}],\label{lenl}
\eeq
with $C$ a constant and $\eta^{-1}=0$.

To prove this, we proceed recursively in the length $l$ of the descent. 
For $l=0$, we have $s\omega^0=0$. Using
the assumption that the $h^0$'s, $\hat h$'s, $e^0$'s and the number 1
provide a basis of $H(s)$,
this gives $\omega^0= C + \sum_{r\geq 0}
\lambda^{i_r}_0 h^0_{i_r} +\sum_{s\geq 0}
\mu^{\alpha_s}_0e^0_{\alpha_s}+\sum_{r\geq 0}\nu^{(0)i_r} \hat
h_{i_r}+s\tilde\eta^0 =\sum_{r\geq 0} \lambda^{i_r}_0 h^0_{i_r} +\sum_{s\geq
0} \mu^{\alpha_s}_0 e^0_{\alpha_s}+s\eta^0 +d(\sum_{r\geq 0}\nu^{(0)i_r}
h^{r}_{i_r})$, 
where $\eta^0=\tilde\eta^0+ \sum_{r\geq 0}\nu^{(0)i_r}
h^{r+1}_{i_r}$. This is (\ref{lenl}) for $l=0$.
 
We assume now that (\ref{lenl}) holds for $l=L$. Then we have for
$l=L+1$, that in $s\omega^{L+1}+d\omega^L=0$, $\omega^L$ is given by
(\ref{lenl}) with $l$ replaced by $L$. 
Using (\ref{prop1}) and (\ref{prop2}), one gets
$d\omega_L= s[-\sum_{0\leq q\leq L}\sum_{r\geq L-q+1}
\lambda^{i_r}_{q}h^{L-q+1}_{i_r}- \sum_{0\leq p\leq l}\sum_{s\geq L-p+1}
\mu^{\alpha_s}_{p}e^{L-p+1}_{\alpha_s} -d\eta^L-\sum_{0\leq q\leq
L}\lambda^{i_{L-q}}_{q} h^{L-q+1}_{i_{L-q}}]+\sum_{0\leq q\leq
L}\lambda^{i_{L-q}}_{q}\hat h_{i_{L-q}} $. Injecting this into
$s\omega^{L+1}+d\omega^L=0$, we find, using the properties of the
basis
and the first relation of (\ref{prop1}), on the one hand that
$\lambda^{i_{L-q}}_{q}=0$, and on the other hand 
that $\omega^{L+1}=C+\sum_{0\leq
q\leq L}\sum_{r\geq L-q+1} \lambda^{i_r}_{q}h^{L-q+1}_{i_r}+ \sum_{0\leq p\leq
l}\sum_{s\geq L-p+1} \mu^{\alpha_s}_{p}e^{L-p+1}_{\alpha_s}+\sum_{r\geq 0}
\lambda^{i_r}_{L+1}h^0_{i_r} +\sum_{s\geq 0}
\mu^{\alpha_s}_{L+1}e^0_{\alpha_s}+s\eta^{L+1}+d[\eta^L+\sum_{r\geq
0}\nu^{(L+1)i_r} h^r_{i_r}]$, which is precisely 
(\ref{lenl}) for $l=L+1$.

\paragraph{The elements of (\ref{bassd}) are cohomologically independent.}

Suppose now that
\beq
C +\sum_{0\leq q\leq l}\sum_{r\geq l-q}
\lambda^{i_r}_{q}h^{l-q}_{i_r} +\sum_{0\leq p\leq l}\sum_{s\geq l-p}
\mu^{\alpha_s}_{p}e^{l-p}_{\alpha_s}=s\tilde\eta^{(l)}+d\eta^{(l)}.
\label{cotr}
\eeq
We have to show that this implies that
$\lambda^{i_r}_{q}=0$, $\mu^{\alpha_s}_{p}=0$ and $C=0$.

Again, we proceed recursively on the length of the descent.
For $l=0$, we have $C+
\sum_{r\geq 0} \lambda^{i_r}_{0}h^{0}_{i_r} +\sum_{s\geq 0}
\mu^{\alpha_s}_{0}e^{0}_{\alpha_s}=s\tilde\eta^{(0)}+d\eta^{(0)}$. Applying
$s$ and using the triviality of the cohomology of $d$, we get that
$\eta^{(0)}$ is an $s$ modulo $d$ cocycle,
$s \eta^{(0)} + d (\cdot) = 0$ (without constant since the
equations of motion are consistent).
Suppose that the descent of
$\eta^{(0)}$ stops after $l^\prime$ steps with $0\leq l^\prime\leq n-1$, i.e.,
$\eta^{(0)}\equiv\eta^{(0)l^\prime}$ with $s\eta^{(0)l^\prime}
+d\eta^{(0)l^\prime-1}=0,\dots,$ $s\eta^{(0)0}=0$ (again no constant here).
It follows that $\eta^{(0)l^\prime}$ is given by (\ref{lenl})
with $l$ replaced by $l^\prime$. Evaluating then $d\eta^{(0)l^\prime}$ in the
equation for $l=0$ and using the properties of the basis implies that
$\lambda^{i_r}_{0}=\mu^{\alpha_s}_{0} = C=0$.
 
Suppose now that the result holds for
$l=L$. If we apply $s$ to (\ref{cotr}) at $l=L+1$ and use the triviality of
the cohomology of $d$  we get $\sum_{0\leq q\leq L}\sum_{r\geq L+1-q}
\lambda^{i_r}_{q}h^{L-q}_{i_r} +\sum_{0\leq p\leq L}\sum_{s\geq L+1-p}
\mu^{\alpha_s}_{p}e^{L-p}_{\alpha_s}=s\eta^{(L+1)}+d(\ )$. The induction
hypothesis implies then that $\lambda^{i_r}_{q}=0=\mu^{\alpha_s}_{p}$ 
for $0\leq
q\leq L$ and $0\leq p\leq L$. 
This implies that the relation at $l=L+1$ reduces
to $\sum_{r\geq 0} \lambda^{i_r}_{L+1}h^{0}_{i_r} +\sum_{s\geq 0}
\mu^{\alpha_s}_{L+1}e^{0}_{\alpha_s}=s\tilde\eta^{(l)}+d\eta^{(l)}$. As we
have already shown, this implies that
$\lambda^{i_r}_{L+1}=0=\mu^{\alpha_s}_{L+1}$. \qed  

\newpage

\mysection{Cohomology in the small algebra}\label{small}

\subsection{Definition of small algebra}

The ``small algebra" $\cB$ is by definition the algebra 
of polynomials in the undifferentiated ghosts
$C^I$, the gauge field 1-forms $A^I$ and their
exterior derivatives $dC^I$ and $dA^I$.
\[
\cB=\{\mbox{polynomials in $C^I,\,A^I,\,dC^I,\,dA^I$}\}.
\]
It is stable under $d$ and $s$
($b\in\cB$ $\Rightarrow$
$db\in\cB$, $sb\in\cB$).
This is obvious for $d$ and holds
for $s$ thanks to
\bea
&s\,C^I=\frac 12\, e\,\f JKI C^KC^J,\quad
s\,A^I=-dC^I-e\,\f JKI C^JA^K,&
\nonumber\\
&s\,dC^I=-e\,\f JKI C^JdC^K,\quad
s\, dA^I=e\,\f JKI (C^KdA^J-A^JdC^K).&
\label{sB}\eea
Accordingly, the cohomological groups $H(s|d,\cB)$
of $s$ modulo $d$ in $\cB$ are well 
defined\footnote{Note that $s$ and $\gamma$ coincide in the small
algebra, which contains no antifield.}.

The small algebra $\cB$ is only
a very small subspace of the complete space
of all local forms
(in fact $\cB$ is finite dimensional
whereas the space of all local forms is infinite
dimensional).
Nevertheless it provides a good deal of the
BRST cohomology in Yang-Mills theories, in that it contains 
all the antifield-independent 
solutions of the consistency condition $sa + db =0$
that descend non-trivially,
in a sense to be made precise in section \ref{solution}.
Furthermore, it will also
be proved there that the representatives
of $H(s|d,\cB)$ remain
nontrivial in the full cohomology, with only very few possible
exceptions.

For this reason, the calculation of $H(s|d,\cB)$ is an essential
part of the calculation of $H(s|d)$ in the full algebra.
This calculation was done first in 
\cite{Dubois-Violette:1985hc,Dubois-Violette:1985jb}
(in fact in the universal algebra defined below).
 
Let us briefly outline the construction of
$H(s|d,\cB)$ before going into the details.
It is based on an analysis of the
descent equations described in Section \ref{Des}.
The central task in this approach is the explicit
construction of a particular basis of
$H(s,\cB)$ as in theorem \ref{descth1} which provides
$H(s|d,\cB)$ by theorem \ref{descth2}.
Technically it is of great help that 
the essential steps of this construction can be carried out
in a free differential algebra
associated with $\cB$. This is shown first.

\subsection{Universal algebra}

The free differential algebra
associated with $\cB$ is denoted by $\cA$. 
It was called the ``universal algebra" in 
\cite{Dubois-Violette:1985hc,Dubois-Violette:1985jb}. 
It has the same set of generators as $\cB$, but these are
not constrained by the condition coming from the
spacetime dimension that there is no exterior form
of form-degree higher than $n$.

Explicitly, $\cA$ is generated by
anticommuting variables $C_\cA^I$, $A_\cA^I$ and commuting variables
$(dC)_\cA^I$, $(dA)_\cA^I$ which correspond to 
$C^I$, $A^I$, $dC^I$ and $dA^I$ respectively,  
$\cA$ is the space of polynomials in these
variables,
\[
\cA=\{\mbox{polynomials in $C_\cA^I,\,A_\cA^I,\,
(dC)_\cA^I,\,
(dA)_\cA^I$}\}.
\] 
These variables
are subject only to the commutation/anticommutation relations
$C_\cA^I C_\cA^J = - C_\cA^J C_\cA^I$ etc but are not constrained
by the further condition that the forms are zero whenever their
form-degree exceeds $n$.  Thus, the free differential algebra
$\cA$ is independent of the spacetime dimension; hence its name
``universal".

The fundamental difference between $\cA$ and $\cB$ is that
the $A_\cA^I$,
$(dC)_\cA^I$ and $(dA)_\cA^I$ are variables by themselves,
whereas their counterparts $A^I$, $dC^I$ and $dA^I$
are composite objects
containing the differentials $dx^\mu$ and
jet space variables (fields and their derivatives),
$A^I=dx^\mu A_\mu^I$, $dC^I=dx^\mu \6_\mu C^I$, 
$dA^I=dx^\mu dx^\nu \6_\mu A_\nu^I$.

The universal algebra $\cA$ is infinite dimensional, whereas $\cB$
is finite dimensional since
the spacetime dimension bounds the form-degree of
elements in $\cB$. By contrast,
$\cA$ contains elements
with arbitrarily high degree in the $(dC)_\cA^I$
and $(dA)_\cA^I$. 

The usefulness
of $\cA$ for the computations in the small algebra
rests on the fact that 
$\cA$ and $\cB$ are isomorphic
at all form-degrees smaller than or equal to
the spacetime dimension $n$. 
To make this statement precise,
we first define a "$p$-degree" which is
the form-degree in $\cA$,
\[
p=A_\cA^I\, \frac{\6}{\6A_\cA^I}+
(dC)_\cA^I\, \frac{\6}{\6(dC)_\cA^I}+
2(dA)_\cA^I\, \frac{\6}{\6(dA)_\cA^I}\ ,
\]
and clearly coincides with the form-degree
in $\cB$ for the corresponding objects.
The $p$-degree is indicated by a superscript.
$\cA$ and $\cB$ decompose into subspaces
with definite $p$-degrees,
\[
\cB=\bigoplus_{p=0}^n \cB^p\quad ,\quad 
\cA=\bigoplus_{p=0}^\infty \cA^p\ .
\]
We then introduce the natural mappings 
$\pi^p$ from $\cA^p$ to $\cB^p$ which
just replace each generator $C_\cA^I$, $A_\cA^I$,
$(dC)_\cA^I$, $(dA)_\cA^I$ by its counterpart in
$\cB$,
\beq
\pi^p:
\left\{
\ba{rcl}
\cA^p&\longrightarrow& \cB^p\\[4pt]
a(C_\cA,A_\cA,(dC)_\cA,(dA)_\cA)&\longmapsto& 
a(C,A,dC,dA)\ .
\ea
\right.
\label{map}
\eeq
These mappings are defined for all $p$, with
$\cB^p\equiv 0$ for $p>n$.
Evidently they are surjective (each
element of $\cB^p$ is in the image of $\pi^p$).
The point is that they are also injective for
all $p\leq n$. Indeed, the kernel of $\pi^p$
is trivial for $p\leq n$ by the following lemma:
\begin{lemma}\label{free}
For all $p\leq n$, 
the image of $a^p\in\cA^p$ under $\pi^p$
vanishes if and only if
$a^p$ itself vanishes,
\beq
\forall\,p\leq n:\quad
\pi^p(a^p)=0\quad \Leftrightarrow\quad a^p=0.
\eeq
\end{lemma}

Lemma \ref{free} holds due to the fact that
the components $A_\mu^I$,
$\6_{\mu}C^I$, $\6^{}_{[\mu}A_{\nu]}^I$ of
$A^I$, $dC^I$, $dA^I$ are algebraically independent variables
in the jet space (see section \ref{Loc})
and occur in elements of $\cB$ always together with the corresponding 
differential(s).
For instance, consider 
$a^p=k_{I_1\dots I_p}A_\cA^{I_1}\dots A_\cA^{I_p}$
where the $k_{I_1\dots I_p}$ are 
constant coefficients. Without loss of generality
the $k_{I_1\dots I_p}$ are antisymmetric since the
$A_\cA^{I}$ anticommute. Hence, one has
$a^p=p!\sum_{I_i<I_{i+1}}k_{I_1\dots I_p}A_\cA^{I_1}\dots A_\cA^{I_p}$
and
$\pi^p(a^p)=p!\sum_{I_i<I_{i+1}}k_{I_1\dots I_p}A^{I_1}\dots A^{I_p}$.
Vanishing of $\pi^p(a^p)$ in the jet space requires in particular
that the coefficients of $(dx^1 A_1^{I_1})\dots (dx^p A_p^{I_p})$
vanish, for all sets $\{I_1,\dots,I_p:I_i<I_{i+1}\}$.
This requires vanishing of all coefficients $k_{I_1\dots I_p}$,
and thus $a^p=0$. The general case
is a straightforward extension of this example.

We can thus conclude:
\begin{corollary}\label{iso}
The mappings $\pi^p$ are bijective for all $p\leq n$ and
establish thus isomorphisms between
$\cA^p$ and $\cB^p$ for all $p\leq n$.
\end{corollary}

In order to use these isomorphisms,
$\cA$ is equipped with differentials $s_\cA$ and $d_\cA$
which are the counterparts of $s$ and $d$.
Accordingly,  $s_\cA$ and $d_\cA$ are defined
on the generators of $\cA$ as follows:
\beq
\ba{c||c|c}
Z_\cA & s_\cA\,Z_\cA &  d_\cA\,Z_\cA 
\\
\hline\rule{0em}{3ex}
C_\cA^I & \frac 12\, e\,\f JKI C_\cA^KC_\cA^J & (dC)_\cA^I
\\
\rule{0em}{3ex}
A_\cA^I & -(dC)_\cA^I-e\,\f JKI C_\cA^JA_\cA^K & (dA)_\cA^I
\\
\rule{0em}{3ex}
(dC)_\cA^I & -e\,\f JKI C_\cA^J(dC)_\cA^K & 0
\\
\rule{0em}{3ex}
(dA)_\cA^I & e\,\f JKI (C_\cA^K(dA)_\cA^J-A_\cA^J(dC)_\cA^K) & 0
\ea
\label{7sd}
\eeq
The definition of $s_\cA$ is extended
to all polynomials in the generators by the
graded Leibniz rule,
$s_\cA\,(ab)=
(s_\cA\, a) b+(-)^{\epsilon_a}a (s_\cA\, b)$,
and the definition of $d_\cA$ is analogously extended.
With these definitions, $s_\cA$ and $d_\cA$ 
are anticommuting differentials in $\cA$,
\[
s_\cA^2=d_\cA^2=s_\cA d_\cA+d_\cA s_\cA=0.
\]
One can therefore define $H(d_\cA,\cA)$ and $H(s_\cA,\cA)$ (and
$H(s_\cA|d_\cA,\cA)$ as well).

By construction one has
\beq
\forall\,p:\quad
\pi^p\circ s_\cA=s\circ \pi^p,\quad
\pi^{p+1}\circ d_\cA=d\circ \pi^p\quad (\mbox{on}\ \cA^p).
\label{isos}
\eeq
This just means that $\pi^p$ and $\pi^{p+1}$
map $s_\cA a^p$ and $d_\cA a^p$
to $s \pi^p(a^p)$ and $d \pi^p(a^p)$ respectively, for every
$a^p\in\cA^p$ (note that 
$s \pi^p(a^p)$ and $d \pi^p(a^p)$ vanish
for $p>n$ and $p\geq n$ respectively).

Now, $\pi^p$ can
be inverted for $p\leq n$ since it is bijective
(see corollary \ref{iso}). Hence, (\ref{isos}) gives
\[
\ba{rll}
\forall\,p\leq n:& 
s=\pi^p\circ s_\cA\circ (\pi^p)^{-1} & (\mbox{on}\ \cB^p),\\
& 
d=\pi^{p+1}\circ d_\cA\circ (\pi^p)^{-1} & (\mbox{on}\ \cB^p),\\
&
s_\cA=(\pi^p)^{-1}\circ s\circ \pi^p & (\mbox{on}\ \cA^p),\\
\forall\,p< n:&
d_\cA=(\pi^{p+1})^{-1}\circ d\circ \pi^p & (\mbox{on}\ \cA^p)\\
\ea
\]
where the last relation holds only for $p<n$ (but not for
$p=n$) since $\pi^{n+1}$ has no inverse due to
$\pi^{n+1}(\cA^{n+1})=0$.
We conclude:
\begin{corollary}\label{isoH}
$H(s,\cB^p)$ is isomorphic to
$H(s_\cA,\cA^p)$ for
all $p\leq n$, and $H(d,\cB^p)$ is isomorphic to
$H(d_\cA,\cA^p)$ for all $p< n$.
At the level of the representatives $[b^p]\in\cB^p$ and
$[a^p]\in\cA^p$ of these cohomologies,
the isomorphisms are
given by the mappings (\ref{map}),
\bea
\forall\,p\leq n:&&
H(s,\cB^p)\simeq H(s_\cA,\cA^p),\quad [b^p]=\pi^p([a^p])\ ;
\nonumber\\
\forall\,p< n:&&
H(d,\cB^p)\simeq H(d_\cA,\cA^p),\quad [b^p]=\pi^p([a^p])\ .
\eea
\end{corollary}

In the following this corollary
will be used to deduce the cohomologies
of $s$ and $d$ in $\cB$ (except for $H(d,\cB^n)$)
from their counterparts in $\cA$.

\subsection{Cohomology of $d$ in the small algebra}
\label{cohdsmall}

It is very easy to compute $H(d_\cA,\cA)$
since all generators of $\cA$
group in contractible pairs for $d_\cA$ 
given by $(C_\cA^I,(dC)_\cA^I)$
and $(A_\cA^I,(dA)_\cA^I)$, see (\ref{7sd}). 
One concludes by means of
a contracting homotopy (cf.\ Section \ref{IIBContractible}):
\begin{lemma}\label{aplA}
$H(d_\cA,\cA^p)$ vanishes for all $p>0$ and
$H(d_\cA,\cA^0)$ is represented by the constants (pure numbers),
\beq
H(d_\cA,\cA^p)=\delta^p_0\, \mathbb{R}\ .
\label{tr0a}\eeq
\end{lemma}
By corollary \ref{isoH}, this implies
\begin{corollary}\label{aplB}
$H(d,\cB^p)$ vanishes for $0<p<n$ and
$H(d,\cB^0)$ is represented by the constants,
\beq
H(d,\cB^p)=\delta^p_0\, \mathbb{R}\quad\mbox{for}\quad p<n.
\label{tr0b}\eeq
\end{corollary}

Corollary \ref{aplB} guarantees that
we can apply theorems \ref{descth1} and \ref{descth2}
of Section \ref{Des} to compute
$H(s|d,\cB)$ since (\ref{des2}) holds.

\subsection{Cohomology of $s_\cA$}\label{HsB}

The cohomology of $s_\cA$ can be derived using
the techniques of Section \ref{LieAlgebraCoho}.
One first gets rid of the exterior derivatives
of the ghosts and of the gauge potentials by
introducing a new basis of generators 
of $\cA$, which are denoted by  $\{u^I,v^I,w^i\}$ with 
\bea
&u^I=A^I_\cA\ ,\quad v^I= -(dC)_\cA^I-e\,\f JKI C_\cA^JA_\cA^K\ ,&
\nonumber\\
&\{w^i\}=\{C^I_\cA,F^I_\cA\}\ ,\quad
F^I_\cA=(dA)^I_\cA+\frac 12\,e\,\f JKI A^J_\cA A^K_\cA\ .&
\eea
Note that $v^I$ and $F^I_\cA$ replace the
former generators $(dC)_\cA^I$ and $(dA)^I_\cA$ 
respectively and that $F^I_\cA$ corresponds of course to the
field strength 2-forms $F^I=\frac 12 dx^\mu dx^\nu F_{\mu\nu}^I
=\pi^2(F^I_\cA)$.
Note also that the change of basis preserves the
polynomial structure:
$\cA$ is the space of polynomials in the new generators.

Using (\ref{7sd}), one easily verifies that
$s_\cA$ acts on the new generators according to
\bea
&s_\cA\, u^I=v^I\ ,\quad 
s_\cA\, v^I=0\ ,&
\nonumber\\
&s_\cA\, C_\cA^I =\frac 12\, e\,\f JKI C_\cA^KC_\cA^J\ ,\quad 
s_\cA\, F^I_\cA=-e\,\f JKI C_\cA^J F_\cA^K\ .&
\label{newA}\eea
The $u^I$ and $v^I$ form thus contractible
pairs for $s_\cA$ and drop therefore from
$H(s_\cA,\cA)$. Hence, $H(s_\cA,\cA)$ reduces to
$H(s_\cA,\cA_w)$ where $\cA_w$ is the space of
polynomials in the $C_\cA^I$ and $F^I_\cA$.

From (\ref{newA}) and our discussion in section \ref{LieAlgebraCoho}, 
it follows that $H(s_\cA,\cA_w)$ is nothing
but the Lie algebra cohomology of $\cG$ 
in the representation space of the polynomials in the $F^I_\cA$,
tansforming under the extension of the coadjoint representation. 
As shown in section \ref{LieAlgebraCoho},
it is generated by
the ghost polynomials $\theta_r(C_\cA)$ and by
$\cG$-invariant polynomials in the $F^I_\cA$. 

We can make the description of $H(s_\cA,\cA)$ completely precise
here,  because the space of invariant polynomials in the
$F^I_\cA$ is completely known.
Indeed, the space of
$\cG$-invariant polynomials in the $F^I_\cA$ is generated
by a finite set of such polynomials given by
\beq
f_r(F_\cA)=
\mathrm{Tr}(F_\cA^{\,m(r)})\ ,\quad
F_\cA=F_\cA^IT_I\ ,\quad r=1,\dots,R\ ,\quad R=\mathit{rank}(\cG)
\eeq
where we follow the notations of subsection \ref{primitive5}:
$r$ labels the independent Casimir operators of $\cG$,
$m(r)$ is the order of the $r$th Casimir operator, and
$\{T_I\}$ is the same matrix representation of $\cG$
used for constructing $\theta_r(C)$ in Eq.\ (\ref{thetas'}).
More precisely one has (see e.g. \cite{GHV,kobayashiIIsec12,zelobenko}):

(i) Every $\cG$-invariant polynomial in the $F_\cA^I$ is
a polynomial $P(f_1(F_\cA),\dots,f_R(F_\cA))$ in the $f_r(F_\cA)$.

(ii) A polynomial
$P(f_1(F_\cA),\dots,f_R(F_\cA))$ in the $f_r(F_\cA)$
vanishes as a function of the $F_\cA^I$
if and only if $P(f_1,\dots,f_R)$ vanishes
as a function of commuting independent variables $f_r$ 
(i.e., in the free
differential algebra of variables $f_1,\dots,f_R$).

Note that (i) states the completeness of $\{f_r(F_\cA)\}$
while (ii) states that the $f_r(F_\cA)$ are algebraically
independent. Analogous properties hold for the $\theta_r(C_\cA)$
(they generate the space of $\cG$-invariant polynomials in
the anticommuting variables $C_\cA^I$). In particular, 
the algebraic independence of the $\theta_r(C_\cA)$
and $f_r(F_\cA)$ implies that
a polynomial in the  $\theta_r(C_\cA)$ and $f_r(F_\cA)$
vanishes in $\cA$ if and only if it vanishes
already as a polynomial in the free differential algebra
of anticommuting variables $\theta_r$ and commuting variables
$f_r$. Summarizing, we have:

\begin{lemma}\label{sA}
$H(s_\cA,\cA)$ is freely generated by the $\theta_r(C_\cA)$ and
$f_r(F_\cA)$:

(i) Every $s_\cA$-closed element of $\cA$ is
a polynomial in the $\theta_r(C_\cA)$ and
$f_r(F_\cA)$ up to an $s_\cA$-exact element of $\cA$,
\bea
& s_\cA\, a=0,\quad a\in\cA\quad
\Leftrightarrow &
\nonumber\\
&
a=P(\theta_1(C_\cA),\dots,\theta_R(C_\cA),f_1(F_\cA),\dots,f_R(F_\cA))
+s_\cA\, a',\quad  a'\in\cA\ .
&
\eea

(ii) No nonvanishing polynomial
$P(\theta_1,\dots,\theta_R,f_1,\dots,f_R)$ 
gives rise to an $s_\cA$-exact
polynomial in $\cA$,
\bea
&
P(\theta_1(C_\cA),\dots,\theta_R(C_\cA),f_1(F_\cA),\dots,f_R(F_\cA))
=s_\cA\, a,\quad
a\in\cA\quad
\then
&
\nonumber\\
&
P(\theta_1,\dots,\theta_R,f_1,\dots,f_R)=0\ .
&
\eea
\end{lemma}

By lemma \ref{sA}, a basis of $H(s_\cA,\cA)$ is
obtained from a basis of all polynomials in anticommuting 
variables $\theta_r$ and commuting
variables $f_r$, $r=1,\dots,R$. Such a basis
is simply given by all monomials of the following form:
\beq
\theta_{r_1}\cdots \theta_{r_K}\,f_{s_1}\cdots f_{s_N}:\quad
K,N\geq 0\ ,\quad
r_i<r_{i+1}\ ,\quad
s_i\leq s_{i+1}\ .
\label{basis1}\eeq
Here it is understood that
$\theta_{r_1}\cdots \theta_{r_0}\equiv 1$ if $K=0$,
and $f_{s_1}\cdots f_{s_0}\equiv 1$ if $N=0$.
The requirements $r_i<r_{i+1}$ and
$s_i\leq s_{i+1}$ take the commutation relations
(Grassmann parities)
of the variables into account.

\subsection{Cohomology of $s$ in the small algebra}

(\ref{basis1}) induces  a basis
of $H(s,\cB)$ thanks to corollary \ref{isoH}.
However, this basis is not best suited
for our ultimate goal, the determination of $H(s|d,\cB)$, because
it is not split into what we called $h^0_{i_r}$, $e^0_{\alpha_s}$ and
$\hat h_{i_r}$ in theorem \ref{descth1}. 
Therefore we will now construct
a better suited basis, using
(\ref{basis1}) as a starting point. For this purpose
we order the Casimirs according to their degrees in the $F$'s,
namely, we assume that the Casimir labels
$r=1,\dots,R$ are such that, for any two such labels $r$ and $r'$,
\beq
r<r'\quad\Rightarrow\quad m(r)\leq m(r')\ .
\label{M0}
\eeq
Note that the ordering (\ref{M0}) is ambiguous if two or more
Casimir operators have the same order. This ambiguity
will not matter, i.e., any ordering
that satisfies (\ref{M0}) is suitable for our purposes.

The set of all monomials (\ref{basis1}) is now split
into three subsets. The first subset is just given by
$\{1\}$. 
The second subset contains all those
monomials which have the property that the lowest appearing Casimir label is
carried by a $\theta$. These
monomials are denoted by $M_{r_1\dots r_K|s_1\dots  s_N}(\theta,f)$,
\bea
&
M_{r_1\dots r_K|s_1\dots s_N}(\theta,f)=
\theta_{r_1}\cdots \theta_{r_K}\,f_{s_1}\cdots f_{s_N}\ ,
&
\nonumber\\ 
&
K\geq 1\ ,\ N\geq 0\ ,\ r_i<r_{i+1}\ ,\
s_i\leq s_{i+1}\ ,\ 
r_1=\mathrm{min}\{r_i,s_i\}\ .
&
\label{basis2}
\eea
Note that $r_1=\mathrm{min}\{r_i,s_i\}$ requires
$r_1\leq s_1$ if $N\neq 0$ while it does not
impose an extra condition if $N=0$ (it is already 
implied by $r_i<r_{i+1}$ if $N=0$).

The third subset contains all remaining monomials.
In these monomials at least one of the $f$'s
has a lower Casimir label than all the $\theta$'s.
Denoting the lowest occurring label again by $r_1$,
the monomials of the third set can thus be written as
\bea
&
\5N_{r_1 \dots r_K s_1\dots  s_N}(\theta,f)=
f_{r_1}\,\theta_{r_2}\cdots \theta_{r_K}\,f_{s_1}\cdots f_{s_N}\ ,
&
\nonumber\\
&
K\geq 1\ ,\ N\geq 0\ ,\ r_i<r_{i+1}\ ,\
s_i\leq s_{i+1}\ ,\ 
r_1=\mathrm{min}\{r_i,s_i\}\ .&
\label{basis3}\eea
Thus, for instance, in the case of $G = U(1) \times
SU(2)$, there are two $f$'s and two $\theta$'s, 
namely, $f_1 =
F^{u(1)}$, $f_2 = \mathrm{Tr}_\mathrm{su(2)}F^2$, 
$\theta_1 = C^\mathrm{u(1)}$
and $\theta_2 = \mathrm{Tr}_\mathrm{su(2)} C^3$. 
The monomial $\theta_1 f_2$
belongs to the second subset, i.e., it is an ``$M$", because
the label on $\theta$ is clearly smaller than that
on $f$. The monomial $\theta_2 f_1$, by contrast, belongs to
the third set.  We shall see that $\theta_1 f_2$ is non trivial
in $H(s \vert d)$ -- and can be regarded as an $h^0_{i_r}$ --, while
$\theta_2 f_1$ is $s$-exact modulo $d$ and is a $\hat h_{i_r}$
(it arises as an obstruction in the lift of $\theta_1 \theta_2$). 

We now define the following polynomials:
\beq
N_{r_1\dots r_K s_1\dots s_N}(\theta,f)=
\sum_{r:\,m(r)=m(r_1)}f_r\
\frac{\6M_{r_1\dots r_K|s_1\dots s_N}(\theta,f)}{\6\theta_r}\ .
\label{basis4}\eeq
$N_{r_1\dots r_K s_1\dots s_N}$ is
the sum
of $\5N_{r_1 \dots r_K s_1\dots  s_N}$ and 
a linear combination of $M$'s,
\[
N_{r_1\dots r_K s_1\dots s_N}(\theta,f)
=\5N_{r_1 \dots r_K s_1\dots  s_N}(\theta,f)
-\sum_{i:\,i\geq 2,\atop m(r_i)=m(r_1)}(-)^{i}
M_{r_1\dots \7r_i\dots r_K|r_is_1\dots  s_N}(\theta,f)
\]
where $\7r_i$ means omission of $r_i$.
This shows that the set of all $M$'s and $N$'s, supplemented
by the number 1, is a basis of polynomials in the 
$\theta_r$ and $f_r$ (as the same holds for the
$M$'s and $\5N$'s). Due to lemma \ref{sA}
this provides
also a basis of $H(s_\cA,\cA)$ after substituting
the $\theta_r(C_\cA)$ and $f_r(F_\cA)$ for the $\theta_r$
and $f_r$:
\begin{corollary}\label{sAbasis}
A basis of $H(s_\cA,\cA)$ is given by
\beq
\{B_\alpha\}
=\{1\, ,\, M_{r_1\dots r_K|s_1\dots  s_N}(\theta(C_\cA),f(F_\cA))\, ,\, 
N_{r_1\dots r_Ks_1\dots s_N}(\theta(C_\cA),f(F_\cA)) \}.
\label{Bs}\eeq
\end{corollary}

Again, "basis" is meant here in the cohomological sense:
(i) every $s_\cA$-closed element of $\cA$ is
a linear combination of the $B_\alpha$
up to an $s_\cA$-exact element
($s_\cA\, a=0$ $\Leftrightarrow$ $a=\lambda^\alpha B_\alpha
+s_\cA\, a'$);
(ii) no nonvanishing linear combination of
the $B_\alpha$ is $s_\cA$-exact
($\lambda^\alpha B_\alpha=s_\cA\, a$ $\Rightarrow$
$\lambda^\alpha=0\ \forall\, \alpha$).

Note that each $B_\alpha$ has a definite $p$-degree
$p_\alpha$ which equals twice its degree
in the $F^I_\cA$.
Hence, by corollary \ref{isoH}, those $B_\alpha$ with
$p_\alpha\leq n$ provide a basis of $H(s,\cB)$.
The requirement $p_\alpha\leq n$ selects those
$M_{r_1\dots r_K|s_1\dots  s_N}$ with
$\Sigma_{i=1}^{N}2m(s_i)\leq n$ and those
$N_{r_1\dots r_Ks_1\dots s_N}$ with
$2m(r_1)+\Sigma_{i=1}^N 2m(s_i)\leq n$.
We conclude:
\begin{corollary}\label{sBbasis}
A basis of $H(s,\cB)$ is given by 
$\{1\, ,\, M_a\, ,\, N_i\}$ where
\bea
\{M_a\}&\equiv&\{
M_{r_1\dots r_K|s_1\dots  s_N}(\theta(C),f(F)):
\Sigma_{i=1}^{N}2m(s_i)\leq n
\},
\nonumber\\[4pt]
\{N_i\}&\equiv&\{
N_{r_1\dots r_Ks_1\dots s_N}(\theta(C),f(F)): 
2m(r_1)+\Sigma_{i=1}^N 2m(s_i)\leq n\}.
\label{sB3}
\eea
\end{corollary}

\subsection{Transgression formulae}\label{russian}

To derive $H(s|d,\cB)$ from corollary \ref{sBbasis},
we need to construct the lifts of the elements $M$
of the previous basis.  The most expedient way to achieve this
task is to
use the celebrated ``transgression formula''
(also called Russian formula) \cite{Stora:1976kd,Stora:1983ct,Zumino:1983ew}
\beq
(s+d)\, q_r(C+A,F)=\mathrm{Tr}\left(F^{\,m(r)}\right)
=f_r(F),
\label{f}\eeq
where $C=C^IT_I$, $A=A^IT_I$, $F=F^I T_I$, and 
\beq
q_r(C+A,F)=m(r)\int_0^1 dt\, \mathrm{Tr}
\left((C+A) \left[tF+e(t^2-t)(C+A)^{2}\right]^{m(r)-1}\right).
\label{q}\eeq
Here $t$ is just an integration variable
and should not be confused with the spacetime coordinate $x^0$.
A derivation of Eqs.\ (\ref{f}) and (\ref{q}) is given 
at the end of this subsection.
$q_r(C+A,F)$ is nothing but the Chern-Simons polynomial
$q_r(A,F)$ with $C+A$ substituting for $A$. It 
fulfills $dq_r(A,F)=f_r(F)$, as can be seen from (\ref{f})
in ghost number $0$.

The usefulness of (\ref{f}) for the determination
of $H(s|d,\cB)$ rests on the fact that it
relates $\theta_r(C)$ and $f_r(F)$
via a set of equations obtained
by decomposing (\ref{f}) into parts with
definite form-degrees,
\bea
&d[\theta_r]^{2m(r)-1}=f_r(F),&
\nonumber\\[4pt]
&s[\theta_r]^p+d[\theta_r]^{p-1}=0& \mbox{for}\ 0<p<2m(r),
\nonumber\\[4pt]
&s[\theta_r]^0=0&
\label{qdes}\eea
where $[\theta_r]^p$ is the $p$-form
contained in $q_r(C+A,F)$,
\beq
q_r(C+A,F)=\sum_{p=0}^{2m(r)-1}[\theta_r]^p\ .
\label{qdec}
\eeq
The 0-form contained in $q_r(C+A,F)$ is nothing
but $\theta_r(C)$,
\bea
[\theta_r]^0 &=&
m(r)\, \mathrm{Tr}(C^{\,2m(r)-1})e^{m(r)-1}
\int_0^1 dt\, (t^2-t)^{m(r)-1}
\nonumber\\
&=& (-e)^{m(r)-1}\, \frac{m(r)!(m(r)-1)!}{(2m(r)-1)!}\,
\mathrm{Tr}(C^{\,2m(r)-1})
=\theta_r(C).
\label{threp}
\eea
Note that
$f_r(F)$ and some of the $[\theta_r]^p$
vanish in sufficiently low
spacetime dimension (when $n<2m(r)$) but that 
$q_r(C+A,F)$ never vanishes completely since it contains 
$\theta_r(C)$.
The same formulae hold in the universal algebra $\cA$, but there,
of course, none of the $f_r$ vanishes.

Consider now the polynomials 
$M_{r_1\dots r_K|s_1\dots  s_N}(q(C+A,F),f(F))$
arising from the
$M_{r_1\dots r_K|s_1\dots  s_N}(\theta,f)$ in
Eq.\ (\ref{basis2}) by substituting the
$q_r(C+A,F)$ and $f_r(F)$ for the corresponding $\theta_r$ and
$f_r$.
Analogously to (\ref{qdec}),
these polynomials
decompose into pieces of various form-degrees,
\bea
&
M_{r_1\dots r_K|s_1\dots  s_N}(q(C+A,F),f(F))
=
\sum_{p=\uns}^{\5p}[M_{r_1\dots r_K|s_1\dots  s_N}]^p\ ,&
\nonumber\\[4pt]
&\uns=\Sigma_{i=1}^K 2m(s_i),\quad
\5p=\uns+\Sigma_{i=1}^K (2m(r_i)-1)&
\label{tr11}
\eea
where some or all $[M_{r_1\dots r_K|s_1\dots  s_N}]^p$
may vanish in sufficiently low spacetime dimension.
Due to (\ref{threp}), one has
\beq
[M_{r_1\dots r_K|s_1\dots  s_N}]^\uns
=M_{r_1\dots r_K|s_1\dots s_N}(\theta(C),f(F)).
\label{tr12}
\eeq

The polynomials (\ref{tr11}) give rise to transgression equations
that generalize Eqs.\ (\ref{qdes}). These equations
are obtained by evaluating
$(s+d)M_{r_1\dots r_K|s_1\dots  s_N}(q(C+A,F),f(F))$:
one gets a sum of terms in which one of the
$q_r(C+A,F)$ in $M$ is replaced by the corresponding $f_r(F)$
as a consequence of (\ref{f}) (note that one has
$(s+d)f_r(F)=(s+d)^2q_r(C+A,F)=0$
due to $(s+d)^2=0$, i.e., $(s+d)$ acts nontrivially only
on the $q$'s contained in $M$). Hence,
$(s+d)M$ is obtained by applying the
operation $\sum_r f_r\6/\6q_r$ to $M$.
This makes it easy to identify 
the part of lowest form-degree contained in the
resulting expression:
it is
$N_{r_1\dots r_Ks_1\dots  s_N}(\theta(C),f(F))$
given in Eq.\ (\ref{basis4})
thanks to the ordering (\ref{M0}) of the
Casimir labels. One thus gets
generalized transgression equations
\bea
&s[M_{r_1\dots r_K|s_1\dots s_N}]^{\uns+2m(r_1)}
+d[M_{r_1\dots r_K|s_1\dots s_N}]^{\uns+2m(r_1)-1}=
N_{r_1\dots r_Ks_1\dots s_N}(\theta(C),f(F)),&
\nonumber\\[4pt]
& s[M_{r_1\dots r_K|s_1\dots s_N}]^{\uns+q}
+d[M_{r_1\dots r_K|s_1\dots s_N}]^{\uns+q-1}=0\quad
\mbox{for}\quad 0<q<2m(r_1),&
\nonumber\\[4pt]
& s[M_{r_1\dots r_K|s_1\dots s_N}]^\uns=0.&
\label{tr13}\eea
Note that (\ref{qdes}) is just a special case of (\ref{tr13}),
arising for $M_{r_1\dots r_K|s_1\dots s_N}(\theta,f)\equiv \theta_r$.

\paragraph{Derivation of Eqs.\ (\ref{f}) and
(\ref{q}).} The derivation is performed in the free
differential algebra $\cA$. 

As shown in subsection \ref{cohdsmall}, the cohomology of $d_\cA$ is 
trivial in the algebra ${\cal A }$ 
of polynomials in $A_\cA^I,(dA)^I_\cA,C^I_\cA,(dC)^I_\cA$. 
The contracting homotopy is explicitly given by 
$\rho=A_\cA^I\frac{\partial}{\partial (dA)^I_\cA}+
C^I_\cA\frac{\partial}{\partial (dC)^I_\cA}$. For a
$d_\cA$-cocycle $f(A_\cA^I,(dA)^I_\cA,C^I_\cA,(dC)^I_\cA)$ with
$f(0,0,0,0)=0$, 
we get
\beq
f(A_\cA^I,(dA)^I_\cA,C^I_\cA,(dC)^I_\cA)=d_\cA \int^1_0 \frac{dt}{t}\ 
[\rho f](tA_\cA^I,t(dA)^I_\cA,tC^I_\cA,t(dC)^I_\cA). 
\label{hot}
\eeq

We then consider the change of generators 
$A_\cA^I,(dA)^I_\cA\longrightarrow
A_\cA^I,F^I_\cA$ in ${\cal A}$ (while $C^I_\cA,(dC)^I_\cA$ remain
unchanged). The differential $d_\cA$ acts on $A_\cA^I$ and
$F^I_\cA$ according to 
\bea
d_\cA A_\cA^I &=& F^I_\cA-\frac 12\,e\,\f JKI A^J_\cA
A^K_\cA,\nonumber\\
d_\cA\,F_\cA^I&=&-e\, \f JKI A_\cA^JF_\cA^K.\label{ye}
\eea
For a $d_\cA$-cocycle $f(A_\cA^I,F^I_\cA)$ with $f(0,0)=0$,
the homotopy formula (\ref{hot}) gives 
\beq
f(A_\cA^I,F^I_\cA)=d_\cA \Big[ A_\cA^L\int^1_0 dt\ [
\frac{\partial f(A_\cA^I,F_\cA^I)}{\partial F^L_\cA}]
(tA_\cA^I,tF_\cA^I+\sfrac 12\,e\,(t^2-t)\f JKI A_\cA^J A_\cA^K)\Big].
\eeq
 
This formula generalizes straightforwardly to an analogous one for the 
differential 
\bea
\4s_\cA=s_\cA+d_\cA. 
\eea
Indeed, with $\4C_\cA^I=C_\cA^I+A_\cA^I$, one gets
\bea
\4s_\cA\, \4C_\cA^I&=&
\frac 12\,e\, \f JKI\4C_\cA^K\4C_\cA^J+F_\cA^I,
\nonumber\\
\4s_\cA\,F_\cA^I&=&-e\, \f JKI \4C_\cA^JF_\cA^K. 
\label{yet}\eea
Comparing (\ref{ye}) and (\ref{yet}), we 
see that the differential algebras
$(d_\cA,\Lambda(A_\cA^I,F^I_\cA))$ and $(\4s_\cA,\Lambda(
\4C_\cA^I,F_\cA^I))$ are isomorphic.
It follows that 
\beq
f(\4C_\cA^I,F^I_\cA)=\4s_\cA\Big[\4C_\cA^L\int^1_0 dt\ 
[\frac{\partial f(\4C_\cA^I,F_\cA^I)}{\partial F^L_\cA}]
(t\4C_\cA^I,tF_\cA^I+\sfrac 12\,e\,(t^2-t)\f JKI \4C_\cA^J \4C_\cA^K)\Big],
\eeq
for an $\4s_\cA$-cocycle $f(\4C_\cA^I,F^I_\cA)$ with $f(0,0)=0$. Taking
$f=\mathrm{Tr} F^{m(r)}_\cA$, this yields (\ref{f}) and
(\ref{q})
through the mappings (\ref{map}).

\subsection{$H(s|d)$ in the small algebra}
\label{sdsmall}

We are now in the position to determine $H(s|d,\cB)$. 
Namely, Eqs.\ (\ref{tr13}) imply
that the basis of $H(s,\cB)$ given in
corollary \ref{sBbasis} has the properties
described in theorems \ref{descth1} and
\ref{descth2} of section \ref{Des}.
To verify this, consider Eqs.\ (\ref{tr13})
first in the cases $\uns+2m(r_1)\leq n$.
In these cases, Eqs.\ (\ref{tr13}) reproduce Eqs.\
(\ref{prop1}) in theorem \ref{descth1}, with the
identifications 
\beann
\uns+2m(r_1)\leq n:&&
\\
\7h_{i_r}&\equiv&N_{r_1\dots r_Ks_1\dots s_N}(\theta(C),f(F))
\\
h^q_{i_r}&\equiv&[M_{r_1\dots r_K|s_1\dots s_N}]^{\uns+q},
\quad q=0,\dots,r
\\
r&=&2m(r_1)-1.
\eeann
In particular this yields $h^0_{i_r}\equiv
M_{r_1\dots r_K|s_1\dots s_N}(\theta(C),f(F))$
for $\uns+2m(r_1)\leq n$
by Eq.\ (\ref{tr12}).

Next consider Eqs.\ (\ref{tr13}) in the cases
$n-2m(r_1)<\uns\leq n$.
This reproduces Eqs.\
(\ref{prop2}) in theorem \ref{descth1}, with
the identifications 
\beann
n-2m(r_1)<\uns\leq n:&&
\\
e^q_{\alpha_s}&\equiv&[M_{r_1\dots r_K|s_1\dots s_N}]^{\uns+q},
\quad q=0,\dots,s
\\
s&=&n-\uns.
\eeann
In particular this yields
$e^0_{\alpha_s}\equiv
M_{r_1\dots r_K|s_1\dots s_N}(\theta(C),f(F))$
for $n-2m(r_1)<\uns\leq n$.

Hence, the basis of $H(s,\cB)$ given in
corollary \ref{sBbasis} has indeed the properties
described in theorem \ref{descth1}. By 
theorem \ref{descth2}, a basis of
$H(s|d,\cB)$ is thus given by the 
$[M_{r_1\dots r_K|s_1\dots s_N}]^p$ specified in
the above equations. The whole set
of these representatives
can be described more compactly through
$p=\uns,\dots,\overline{m}$ where
$\overline{m}=\mathrm{min}\{\uns+2m(r_1)-1,n\}$.

We can summarize the result in the form of a receipe.
Given the gauge group $G$ and the spacetime
dimension $n$, one obtains $H(s|d,\cB)$ as follows:
\ben
\item 
Specify the independent Casimir operators of $G$ and
label them such that
\beq
r<r'\quad\Rightarrow\quad m(r)\leq m(r')
\label{small1}
\eeq
where $m(r)$ is the order of the $r$th Casimir operator.
\item 
Specify the following monomials:
\bea
&
M_{r_1\dots r_K|s_1\dots s_N}(\theta,f)=
\theta_{r_1}\cdots \theta_{r_K}\,f_{s_1}\cdots f_{s_N}\ :
&
\nonumber\\ 
&
K\geq 1\ ,\ N\geq 0\ ,\ r_i<r_{i+1}\ ,\
s_i\leq s_{i+1}\ ,
&
\nonumber\\ 
&
r_1=\mathrm{min}\{r_i,s_i\}\ ,\
\sum_{i=1}^N2m(s_i)\leq n\ .
&
\label{small2}
\eea
\item
Replace in (\ref{small2})
the $\theta_r$ and $f_r$ by the corresponding
$q_r(C+A,F)$ and $f_r(F)$ given in (\ref{q}) and (\ref{f})
and decompose the resulting polynomials in the
$q_r(C+A,F)$ and $f_r(F)$ into pieces of
definite form-degree,
\beq
M_{r_1\dots r_K|s_1\dots  s_N}(q(C+A,F),f(F))
=
\sum_p\, [M_{r_1\dots r_K|s_1\dots  s_N}]^p\ .
\label{small3}
\eeq
\item
A basis of $H(s|d,\cB)$ is then given by 
the number 1 and the following
$[M_{r_1\dots r_K|s_1\dots  s_N}]^p$:
\bea
[M_{r_1\dots r_K|s_1\dots  s_N}]^p\ :&&
p=\uns,\dots,\overline{m}\ ,
\nonumber\\
&&
\uns=\Sigma_{i=1}^N2m(s_i),
\nonumber\\
&&
\overline{m}=\mathrm{min}\{\uns+2m(r_1)-1,n\}.
\label{small4}
\eea
\een

A similar results hold in the universal algebra $\cA$, but in this
case, there is no $e^0_{\alpha_s}$ but only $h^0_{i_r}$: all
lifts are obstructed at some point.  This implies, in particular,
that any solution of the consistency condition in $\cA$ can
be seen as coming from an obstruction living above.  For instance,
the Adler-Bardeen-Bell-Jackiw anomaly in four
dimensions, which is a four-form, comes from the six-form
$\mathrm{Tr}F^3_\cA$ through the Russian formula.  This makes sense only
in the universal algebra, although the anomaly itself is meaningful
both in $\cA$ and $\cB$.

\subsection{$H^{0,n}(s|d)$ and $H^{1,n}(s|d)$ in the small algebra}
\label{g=01}

Physically important representatives of $H(s|d,\cB)$
are those
with form-degree $n$ and ghost numbers 0 or 1
as they provide possible counterterms and gauge anomalies
respectively. To extract these
representatives from Eqs.\ (\ref{small1}) through (\ref{small4}),
one uses that $q_r(C+A,F)$ and $f_r(F)$ have total degree
(= form-degree + ghost number)
$2m(r)-1$ and $2m(r)$ respectively.
The total
degree of $[M_{r_1\dots r_K|s_1\dots  s_N}]^p$
is thus 
$\sum_{i=1}^K(2m(r_i)-1)+\sum_{i=1}^N2m(s_i)=
\uns+\sum_{i=1}^K(2m(r_i)-1)$.
Hence, representatives with 
form-degree $n$ and ghost number $g$ fulfill
\[
n+g=\uns+\sum_{i=1}^K(2m(r_i)-1).
\]
Furthermore, representatives with 
form-degree $n$ fulfill
\[ n\leq \uns+2m(r_1)-1\]
because of the requirement $\overline{m}=\mathrm{min}\{\uns+2m(r_1)-1,n\}$
in Eq.\ (\ref{small4}).
Combining these two conditions, one gets
\beq
\sum_{i=2}^K(2m(r_i)-1)\leq g\ .
\label{gcond}
\eeq

Note that here the sum runs from $2$ to $K$, and that
we have $K\geq 1$ by (\ref{small2}).
Hence, for $g=0$,
(\ref{gcond}) selects the value $K=1$.
The representatives of $H^{0,n}(s|d,\cB)$ arise
thus from (\ref{small3}) by setting $K=1$ and selecting
the ghost number 0 part. These representatives are
\beq
[M_{r|s_1\dots  s_N}]^{\uns+2m(r)-1}=
[\theta_r]^{2m(r)-1}f_{s_1}(F)\cdots f_{s_N}(F)\ .
\label{g=0}
\eeq
Note that (\ref{small2}) imposes
$r\leq s_1\leq s_2\leq\dots\leq s_N$ if $N\neq 0$.
$[\theta_r]^{2m(r)-1}$ is nothing but the Chern-Simons form
corresponding to $f_r(F)$, see equation (\ref{qdes}).
(\ref{g=0}) is thus a Chern-Simons form too, corresponding to
$f_r(F)f_{s_1}(F)\cdots f_{s_N}(F)$.
All representatives (\ref{g=0}) have odd form-degree and
occur thus only in odd spacetime dimensions.

For $g=1$,
(\ref{gcond}) leaves two possibilities: $K=1$, or
$K=2$ where the latter case requires in addition $m(r_2)=1$. 
For $K=1$, this yields the following representatives
of $H^{1,n}(s|d,\cB)$,
\beq
[M_{r|s_1\dots  s_N}]^{\uns+2m(r)-2}=
[\theta_r]^{2m(r)-2}f_{s_1}(F)\cdots f_{s_N}(F)\ .
\label{g=1a}
\eeq
Again, (\ref{small2}) imposes
$r\leq s_1\leq s_2\leq\dots\leq s_N$ if $N\neq 0$.
By (\ref{q}) one has
\[
[\theta_r]^{2m(r)-2}=\mathrm{Tr}(CF^{\,m(r)-1}+\dots).
\]
All representatives (\ref{g=1a}) have even
form-degree. They represent the consistent chiral 
gauge anomalies in even spacetime dimensions.

The remaining representatives 
of $H^{1,n}(s|d,\cB)$ have $K=2$
and $m(r_2)=1$. $m(r_2)=1$ requires $m(r_1)=1$
by (\ref{small1}) and (\ref{small2}). 
The Casimir operators of order 1 are the abelian
generators (see Section \ref{primitive5}). 
The corresponding
$q_r(C+A,F)$ coincide with the abelian $C+A$,
\beq
\{q_r(C+A,F):m(r)=1\}=\{\mbox{abelian}\ C^I+A^I\}.
\label{abelq}
\eeq
The representatives of $H^{1,n}(s|d,\cB)$ 
with $K=2$ read
\beq
[M_{IJ|s_1\dots  s_N}]^{\uns+1}=
(C^I A^J-C^J A^I) f_{s_1}(F)\cdots f_{s_N}(F)\quad
(\mbox{abelian}\, I,J).
\label{g=1b}
\eeq
(\ref{small2}) imposes $I<J$, $s_i\leq s_{i+1}$ and that 
$f_{s_1}(F)$ is not an abelian $F^K$ with $K<I$.
Note that the representatives (\ref{g=1b}) have odd form-degree and
are only present if the gauge group contains at least two
abelian factors.
They yield candidate gauge anomalies in odd spacetime dimensions.

\subsection{Examples}

To illustrate the results, we
shall now spell out $H^{0,n}(s|d,\cB)$ and $H^{1,n}(s|d,\cB)$
for specific choices of $n$ and $G$. We list
those $\theta_r(C)$ (up to the normalization factor)
and $f_r(F)$ needed to construct  
$H^{0,n}(s|d,\cB)$ and $H^{1,n}(s|d,\cB)$ and give a complete
set of the
inequivalent representatives (``Reps.'')
of these cohomological groups and 
the corresponding obstructions (``Obs.'') in the
universal algebra $\cA$ (except that in the last example
we leave it to the reader to spell out the obstructions
as it is similar to the second example).
The inclusion of $SO(1,9)$ and $SO(1,10)$ in the last two
examples is relevant in the 
gravitational context because the Lorentz group plays
a r\^ole similar to the Yang-Mills gauge group when one
includes gravity in the analysis.
\smallskip

\underline{$n=4$, $G=U(1)\times SU(2)\times SU(3)$}
\[
\ba{c||c|c|c|c}
r    & 1 & 2 & 3 & 4 \\
\hline\rule{0em}{3ex}
m(r) & 1 & 2 & 2 & 3 \\
\rule{0em}{3ex}
\theta_r(C) & C^\mathrm{u(1)} & \mathrm{Tr}_\mathrm{su(2)}C^3 &
\mathrm{Tr}_\mathrm{su(3)}C^3 & \mathrm{Tr}_\mathrm{su(3)}C^5 \\
\rule{0em}{3ex}
f_r(F) & F^\mathrm{u(1)} & \mathrm{Tr}_\mathrm{su(2)}F^2 &
\mathrm{Tr}_\mathrm{su(3)}F^2 & 0
\ea
\]
\beann
H^{0,4}(s|d,\cB):&& \mbox{empty}
\\[4pt]
H^{1,4}(s|d,\cB):&& 
\ba[t]{c||c|c|c|c}
\mathrm{Reps.} &
C^\mathrm{u(1)}(F^\mathrm{u(1)})^2 &
C^\mathrm{u(1)}f_2(F) & 
C^\mathrm{u(1)}f_3(F) & 
[\theta_3]^4
\\
\hline\rule{0em}{3ex}
\mathrm{Obs.} &
(F_\cA^\mathrm{u(1)})^3&
F_\cA^\mathrm{u(1)}f_2(F_\cA)&
F_\cA^\mathrm{u(1)}f_3(F_\cA)&
f_4(F_\cA)
\ea
\\[6pt]
\mbox{where}&&  
[\theta_3]^4=\mathrm{Tr}_\mathrm{su(3)}[Cd(AdA+\sfrac 12 
e_\mathrm{su(3)} A^3)],\
f_4(F_\cA)=\mathrm{Tr}_\mathrm{su(3)}(F_\cA)^3
\eeann
\smallskip

\underline{$n=10$, $G=SO(32)$}
\[
\ba{c||c|c|c}
r    & 1 & 2 & 3 \\
\hline\rule{0em}{3ex}
m(r) & 2 & 4 & 6 \\
\rule{0em}{3ex}
\theta_r(C) & \mathrm{Tr}C^3 & 
              \mathrm{Tr}C^7 & 
              \mathrm{Tr}C^{11} \\
\rule{0em}{3ex}
f_r(F) &
              \mathrm{Tr}F^2 & 
              \mathrm{Tr}F^4 & 
              0
\ea
\]
\beann
H^{0,10}(s|d,\cB):&& \mbox{empty}
\\[4pt]
H^{1,10}(s|d,\cB):&&
\ba[t]{c||c|c|c}
\mathrm{Reps.} &
[\theta_1]^2 (f_1(F))^2 & 
[\theta_1]^2 f_2(F)&
[\theta_3]^{10}
\\
\hline\rule{0em}{3ex}
\mathrm{Obs.} &
(f_1(F_\cA))^3 &
f_1(F_\cA)f_2(F_\cA) &
f_3(F_\cA)
\ea
\\[6pt]
\mbox{where}&&
[\theta_1]^2=\mathrm{Tr}(CdA),\ 
[\theta_3]^{10}=\mathrm{Tr}[C(dA)^5+\dots],\\
&&
f_3(F_\cA)=\mathrm{Tr}(F_\cA)^6
\eeann
\smallskip

\underline{$n=11$, $G=SO(1,10)$}
\[
\ba{c||c|c|c}
r    & 1 & 2 & 3 \\
\hline\rule{0em}{3ex}
m(r) & 2 & 4 & 6 \\
\rule{0em}{3ex}
\theta_r(C) & \mathrm{Tr}C^3 & 
              \mathrm{Tr}C^7 & 
              \mathrm{Tr}C^{11} \\
\rule{0em}{3ex}
f_r(F) &
              \mathrm{Tr}F^2 & 
              \mathrm{Tr}F^4 & 
              0
\ea
\]
\beann
H^{0,11}(s|d,\cB):&& 
\ba[t]{c||c|c|c}
\mathrm{Reps.} &
[\theta_1]^3 (f_1(F))^2 &
[\theta_1]^3 f_2(F) &
[\theta_3]^{11}
\\
\hline\rule{0em}{3ex}
\mathrm{Obs.} &
(f_1(F_\cA))^3 &
f_1(F_\cA)f_2(F_\cA) &
f_3(F_\cA)
\ea
\\[6pt]
\mbox{where}&&
[\theta_1]^3=\mathrm{Tr}( AdA+\sfrac 13 e A^3),\ 
[\theta_3]^{11}=\mathrm{Tr}[A(dA)^5+\dots],
\\
&& f_3(F_\cA)=\mathrm{Tr}(F_\cA)^6
\\[6pt]
H^{1,11}(s|d,\cB):&&  \mbox{empty}
\eeann
\smallskip

\underline{$n=10$, $G=SO(1,9)\times SO(32)$} 
\[
\ba{c||c|c|c|c|c|c}
r    & 1 & 2 & 3 & 4 & 6 & 7 \\
\hline\rule{0em}{3ex}
m(r) & 2 & 2 & 4 & 4 & 6 & 6 \\
\rule{0em}{3ex}
\theta_r(C) & \mathrm{Tr}_\mathrm{so(1,9)}C^3 & 
              \mathrm{Tr}_\mathrm{so(32)}C^3 &
              \mathrm{Tr}_\mathrm{so(1,9)}C^7 & 
              \mathrm{Tr}_\mathrm{so(32)}C^7 &
              \mathrm{Tr}_\mathrm{so(1,9)}C^{11} & 
              \mathrm{Tr}_\mathrm{so(32)}C^{11} \\
\rule{0em}{3ex}
f_r(F) &
              \mathrm{Tr}_\mathrm{so(1,9)}F^2 & 
              \mathrm{Tr}_\mathrm{so(32)}F^2 &
              \mathrm{Tr}_\mathrm{so(1,9)}F^4 & 
              \mathrm{Tr}_\mathrm{so(32)}F^4 &
              0 & 0
\ea
\]
\beann
H^{0,10}(s|d,\cB):&& \mbox{empty}
\\[4pt]
H^{1,10}(s|d,\cB)\ (\mbox{Reps.}):&&
[\theta_1]^2 (f_1(F))^2,\ 
[\theta_1]^2 f_1(F)f_2(F),\ 
[\theta_1]^2 (f_2(F))^2,
\\
&&
[\theta_1]^2 f_3(F),\ [\theta_1]^2 f_4(F),
\\
&&
[\theta_2]^2 (f_2(F))^2,\
[\theta_2]^2 f_3(F),\ [\theta_2]^2 f_4(F),
\\
&&
[\theta_6]^{10},\ [\theta_7]^{10}
\eeann
Remark: the Pfaffian of $SO(1,9)$ yields
$f_5(F)$ with $m(5)=5$; however it
does not contribute to
$H^{1,10}(s|d,\cB)$ in this case (it would contribute
through $C^\mathrm{u(1)}f_5(F)$
if $G$ contained in addition a $U(1)$).

\newpage

\mysection{General solution of the consistency condition in
Yang-Mills type theories}\label{solution}
\subsection{Assumptions} \label{AssUMPtions}

We shall now put the pieces together and determine
$H(s|d,\Omega)$
completely in Yang-Mills 
type theories for two cases:
\bea
\mbox{Case I:}&&\Omega=\{\mbox{all local forms}\},
\nonumber\\
\mbox{Case II:}&&\Omega=\{\mbox{Poincar\'e-invariant local forms}\},
\label{I+II}
\eea
on the following assumptions:

(a) the gauge group does not contain
abelian gauge symmetries under which
all matter fields are uncharged -- we shall call such
special abelian symmetries ``free abelian gauge
symmetries'' in the following;

(b) the spacetime dimension (denoted by $n$) is larger than 2;

(c) the theory is normal and the regularity conditions hold;

(d) in case II it is assumed that the Lagrangian itself is 
Poincar\'e-invariant
(in case I it need not be Poincar\'e-invariant).

Assumption (c) is a technical one and has been
explained in Sections \ref{KoszulSection} and \ref{Koszul2Section}.
Assumptions (a) and (b) reflect special properties of
free abelian gauge symmetries and 2-dimensional (pure) Yang-Mills 
theory which complicate somewhat the general analysis. 
These special properties of free abelian gauge symmetries
are illustrated and dealt with in Section \ref{Free} where we compute
the cohomology for a set of
free abelian gauge fields.
Two-dimensional pure Yang-Mills theory is treated separately in
the appendix to this Section. 
Note that assumption (a) does not exclude
abelian gauge symmetries. It excludes only
the presence of free abelian gauge fields, or of
abelian gauge fields that couple exclusively non-minimally to matter
or gauge fields (i.e., through the field strengths and their derivatives
only). 

The space of all local forms is the direct product 
$\cP\otimes \Omega(\mathbb{R}^n)$ where $\cP$ is the
space of local functions of
the fields, antifields, and all their derivatives,
while $\Omega(\mathbb{R}^n)$ is the space of ordinary differential
forms $\omega(x,dx)$ in $\mathbb{R}^n$.
Depending on the context and Lagrangian, $\cP$
can be, for instance, the space of polynomials
in the fields, antifields, and all their derivatives
(when the Lagrangian is polynomial too), or
it can be the space of local forms that depend 
polynomially
on the derivatives of the fields and antifields but may depend
smoothly on (some of) the undifferentiated fields (when
the Lagrangian has the same property). The latter would be the case 
for instance for Yang-Mills theory coupled to a dilaton. It can also
be the space of formal power series in the
fields, antifields, and all their derivatives, with coefficients
that depend on the coupling constants (in the case of effective theories).
More generally speaking, the results and their derivation apply whenever
the various cohomological results derived and discussed in the previous
sections (especially
those on $H(s)$ and $H_\mathrm{char}(d)$)
hold, since these
will be used within the computation below.

The space of Poincar\'e-invariant local forms is
a subspace of $\cP\otimes \Omega(\mathbb{R}^n)$.
It contains only those local forms which do not depend explicitly
on the spacetime coordinates $x^\mu$ and are Lorentz-invariant.
Lorentz-invariance requires here simply that all Lorentz indices
(including the indices of derivatives and differentials, and
the spinor indices of spacetime fermions) are contracted in an
$SO(1,n-1)$-invariant manner. 

Two central ingredients in our computation of $H(s \vert d)$
are the results on the characteristic
cohomology of $d$ in Section \ref{Koszul2Section} and
on the cohomology of $s$ in Section \ref{LieAlgebraCoho}. 
We shall repeat them here, for case I and case II,
and reformulate the result on $H(s)$ since we shall frequently use
it in that formulation.
For the characteristic cohomology of $d$ we have:
\begin{corollary}
\label{LEM0}
$H^p_\mathrm{char}(d,\Omega)$ vanishes at all form-degrees $0<p<n-1$
and is at form-degree 0 represented by the constants,
\bea
0<p<n-1:&& d\omega^p\approx 0,\ \omega^p\in\Omega\quad
\LRA\quad \omega^p\approx d\omega^{p-1},\ \omega^{p-1}\in\Omega\ ;
\nonumber\\
p=0:&& d\omega^0\approx 0,\ \omega^0\in\Omega\quad
\LRA\quad \omega^0\approx \mathit{constant}.
\eea
\end{corollary}
In case I this follows directly from the results
of Section \ref{Koszul2Section} thanks to assumptions (a), (b) and (c),
where assumptions (a) and (b) are only needed for the
vanishing of $H^{n-2}_\mathrm{char}(d,\Omega)$ 
($H^{n-2}_\mathrm{char}(d,\Omega)$ does not vanish when
free abelian gauge symmetries are present, and is
not exhausted by the constants in 2-dimensional
pure Yang-Mills theory; this makes these two cases special).
The analysis and results of Section \ref{Koszul2Section}
extend to case II 
because both $d$ and $\delta$ 
are Lorentz invariant according to the definition
of Lorentz invariance used here (i.e., $d$ and $\delta$ commute
in $\Omega$ with $SO(1,n-1)$-rotations of
all Lorentz indices).
This is obvious for $d$ and holds for $\delta$
thanks to assumption (d) because that assumption
guarantees the Lorentz-covariance of the equations of motion. 

The results on $H(s)$ in section \ref{LieAlgebraCoho} can
be reformulated as follows:
\begin{corollary}
\label{LEM1}
Description of $H(s,\Omega)$:
\bea
& s\omega=0,\ \omega\in\Omega\quad \LRA\quad 
\omega=I^\tin \Theta_\tin+s\eta,
\ I^{\tin} \in\cI, \ \eta\in\Omega\ ; &
\label{LEM1A}\\[4pt]
& I^\tin \Theta_\tin=s\omega,\ I^{\tin} \in\cI,\ \omega\in\Omega
\quad \LRA\quad I^\tin\approx 0\ \forall\tin. &
\label{LEM1B}\eea
\end{corollary}
Here, $\{\Theta_\tin\}$ is a basis of all polynomials
in the $\theta_r(C)$,
\beq
\{\Theta_\tin\}=\{1, \prod_{i=1,\dots,K \atop r_i< r_{i+1}}\theta_{r_i}(C):
K=1,\dots,\mathit{rank}(\cG)\},
\label{Theta}
\eeq 
and
$\cI$ is the antifield independent
gauge invariant subspace of $\Omega$ given in the
two cases under study respectively by
\bea
\mbox{Case I:}&&\cI=\{\mbox{$\cG$-invariant local functions of
$F_{\mu\nu}^I$, $\psi^i$, $D_\rho F_{\mu\nu}^I$, $D_\rho\psi^i$, \dots}\}
\otimes \Omega(\mathbb{R}^n)
\nonumber\\
\mbox{Case II:}&&\cI=\{\mbox{$\cG$-invariant and
Lorentz-invariant
local functions of}
\nonumber\\
&& \phantom{\cI=\{}
\mbox{$dx^\mu$, $F_{\mu\nu}^I$, $\psi^i$, $D_\rho F_{\mu\nu}^I$,
$D_\rho\psi^i$, \dots}\}.
\label{I}
\eea 
Again, the precise definition of ``local functions''
depends on the context and Lagrangian.

Let us now briefly explain how and why corollary \ref{LEM1} 
reformulates the results in Section \ref{LieAlgebraCoho}.
We first treat case I. The results in Section \ref{LieAlgebraCoho} 
give (using a notation as in that section)
$H(s,\Omega)=\Omega(\mathbb{R}^n) \otimes V^X_{\rho=0} \otimes
\Lambda(C)_{\rho^C=0}$ where $V^X_{\rho=0}$
is the space of $\cG$-invariant local functions of the $X_A^u$
specified in Section \ref{Xvariables},
and $\Lambda(C)_{\rho^C=0}$ is the space of polynomials in the
$\theta_r(C)$. Now, one has
$\Omega(\mathbb{R}^n) \otimes V^X_{\rho=0}\subset \cI$
and thus $\Omega(\mathbb{R}^n) \otimes V^X_{\rho=0} \otimes
\Lambda(C)_{\rho^C=0} \subset \cI\otimes \Lambda(C)_{\rho^C=0}$.
This yields the implication $\then$ in (\ref{LEM1A}).
The implication $\Leftarrow$ holds because
all elements
of $\cI\otimes \Lambda(C)_{\rho^C=0}$ are $s$-closed.
Furthermore, since $\Omega(\mathbb{R}^n) \otimes V^X_{\rho=0}$ is  
only a subspace of $\cI$, one has $\cI=
(\Omega(\mathbb{R}^n) \otimes V^X_{\rho=0})\oplus 
(\Omega(\mathbb{R}^n) \otimes V^X_{\rho=0})^\top$. 
By the results in Section 
\ref{LieAlgebraCoho}, all nonvanishing elements of 
$\Omega(\mathbb{R}^n) \otimes V^X_{\rho=0}\otimes \Lambda(C)_{\rho^C=0}$ 
are nontrivial in $H(s,\Omega)$ whereas all elements of 
$(\Omega(\mathbb{R}^n) \otimes V^X_{\rho=0})^\top
\otimes \Lambda(C)_{\rho^C=0}$ are trivial in $H(s,\Omega)$
and vanish on-shell. This gives (\ref{LEM1B}).

We now turn to case II. Thanks to assumption (d), both
$\delta$ and $\gamma$ commute
with Lorentz transformations. Therefore the results of 
Section \ref{LieAlgebraCoho} hold analogously in the
Lorentz-invariant subspace of $\cP\otimes \Omega(\mathbb{R}^n)$.
This gives in case II
$H(s,\Omega)=V^{X,dx}_{\rho=0} \otimes
\Lambda(C)_{\rho^C=0}$ where $V^{X,dx}_{\rho=0}$ is the
space of Lorentz-invariant and $\cG$-invariant local functions 
of the $X_A^u$ and $dx^\mu$. Corollary \ref{LEM1}
holds now by reasons analogous to case I.

Finally we shall often use the following immediate consequence
of the isomorphism (\ref{74}) (cf.\ proof of that isomorphism):
\begin{corollary}
\label{LEM2} 
An $s$-cocycle with nonnegative
ghost number is $s$-exact whenever 
its antifield independent part vanishes on-shell,
\[
s\omega=0,\ \omega\in\Omega,\ \gh(\omega)\geq 0,\
\omega_0\approx 0\quad \then\quad \omega=s K,\ K\in\Omega
\]
where $\omega_0$ is the antifield independent part of $\omega$.
\end{corollary}

\subsection{Outline of the derivation and result}\label{Strategy}

We shall determine the
general solution of the consistency condition
\beq
s\omega^p+d\omega^{p-1}=0,\quad \omega^p,\omega^{p-1}\in\Omega
\label{CC}\eeq
for all values of the form-degree $p$ and of the ghost number
(the ghost number will not be made explicit throughout this section,
contrary to the form-degree).
Since the precise formulation and derivation of the result are
involved, we shall first outline
the crucial steps of the computation and describe the
various nontrivial solutions. 
The precise formulation of the result and its proof
will be given in the following Section \ref{Main}.

The computation relies on 
the descent equation technique described in Section \ref{Des},
which can be used because
$d$ has trivial cohomology
at all form-degrees different from 0 and $n$,
\beq
H^p(d,\Omega)=\delta^p_0\, \mathbb{R}\quad \mbox{for}\quad p<n. 
\label{APLomega}
\eeq
This holds in the space of all local forms (case I) by
the algebraic Poincar\'e lemma (theorem \ref{Loct2}). It also
holds in the space of Poincar\'e invariant local forms
because $d$ is Lorentz-invariant (cf.\ text after corollary \ref{LEM0}).
In fact, one may even
deduce this directly from the proof
of the algebraic Poincar\'e lemma given in Section \ref{Loc}
because the operators $\rho$ and $P_m$ used there 
are manifestly Lorentz-invariant,
and because there are no Lorentz-invariant constant forms
$c_{\mu_1\dots {\mu_p}}dx^{\mu_1}\dots dx^{\mu_p}$ with
form degree $0<p<n$.
In fact, that proof of the algebraic Poincar\'e lemma
is not just an existence proof but
provides an explicit construction
of $\omega^{p-1}$ for given $d$-closed $\omega^p$
such that $\omega^p=d\omega^{p-1}$, both in case I and
case II.

One distinguishes between
solutions with a trivial descent
and solutions with a nontrivial descent.

\subsubsection{Solutions with a trivial descent.} 
These are solutions
to (\ref{CC}) that
can be redefined by the addition of trivial solutions such that
they solve $s\omega^p=0$. 
The result on $H(s)$ (corollary \ref{LEM1}) implies then 
that these
solutions have the form
\beq
\omega^p=I^{p\,\tin} \Theta_\tin\ ,\quad I^{p\,\tin} \in\cI
\label{(a)}\eeq
modulo trivial solutions. We note that (\ref{(a)}) can be trivial in
$H(s|d)$ even if $I^{p\,\tin}\not\approx 0$. 

\subsubsection{Lifts and equivariant characteristic cohomology}
The solutions with a non trivial descent are those for which it is 
impossible to make $\omega^{p-1}$
vanish in (\ref{CC}) through the addition of trivial solutions.
Their determination is more difficult. In particular it 
calls for the solution of a cohomological problem
that we have not discussed so far.
Namely, one has to determine
the characteristic cohomology in the space $\cI$ defined in (\ref{I}).
This cohomology is well defined because $d\cI\subset \cI$. Indeed, 
for  $I\in \cI$, one has $d I=DI$, where 
$D=dx^\mu D_\mu$, with $D_\mu$ the covariant derivative on the fields
and $D_\mu x^\nu=\delta^\nu_\mu$. Furthermore 
$DI$ is $\cG$-invariant.
We call this cohomology the ``equivariant characteristic cohomology''
and denote it by $H_\mathrm{char}(d,\cI)$.
It is related to, but different from
the ordinary characteristic cohomology 
discussed in Section \ref{Koszul2Section}. 

To understand the difference,
assume that $I\in\cI$ is weakly $d$-exact in $\Omega$, i.e.,
$I\approx d\omega$ for some $\omega\in\Omega$. The 
equivariant characteristic cohomology poses the following question:
is it possible to choose $\omega\in\cI$ ? We shall
answer this question in the affirmative 
with the exception when $I$ contains
a ``characteristic class''. Abusing slightly
standard terminology, a characteristic class is in this context
a $\cG$-invariant polynomial in the curvature 2-forms $F^I$
and is thus, in particular, an element of the small algebra.%
\footnote{It is a semantic coincidence that
the word ``characteristic'' is used in the literature
both in the context of the
polynomials $P(F)$ and to term the weak cohomology of $d$.
The invariant cohomology of $d$ without antifields and use
of the equations of motion has been investigated in
\cite{Brandt:1989rd,Brandt:1990gy,Dubois-Violette:1992ye}.
The fact that it contains only the characteristic classes
has been called the ``covariant
Poincar\'e lemma" in \cite{Brandt:1989rd,Brandt:1990gy}.}
Furthermore, we shall show that no characteristic class 
with form-degree $<n$ is trivial
in $H_\mathrm{char}(d,\cI)$ (at form-degree $n$ there may be
exceptions).

Hence,
$H_\mathrm{char}(d,\cI)$ is the sum of a subspace of 
$H_\mathrm{char}(d,\Omega)$ (given by
$H_\mathrm{char}(d,\Omega)\cap \cI$) and of the space of 
characteristic classes.
Using corollary \ref{LEM0},
one thus gets that $H_\mathrm{char}(d,\cI)$ is at all
form-degrees $<n-1$ solely represented by characteristic classes
(at form-degree $n-2$ this is due to assumptions (a) and (b)).
In contrast, there are in general
additional nontrivial representatives of $H_\mathrm{char}(d,\cI)$ at
form-degrees $n$ and $n-1$.%
\footnote{An exception is 3-dimensional pure 
Chern-Simons theory (with semisimple gauge group)
where $H_\mathrm{char}(d,\cI)$ vanishes even in form-degrees
$n=3$ and $n-1=2$, see Section \ref{CS}.}
At form-degree $n$ they are present because
an $n$-form is automatically $d$-closed but not necessarily weakly
$d$-exact. The additional representatives
at form-degree $n-1$ are gauge-invariant nontrivial
Noether currents written as $(n-1)$-forms. 

The result on
$H_\mathrm{char}(d,\cI)$ is interesting 
in itself and a cornerstone
of the local BRST cohomology in Yang-Mills type theories.
The technical assumption of ``normality" assumed throughout the
calculation is made in order to be able to characterize
completely $H_\mathrm{char}(d,\cI)$.
The properties of $H_\mathrm{char}(d,\cI)$
are at the origin of the importance of the small
algebra for the cohomology, as we now explain.

The equivariant characteristic cohomology arises as follows when
discussing the descent equations.
One has
\[ s\omega^p+d\omega^{p-1}=0\ ,\ 
s\omega^{p-1}+d\omega^{p-2}=0\ ,\ \dots\ ,\ 
s\omega^{\um}=0.\]
Without loss of generality, one can assume that
the bottom form $\omega^\um$ is a nontrivial solution
of the consistency condition. It is thus a solution with
a trivial descent and can be taken of the form
\beq
\omega^\um=I^{\um\,\tin} \Theta_\tin\ ,\quad I^{\um\,\tin} \in\cI\ .
\label{rough1}\eeq 
In the case of a nontrivial descent, $\omega^\um$
satisfies additionally
\beq
s\omega^{\um+1}+d\omega^\um=0
\label{rough2}\eeq 
which is the last but one descent equation. It turns
out that this equation is a very restrictive condition
on the bottom $\omega^{\um}$.  Few bottoms can be lifted
at least once.  In order to be ``liftable", all $I^{\um\,\tin}$
must be representatives of the equivariant characteristic
cohomology.  

Indeed, one has
\beq
d(I^{\um\,\tin} \Theta_\tin)=
(dI^{\um\,\tin}) \Theta_\tin-s(I^{\um\,\tin} [\Theta_\tin]^1)
\label{rough3}\eeq
where we used $sI^{\um\,\tin}=0$
(which holds due to $I^{\um\,\tin}\in\cI$) and
\beq
d\Theta_\tin+s[\Theta_\tin]^1=0,\quad
[\Theta_\tin]^1=A^I\,\frac{\6\Theta_\tin}{\6 C^I}\ .
\label{rough0}\eeq
(\ref{rough0}) is nothing but the
equation with $q=1$ contained in Eqs.\ (\ref{tr13}), for the
particular case
$M_{r_1\dots r_K|s_1\dots s_N}\equiv
M_{r_1\dots r_K}$.

Now, (\ref{rough1}) through (\ref{rough3}) imply
$(dI^{\um\,\tin}) \Theta_\tin=s(\dots)$. By corollary \ref{LEM1} this
implies that all $dI^{\um\,\tin}$ vanish weakly,
\beq
\forall\,\tin:\quad dI^{\um\,\tin}\approx 0.
\label{rough4}\eeq
Furthermore, if $I^{\um\,\tin}\approx dI^{\um-1}$ for some $I^{\um-1}\in\cI$
and some $\tin$, the piece $I^{\um\,\tin} \Theta_\tin$ (no sum over
$\tin$ here) can be removed by subtracting a
trivial term from $\omega^\um$. Indeed, $I^{\um\,\tin}\approx dI^{\um-1}$
implies that $I^{\um\,\tin}= dI^{\um-1}+sK^\um$ for some $K^\um$ by
corollary \ref{LEM2} and thus that $I^{\um\,\tin} \Theta_\tin$ is trivial,
$I^{\um\,\tin} \Theta_\tin
=d(I^{\um-1}\Theta_\tin)+s(I^{\um-1}[\Theta_\tin]^1+K^\um \Theta_\tin)$ 
(no sum over $\tin$).
Hence, without loss of generality one can assume that
\beq
\forall\,\tin:\quad 
I^{\um\,\tin}\not\approx dI^{\um-1\,\tin},\quad  
I^{\um-1\,\tin}\in\cI\ .
\label{rough5}\eeq

By Eqs.\ (\ref{rough4}) and (\ref{rough5}), 
every $I^{\um\,\tin}$ is a nontrivial representative of the
equivariant characteristic cohomology.
This cohomology
qualifies thus to some extent the
bottom forms that appear in nontrivial descents:
all those $I^{\um\,\tin}$ with $\um<n-1$ can be assumed to
be characteristic classes $P(F)$, while those with $\um=n-1$ can
additionally contain nontrivial gauge-invariant
Noether currents (in the special case $n=2$ or when
free gauge symmetries are present, there can be
bottom forms of yet another type with form-degree $n-2$).

\subsubsection{Solutions with a nontrivial descent}
Of course, the previous discussion gives 
not yet a complete characterization of the bottom forms
because
$I^{\um\,\tin}\Theta_\tin$ may be trivial in $H(s|d,\Omega)$ even 
when all $I^{\um\,\tin}$ represent nontrivial classes of
the equivariant characteristic cohomology
(the nontriviality of $I^{\um\,\tin}$ in 
$H_\mathrm{char}(d,\cI)$ is necessary but not sufficient
for the nontriviality of $I^{\um\,\tin}\Theta_\tin$). 
Furthermore one still
has to investigate how far the nontrivial bottom forms
can be maximally lifted (so far we have only discussed
lifting bottom forms once).
Nevertheless the above discussion gives already an idea
of the result. Namely,
one finds ultimately that the consistency condition
has at most three types of solutions with a nontrivial descent:
\ben
\item
Solutions which lie in the small algebra $\cB$.
These solutions are linear combinations of
those $[M_{r_1\dots r_K|s_1\dots  s_N}]^p$
in (\ref{small4}) with
$\sum_{i=1}^N2m(s_i) <p$
(the solutions with $\sum_{i=1}^N2m(s_i)=p$ are
$s$-closed and have thus a trivial descent). 
Here and in the following,
such linear combinations are denoted
by $B^p$,
\bea
B^p=\sum_{\uns= p-2m(r_1)+1}^{p-1}
\lambda^{r_1\dots r_K|s_1\dots  s_N}
[M_{r_1\dots r_K|s_1\dots  s_N}]^p,\quad
\uns=\sum_{i=1}^N2m(s_i)
\label{Bp}
\eea
where the $\lambda^{r_1\dots r_K|s_1\dots  s_N}$ 
are constant coefficients. We shall prove that
no nonvanishing $B^p$ is trivial in $H(s|d,\Omega)$. 
In other words: $B^p$ is trivial only if all coefficients
$\lambda^{r_1\dots r_K|s_1\dots  s_N}$ vanish (because 
the $[M_{r_1\dots r_K|s_1\dots  s_N}]^p$ are linearly independent,
see Section \ref{small}).
The solutions $B^p$ descend to bottom forms
involving characteristic classes $P(F)$.
\item
Antifield dependent solutions which involve
nontrivial global symmetries corresponding to 
gauge invariant nontrivial Noether currents.
These solutions cannot be given explicitly in a model
independent manner because the
set of global symmetries is model dependent.
To describe them we introduce the notation $\{j^\mu_\Delta\}$
for a basis of those Noether
currents which can be 
brought to a form such that
the corresponding $(n-1)$-forms are elements of $\cI$
(possibly by the addition of
trivial currents, see Section \ref{Koszul2Section}),
\beq
j_\Delta=\frac{1}{(n-1)!}\,
dx^{\mu_1}\dots dx^{\mu_{n-1}}
\epsilon_{\mu_1\dots\mu_n}\,j^{\mu_n}_\Delta\in\cI\ ,
\quad d j_\Delta\approx 0.
\label{Delta5}\eeq
``Basis'' means here that (a) 
every Noether current which has a representative in $\cI$
is a linear combination of the $j^\mu_\Delta$ up
to a current which is trivial in $\Omega$, and 
(b) no nonvanishing linear combination of
the $j_\Delta$ is trivial in $H_\mathrm{char}(d,\Omega)$,
\bea
&&dI^{n-1}\approx 0,\quad I^{n-1}\in\cI\quad \then\quad 
I^{n-1}\approx \lambda^\Delta j_\Delta+d\omega^{n-2}\ ;
\label{jbasis1}\\
&& \lambda^\Delta j_\Delta\approx d\omega^{n-2}
\quad \then\quad \lambda^\Delta=0\quad\forall\,\Delta \ .
\label{jbasis2}\eea
Note that, in general, $\{j_\Delta\}$ does not represent a complete basis of
$H^{n-1}_\mathrm{char}(d,\cI)$ because our definition does not 
use the coboundary condition in $H^{n-1}_\mathrm{char}(d,\cI)$
($\{j_\Delta\}$ does not contain
characteristic classes $P(F)$ while $H^{n-1}_\mathrm{char}(d,\cI)$
may contain characteristic classes when $n$ is odd).
Note also that, in general,
$\{j^\mu_\Delta\}$ differs in case I and II.
For instance,
the various components ${T_0}^\mu$, \dots , ${T_{n-1}}^\mu$
of the energy momentum tensor ${T_\nu}^\mu$ can normally
be redefined such that they are gauge invariant\footnote{An
example where ${T_0}^\mu$ cannot be made gauge invariant
is example 3 in Section \ref{examplecurrents}.} and provide
then $n$ elements of $\{j_\Delta\}$ in case I; however,
they are not contravariant Lorentz-vectors
and therefore
they do not provide elements of $\{j_\Delta\}$
in case II. Similarly,
in globally supersymmetric Yang-Mills models, the supersymmetry 
currents provide normally elements of $\{j_\Delta\}$
in case I, but not in case II.

Since $j_\Delta$ is gauge invariant, one has $d j_\Delta=D j_\Delta$
and therefore $d j_\Delta$ is gauge invariant and
$s$-invariant too.
By corollary \ref{LEM2}, $d j_\Delta\approx 0$ implies thus the existence
of a volume form $K_\Delta$ such that
\beq
sK_\Delta+dj_\Delta=0.
\label{Delta3ex}\eeq
$K_\Delta$ encodes the global symmetry corresponding to $j_\Delta$
(see Section \ref{Koszul2Section} 
and also Section \ref{homological}
for the connection between $s$ and $\delta$).
We now define the $n$-forms
\beq
V_{\Delta \tin}= K_\Delta \Theta_\tin
+j_\Delta [\Theta_\tin]^1
\label{V}\eeq
with $[\Theta_\tin]^1$ as in (\ref{rough0}).
These forms solve the consistency condition. Indeed,
Eqs.\ (\ref{rough0}) and (\ref{Delta3ex}) give immediately
\beq
sV_{\Delta \tin}+d\, [j_\Delta\Theta_\tin]=0.
\label{Delta3}
\eeq 
This also shows that $j_\Delta\Theta_\tin$ is a
bottom form corresponding to $V_{\Delta \tin}$ (as one
has $s(j_\Delta\Theta_\tin)=0$ due to $j_\Delta\in\cI$).
\item
In special cases (i.e., for special Lagrangians), there are
additional nontrivial antifield dependent solutions. They
emerge from nontrivial conserved currents which
can {\em not} be made gauge invariant
by the addition of trivial currents.
These ``accidental" solutions complicate the derivation
of the general solution of the consistency condition but,
even though exceptional, they must be covered since we do not make
restrictions on the Lagrangian besides the technical ones
explained above.

We shall prove that, if $n>2$ and free abelian gauge symmetries
are absent, such currents exist if and only if
characteristic classes with maximal form-degree $n$ 
are trivial in $H_\mathrm{char}(d,\cI)$
(examples are given in Section \ref{DisDisDis}).
Assume that $\{P_A(F)\}$ is a basis for
characteristic classes of this type, i.e., assume that
every characteristic class with
form-degree $n$ which is trivial in $H_\mathrm{char}(d,\cI)$
is a linear combination of the $P_A(F)$ and the $P_A(F)$ are
linearly independent,
\bea
&&
P_A(F)\approx dI^{n-1}_A,\quad  I^{n-1}_A\in\cI\ ;
\label{P_A1}\\
&&
P^n(F)\approx dI^{n-1},\quad I^{n-1}\in\cI\quad
\then\quad P^n=\lambda^A P_A(F)\ ;
\label{P_A2}\\
&&
\lambda^A P_A(F)=0\quad\then\quad \lambda^A=0\quad \forall\,A\ .
\label{P_A3}
\eea
Note that $P_A(F)\approx dI^{n-1}_A$ implies 
$d(I^{n-1}_A-q_A^{n-1})\approx 0$
where $q_A^{n-1}$ is a Chern-Simons $(n-1)$-form fulfilling
$P_A(F)=dq_A^{n-1}$. The $I^{n-1}_A-q_A^{n-1}$ are thus conserved
$(n-1)$-forms; they are the afore-mentioned
Noether currents that cannot be
made gauge invariant.
$P_A(F)-dI^{n-1}_A$ is $s$-invariant (as it is in
$\cI$) and vanishes weakly; hence,
corollary \ref{LEM2} guarantees the existence
of a volume form $K_A$ such that
\beq
P_A(F)=dI^{n-1}_A+sK_A\ .
\label{P_A7}\eeq
This gives $sK_A+d(I^{n-1}_A-q_A^{n-1})=0$,
i.e. $K_A$ is a cocycle of $H^{-1,n}(s|d)$ 
(it contains the global symmetry corresponding to
the Noether current $I^{n-1}_A-q_A^{n-1}$).

Every linear combination of $n$-forms $P_A(F)\Theta_\tin$ belongs 
to the cohomology of 
$s$ in the small algebra and can therefore be expanded in the 
$N_i$ and $M_a$ in (\ref{sB3}). We now consider only those linear 
combinations which can be written solely in terms of $N_i$
(which must have form degree $n$).
A basis for these linear combinations is denoted by $\{N_\Gamma\}$:
\bea
&&
N_\Gamma=k^i_\Gamma N_i=
k^{A\tin}_\Gamma\, P_A(F)\, \Theta_\tin\ ;
\label{P_A4}\\
&&
\lambda^i N_i=\lambda^{A\tin}\, P_A(F)\, \Theta_\tin\quad
\LRA\quad \lambda^i N_i=\lambda^\Gamma N_\Gamma\ ;
\label{P_A5}\\
&&
\lambda^\Gamma N_\Gamma=0\quad\then\quad \lambda^\Gamma=0
\quad\forall\,\Gamma\ .
\label{P_A6}
\eea
In particular one can choose the basis $\{N_\Gamma\}$ 
such that it contains $\{P_A(F)\}$ (since
$\{N_i\}$ contains a basis of all $P(F)$).

Now, on the one hand, one gets
\bea
N_\Gamma&=&k^{A\tin}_\Gamma\, (dI^{n-1}_A+sK_A)\, \Theta_\tin
\nonumber\\[4pt]
&=&k^{A\tin}_\Gamma\, \left[ d(I^{n-1}_A\Theta_\tin)
+s(I^{n-1}_A[\Theta_\tin]^1+K_A \Theta_\tin)\right]
\label{P_A8}\eea
where we used (\ref{P_A7}) and (\ref{rough0}).
On the other hand, $N_\Gamma$ is a linear combination of those
$N_i$ with form-degree $n$ and thus of the form
\beq
N_\Gamma=sb^n_\Gamma+dB^{n-1}_\Gamma\ ,\quad 
b^n_\Gamma\, ,\, B^{n-1}_\Gamma\in\cB
\label{P_A9}\eeq
by the first equation in (\ref{tr13})
[$b^n_\Gamma$ is a linear combination of 
the $[M_{r_1\dots r_K|s_1\dots s_N}]^{n}$
with $n=2m(r_1)+\sum_{i=1}^N2m(s_i)$ and is therefore not
of the form (\ref{Bp}), in contrast to
$B^{n-1}_\Gamma$].
Subtracting (\ref{P_A8}) from (\ref{P_A9}), one gets
\beq
sW_\Gamma+d[B^{n-1}_\Gamma-k^{A\tin}_\Gamma\, I^{n-1}_A\Theta_\tin]=0
\label{QGamma2}\eeq
where
\beq
W_\Gamma=
b^n_\Gamma-k^{A\tin}_\Gamma\, (K_A\, \Theta_\tin
+I^{n-1}_A[\Theta_\tin]^1)\ .
\label{VGamma}\eeq
$W_\Gamma$ descends to a bottom form in the small algebra which
involves characteristic classes $P(F)$, namely to the same bottom form
to which $B^{n-1}_\Gamma$ descends.
From the point
of view of the descent equations this means the following.
In the small algebra, the bottom-form corresponding to $B^{n-1}_\Gamma$
can only be lifted to form-degree $(n-1)$; there is
no way to lift it to an $n$-form in the small algebra because 
this lift
is obstructed by $N_\Gamma$. However, 
there is no such obstruction in the full algebra
because the characteristic
classes contained in $N_\Gamma$ are trivial
in $H^n_\mathrm{char}(d,\cI)$ ($W_\Gamma$ is not entirely
in the small algebra: in particular it contains antifields
through the $K_A$).

\een

\subsection{Main result and its proof}\label{Main}

According to the discussion in Section \ref{Strategy} it may appear natural
to determine first $H_\mathrm{char}(d,\cI)$ and afterwards
$H(s|d,\Omega)$. In fact that strategy was followed in
previous computations \cite{Barnich:1995db,Barnich:1995mt}.
However, it is more efficient to determine
$H_\mathrm{char}(d,\cI)$ and $H(s|d,\Omega)$ at a stroke.
The reason is that these two cohomologies
are strongly interweaved. In fact, not only does
one need $H_\mathrm{char}(d,\cI)$ to compute
$H(s|d,\Omega)$; but also,
$H_\mathrm{char}(d,\cI)$ at form-degree $p$
can be computed by means of $H(s|d,\Omega)$ at
lower former degrees using descent equation techniques. 
That makes it possible
to determine both cohomologies simultaneously in a recursive
manner, 
starting at form-degree 0 where $H(s|d,\Omega)$
reduces to $H(s,\Omega)$,
and then proceeding successively to higher form-degrees.
This strategy streamlines the derivation as compared to previously
used approaches (but reaches of course identical 
conclusions!), and is reflected
in the formulation and proof of the theorem given below.
The theorem is formulated such that it
applies both to case I and to case II; however, 
the different
meaning of $\Omega$, $\cI$ (and thus also $j_\Delta$ and $V_{\Delta\tin}$) 
in these cases should be kept in mind.

To formulate and prove the theorem, we use the same notation as in
Section \ref{Strategy}. In addition we introduce, similarly
to (\ref{Bp}), the notation
$M^p$, $N^p$ and $b^p$ for linear combinations of the
$[M_{r_1\dots r_K|s_1\dots s_N}]^{\uns}$,
$N_{r_1\dots r_Ks_1\dots s_N}(\theta(C),f(F))$,
and $[M_{r_1\dots r_K|s_1\dots s_N}]^{\uns+2m(r_1)}$
with form-degree $p$ respectively,
\bea
M^p&\equiv&
\sum_{\uns=p}\lambda^{r_1\dots r_K|s_1\dots  s_N}
M_{r_1\dots r_K|s_1\dots  s_N}(\theta(C),f(F))\ ,
\label{Mp}\\
N^p&\equiv&
\sum_{\uns+2m(r_1)=p}\lambda^{r_1\dots r_K s_1\dots  s_N}
N_{r_1\dots r_K s_1\dots  s_N}(\theta(C),f(F))
\label{Np}\\
b^p&\equiv&
\sum_{\uns+2m(r_1)=p}\lambda^{r_1\dots r_K| s_1\dots  s_N}
[M_{r_1\dots r_K|s_1\dots s_N}]^{\uns+2m(r_1)}
\label{bp}\eea
where we used once again the notation
$\uns=\sum_{i=1}^N2m(s_i)$.
Note that we have, for every $B^p$ as in (\ref{Bp}),
\bea
sB^p=-d(B^{p-1}+M^{p-1}),\quad dB^p=-s(B^{p+1}+b^{p+1})+N^{p+1}
\label{sdB}
\eea
for some
$B^{p-1}$, $M^{p-1}$, $B^{p+1}$, $b^{p+1}$ and $N^{p+1}$
by Eqs.\ (\ref{tr13}).

We can now formulate the result as follows.

\begin{theorem}\label{main}
Let $\omega^p\in\Omega$ with $\Omega$ as in (\ref{I+II}),
$I^p,I^{p\,\tin}\in\cI$ with $\cI$ as in (\ref{I}).
Let $P^p(F)$ denote characteristic classes 
($p$ indicating the form-degree respectively),
and $\sim$ denoting equivalence in $H(s|d,\Omega)$
(i.e. $\omega'{}^p\sim \omega^p$ means $\omega'{}^p=
\omega^p+s\eta^p+d\eta^{p-1}$ for some $\eta^p,\eta^{p-1}\in\Omega$).
For Yang-Mills type theories without free
abelian gauge symmetries,
the following statements hold in all spacetime dimensions
$n>2$:

(i) At all form-degrees $p<n$,
the general solution of the consistency condition 
is given, up to trivial solutions,
by the sum of a term $I^{\tin}\Theta_\tin$
and a solution in the small algebra
as in Eq.\ (\ref{Bp});
at form-degree $p=n$ it
contains in addition 
a linear combination of the $V_{\Delta \tin}$ and
$W_\Gamma$ given in (\ref{V}) and (\ref{VGamma}) respectively,
\beq
s\omega^p+d\omega^{p-1}=0\ \LRA\ 
\omega^p\sim I^{p\,\tin}\Theta_\tin+B^p
+\delta^p_n\,(\lambda^{\Delta \tin}V_{\Delta \tin}
+\lambda^\Gamma W_\Gamma).
\label{soli}
\eeq

(ii) $I^{p\,\tin}\Theta_\tin+B^p
+\delta^p_n\,(\lambda^{\Delta \tin}V_{\Delta \tin}
+\lambda^\Gamma W_\Gamma)$ 
is trivial in $H(s|d,\Omega)$ if and only
if $B^p$ and all coefficients $\lambda^{\Delta \tin}$,
$\lambda^\Gamma$ vanish and
$I^{p\,\tin}\Theta_\tin$
is weakly equal to
$N^p+(dI^{p-1\,\tin})\Theta_\tin$ for some $N^p$
and $I^{p-1\,\tin}$,%
\footnote{Note that
this requires $I^{p\,\tin}\approx dI^{p-1\,\tin}
+P^{p\,\tin}(F)$ with $P^{p\,\tin}(F)$ such that
$P^{p\,\tin}(F)\Theta_\tin=N^p$.}
\bea
&&
I^{p\,\tin}\Theta_\tin+B^p
+\delta^p_n\,(\lambda^{\Delta \tin}V_{\Delta \tin}
+\lambda^\Gamma W_\Gamma)\sim 0
\nonumber\\[4pt]
&\LRA &
B^p=0,\
\lambda^{\Delta \tin}=0\ \forall\, (\Delta,\alpha),\
\lambda^\Gamma=0\ \forall\,\Gamma,\
%\nonumber\\[4pt]
%&&
I^{p\,\tin}\Theta_\tin\approx N^p+(dI^{p-1\,\tin})\Theta_\tin .
\quad
\label{solii}
\eea

(iii) If $I^p\in\cI$ ($p>0$) is trivial in $H^p_\mathrm{char}(d,\Omega)$
then it is the sum of a characteristic class
and a piece which is trivial in $H^p_\mathrm{char}(d,\cI)$,
\beq
p>0:\quad I^p\approx d\omega^{p-1}\ \LRA\ I^p\approx 
P^p(F)+d I^{p-1}.
\label{soliii}
\eeq

(iv) No nonvanishing characteristic class with form-degree
$p<n$ is trivial in $H_\mathrm{char}(d,\cI)$,
\beq
p<n:\quad P^p(F)\approx d I^{p-1}\quad\then\quad
P^p(F)=0.
\label{soliv}
\eeq
\end{theorem}

\noindent
{\bf Proof.} 

Step 1. To prove the theorem, we first verify that
(i), (ii) and (iv) hold at form-degree 0. 
There are no $B^0$ or $N^0$ since
the ranges of values for $\uns$ in the sums in Eqs.\ (\ref{Bp})
and (\ref{Np}) are empty for $p=0$. 
Hence, for $p=0$, (i) and (ii) reduce to
$s\omega^0=0\LRA\omega^0= I^{0\tin}\Theta_\tin+s\eta^0$ and
$I^{0\tin}\Theta_\tin= s\eta^0\LRA
I^{0\tin}\approx 0$ respectively 
and hold by the results on $H(s)$ (corollary \ref{LEM1}).
(iv) reduces for $p=0$ to 
$\mathit{constant}\approx 0\LRA \mathit{constant}=0$
which holds for every meaningful Lagrangian (if it would not
hold then the equations of motion were inconsistent, see
Section \ref{Des}).

Step 2. In the second (and final) step we show that (i) through (iv)
hold for $p=m$ if they hold for $p=m-1$, excluding
$m=n$ in the case (iv). 
\ben
\item[(iv)] 
By corollary \ref{LEM2},
$P^m(F)\approx dI^{m-1}$ implies $P^m(F)=dI^{m-1}+sK^m$
for some local $K^m$.
On the other hand one has $P^m(F)=dq^{m-1}$ for some Chern-Simons 
$(m-1)$-form $q^{m-1}$ which we choose
to be the $B^{m-1}$ corresponding to $P^m(F)$.\footnote{Note that 
there can be an ambiguity in the choice
of Chern-Simons forms. Indeed, consider $P^m(F)=f_1(F)f_2(F)$.
One has $P^m(F)=d[\alpha q_1(A,F)f_2(F)+(1-\alpha)q_2(A,F)f_1(F)]$
where $\alpha$ is an arbitrary number.
Our prescription in Section \ref{small} selects 
$B^{m-1}=q_1(A,F)f_2(F)$ ($\alpha=1$). 
This extends to
all characteristic classes $P^m(F)$: our prescription 
selects precisely one $B^{m-1}$
among all Chern-Simons forms corresponding to $P^m(F)$.}
This gives $sK^m+d(I^{m-1}-q^{m-1})=0$,
i.e., $K^m$ is a cocycle of $H^{-1,m}(s|d,\Omega)$.
One has $H^{-1,m}(s|d,\Omega)\simeq H^{m}_1(\delta|d,\Omega)
\simeq H^{m-1}_0(d|\delta,\Omega)=H^{m-1}_\mathrm{char}(d,\Omega)$
for $m>1$ and analogously $H^{-1,1}(s|d,\Omega)\simeq 
H^{0}_\mathrm{char}(d,\Omega)/\mathbb{R}$
by theorems \ref{Loctdec} and \ref{Loct6}.
By corollary \ref{LEM0} this gives $H^{-1,m}(s|d,\Omega)=0$
(since we are assuming $0<m<n$) and thus
$K^m\sim 0$. This implies $I^{m-1}-q^{m-1}\sim 0$ by the standard
properties of the descent equations\footnote{\label{foot1}If 
one of the forms
in the descent equations is trivial, then all its descendants
are trivial too, see Section \ref{Des}.}. 
Now, since we assume that (ii) holds for $p=m-1$,
we conclude from $I^{m-1}-q^{m-1}\sim 0$ in particular that $q^{m-1}=0$
(since $q^{m-1}$ is a $B^{m-1}$)
and thus that $P^m(F)=dq^{m-1}=0$ which is (iv) for $p=m<n$.
\item[(i)]
$s\omega^{m}+d\omega^{m-1}=0$ implies descent equations 
(Section \ref{Des}).
In particular there is some $\omega^{m-2}$ such that
$s\omega^{m-1}+d\omega^{m-2}=0$.
Since we assume that (i) holds for $p=m-1$,
we conclude
\beq
\omega^{m-1}=I^{m-1\,\tin}\Theta_\tin+B^{m-1}
\label{soli0}
\eeq
for some
$I^{m-1\,\tin}$ and $B^{m-1}$ (without loss of generality,
since trivial contributions to any form in the descent equations
can be neglected, see Section \ref{Des}).
By (\ref{sdB}) we have
\beq
dB^{m-1}=-s(\7B^m+b^m)+N^m
\label{soli1a}
\eeq
for some $\7B^m$, $b^{m}$, $N^m$.
Using in addition (\ref{rough0}), we get
\[
d\omega^{m-1}=(dI^{m-1\,\tin})\Theta_\tin
-s(I^{m-1\,\tin}[\Theta_\tin]^1+\7B^{m}+b^{m})+N^{m}.
\]
Inserting this in $s\omega^{m}+d\omega^{m-1}=0$,
we obtain 
\beq
s(\omega^{m}-I^{m-1\,\tin}[\Theta_\tin]^1
-\7B^{m}-b^{m})+[dI^{m-1\,\tin}+P^{m\, \tin}(F)]\Theta_\tin=0
\label{soli1}
\eeq
where we used that 
\beq
N^m=P^{m\, \tin}(F)\Theta_\tin
\label{soli1b}
\eeq
for some $P^{m\, \tin}(F)$.
Using corollary \ref{LEM1}, we conclude
from (\ref{soli1}) that
\beq
dI^{m-1\,\tin}+P^{m\, \tin}(F)\approx 0\quad \forall \tin.
\label{soli2}
\eeq
To go on, we must distinguish the cases $m<n$ and $m=n$.
\medskip

\underline{$m<n$}.
We have just proved that (iv) holds for $p=m$ if $m<n$.
Using this, we conclude from (\ref{soli2}) that
\beq
m<n:\quad P^{m\, \tin}(F)=0\quad \forall\tin\ .
\label{soli3}\eeq
Hence, $N^m$ vanishes, see
(\ref{soli1b}). Therefore
$b^m$ vanishes as well because $b^m$ is present
in Eq.\ (\ref{soli1a}) only if $N^m$ is present too
[using Eqs.\ (\ref{tr13}), one verifies this by
making the linear combinations of the
$[M_{r_1\dots r_K|s_1\dots  s_N}]^p$ and $N_{r_1\dots r_Ks_1\dots  s_N}$ 
explicit that enter
in (\ref{sdB})]. Hence, we have
\beq
m<n:\quad N^m=b^m=0,\quad dB^{m-1}=-s\7B^{m}.
\label{soli4}\eeq
Moreover, using (\ref{soli3}) in (\ref{soli2}), we
get $dI^{m-1\,\tin}\approx 0$. 
Hence, $I^{m-1\,\tin}$ is weakly $d$-closed
and has form-degree $<n-1$ (since
we are discussing the cases $m<n$).
We conclude, using corollary \ref{LEM0},
that $I^{m-1\,\tin}\approx d\omega^{m-2\, \tin}$ for some
$\omega^{m-2\,\tin}$ if $m-1>0$, or
$I^{0\,\tin}=\lambda^\tin$ for some constants $\lambda^\tin\in\mathbb{R}$
if $m-1=0$. 
Since we assume that (iii) holds for $p=m-1$,
we get 
\beq
m<n:\quad
I^{m-1\,\tin}\approx dI^{m-2\,\tin}+P^{m-1\,\tin}(F)\quad 
\forall\tin
\label{soli5}
\eeq
(with $P^{0\,\tin}(F)\equiv\lambda^\tin$ if $m-1=0$).
Using corollary \ref{LEM2} we conclude from (\ref{soli5})
that 
$I^{m-1\,\tin}- dI^{m-2\,\tin}-P^{m-1\,\tin}(F)$ is
$s$-exact,
\beq
m<n:\quad
I^{m-1\,\tin}= sK^{m-1\,\tin}+dI^{m-2\,\tin}+P^{m-1\,\tin}(F)\quad 
\forall\tin\ .
\label{soli6}
\eeq
Using (\ref{soli6}) in (\ref{soli0}), we get
\beq
m<n:\quad
\omega^{m-1}=
[P^{m-1\,\tin}(F)+sK^{m-1\,\tin}+dI^{m-2\,\tin}]\Theta_\tin+B^{m-1}.
\label{soli7}
\eeq
To deal with the first term on the right hand side of (\ref{soli7}),
we use that every $P^{m-1\,\tin}(F)\Theta_\tin$
can be written as
$P^{m-1\,\tin}(F)\Theta_\tin=N^{m-1}+M^{m-1}+\delta^{m-1}_0\7\lambda$ 
for some $N^{m-1}$ and $M^{m-1}$ and some constant $\7\lambda$ which
can only contribute if $m-1=0$.
This is guaranteed because 
$\{1,
M_{r_1\dots r_K|s_1\dots  s_N}(\theta(C),f(F)),
N_{r_1\dots r_Ks_1\dots  s_N}(\theta(C),f(F))\}$ is
a basis of all $P^\tin(F)\Theta_\tin$, see Section \ref{small}
(note that this is the place where we use the completeness 
property of this basis).

Now, $N^{m-1}$ is trivial in $H(s|d,\Omega)$ since each
$N_{r_1\dots r_Ks_1\dots  s_N}(\theta(C),f(F))$ is
trivial, see Eqs.\ (\ref{tr13}). 
Furthermore, $(sK^{m-1\,\tin}+dI^{m-2\,\tin})\Theta_\tin$
is trivial too, due to
\[
(sK^{m-1\,\tin}+dI^{m-2\,\tin})\Theta_\tin
=
s(K^{m-1\,\tin}\Theta_\tin+I^{m-2\,\tin}[\Theta_\tin]^1)
+d(I^{m-2\,\tin}\Theta_\tin)
\]
where we used once again (\ref{rough0}). Since
trivial contributions to $\omega^{m-1}$ can be neglected
(see above),
we can thus assume, without loss of generality,
\beq
m<n:\quad
\omega^{m-1}=M^{m-1}+B^{m-1}+\delta^{m-1}_0\,\7\lambda.
\label{soli8}
\eeq
For every $M^{m-1}$ there is a $\4B^m$ such that
$dM^{m-1}=-s\4B^m$. This holds by Eqs.\ (\ref{tr13})
(more precisely: the equation with $q=1$ there).
Using in addition Eq.\ (\ref{soli4}), we get
\[
m<n:\quad
d\omega^{m-1}=-sB^m,\quad B^m=\4B^m+\7B^m.
\]
Using this in $s\omega^{m}+d\omega^{m-1}=0$, we get
\[
m<n:\quad s(\omega^{m}-B^m)=0.
\]
{}From this we conclude, using corollary \ref{LEM1},
\beq
m<n:\quad
\omega^{m}\sim B^m+I^{m\, \tin}\Theta_\tin\ .
\label{soli9}
\eeq
This proves (\ref{soli}) for $p=m$ if $m<n$.
\medskip

\underline{$m=n$}.
In this case
we conclude from (\ref{soli2}), using
(\ref{P_A1}) and (\ref{P_A2}),
\beq
P^{n \tin}=\lambda^{A \tin}P_A\ ,\quad
d(I^{n-1 \tin}+\lambda^{A \tin}I_A^{n-1})\approx 0
\label{soli10}
\eeq
for some constant coefficents $\lambda^{A \tin}$.
Using (\ref{jbasis1}), we conclude from (\ref{soli10}) that
$I^{n-1 \tin}+\lambda^{A \tin}I_A^{n-1}\approx 
\lambda^{\Delta \tin}j_\Delta+d\omega^{n-2}$ for
some constant coefficients $\lambda^{\Delta \tin}$ and some $\omega^{n-2}$.
Hence, $I^{n-1 \tin}+\lambda^{A \tin}I_A^{n-1} 
-\lambda^{\Delta \tin}j_\Delta\in\cI$ is weakly $d$-exact.
Using (iii) for $p=m-1=n-1$
and then once again corollary \ref{LEM2}, we conclude from (\ref{soli10})
\beq
I^{n-1\, \tin}=-\lambda^{A \tin}I_A^{n-1}
+\lambda^{\Delta \tin}j_\Delta
+P^{n-1\,\tin}(F)+dI^{n-2\, \tin}+sK^{n-1\, \tin}
\label{soli11}
\eeq
for some $K^{n-1\, \tin}$. 
Furthermore, because of
(\ref{P_A5}) and (\ref{soli10}), we have 
$N^n=\lambda^\Gamma N_\Gamma$ in Eq.\ (\ref{soli1b}), for some
$\lambda^\Gamma$ such that $\lambda^{A \tin}=\lambda^\Gamma
k^{A \tin}_\Gamma$.
Now consider
\bea
\7\omega^n&:=&\omega^n-\lambda^{\Delta \tin}V_{\Delta \tin}
-\lambda^\Gamma W_\Gamma
\nonumber\\
\7\omega^{n-1}&:=&\omega^{n-1}-\lambda^{\Delta \tin}j_\Delta\Theta_\tin
-\lambda^\Gamma [B^{n-1}_\Gamma-k^{A\tin}_\Gamma\, I^{n-1}_A\Theta_\tin]
\nonumber\\
&=&[P^{n-1\,\tin}(F)+dI^{n-2\, \tin}+sK^{n-1\, \tin}]\Theta_\tin+\7B^{n-1} 
\label{soli12}\eea
where $\7B^{n-1} =B^{n-1} - \lambda^\Gamma B^{n-1}_\Gamma$.
One has $s\7\omega^n+d\7\omega^{n-1}=0$, due to
(\ref{Delta3}) and (\ref{QGamma2})
(and $s\omega^n+d\omega^{n-1}=0$).
The last line in (\ref{soli12}) is
analogous to (\ref{soli7}).
By the same arguments that have led from
(\ref{soli7}) to (\ref{soli9}), we conclude that
\[
\7\omega^n\sim B^n+I^{n \tin}\Theta_\tin\ .
\]
This yields (\ref{soli}) for $p=m=n$
due to $\omega^n= \7\omega^n+
\lambda^{\Delta \tin}V_{\Delta \tin}
+\lambda^\Gamma W_\Gamma$.
\item[(ii)]
We shall treat the case $m=n$; the proof for $m<n$
is simpler and obtained from the one for $m=n$ by setting
$\lambda^{\Delta \tin}=\lambda^\Gamma=0$ and substituting
$m$ for $n$ in the following formulae.

Consider the $n$-form $\omega^n=I^{n\,\tin}\Theta_\tin+B^n
+\lambda^{\Delta \tin}V_{\Delta \tin}
+\lambda^\Gamma W_\Gamma$. 
Due to (\ref{sdB}), (\ref{Delta3}) and (\ref{QGamma2}), one has
\bea
&&s(I^{n\,\tin}\Theta_\tin)=0
\\
&&sB^n=-d(B^{n-1}+M^{n-1})
\label{soliz}\\
&&sV_{\Delta \tin}=-d\, [j_\Delta\Theta_\tin]
\\
&&sW_\Gamma=-d[B^{n-1}_\Gamma-k^{A\tin}_\Gamma\, I^{n-1}_A\Theta_\tin]
\eea
for some $B^{n-1}$ and $M^{n-1}$.
Hence, one has $s\omega^n+d\omega^{n-1}=0$ where
$\omega^{n-1}=\7B^{n-1}+\7I^{n-1\,\tin}\Theta_\tin$ with
\bea
&\7B^{n-1}=B^{n-1}+\lambda^\Gamma B^{n-1}_\Gamma&
\\
& \7I^{n-1\,\tin}\Theta_\tin=M^{n-1}
+\lambda^{\Delta \tin}j_\Delta\Theta_\tin
-\lambda^\Gamma k^{A\tin}_\Gamma\, I^{n-1}_A\Theta_\tin&
\label{soli13}
\eea
We assume now that $\omega^n$ is trivial,
\beq
I^{n\,\tin}\Theta_\tin+B^n
+\lambda^{\Delta \tin}V_{\Delta \tin}
+\lambda^\Gamma W_\Gamma\sim 0.
\label{soli14}
\eeq
Then $\omega^{n-1}$ is trivial too (see footnote \ref{foot1}),
\beq
\7B^{n-1}+\7I^{n-1\,\tin}\Theta_\tin\sim 0.
\label{soli15}\eeq
Since we assume that (ii) holds for $p=n-1$,
we conclude from Eq.\ (\ref{soli15})
\bea
& B^{n-1}=-\lambda^\Gamma B^{n-1}_\Gamma\quad (\LRA\ \7B^{n-1}=0) &
\label{soli18}\\
&\7I^{n-1\,\tin}\Theta_\tin\approx N^{n-1}
+(dI^{n-2\,\tin})\Theta_\tin \ .&
\label{soli19}
\eea
(\ref{soli18})
implies that both $\lambda^\Gamma B^{n-1}_\Gamma$ and $B^{n-1}$
vanish. This is trivial if $n$ is odd because then
no $N_\Gamma$ 
is present (recall that $N_\Gamma$ is a polynomial
in the $C^I$ and $F^I$ and has thus
even form-degree). If $n$ is even, then
no $M^{n-1}$ can be present in Eq.\ (\ref{soliz}) (as $M^{n-1}$
is a polynomial in the $C^I$ and $F^I$ too).
By (\ref{P_A9}), we have
$d(\lambda^\Gamma B^{n-1}_\Gamma)=-s(\lambda^\Gamma b^n_\Gamma)
+\lambda^\Gamma N_\Gamma$. Using this and (\ref{soli18}) in
Eq.\ (\ref{soliz}), for $n$ even, one gets
$s(B^n+\lambda^\Gamma b^n_\Gamma)=\lambda^\Gamma N_\Gamma$, i.e., 
$\lambda^\Gamma N_\Gamma$ is $s$-exact
in the small algebra. This implies
$\lambda^\Gamma N_\Gamma=0$ because
$\lambda^\Gamma N_\Gamma$ is a linear combination of
nontrivial representatives of $H(s,\cB)$ by construction
(recall that it is a linear combination of the $N_i$ in
corollary \ref{sBbasis})
and is thus $s$-exact in $\cB$ only if it vanishes.
$\lambda^\Gamma N_\Gamma=0$
implies that all coefficients
$\lambda^\Gamma$ vanish because the $N_\Gamma$
are linearly independent by assumption, see Eq.\ (\ref{P_A6}).
Hence, we get indeed
\beq
\lambda^\Gamma=0\quad \forall\,\Gamma
\label{soli20}\eeq
and thus also, by Eq.\ (\ref{soli18}),
\beq
B^{n-1}=0.
\label{soli20a}
\eeq
Using $ \lambda^\Gamma=0$, Eqs.\ 
(\ref{soli13}) and (\ref{soli19}) give
\beq
N^{n-1}-M^{n-1}\approx 
\lambda^{\Delta \tin}j_\Delta\Theta_\tin
-(dI^{n-2\,\tin})\Theta_\tin\ .
\label{soli21}
\eeq
We have
\beq
N^{n-1}-M^{n-1}=P^{n-1\, \tin}(F)\Theta_\tin(C)
\eeq
for some $P^{n-1\, \tin}(F)$. 
By assumption no nonvanishing linear combination of the
$j_\Delta$ is weakly $d$-exact, see Eq.\ (\ref{jbasis2}).
Since each $P^{n-1\, \tin}(F)$ is $d$-exact,
(\ref{soli21}) implies
\beq
\lambda^{\Delta \tin}=0\quad \forall\,(\Delta,\tin).
\label{soli22}
\eeq
Since we assume that (iv) holds for $p=n-1$,
we conclude from (\ref{soli21}) through (\ref{soli22}) also
that all $P^{n-1\, \tin}(F)$ vanish and thus that
$N^{n-1}-M^{n-1}=0$. 
The latter implies that $N^{n-1}$ and $M^{n-1}$ vanish separately
because they contain independent 
representatives of $H(s,\cB)$,
\[
N^{n-1}=0,\quad M^{n-1}=0.
\]
Using this and Eq.\ (\ref{soli20a}) in
(\ref{soliz}), the latter turns into $sB^n=0$.
By the very definition (\ref{Bp}), $B^n$ is a linear
combination of terms with nonvanishing and linearly independent
$s$-transformations. Hence, $sB^n=0$ holds if and only if
$B^n$ itself vanishes. We conclude
\beq
B^n=0.
\label{soli23}
\eeq

(\ref{soli20}), (\ref{soli22}) and (\ref{soli23})
provide already the assertions for 
$\lambda^\Gamma$, $\lambda^{\Delta \tin}$ and $B^n$
in part (ii) of the theorem. 
We still have to prove those for $I^{n\,\tin}\Theta_\tin$.
Using $\lambda^\Gamma=\lambda^{\Delta \tin}=B^n=0$,
(\ref{soli14}) reads
\beq
I^{n\,\tin}\Theta_\tin=s\eta^n+d\eta^{n-1}
\label{soli25}\eeq
where we made the trivial terms explicit.
Acting with $s$ on this equation gives $d(s\eta^{n-1})=0$
and thus 
\beq
s\eta^{n-1}+d\eta^{n-2}=0
\label{soli24}
\eeq
for some $\eta^{n-2}$, thanks to the algebraic Poincar\'e
lemma. Since we assume that (i) holds for $p=n-1$, we conclude
from (\ref{soli24}) that
\beq
\eta^{n-1}=\4B^{n-1}+\4I^{n-1\, \tin}\Theta_\tin
+s\4\eta^{n-1}+d\4\eta^{n-2}.
\eeq
As above, we have
\beann
&& d\4B^{n-1}=-s(\4B^{n}+\4b^n)+\4N^n
\\
&& d(\4I^{n-1\, \tin}\Theta_\tin)=
(d\4I^{n-1\, \tin})\Theta_\tin
-s(\4I^{n-1\, \tin}[\Theta_\tin]^1).
\eeann
Using this in (\ref{soli25}), we get
\beq
[I^{n\,\tin}-d\4I^{n-1\, \tin}-\4P^{n\,\tin}]\Theta_\tin
=s(\eta^n-\4B^{n}-\4b^n-d\4\eta^{n-1}-\4I^{n-1\, \tin}[\Theta_\tin]^1),
\label{soli26}
\eeq
where
\beq
\4P^{n\,\tin}\Theta_\tin=\4N^n\ .
\label{soli27}
\eeq
Using corollary \ref{LEM1} we conclude from (\ref{soli26}) that
\beq
I^{n\,\tin}-d\4I^{n-1\, \tin}-\4P^{n\,\tin}\approx 0.
\label{soli28}
\eeq
(\ref{soli27}) and (\ref{soli28}) complete the
demonstration of (ii).
\item[(iii)]
$I^m\approx d\omega^{m-1}$
implies $I^m\sim 0$, i.e.
$I^m$ is trivial in $H(s|d,\Omega)$.
Indeed, $I^m\approx d\omega^{m-1}$ means that
$I^m= \delta\omega^m+d\omega^{m-1}$ for some
$\omega^m$ with antifield number 1. Hence, $I^m$ is a cocycle
of $H(s|d,\Omega)$ (since it is $s$-closed due to $I^m\in\cI$)
and trivial in $H(\delta|d,\Omega)$. It is therefore
also trivial in $H(s|d,\Omega)$
by theorem \ref{Loctdec} (cf.\ proof of (\ref{75})).
Now, $I^m\sim 0$ is just a special case of 
$I^{p\,\tin}\Theta_\tin\sim 0$ (due to $1\in\{\Theta_\tin\}$).
Hence, using (ii) for $p=m$ (which we have already proved),
we conclude $I^m\approx dI^{m-1}+P^m(F)$ for some
$I^{m-1}\in\cI$ and some $P^m(F)$. Conversely, if
$m>0$, we have $P^m(F)=dq^{m-1}$ for some Chern-Simons
form $q^{m-1}$ and thus $I^m\approx dI^{m-1}+P^m(F)$
implies $I^m\approx d\omega^{n-1}$ 
with $\omega^{n-1}=I^{m-1}+q^{m-1}$. \qed
\een

\subsection{Appendix \ref{solution}.A: 2-dimensional 
pure Yang-Mills theory}
\label{2dYM}

Pure 2-dimensional Yang-Mills theory needs a special
treatment because $H^{n-2}_{char}(d, \Omega)
\equiv H^0_{char}(d, \Omega)$ is not given by the
global reducibility identities associated with
the abelian gauge symmetries (theorem \ref{Loct12}),
but is much bigger and in fact infinite-dimensional
(see explicit description at the end of the appendix).
This feature disappears if one couples coloured
matter fields.
We discuss the pure Yang-Mills case for 
the sake of completeness contenting ourselves
with case I, i.e., with the solution
of the consistency condition $s\omega^p+d\omega^{p-1}=0$
in the space of all local forms.

We consider the standard Lagrangian
\[
L=-\sfrac 14\, F_{\mu\nu}^I F^{\mu\nu}_I
\]
where $F^{\mu\nu}_I=g_{IJ} F^{\mu\nu J}$ involves an
invertible $\cG$-invariant symmetric tensor $g_{IJ}$.
The gauge group may contain abelian factors.

Due to $n=2$, we have $F^I_{01}=-F^I_{10}=
(1/2)\epsilon^{\mu\nu}F^I_{\mu\nu}=\star F^I$ and the equations of motion
set all covariant derivatives of $ F_{01I}$ to zero.
The result for $H(s)$ (corollary \ref{LEM1})
implies thus immediately
that
\beq
\omega^0=I^{\tin}(x,F_{01})\Theta_\tin+s\eta^0
\label{2d0}
\eeq
where the $I^{\tin}(x,F_{01})$ are arbitrary $\cG$-invariant
local functions of
the $ F_{01I}$ and the $x^\mu$ 
(the latter can occur because we are discussing case I).
(\ref{2d0}) is therefore
the general solution of the consistency condition
for $p=0$.

To find the general solutions with $p=1$ and $p=2$, we use
the descent equations and examine
whether an $\omega^0$ as in Eq.\ (\ref{2d0})
can be lifted to solutions of the consistency condition with
form-degree 1 or 2. In order to lift $\omega^0$
to form-degree 1, it is necessary
and sufficient that the $dI^{\tin}(x,F_{01})$ vanish weakly, for all
$\tin$ (see Section \ref{Strategy}).
Since the $I^{\tin}(x,F_{01})$ are $\cG$-invariant, we have
\[
dI^{\tin}(x,F_{01})=dx^\mu \left[
\frac{\6I^{\tin}(x,F_{01})}{\6x^\mu}+
(D_\mu F_{01I})\,\frac{\6I^{\tin}(x,F_{01})}{\6 F_{01I}}
\right]
\approx
dx^\mu\,
\frac{\6I^{\tin}(x,F_{01})}{\6x^\mu}
\]
where we have used $D_\mu  F_{01I}\approx 0$.
No nonvanishing function of the undifferentiated 
$ F_{01I}$ is weakly zero since the equations of motion
contain derivatives of $ F_{01I}$. Hence,
$\omega^0$ can be lifted to form-degree 1 if and only if 
$\6I^\tin/\6x^\mu=0$, i.e.,
the $I^{\tin}$ must not
depend explicitly on the $x^\mu$.
It turns out that this also suffices to lift $\omega^0$ to
form-degree 2.
To show this, we introduce
\[
\star\4C^*_I:=\star C^*_I+\star A^*_I+\star F_I
\]
where $\star C^*_I=d^2x\, C^*_I$,
$\star A^*_I=dx^\mu \epsilon_{\mu\nu}A^{*\nu}_I$ and 
$\star F_I=\sfrac 12\epsilon_{\mu\nu}F^{\mu\nu}_I$. 
One has
\[
(s+d)\star\4C^*_I=(C^J+A^J)e\f JIK \star\4C^*_K\ ,
\]
i.e., the $\star\4C^*_I$ transform under $(s+d)$ according
to the adjoint representation of $\cG$ 
with ``$(s+d)$-ghosts $(C^I+A^I)$''. 
$\cG$-invariant functions of the $\star\4C^*_I$ are 
thus $(s+d)$-closed,
\[
(s+d)I^{\tin}(\star \4C^*)=0.
\]
Recall that the $\Theta_\tin$ are polynomials in the
$\theta_r(C)$ and that the latter are related to $\cG$-invariant
polynomials $f_r(F)$ via the transgression formula (\ref{f})
which decomposes into Eqs.\ (\ref{qdes}). In two dimensions,
all $f_r(F)$ with degree $m(r)>1$ in the $F^I$ vanish.
The transgression formula gives thus
\[
m(r)>1:\quad 
(s+d)q_r(C+A,F)=0,\quad q_r(C+A,F)=[\theta_r]^0+[\theta_r]^1+[\theta_r]^2.
\]
For $m(r)=1$ one gets $(s+d)(C^I+A^I)=F^I$ where 
$C^I$, $A^I$ and $F^I$ are abelian. One has
\beann
\mbox{abelian $F^I$:}\quad
F^I&=&\sfrac 12\, dx^\mu dx^\nu F_{\mu\nu}^I
\\
&=& d(\sfrac 12\, x^\mu dx^\nu F_{\mu\nu}^I)+
\sfrac 12\, x^\mu dx^\nu dx^\rho\6_\rho F_{\mu\nu}^I
\\
&=&
d(\sfrac 12\, x^\mu dx^\nu F_{\mu\nu}^I)-
s(\sfrac 12\, d^2x\, x^\mu \epsilon_{\mu\nu} A^{*\nu I})
\\
&=&(s+d)(\sfrac 12\, x^\mu dx^\nu F_{\mu\nu}^I
-\sfrac 12\, d^2x\, x^\mu \epsilon_{\mu\nu} A^{*\nu I})
\eeann
where we have used that one has
$\6_\mu F_{01}^I=s(\epsilon_{\mu\nu} A^{*\nu I})$
for abelian $F^I$. Hence we have two different
quantities whose $(s+d)$-transformation equals
$F^I$ in the abelian case ($C^I+A^I$ and the
quantity in the previous equation). The difference
of these quantities is thus an $(s+d)$-closed 
extension of the abelian ghosts.
Hence, we can complete every $\theta_r(C)$, whether nonabelian
or abelian, to an $(s+d)$-invariant quantity $\4q_r$,
\beann
m(r)>1:&& \4q_r=q_r(C+A,F)=[\theta_r]^0+[\theta_r]^1+[\theta_r]^2
\\
m(r)=1:&& \4q_r=C^I+A^I-\sfrac 12\,(x^\mu dx^\nu F_{\mu\nu}^I
-d^2x\, x^\mu \epsilon_{\mu\nu} A^{*\nu I})\quad
(\mbox{abelian $I$}).
\eeann
Due to $(s+d)\4q_r=0$ and $(s+d)I^{\tin}(\star \4C^*)=0$, we 
have
\beq
(s+d)\left[I^{\tin}(\star \4C^*)\Theta_\tin(\4q)\right]=0,
\label{2ds+d}
\eeq
where $\Theta_\tin(\4q)$ arises from $\Theta_\tin$ by substituting
the $\4q_r$ for the $\theta_r(C)$. 
The decomposition of
(\ref{2ds+d}) into pieces with definite form-degree reads
\beann
&s[I^{\tin}\Theta_\tin]^2+
d[I^{\tin}\Theta_\tin]^1=0,&
\\
&s[I^{\tin}\Theta_\tin]^1+
d[I^{\tin}\Theta_\tin]^0=0,&
\\
& s[I^{\tin}\Theta_\tin]^0=0,&
\eeann
where $[I^{\tin}\Theta_\tin]^p$ is the $p$-form contained
in $I^{\tin}(\star \4C^*)\Theta_\tin(\4q)$,
\[
I^{\tin}(\star \4C^*)\Theta_\tin(\4q)=
\sum_{p=0}^2 [I^{\tin}\Theta_\tin]^p.
\]
Every $I^{\tin}(F_{01})\Theta_\tin=[I^{\tin}\Theta_\tin]^0$
can thus indeed be lifted to solutions of the consistency condition
with form-degrees 1 and 2.

It is now easy to complete the analysis. 
$s\omega^1+d\omega^0=0$ yields $s(\omega^1-[I^{\tin}\Theta_\tin]^1
-d\eta^0)=0$. By the result on $H(s)$, the general
solution of $s\omega^1+d\omega^0=0$ is accordingly 
\beq
\omega^1=[I^{\tin}\Theta_\tin]^1
+dx^\mu I^\tin_\mu(x,F_{01})\Theta_\tin
+s\eta^1+d\eta^0
\label{2d1}
\eeq
where the $I^\tin_\mu(x,F_{01})$ are arbitrary $\cG$-invariant
local functions of the $ F_{01I}$ and the $x^\mu$.

We know already that every $[I^{\tin}\Theta_\tin]^1$ can be lifted
to $[I^{\tin}\Theta_\tin]^2$. In order to lift
an $\omega^1$ as in Eq.\ (\ref{2d1}), it is therefore necessary
that the piece $\7\omega^1:=
dx^\mu I^\tin_\mu(x,F_{01})\Theta_\tin$ can be lifted
too. By arguments analogous to those used above, this
requires $d_x \7\omega^1 =0$ where $d_x=dx^\mu \6/\6x^\mu$.
Since $H^1(d_x)$ is trivial (ordinary Poincar\'e lemma in
$\mathbb{R}^2$), this gives $\7\omega^1=d_x J^\tin(x,F_{01})\Theta_\tin$ 
for some $\cG$-invariant local functions $J^\tin(x,F_{01})$.
This implies that $\7\omega^1$ is trivial in $H(s|d)$. Indeed,
using $D_\mu F_{01I}=\delta(\epsilon_{\mu\nu}A^{*\nu}_I)$,
the $\cG$-invariance of $I^{\tin}$ and $J^\tin$, and Eq.\ (\ref{rough0}), 
one obtains
\beann
d_x J^\tin(x,F_{01})\Theta_\tin=d[J^\tin(x,F_{01})\Theta_\tin]
-(dx^\mu D_\mu F_{01I})\, 
\frac{\6 J^\tin(x,F_{01})}{\6 F_{01I}}\,\Theta_\tin
-J^\tin(x,F_{01})d\Theta_\tin
\\
=d[J^\tin(x,F_{01})\Theta_\tin]+s\left[
\star A^*_I\, \frac{\6 J^\tin(x,F_{01})}{\6 F_{01I}}\,\Theta_\tin
+J^\tin(x,F_{01})A^I\,\frac{\6\Theta_\tin}{\6C^I}
\right].
\eeann
Hence, those solutions $\omega^1$ which can be lifted
are of the form $[I^{\tin}\Theta_\tin]^1+s\eta^{\prime 1}
+d\eta^{\prime 0}$. 
Inserting this in $s\omega^2+d\omega^{1}=0$ yields
$s(\omega^2-[I^{\tin}\Theta_\tin]^2-d\eta^{\prime 1})=0$.
Every element of $H(s)$ with form-degree 2 is $d_x$-closed
and thus $d_x$-exact, due to $H^2(d_x)=0$. Using arguments
as before, one concludes that the general solution
of $s\omega^2+d\omega^1=0$ is
\beq
\omega^2=[I^{\tin}\Theta_\tin]^2
+s\eta^{\prime\prime2}+d\eta^{\prime\prime 1}.
\label{2d2}
\eeq

{\sl Remark.}
Using the isomorphism 
$H^0_\mathrm{char}(d,\Omega)\simeq H^2_2(\delta \vert d,\Omega)
\oplus \mathbb{R} \simeq H^{-2,2}(s|d,\Omega) \oplus \mathbb{R}$
(see theorems \ref{Loct6} and \ref{Loctdec}), 
one deduces from the above result that
$H^0_\mathrm{char}(d,\Omega)$
and $H^0_\mathrm{char}(d,\cI)$ are
represented by arbitrary $\cG$-invariant
polynomials in the $ F_{01I}$. These cohomological
groups are thus infinite dimensional. 
This explains the
different results as compared to higher dimensions where
the nontrivial representatives of
$H^{n-2}_\mathrm{char}(d,\Omega)$ correspond one-to-one
to the free abelian gauge symmetries.

\newpage

\mysection{Discussion of the results for Yang-Mills type theories}
\label{DisDisDis}

Theorem \ref{main} gives the general solution of the
consistency condition $sa + db = 0$
at all form-degrees and ghost numbers
for theories of the Yang-Mills type without free abelian
gauge symmetries (in the sense of subsection \ref{AssUMPtions})
and in spacetime dimensions greater
than $2$. The case of free abelian symmetries is treated
in the next section.  In this section,
we spell out the physical implications of the 
theorem by expliciting the results in the relevant
ghost numbers.

To that end
we shall use the notation $f(\cov)$ 
for functions that depend only on the Yang-Mills field strengths,
the matter fields and their covariant derivatives,
\beann
f(\cov)\equiv
f(F_{\mu\nu}^I,D_\rho F_{\mu\nu}^I,D_\rho D_\sigma F_{\mu\nu}^I,
\dots,\psi^i,D_\mu\psi^i,D_\mu D_\nu\psi^i,\dots).
\eeann

We recall that the results are valid for general Lagrangians
of the Yang-Mills type, provided these fulfill the technicality
assumptions of ``regularity" and ``normality" explained above.
The results cover in particular the standard model and effective
gauge theories.

\subsection{$H^{-1,n}(s|d)$: Global symmetries and Noether currents}

\subsubsection{Solutions of the consistency condition at negative
ghost number}

We start with the discussion of the results at negative ghost
number.
First, we recall that the groups $H^{-q,n}(s|d)$ are trivial for
$q>1$.  This implies that there is no characteristic cohomology
in form degree $<n-1$, i.e., no non trivial higher order
conservation law.  Any conserved local antisymmetric tensor $A^{\mu_1
\cdots \mu_q}$ ($q>1$) is trivial, i.e., equal on-shell to the
divergence of a {\em local} antisymmetric tensor with
one more index,
\beann 
q>1:&&
\partial_{\mu_1} A^{\mu_1 \dots \mu_q}
\approx 0 , \
A^{\mu_1 \dots \mu_q} = A^{[\mu_1 \dots \mu_q]}
\\
&\Rightarrow& A^{\mu_1 \dots \mu_q} \approx
\partial_{\mu_0} B^{\mu_0 \mu_1 \dots \mu_q}, \;
B^{\mu_0 \mu_1 \dots \mu_q} = B^{[\mu_0 \mu_1 \dots \mu_q]}.
\eeann
We stress again that the important point in this statement is that the
$B^{\mu_0 \mu_1 \cdots \mu_q}$ are  local functions; 
the statement would otherwise be
somewhat empty due to the ordinary Poincar\'e lemma for ${\mathbb R}^n$.

The only non-vanishising group at negative ghost number is
$H^{-1,n}(s|d,\Omega)$.
The nontrivial representatives of $H^{-1,n}(s|d,\Omega)$
are the generators of the nontrivial global symmetries,
denoted by $K_\Delta$ and $K_A$
in the
previous section. Indeed, one must 
set $\Theta_\alpha= 1$ in order 
that Eqs.\ (\ref{V}) and (\ref{VGamma}) yield solutions with
ghost number $-1$.
The general solution of the consistency condition
with ghost number $-1$ is thus
\[
\omega^{-1,n}\sim \lambda^\Delta K_\Delta+\lambda^A K_A,
\]
in form-degree $n$, where $K_\Delta$ and $K_A$ are related to the
gauge invariant conserved currents $j_\Delta$ and to the
characteristic classes $P_A(F)$ respectively, through
\beann
& sK_\Delta+dj_\Delta=0,\quad j_\Delta\in\cI\ ,&
\\
& sK_A+dI^{n-1}_A=P_A(F),\quad I^{n-1}_A\in\cI\ .&
\eeann
That is, the coefficients of the antifields $A^{*\mu}_I$ and
$\psi^*_i$ in the $K$'s determine the transformations of the
corresponding field in the global symmetry associated with the
conserved Noether currents.

\subsubsection{Structure of global symmetries and conserved currents.}

A determination of a complete set of gauge invariant nontrivial
conserved currents $j_\Delta$ depends on the specific
model under study. It also depends on the detailed form of 
the Lagrangian whether or not invariants $I^{n-1}_A$
exist which are related to characteristic classes
by Eq.\ (\ref{P_A1}).
However,
we can make the description of the $K_\Delta$ and $K_A$
a little more precise without specifying $L$.
Namely, as we prove in the appendix to this section, one can always
choose the $j_\Delta$ and $I^{n-1}_A$ such
that all $K_\Delta$ and $K_A$ take the form
\bea
K_\Delta&=&d^nx\, \left[A^{*\mu}_I Q_{\Delta\mu}^I(x,\cov)
              +\psi^*_i Q_{\Delta}^i(x,\cov)\right]
\label{dis1}\\
K_A&=&d^nx\, \left[A^{*\mu}_I Q_{A\mu}^I(x,\cov)
              +\psi^*_i Q_A^i(x,\cov)\right]
\label{dis2}\eea
where the
$Q_{\Delta\mu}^I(x,\cov)$
transform under $\cG$ according to the coadjoint representation
and the $Q_{\Delta}^i(x,\cov)$ according
to the same representation as the $\psi^i$.
We assume here that we work in the space of all
local forms (case I). An analogous statement holds in the space of
Poincar\'e invariant local forms (case II) where
(\ref{dis1}) and (\ref{dis2})
hold with Poincar\'e invariant $K$'s.

This result, and the relationship between
the $K$'s and the conserved currents, enables
us to draw the following conclusions 
(in Yang-Mills type theories without free abelian gauge symmetries,
when the spacetime dimension exceeds 2):
\ben
\item
In odd dimensional spacetime,
every nontrivial conserved current is equivalent
to a gauge invariant conserved current,
\[
n=2k+1:\quad
\6_\mu j^\mu\approx 0\ \then\
j^\mu\sim j^\mu_\mathrm{inv}(x,\cov)
\]
where $j^\mu_\mathrm{inv}(x,\cov)$ 
is $\cG$-invariant and $\sim$ means ``equal modulo trivial
conserved currents'',
\[
j^\mu\sim h^\mu\ :\LRA\ 
j^\mu\approx h^\mu+\6_\nu m^{[\nu\mu]}.
\]
\item
In even dimensional spacetime
a nontrivial conserved current is either equivalent
to a completely gauge invariant current or to a current
that is gauge invariant except for 
a Chern-Simons term,
\[
n=2k:\quad
\6_\mu j^\mu\approx 0\ \then\ 
j^\mu\sim
\left\{
\ba{cc}
j^\mu_\mathrm{inv}(x,\cov)
\\
\mbox{or}
\\
I^\mu_\mathrm{inv}(x,\cov)+q_\mathrm{CS}^\mu(A,\6A)
\ea
\right.
\]
where $j^\mu_\mathrm{inv}(x,\cov)$ and
$I^\mu_\mathrm{inv}(x,\cov)$ are $\cG$-invariant, and
$q_\mathrm{CS}^\mu(A,\6A)$ is dual to a Chern-Simons $(n-1)$-form, i.e.,
\[
q_\mathrm{CS}^\mu(A,\6A)=\epsilon^{\mu\nu_1\mu_2\nu_2\dots\mu_k\nu_k}
d_{I_1\dots I_k}A_{\nu_1}^{I_1}
\6_{\mu_2}A_{\nu_2}^{I_2}\dots \6_{\mu_k}A_{\nu_k}^{I_k}+\dots
\]
One can choose the basis of inequivalent conserved currents such that
those currents which contain Chern-Simons terms correspond one-to-one
to the characteristic classes $P_A(F)$ which
are trivial in the equivariant characteristic cohomology.
In particular, all conserved currents can be made strictly gauge invariant
when no characteristic class is trivial in the equivariant 
characteristic cohomology.%
\footnote{To our knowledge, it is still an open problem whether
in standard Yang-Mills theory characteristic classes $P(F)$ 
with form-degree $n$ can be trivial in $H_\mathrm{char}(d,\cI)$. 
(The problem occurs only in the space of forms with explict 
$x^\mu$-dependence.)
In Section 13 of \cite{Barnich:1995db}, we have claimed that the answer is 
negative for a polynomial dependence on $x^\mu$. However, the proof
of the assertion given there is incorrect because the
$sl(n)$-decomposition of the equations of motion used there
does not yield pieces which are all weakly zero 
separately. If the answer were positive (contrary to our expectations), 
it would mean that non covariantizable currents could occur in standard
Yang-Mills theory, contrary to the claim in theorem 2 in 
\cite{Barnich:1995cq}\label{cmperratum}.}

\item 
Every nontrivial global symmetry can be brought
to a gauge covariant form. More precisely,
let $\delta_Q$ be the generator of a global
symmetry whose characteristics $\delta_Q A_\mu^I=Q_\mu^I$
and $\delta_Q \psi^i=Q^i$ are local functions of the fields
($Q_\mu^I$ or $Q^i$ 
may depend explicitly on the $x^\mu$). Then one can bring
$\delta_Q$ to a form (by subtracting trivial symmetries if necessary)
such that the characteristics depend only on the $x^\mu$,
the Yang-Mills field strengths and their covariant
derivatives, and the matter fields and their covariant
derivatives, 
\[
\delta_Q A_\mu^I=Q_\mu^I(x,\cov),\quad
\delta_Q \psi^i=Q^i(x,\cov),
\]
where the $Q_\mu^I$ transform under the 
coadjoint representation of $\cG$
and the $Q^i$ transform under the same representation of $\cG$ as
the $\psi^i$. 
\een
Note that the gauge covariant
global symmetries commute with the gauge transformations 
(\ref{gauge007}): for instance, one has
\beann
{}[\delta_Q,\delta_\epsilon] A_\mu^I&=&
\delta_Q(\6_\mu \epsilon^I + 
e\, \f JKI A_\mu^J \epsilon^K)-
\delta_\epsilon Q_\mu^I
\\
&=&e\, \f JKI Q_\mu^J \epsilon^K
-(-e\,\epsilon^K\f KJI Q_\mu^J)=0.
\eeann
Here we used $\delta_Q \epsilon^I=0$ where $\epsilon^I$ are
arbitrary fields. Of course,
in general $\delta_Q$ would not commute with a special gauge 
transformation obtained by substituting functions
of the $A_\mu^I$, $\psi^i$ and their derivatives for
$\epsilon^I$. 

\subsubsection{Examples.}\label{examplecurrents}
\ben
\item
It should be noted that the gauge covariant form of
a global symmetry is not always its most familiar version.
In order to make a global symmetry gauge covariant, it may
be necessary to add a trivial symmetry to it.
We illustrate this feature now for conformal transformations.
Consider 4-dimensional massless scalar electrodynamics,
\[
n=4,\quad L=-\sfrac 14\, F_{\mu\nu} F^{\mu\nu}
- \sfrac 12\, (D_\mu\varphi) D^\mu\5\varphi
\]
where $\varphi$ is a complex scalar field, $\5\varphi$ is the
complex conjugate of $\varphi$, and
\[
F_{\mu\nu}=\6_\mu A_\nu-\6_\nu A_\mu\ ,\ 
D_\mu\varphi=\6_\mu \varphi+\Ii e A_\mu \varphi\ ,\ 
D_\mu\5\varphi=\6_\mu \5\varphi-\Ii e A_\mu \5\varphi\ .
\]
The action is invariant under the following 
infinitesimal conformal transformations
\beann
\delta_\mathrm{conf} A_\mu&=&\xi^\nu \6_\nu A_\mu +(\6_\mu\xi^\nu) A_\nu\ ,
\\
\delta_\mathrm{conf} \varphi&=&\xi^\nu \6_\nu \varphi
+\sfrac 14\,(\6_\nu\xi^\nu)\varphi\ ,
\\
\xi^\mu&=& a^\mu+\omega^{[\mu\nu]} x_\nu+\lambda x^\mu + b^\mu x_\nu x^\nu
-2x^\mu b^\nu x_\nu
\eeann
where the $a^\mu$, $\omega^{[\mu\nu]}$, $\lambda$ and $b^\mu$ are
constant parameters of conformal transformations 
and $x_\mu=\eta_{\mu\nu}x^\nu$.
$\delta_\mathrm{conf}$ is not gauge covariant. To make it
gauge covariant we add a trivial global symmetry to it, namely
a special gauge transformation with ``gauge parameter''
$\epsilon=-\xi^\nu A_\nu$. This
special gauge transformation is
$\delta_\mathrm{trivial} A_\mu=\6_\mu (-\xi^\nu A_\nu)$, 
$\delta_\mathrm{trivial}\varphi=\Ii e \xi^\nu A_\nu \varphi$. 
$\7\delta_\mathrm{conf}=\delta_\mathrm{conf}+\delta_\mathrm{trivial}$ 
is gauge covariant,
\[
\7\delta_\mathrm{conf} A_\mu=\xi^\nu F_{\nu\mu}\ ,\
\7\delta_\mathrm{conf} \varphi=\xi^\nu D_\nu \varphi
+\sfrac 14\,(\6_\nu\xi^\nu)\varphi\ .
\]
Since $\7\delta_\mathrm{conf}$ and $\delta_\mathrm{conf}$ 
differ only by a special gauge
transformation, they are equivalent and yield the same variation
of the Lagrangian,
\[
\7\delta_\mathrm{conf} L=\delta_\mathrm{conf} L
=\6_\mu (\xi^\mu L-b^\mu\varphi\5\varphi).
\] 
The Noether current corresponding to 
$\7\delta_\mathrm{conf}$ is gauge invariant,
\[
j^\mu_\mathrm{inv}(x,[F,\varphi,\5\varphi]_D)=
\sum_{\Phi=A_\nu,\varphi,\5\varphi}(\7\delta_\mathrm{conf} \Phi)\, 
\frac{\6L}{\6(\6_\mu\Phi)}-\xi^\mu L+b^\mu\varphi\5\varphi\ .
\]
\item 
We shall now illustrate the unusual situation in which 
a nontrivial Noether current contains a Chern-Simons term.
As a first example we consider 4-dimensional Yang-Mills theory 
with gauge group $SU(2)$ coupled nonminimally 
to a real $SU(2)$-singlet
scalar field $\phi$ via the following Lagrangian,
\[
n=4,\quad
L=-\sfrac 14 F_{\mu\nu}^I F^{\mu\nu J}\delta_{IJ}
-\sfrac 12\,(\6_\mu\phi)\6^\mu\phi+\sfrac 14\,\phi 
\epsilon^{\mu\nu\rho\sigma}F_{\mu\nu}^I F_{\rho\sigma}^J\delta_{IJ}
\]
where
\[
F_{\mu\nu}^I=\6_\mu A_\nu^I-\6_\nu A_\mu^I+e\,\epsilon_{IJK}A_\mu^JA_\nu^K\ .
\]
The action is invariant under constant shifts of $\phi$,
\beann
&\delta_\mathrm{shift}\phi=-1\quad ,\quad 
\delta_\mathrm{shift} A_\mu^I=0\quad \then\quad
\delta_\mathrm{shift} L=-\6_\mu q^\mu_\mathrm{CS}(A,\6A),&
\\
&q^\mu_\mathrm{CS}(A,\6A)=\epsilon^{\mu\nu\rho\sigma}
(\delta_{IJ} A_\nu^I\6_\rho A_\sigma^J
+\sfrac 13 e\,\epsilon_{IJK}A_\nu^I A_\rho^J A_\sigma^K).&
\eeann
$\delta_\mathrm{shift}$ is obviously nontrivial and gauge covariant.
The corresponding Noether current contains the Chern-Simons
term $q^\mu_\mathrm{CS}(A,\6A)$ and is otherwise gauge invariant,
\[
j^\mu=\6^\mu\phi+q^\mu_\mathrm{CS}(A,\6A)\ .
\]
\item 
A variant of the previous example arises when one replaces
the scalar field by the time coordinate $x^0$,
\[ 
n=4,\quad
L=-\sfrac 14 F_{\mu\nu}^I F^{\mu\nu J}\delta_{IJ}
+\sfrac 14\,x^0 
\epsilon^{\mu\nu\rho\sigma}F_{\mu\nu}^I F_{\rho\sigma}^J\delta_{IJ}
\]
with $F_{\mu\nu}^I$ as in the previous example.
This example breaks of course Lorentz invariance 
and is given only for illustrative purposes.
One can get rid of $x^0$ by integrating by parts
the last term, at the price of introducing an undifferentiated
$A_\mu$.
The action is therefore invariant under temporal translations,
\[
\delta_\mathrm{time} A_\mu^I=F_{0\mu}^I\quad \then
\quad \delta_\mathrm{time} L=\6_0L-\6_\mu q^\mu_\mathrm{CS}(A,\6A)
\]
with $q^\mu_\mathrm{CS}(A,\6A)$ as in the previous example. 
$\delta_\mathrm{time} A_\mu^I= F_{0\mu}^I$ is
already the gauge covariant version of temporal translations (one
has $\delta_\mathrm{time}A_\mu^I=\6_0A_\mu^I+\delta_\mathrm{trivial}A_\mu^I$
where $\delta_\mathrm{trivial}A_\mu^I$ is a special gauge
transformation with $\epsilon^I=-A_0^I$). Note that
we used $\delta_\mathrm{time} x^0=0$, i.e., we transformed only the fields.
The conserved Noether current corresponding to
$\delta_\mathrm{time}$ is the component $\nu=0$ of the energy momentum
tensor ${T_\nu}^\mu$. In the present case,
it cannot be made fully gauge invariant but
contains the Chern-Simons term $q^\mu_\mathrm{CS}(A,\6A)$,
\[
{T_0}^\mu= F_{0\nu}^I\, 
\frac{\6L}{\6(\6_\mu A_\nu^I)}-\delta^\mu_0 L
+q^\mu_\mathrm{CS}(A,\6A).
\]
\een

\subsection{$H^{0,n}(s|d)$: Deformations and BRST-invariant counterterms}

We now turn to the local BRST cohomology at ghost 
number zero.  This case covers
deformations of the action and controls therefore the stability
of the theory.
We first make the results of theorem
\ref{main}  more explicit for the particular value $0$ of the
ghost number; we then discuss
the implications.

The most general solution of
the consistency condition with ghost number 0 and
form-degree $n$ is
\[
\omega^{0,n}\sim I^n+B^{0,n}+V^{0,n}+W^{0,n}
\]
where:
\ben
\item
$I^n\in\cI$, i.e., $I^n$ is a strictly gauge invariant $n$-form,
\beann
\mbox{Case I:}&& 
I^n=d^nx\, I_\mathrm{inv}(x,\cov)
\\
\mbox{Case II:}&& 
I^n=d^nx\, I_\mathrm{inv}(\cov).
\eeann
(we recall that in case I, one computes the cohomology
in the algebra of 
forms having a possible explicit $x$-dependence; while
the forms in case II have no explicit $x$-dependence
and are Lorentz-invariant)
\item
$B^{0,n}$ is a linear combination of the independent
Chern-Simons $n$-forms, see
Eq.\ (\ref{g=0}). Solutions
$B^{0,n}$ can thus only exist
in odd spacetime dimensions.
\item
$V^{0,n}$ are linear combinations of the solutions
$V_{\Delta\tin}$ (\ref{V}) related to global
symmetries. In order 
that $V_{\Delta\tin}$ has
ghost number 0, the $\Theta_\tin$ which appears in it
must have ghost number 1.  There are such $\Theta_\tin$
only when the gauge group has abelian factors, in which case
the $\Theta_\tin$ are the abelian ghosts.  In the
absence of abelian factors, there are thus no $V_{\Delta\tin}$
at ghost number zero.
Using Eq.\ (\ref{dis1}), one gets explicitly
\bea
V^{0,n}&=&\sum_{I:\mathrm{abelian}} \lambda^\Delta_I\, 
(K_\Delta C^I+j_\Delta A^I)
\nonumber\\
&=&d^nx\sum_{I:\mathrm{abelian}} \lambda^\Delta_I
\left[A^{*\mu}_J Q_{\Delta\mu}^J C^I
              +\psi^*_i Q_{\Delta}^i C^I
-(-)^{\epsilon_\Delta}j_\mathrm{\Delta}^\mu A_\mu^I\right]
\label{V0n}
\eea
where the $j_\mathrm{\Delta}^\mu$ are the nontrivial
gauge invariant Noether currents,
$Q_{\Delta\mu}^J$ and $Q_{\Delta}^i$ are the 
corresponding gauge covariant symmetries, and
$\epsilon_\Delta$ is the parity of $j_\mathrm{\Delta}^\mu$
(e.g., $\epsilon_\Delta=1$ when $j_\mathrm{\Delta}^\mu$
is the conserved current of a global supersymmetry).
We stress again that the sets of $Q$'s and $j$'s
are different in
case I and case II, see text after Eq.\ (\ref{jbasis2}).
\item
$W^{0,n}$ are linear combinations of solutions $W_\Gamma$
(\ref{VGamma}) with ghost number 0; such solutions exist only for peculiar
choices of Lagrangians discussed below - and again only when there
are abelian factors.
\een

\paragraph{Nontriviality of the solutions.}
A solution 
$I^n+B^{0,n}+V^{0,n}+W^{0,n}$
is only trivial when $B^{0,n}$, $V^{0,n}$ and $W^{0,n}$
all vanish and $I^n$ is weakly $d$-exact,
\[
I^n+B^{0,n}+V^{0,n}+W^{0,n}\sim 0
\ \LRA\ 
B^{0,n}=V^{0,n}=W^{0,n}=0,\ I^n\approx d\omega^{n-1}.
\]
$I^n\approx d\omega^{n-1}$ is equivalent to
$I^n\approx dI^{n-1}+P(F)$ for some
$I^{n-1}\in\cI$ and some characteristic class $P(F)$.

\paragraph{Semisimple gauge group.}
When the gauge group $G$ is semisimple there are no solutions
$V^{0,n}$ or $W^{0,n}$ at all because all these solutions
require the presence of abelian gauge symmetries.
Hence, when $G$ is semisimple,
all representatives of $H^{0,n}(s|d)$ can be taken to be
strictly gauge invariant except for the Chern-Simons
forms in odd spacetime dimensions.
In particular, the antifields can then be removed from all BRST-invariant
counterterms and integrated composite operators
by adding cohomologically trivial terms, and
the gauge transformations are stable, i.e.,
they cannot be deformed in a continuous and nontrivial manner.
This result implies, in particular, the structural stability
of effective Yang-Mills theories in the sense of
\cite{Gomis:1995he}.

\paragraph{Comment.}
We add a comment on Chern-Simons forms which should
also elucidate a bit the distinction between case I and case II. 
Chern-Simons forms $B^{0,n}$
are Lorentz-invariant in $n$-dimensional spacetime 
and occur thus among the
solutions both in case I and in case II.
However, these are not the only solutions constructible
out of Chern-Simons forms.
For instance, in 4 dimensions there is the solution
$\omega^{0,4}= B^{0,3}dx^0$ where $B^{0,3}$ is a Chern-Simons
3-form. This solution is not Lorentz-invariant and is thus present
only in case I but not in case II. 
Have we missed this solution? The answer is ``no'' 
because it is equivalent to the solution $I^4=x^0P(F)$
where $P(F)=dB^{0,3}$. Namely we have
$B^{0,3}dx^0=d(-x^0 B^{0,3})+x^0dB^{0,3}$ and thus indeed
$B^{0,3}dx^0\sim x^0P(F)$. 
Note that in order
to establish this equivalence it is essential that we work in
the space of local forms that may depend explicitly on the $x^\mu$.

\paragraph{The exceptional solutions {\boldmath $W^{0,n}$}.}
The existence of a solution $W^{0,n}$ requires
a relation $k^iN_i=k^{A\tin}P_A(F)\Theta_\tin$ at ghost number 1,
cf.\ Eq.\ (\ref{P_A4}).
The $N_i$ with ghost number
number 1 are linear combinations of terms $(C^IF^J-C^JF^I)P(F)$ where
$C^I$, $C^J$, $F^I$, $F^J$ are abelian and $P(F)$ is
some characteristic class. The $\Theta_\tin$ with ghost number
1 are the abelian ghosts, and the
$P_A(F)$ are characteristic classes which are trivial
in the equivariant characteristic cohomology $H^n_\mathrm{char}(d,\cI)$.
Hence, in order that a solution $W^{0,n}$ exists, a nonvanishing 
linear combination
of terms $(C^IF^J-C^JF^I)P(F)$ must be equal to a linear combination
of the $P_A(F)C^I$, where $C^I$, $C^J$, $F^I$, $F^J$ are abelian.
The gauge group must therefore contain
at least two abelian factors
and, additionally, there must be at least 
two different $P_A(F)$ containing abelian field strengths.
This is really a very special situation not
met in practice (to our
knowledge), which must be included in the discussion because
we allow here for general Lagrangians. We
illustrate it with a simple example:
\[
n=4,\quad
L=\sum_{I=1}^2 \left[-\sfrac 14\,F_{\mu\nu}^I F^{\mu\nu I}
-\sfrac 12\,(\6_\mu\phi^I)\6^\mu\phi^I+\sfrac 14\,\phi^I
\epsilon^{\mu\nu\rho\sigma}F_{\mu\nu}^I F_{\rho\sigma}^2\right]
\]
where $F_{\mu\nu}^I=\6_\mu A_\nu^I-\6_\nu A_\mu^I$ are abelian
field strengths and
$\phi^1$ and $\phi^2$ are real scalar fields.
In this case we have $\{P_A(F)\}=\{F^1F^2,F^2F^2\}$ with
corresponding $\{-I^3_A\}=\{\star d\phi^1,\star d\phi^2\}$
and
$\{K_A\}=\{\star\phi^*_1,\star\phi^*_2\}$. Furthermore we have
one $N_\Gamma$ with ghost number 1 given by
$\epsilon_{JI}C^IF^JF^2$ and corresponding $b^4_\Gamma$ given by
$(1/2)\epsilon_{IJ}A^IA^JF^2$ ($\epsilon_{IJ}=-\epsilon_{JI}$).
(\ref{VGamma}) gives now the following solution:
\beq
W^{0,4}= d^4x \sum_{I,J=1}^2\epsilon_{IJ}[
\sfrac 14\,\epsilon^{\mu\nu\rho\sigma}A_\mu^I A_\nu^J F_{\rho\sigma}^2
+A_\mu^I\6^\mu\phi^J-\phi^*_IC^J].
\label{W04}
\eeq

\subsection{$H^{1,n}(s|d)$: Anomalies}

We know turn to $H^{1,n}(s|d)$, i.e., to anomalies.
The most general solution of
the consistency condition with ghost number 1 and
form-degree $n$ is
\[
\omega^{1,n}\sim \sum_{I:\mathrm{abelian}} C^I I^n_I
+B^{1,n}+V^{1,n}+W^{1,n}
\]
where:
\ben
\item
$I_I^n\in\cI$, i.e., $\{I_I^n\}$ is a set of strictly gauge 
invariant $n$-forms.
\item
When $n$ is even,
$B^{1,n}$ is a linear combination of the celebrated
chiral anomalies listed in Eq.\ (\ref{g=1a}), except
for those which contain abelian ghosts (the chiral anomalies
with abelian ghosts are already included
in $\sum_{I:\mathrm{abelian}} C^I I^n_I$).
When $n$ is odd, $B^{1,n}$ is a linear combination of
the solutions (\ref{g=1b}) which exist only
when the gauge group contains at least two abelian factors.
\item
$V^{1,n}$ are linear combinations of the solutions
$V_{\Delta\tin}$ (\ref{V}) with ghost number 1
related to global symmetries;
in order that $V_{\Delta\tin}$ has
ghost number 1, the $\Theta_\tin$ which appears in it
must have ghost number 2 and must thus
be a product of two abelian ghosts. Hence, solutions
$V^{1,n}$ exist only if the gauge group contains
at least two abelian factors. They are given by
\beq
V^{1,n}=\sum_{I,J:\mathrm{abelian}} \lambda^\Delta_{IJ}
\left[ K_\Delta C^IC^J+j_\Delta (A^IC^J-A^JC^I)\right].
\label{V1n}
\eeq
Using (\ref{dis1}), the antifield dependence
of $V^{1,n}$ can be made explicit, analogously to (\ref{V0n}).
\item
A discussion of Eq.\ (\ref{P_A4}) similar to the
one performed for $W^{0,n}$ shows that the
solutions $W^{1,n}$ are even more exceptional
than their counterparts in ghost number zero;
they exist only in the following situation:
the gauge group must contain
at least three abelian factors
and, additionally, there must be at least 
three different $P_A(F)$ containing abelian field strengths.
An example is the following:
\[
n=4,\quad
L=\sum_{I=1}^3 \left[-\sfrac 14\,F_{\mu\nu}^I F^{\mu\nu I}
-\sfrac 12\,(\6_\mu\phi^I)\6^\mu\phi^I+\sfrac 14\,\phi^I
\epsilon^{\mu\nu\rho\sigma}F_{\mu\nu}^I F_{\rho\sigma}^3\right]
\]
where $F_{\mu\nu}^I=\6_\mu A_\nu^I-\6_\nu A_\mu^I$ are abelian
field strengths, and
$\phi^I$ are real scalar fields. A solution $W^{1,4}$ is
\beq
W^{1,4}= d^4x \sum_{I,J,K=1}^3\epsilon_{IJK}[
\sfrac 14\,\epsilon^{\mu\nu\rho\sigma}C^I A_\mu^J A_\nu^K F_{\rho\sigma}^3
+C^I A_\mu^J\6^\mu\phi^K-\sfrac 12\,C^IC^J\phi^*_K].
\label{W14}
\eeq
\een

\paragraph{Nontriviality of the solutions.}
A solution 
$\sum_{I:\mathrm{abelian}} C^I I^n_I+B^{1,n}+V^{1,n}+W^{1,n}$
is only trivial when $B^{1,n}$, $V^{1,n}$ and $W^{1,n}$
all vanish and, additionally,
\[
\sum_{I:\mathrm{abelian}} C^I I^n_I
\approx \sum_{I:\mathrm{abelian}} C^IdI_I^{n-1}
+\sum_{I,J:\mathrm{abelian}}(C^IF^J-C^JF^I)P_{IJ}(F)
\]
for some
$I_I^{n-1}\in\cI$ and some characteristic classes $P_{IJ}(F)$.

\paragraph{Semisimple gauge group.}
Note that all nontrivial solutions $\omega^{1,n}$
involve abelian ghosts, except for the solutions $B^{1,n}$
in even dimensions. 
Hence,
when the gauge group is semisimple, the candidate gauge anomalies are
exhausted by the well-known nonabelian chiral anomalies
in even dimensions.  These live in the small algebra and can be
obtained from the characteristic classes living in two dimensions
higher through the Russian formula of section \ref{russian}.
Furthermore, these anomalies are in finite number (independently
of power counting arguments) and do not
depend on the specific form of the Lagrangian.  
For some groups, there
may be none (``anomaly-safe groups''), in which case the
consistency condition implies absence of anomalies, for
any Lagrangian.

In $4$ dimensions, $B^{1,4}$ is
the non abelian gauge anomaly \cite{Bardeen:1969md}:
\beq
B^{1,4}=d_{IJK} C^Id[A^JdA^K+\sfrac{1}{4}\,e\f LMJ A^KA^LA^M],
\eeq
where
$d_{IJK}$ is the general symmetric $\cG$-invariant tensor. 
Hence, the gauge group is anomaly safe for any Lagrangian in 4 dimensions
if there is no $d_{IJK}$-tensor.

\subsection{The cohomological groups $H^{g,n}(s|d)$ with $g>1$}

The results for ghost numbers $g>1$ are similiar to those
for $g=0$ and $g=1$; one gets
\[
\omega^{g,n}\sim  I^{n\,\tin_g}\Theta_{\tin_g}
+B^{g,n}+V^{g,n}+W^{g,n}
\]
where $\{\Theta_{\tin_g}\}$ is the subset of $\{\Theta_{\tin}\}$
containing those $\Theta$'s with ghost number $g$.
Of course, this subset depends on the gauge group $G$.
It depends both on $G$ and on the spacetime
dimension which solutions are present for given $g$.
For instance, for $G=SU(2)$ and $n=4$, one
has $\omega^{2,4}\sim V^{2,4}$.
The highest ghost number for which
nontrivial solutions exist is $g=\mathrm{dim}(G)$ because
this is the ghost number of the product of all
$\theta_r(C)$. 
Antifield dependent solutions
$V^{g,n}$ exist up to ghost number $g=\mathrm{dim}(G)-1$.
As we have mentioned already several times, solutions
$W^{g,n}$ exist only for exceptional Lagrangians.

\subsection{Appendix \ref{DisDisDis}.A: 
Gauge covariance of global symmetries}

(\ref{dis1}) and (\ref{dis2}) can be achieved because the
equations of motion are gauge covariant.
This is seen as follows. Consider an $n$-form
$I\in\cI$ which vanishes weakly, $I\approx 0$.
This is equivalent to
$I =\delta \7K$ for some $n$-form $\7K$.
The equations of motion are gauge covariant in the sense that
one has $\delta A_I^{*\mu}=L_I^\mu(x,\cov)$ and
$\delta \psi^*_i=L_i(x,\cov)$ where
the $L_I^\mu(x,\cov)$ are in
the adjoint representation of $\cG$ and
the $L_i(x,\cov)$ in the representation
dual to the representation of the $\psi^i$ (cf.\ section \ref{AdCelim}).
In particular, $\delta$ is stable
in the space of $\cG$-invariant functions $f_\mathrm{inv}(x,[F,\psi,
A^*,\psi^*,C^*]_D)$, i.e., it maps this space into itself.
We can thus choose
\beann
\7K&=&d^nx
\left[
A_I^{*\mu} \7Q^{I}_{\mu}(x,\cov)+
\psi^*_i\7Q^{i}(x,\cov)
\right.
\\
&&\phantom{d^nx\left[\right.}\left.
+(D_\nu A_I^{*\mu}) \7Q^{I\nu}_{\mu}(x,\cov)+
(D_\nu \psi^*_i)\7Q^{i\nu}(x,\cov)+\dots
\right]
\eeann
where $\7K$ is $\cG$-invariant.
Note that $\7K$ contains in general covariant derivatives
of antifields. To deal with these terms, we write
\beann
\7K&=&d^nx\, [A_I^{*\mu} Q^{I}_{\mu}+\psi^*_iQ^{i}
+D_\nu R^\nu]\ ,
\\
Q^{I}_{\mu}&=&\7Q^{I}_{\mu}
-D_\nu\7Q^{I\nu}_{\mu}+\dots\ ,
\\
Q^{i}&=&\7Q^{i}-D_\nu \7Q^{i\nu}+\dots\ ,
\\
R^\nu&=&A_I^{*\mu} \7Q^{I\nu}_{\mu}+
\psi^*_i\7Q^{i\nu}+\dots\ .
\eeann
Since $R^\nu$ is $\cG$-invariant, we have
$D_\nu R^\nu=\6_\nu R^\nu$ and thus
\beann
\7K&=&K+d R\ ,
\\
K&=&d^nx\, [A_I^{*\mu} Q^{I}_{\mu}+\psi^*_iQ^{i}]\ ,
\\
R&=&(-)^{n}\sfrac 1{(n-1)!}\, dx^{\mu_1}\dots dx^{\mu_{n-1}}
\epsilon_{\mu_1\dots\mu_n} R^{\mu_n}.
\eeann
Using this in $I=\delta \7K$, we get
\[
I+d\delta R=s K
\]
because the $\cG$-invariance of $K$ implies
$\delta K=sK$.

(\ref{dis1}) follows by setting
$I=-dj$ where $j\in\cI$ is a conserved current
(we have $dj=Dj\in\cI$).
Namely the above formula gives in this case $d(j-\delta R)=-sK$.
Note that $j-\delta R$ is equivalent to $j$
and gauge invariant (due to $\delta R\approx 0$ and $\delta R\in\cI$).
Hence we can indeed choose the basis $\{j_\Delta\}$ of the
inequivalent gauge invariant currents such that
$dj_\Delta=-sK_\Delta$ with $K_\Delta$ as in (\ref{dis1}).

Now consider the equation
$P(F)\approx dI^{n-1}$ with $I^{n-1}\in\cI$. Setting
$I=P(F)-dI^{n-1}$, the above formula gives
$P(F)-d(I^{n-1}-\delta R)=sK$.
Hence, we can choose all $I^{n-1}_A$ such that
$P_A(F)=sK_A+dI^{n-1}_A$ with $K_A$ as in (\ref{dis2}).

\newpage

\mysection{Free abelian gauge fields}\label{Free}

\subsection{Pecularities of free abelian gauge fields}

We now compute the local BRST cohomology
for a set of  $R$ abelian gauge fields
with a free Lagrangian of the Maxwell type,
\beq
L=-\frac 14\, \sum_{I=1}^R F_{\mu\nu}^I F^{\mu\nu I},\quad
F_{\mu\nu}^I=\6_\mu A_\nu^I-\6_\nu A_\mu^I\ .
\label{free1}
\eeq
This question is relevant for determining the possible 
consistent interactions
that can be defined among massless vector particles, where
both groups $H^{0,n}$ and $H^{1,n}$ play a r\^ole, as we shall
discuss in subsection \ref{disfreeab} below.

As we have already mentioned, theorem \ref{main}
does not hold for (\ref{free1}). The reason is that
the characteristic cohomology group $H^{n-2}_\mathrm{char}(d,\Omega)$ 
does not vanish in the free model. 
Rather, by theorem \ref{Loct12}, 
this cohomological group
is represented in all spacetime dimensions $n>2$
by the Hodge-duals
of the abelian curvature 2-forms $F^I=dA^I$,
\beq
\star F^I=\frac 1{(n-2)!\,2}\, dx^{\mu_1}\dots dx^{\mu_{n-2}}
\epsilon_{\mu_1\dots\mu_n}F^{\mu_{n-1}\mu_n I}\ .
\label{starF}
\eeq
This modifies the results for
the form-degrees $p= n-1$ and $p=n$ as compared to
theorem \ref{main}, by allowing solutions of a new
type. These solutions are precisely those that appear
in the non Abelian deformation of (\ref{free1}).
 
In contrast, the results
for lower form-degrees remain valid as an
inspection of the proof of the theorem shows since
one still has $H^{p}_\mathrm{char}(d,\Omega)=\delta^p_0\mathbb{R}$ 
for $p<n-2$.

The discussion of this section applies also to abelian gauge
fields with self-couplings involving the curvature only
(like in the Born-Infeld Lagrangian), or in the
case of non-minimal interactions with matter through
terms involving only the field strength (e.g., $F_{\mu \nu} \,
\bar{\psi} \gamma^{[\mu}\gamma^{\nu]}\psi$, where $\psi$ is
a Dirac spinor). 
In that case, the matter fields do not transform under the
abelian gauge symmetry so that the global reducibility
identities behind theorem \ref{Loct12} are still
present.

\subsection{Results}

We shall now work out the modifications for
form-degrees $p=n-1$ and $p=n$, assuming the spacetime
dimension $n$ to be greater than $2$.

\subsubsection{Results in form-degree $p=n-1$} 
Let $\omega^{n-1}$
be a cocycle of 
$H^{*,n-1}(s|d,\Omega)$,
\beq
s\omega^{n-1}+d\omega^{n-2}=0.
\label{free4}
\eeq
The same arguments as in the proof
of theorem \ref{main} until Eq.\ (\ref{soli4}) included
yield $\omega^{n-2}=I^{n-2\,\tin}\Theta_\tin$
where (i) $I^{n-2\,\tin}$ is gauge-invariant
and fulfills $dI^{n-2\,\tin}\approx 0$; and (ii) the $\Theta_\tin$
form a basis of polynomials in the undifferentiated ghosts 
(in the purely abelian
case each ghost polynomial is invariant; furthermore, Eq.\ (\ref{soli4})
has no solution $B^{n-2}$ in the purely abelian
case because the $B$'s cannot be lifted, see below).
The condition  $dI^{n-2\,\tin}\approx 0$ gives now
$I^{n-2\,\tin}\approx \lambda_I^\tin \star F^I+P^{n-2\,\tin}(F)+dI^{n-3\,\tin}$
where the linear combination $\lambda_I^\tin \star F^I$
of the $\star F^I$ comes
from $H^{n-2}_\mathrm{char}(d,\Omega)$.  It is here that
the extra characteristic cohomology enters and gives extra terms
in $I^{n-2\,\tin}$ compared with
Eq.\ (\ref{soli5}). These extra terms fulfill 
\beq
d\star F^I=-s\star A^{*I}
\label{free5}
\eeq
where $\star A^{*I}$ is the
antifield dependent $(n-1)$-form
\beq
\star A^{*I}=\frac 1{(n-1)!}\, dx^{\mu_1}\dots dx^{\mu_{n-1}}
\epsilon_{\mu_1\dots\mu_n}A^{*\mu_n}_J\delta^{JI}.
\label{starA*}
\eeq
It is then straightforward to
adapt Eqs.\ (\ref{soli6}) through (\ref{soli9}). This gives,
instead of Eq.\ (\ref{soli9}),
\[
\omega^{n-1}\sim 
\lambda_I^\tin (\star A^{*I}\Theta_\tin+(\star F^I)[\Theta_\tin]^1)
+B^{n-1}+I^{n-1\,\tin}\Theta_\tin\ .
\]
Since we are dealing with a purely abelian case, the $\Theta_\tin$
are just products of the undifferentiated ghosts.
We can therefore write the result,
up to trivial solutions, as
\bea
\omega^{n-1}\sim
\star A^{*I}P_I(C)+(\star F^I)A^J\6_JP_I(C)
+B^{n-1}+I^{n-1\,\tin}P_\tin(C)
\label{free8}
\eea
where $P_I(C)$ and $P_\tin(C)$ are arbitrary 
polynomials in the undifferentiated ghosts, and
\[ 
\6_I\equiv \frac{\6}{\6C^I}\ .
\]
Furthermore, the descent is particularly simple in the small algebra
because a non trivial bottom can be lifted only once; at the next
step, one meets an obstruction.  This implies
that the solutions $B^{n-1}$ can occur in (\ref{free8}) only when
the spacetime dimension $n$ is even: 
Eq.\ (\ref{Bp}) gives in the purely abelian case only
solutions with odd form-degrees, which are linear in the
one-forms $A^I$,
\bea
B^{2N+1}&=&\sum_{i=1}^{K}\lambda_{I_1\dots I_K J_1\dots J_N}
C^{I_1}\cdots C^{I_{i-1}}A^{I_i}
C^{I_{i+1}}\cdots C^{I_K}F^{J_1}\cdots F^{J_{N}}
\nonumber\\
B^{2k}&=&0
\label{freeBp}\eea
where $I_i<I_{i+1}$, $J_i\leq J_{i+1}$, and (if $N>0$) $I_1\leq J_1$.
These solutions descend on the gauge-invariant
term $\lambda_{I_1\dots I_K J_1\dots J_N}
C^{I_1}\cdots C^{I_K}F^{J_1}\cdots F^{J_{N}}$.

Eq. (\ref{free8}) gives the general solution of
(\ref{free4}), both in the space of all local forms
(case I) and in the space of Poincar\'e invariant local forms
(case II), with $I^{n-1\,\tin}\in\cI$ where $\cI$ is the respective
gauge invariant subspace of local forms,
\bea
\mbox{Case I:}&& \cI=\{\mbox{local functions of
$F_{\mu\nu}^I$, $\6_\rho F_{\mu\nu}^I$, \dots}\}
\otimes \Omega(\mathbb{R}^n)
\label{free22}\\
\mbox{Case II:}&&
\cI=\{\mbox{Lorentz-invariant local functions of
$dx^\mu$, $F_{\mu\nu}^I$, $\6_\rho F_{\mu\nu}^I$, \dots}\}.
\quad
\label{free17}
\eea
[As before, $\Omega(\mathbb{R}^n)$ denotes the space of ordinary 
differential forms in $\mathbb{R}^n$.]

\subsubsection{Results in form-degree $p=n$} 
Let $\omega^{n}$ be a cocycle of 
$H^{*,n}(s|d,\Omega)$, 
\beq
s\omega^{n}+d\omega^{n-1}=0 .
\label{free10}
\eeq
By the standard arguments of the descent equation technique,
$\omega^{n-1}$ is a cocycle of $H^{*,n-1}(s|d,\Omega)$
and trivial contributions to $\omega^{n-1}$ can be neglected
without loss of generality. Hence, $\omega^{n-1}$ can be assumed
to be of the form (\ref{free8}) and we have to analyse the
restrictions imposed on it by the fact that it can be lifted
once to give $\omega^{n}$ through (\ref{free10}).
To this end we compute $d\omega^{n-1}$.

To deal with the first two terms in (\ref{free8}),
we use once again (\ref{free5}) as well as
\beq
d \star A^{*I}=-s\star C^{*I}
\label{free11}
\eeq
where
\beq
\star C^{*I}=d^nx\, C^{*}_J\,\delta^{JI}.
\label{starC*}
\eeq
This yields\footnote{Eq.\ (\ref{free12}) can be elegantly
derived using the quantities 
$\star \4C^{*I}=\star C^{*I}+\star A^{*I}+\star F^I$ and
$\4C^I=C^I+A^I$. One has
$(s+d)\star \4C^{*I}=0$ and $(s+d)\4C^I=F^I$. This
implies $(s+d)[\star \4C^{*I}P_I(\4C)]=(-)^n(\star \4C^{*I})F^J\6_JP_I(C)$
whose $n$-form part is Eq.\ (\ref{free12}).}
\bea
&d [\star A^{*I}P_I(C)+(\star F^I)A^J\6_JP_I(C)]
=(-)^n(\star F^I)F^J\6_JP_I(C)&
\nonumber\\
&-s[\star C^{*I}P_I(C)+(\star A^{*I})A^J\6_JP_I(C)
+\sfrac 12 (\star F^I)A^JA^K\6_K \6_JP_I(C)].&
\label{free12}
\eea
The remaining terms in (\ref{free8}) are dealt with as in the
proof of part (i) of theorem \ref{main}. One gets
\beq
d[B^{n-1}+I^{n-1\,\tin}P_\tin(C)]
=-s[b^n+I^{n-1\,\tin}A^I\6_I P_\tin(C)]+N^n+(dI^{n-1\,\tin})P_\tin(C)
\label{free13}
\eeq
where $N^n$ is in the small algebra (it is an obstruction to a lift
in the small algebra, see corollary \ref{sAbasis} and theorem \ref{main}).
Using Eqs.\ (\ref{free12}) and (\ref{free13}) in Eq.\ (\ref{free10}),
one obtains
\beq
s(\omega^{n}-\dots)=(-)^nF^J\star F^I\6_JP_I(C)+N^n
+(dI^{n-1\,\tin})P_\tin(C).
\label{free14}
\eeq
The right hand side of (\ref{free14}) has zero antighost number
and does not contain the derivatives of the ghosts. 
Due to $\gamma A^I_\mu=\6_\mu C^I$ and
$\gamma C^I=\gamma A^{*\mu}_I=\gamma C^*_I=0$,
$\gamma (\omega^{n}-\dots)$ is a sum of field monomials each of which
contains derivatives of the
ghosts (unless it vanishes). Hence,
(\ref{free14}) implies that the term on the right hand side is 
$\delta$-exact, i.e., 
that it vanishes  weakly,
\beq
(-)^n(\star F^I)F^J\6_JP_I(C)+N^n
+(dI^{n-1\,\tin})P_\tin(C)\approx 0.
\label{free15}
\eeq
To analyse this condition, we must distinguish cases I and II.

We treat first case II, i.e.\ the space of 
Poincar\'e-invariant local forms, for which the
analysis can be pushed to the end. In this case we have
$I^{n-1\,\tin}\in\cI$ with $\cI$ as in (\ref{free17}).
Hence,
$dI^{n-1\,\tin}$ is a sum of field monomials each of which 
contains a first
or higher order derivative of at least 
one of the $F_{\mu\nu}^I$. Furthermore
the equations of motion contain at least first order derivatives
of the $F_{\mu\nu}^I$. Hence, in case II,
the part of Eq.\ (\ref{free15}) which
contains only undifferentiated $F_{\mu\nu}^I$ reads 
$(-)^nF^J\star F^I\6_JP_I(C)+N^n=0$.%
\footnote{The same argument yields 
$P(F)\approx dI,I\in\cI\ \then\ P(F)=0$ in case II.\label{footP}}
This implies that both $F^J\star F^I\6_JP_I(C)$ and $N^n$
vanish
since $N^n$ contains the $F_{\mu\nu}^I$ only
via wedge products of the $F^I$ (all wedge products of the $F^I$
are total derivatives while no $F^J\star F^I$ is a total derivative).
Eq.\ (\ref{free15}) yields thus
\beq
\mbox{Case II:}\quad F^J\star F^I\6_JP_I(C)=0,\quad
N^n=0,\quad (dI^{n-1\,\tin})P_\tin(C)\approx 0.
\label{free18}
\eeq
Since $F^J\star F^I=-\frac 12 d^nx F_{\mu\nu}^JF^{\mu\nu I}$
is symmetric in $I$ and $J$, the first condition in (\ref{free18})
gives
\beq
\mbox{Case II:}\quad \6_IP_J(C)+\6_JP_I(C)=0.
\label{free19}
\eeq
The general solution of Eq.\ (\ref{free19}) is obtained
from the cohomology $H(D,\cC)$ of the differential $D=\xi^I\6_I$ in
the space $\cC$ of polynomials in commuting extra variables $\xi^I$ and
anticommuting variables $C^I$. Indeed, by contracting
Eq.\ (\ref{free19}) with $\xi^I\xi^J$, it reads
$Da=0$ where $a=\xi^I P_I(C)$. Using the contracting homotopy
$\varrho=C^I\6/\6\xi^I$ (see appendix \ref{IIBContractible}),
one easily proves that $H(D,\cC)$ is represented
solely by pure numbers (``Poincar\'e lemma for $D$''; note that $D$ is
similar to $dx^\mu\6/\6x^\mu$ except that the ``differentials'' $\xi^I$
commute while the ``coordinates'' $C^I$ anticommute). 
In particular this implies that $Da=0\then a=DP(C)$
for $a=\xi^I P_I(C)$, i.e.,
\beq
\mbox{Case II:}\quad P_I(C)=\6_I P(C)
\label{free20}
\eeq
for some polynomial $P(C)$ in the $C^I$. 
The analysis of Eq.\ (\ref{free10})
can now be finished along the lines of the proof of
theorem \ref{main}. One obtains in case II that the general solution
of Eq.\ (\ref{free10}) is, up to trivial solutions, given by
\bea
\mbox{Case II:}
&\omega^n=
[\star C^{*I}\6_I+(\star A^{*I})A^J\6_J\6_I
+\sfrac 12 (\star F^I)A^JA^K\6_K \6_J\6_I] P(C)&
\nonumber\\
&+B^n+I^{n\,\tin}P_\tin(C)
+[K_\Delta+j_\Delta A^I\6_I] P^\Delta(C)&
\label{free21}
\eea
where $P(C)$, $P_\tin(C)$ and $P^\Delta(C)$
are arbitrary polynomials in the undifferentiated ghosts,
$B^n$ occurs only in odd dimensional spacetime
due to (\ref{freeBp}), $j_\Delta\in\cI$ are
gauge-invariant and Poincar\'e-invariant conserved
$(n-1)$-forms, see Eqs. (\ref{Delta5}) through (\ref{jbasis2}),
and $K_\Delta$ contains the global symmetry corresponding
to $j_\Delta$ and satisfies $sK_\Delta+dj_\Delta=0$.
There are no solutions $W_\Gamma$ because
all characteristic classes $P(F)$, including those with form-degree $n$,
are nontrivial in the equivariant characteristic cohomology
$H_\mathrm{char}(d,\cI)$ with $\cI$ as in (\ref{free17}) (cf.\ footnote 
\ref{footP}).

Let us finally discuss case I, i.e., the space of all local forms
(with a possible, explicit $x$-dependence).
In this case Eq.\ (\ref{free15}) holds for
$I^{n-1\,\tin}\in\cI$ with $\cI$ as in (\ref{free22}).
In contrast to case II, $dI^{n-1\,\tin}$ may thus 
contain field monomials which involve only undifferentiated $F_{\mu\nu}^I$
because $I^{n-1\,\tin}$ may depend explicitly on the 
spacetime coordinates $x^\mu$. Therefore
the arguments that have led us to Eq.\ (\ref{free20}) do
not apply in case I. In fact one finds in all 
spacetime dimensions $n\neq 4$ that Eq.\ (\ref{free20})
need not hold in case I. Rather, if $n\neq 4$,
(\ref{free10}) does not impose any
restriction on the $P_I(C)$ at all, i.e., the
terms in $\omega^{n-1}$ related to the $P_I(C)$ can be lifted
to a solution $\omega^n$ for any set $\{P_I(C)\}$.
This solution is $d^nx\, a$ where
\bea
a&=&
(C^{*I}+A^{*\mu I}A_\mu^J\6_J
-\sfrac 12\, F^{\mu\nu I}A_\mu^J A_\nu^K \6_K \6_J) P_I(C)
\nonumber\\
&&+\sfrac {2}{n-4}\, 
F^I_{\mu\nu}(x^\mu A^{*\nu J}+x^\mu F^{\rho \nu J}A_\rho^K\6_K
+\sfrac 14F^{\mu\nu J}x^\rho A_\rho^K\6_K)
\6_{(I}P_{J)}(C).\quad
\label{free23}
\eea
$d^nx\, a$ fulfills (\ref{free20}), i.e., $s(d^nx\, a)+da^{n-1}=0$
where $a^{n-1}$ is indeed of the form (\ref{free8}),
\beann
a^{n-1}&=&[\star A^{*I}+(\star F^I)A^J\6_J]P_I(C)
+\sfrac 1{(n-1)!}\,
dx^{\mu_1}\dots dx^{\mu_{n-1}}\epsilon_{\mu_1\dots\mu_n}
I^{\mu_n},
\\
I^\mu&=&\sfrac {2}{n-4}\left(\sfrac 14\, x^\mu F_{\nu\rho}^IF^{\nu\rho J}+
F^{\mu\nu I}F_{\nu\rho}^J x^\rho\right)\6_{(I}P_{J)}(C).
\eeann
Note that both $a$ and $a^{n-1}$ depend explicitly on $x^\mu$
and are therefore present only in case I but not in case II,
except when $\6_{(I}P_{J)}(C)=0$ (then $d^nx\, a$ reproduces
the first line of (\ref{free21}). 
When one multiplies $a$ by $n-4$, one gets solutions
for all $n$. For $n=4$, they become solutions of the form
$[K_\Delta+j_\Delta A^I\6_I] P^\Delta(C)$ with gauge invariant
Noether currents $j_\Delta\in\cI$ involving
explicitly the $x^\mu$. One may now proceed
along the previous lines. 
However, two questions remain open in case I: Does
(\ref{free10}) impose restrictions on the $P_I(C)$ 
when $n=4$ ? Are there characteristic classes $P(F)$ with 
form-degree $n$ which
are trivial in the equivariant characteristic cohomology
$H_\mathrm{char}(d,\cI)$ with $\cI$ as in (\ref{free22}) ?
(See also footnote \ref{cmperratum}.) 

\subsection{Uniqueness of Yang-Mills cubic vertex}
\label{disfreeab}

We now use the above results to discuss the consistent deformations 
of the action (\ref{free1}). Requiring that the interactions
be Poincar\'e invariant, the relevant results are those
of case II.

As shown in \cite{Barnich:1993vg}, the consistent deformations
of an action are given,
to first order in the deformation parameter,
by the elements of $H^{0,n}(s \vert d)$, i.e.,
here, from (\ref{free21}),
\bea
&\omega^{0,n}=
[\star C^{*I}\6_I+(\star A^{*I})A^J\6_J\6_I
+\sfrac 12 (\star F^I)A^JA^K\6_K \6_J\6_I] P(C)&
\nonumber\\
&+B^n+I^{n}
+[K_\Delta+j_\Delta A^I\6_I] P^\Delta(C)&
\label{free40}
\eea  
where $P(C)$ has ghost number $3$ and $P^\Delta(C)$
ghost number one ($P_\tin(C)$ in (\ref{free21}) has ghost number
zero and thus is a constant; this has been taken into account
in (\ref{free40})).  

The term $B^n$ is the familiar Chern-Simons term \cite{Deser:1982vy}, 
and exists
only in odd dimensions.  It belongs to the small algebra and is 
of the form $A F\dots F$.
The term $I^{n}$ is strictly gauge-invariant and thus
involves the abelian field strengths and their derivatives.
Born-Infeld or Euler-Heisenberg deformations are of this type.
Since these terms are well understood and do not
affect the gauge symmetry, we shall drop them from
now on and focus on the other two terms, which
are,
\beq
[\star C^{*I}\6_I+(\star A^{*I})A^J\6_J\6_I
+\sfrac 12 (\star F^I)A^JA^K\6_K \6_J\6_I] P(C)
\label{NonAbDef}
\eeq
and
\beq
[K_\Delta+j_\Delta A^I\6_I] P^\Delta(C)
\label{FTvert}
\eeq
Expression (\ref{NonAbDef}) involves the antifields
conjugate to the ghosts, while (\ref{FTvert}) involves only
the antifields conjugate to $A^I$.

Now, it has also been shown in \cite{Barnich:1993vg} (see
\cite{Henneaux:1997bm} for further details) that deformations involving
nontrivially the antifields do deform the gauge symmetries.  Those
that involve the antifields conjugate to the ghosts
deform not only the gauge transformations but also their
algebra; while those that involve only $A^*_I$ modify
the gauge transformations but leave
the gauge algebra unchanged (at least to first order in
the deformation parameter).

Writing $P(C) = (1/3!) f_{IJK} C^I C^J C^K$ (with
$f_{IJK}$ completely antisymmetric), one gets from
(\ref{NonAbDef}) that the deformations of the theory
that deform the gauge algebra are given by
\bea
&\sfrac 12\star C^{*I} f_{IJK} C^J C^K
+(\star A^{*I})A^J f_{IJK} C^K
+\sfrac 12 (\star F^I)A^JA^Kf_{IJK} 
&
\nonumber\\
&
= -d^nx\, (\sfrac 12 f_{IJK} C^KC^JC^{*I}+f_{IJK} A_\mu^J C^K A^{*\mu I}
+\sfrac 12 f_{IJK} F^{\mu\nu I} A_\mu^J A_\nu^K).
&
\label{NonAbDef'}
\eea   
The term independent of the antifields is the 
first order deformation of the action,
and one recognizes the standard Yang-Mills cubic vertex -- except
that the $f_{IJK}$ are not subject to the Jacobi identity at this stage.
This condition arises, however, when one investigates consistency
of the deformation to second order: the deformation is obstructed
at second order by a non trivial element of $H^{1,n}(s \vert d)$
unless 
$\sum_K (f_{IJK}f_{KLM}+f_{JLK}f_{KIM}+f_{LIK}f_{KJM})=0$
\cite{Barnich:1994pa}.  The obstruction
is precisely of the type (\ref{NonAbDef}), 
with $P(C)=-(1/36) \sum_K f_{IJK}f_{KLM}C^IC^JC^LC^M$.
The field monomials in front of the
antifields in (\ref{NonAbDef'})
give the deformations of the BRST transformations of the
ghosts and gauge fields respectively (up to a minus sign) 
and provide thus
the nonabelian extension of the abelian gauge transformations
and their algebra.
Thus, one recovers the known fact that the Yang-Mills
construction provides the only deformation of the
action for a set of free abelian gauge fields that deforms
the algebra of the gauge transformations at first order
in the deformation parameter.  Any 
consistent interaction which deforms nontrivially the
gauge algebra at first order
contains therefore the Yang-Mills vertex.  Furthermore, this
deformation automatically incorporates the Lie algebra structure
underlying the Yang-Mills theory, without having to postulate it
a priori.  This result has been derived recently in
\cite{wald1}, along different lines und under stronger assumptions
on the form of the new gauge symmetries.
These extra assumptions are in fact not necessary as the
cohomological derivation shows.

Having dealt with the deformations (\ref{NonAbDef}), we can turn to
the deformations (\ref{FTvert}), 
which do not deform the (abelian) gauge algebra at first order
although they do deform the gauge transformations.  
These involve Lorentz covariant and gauge
invariant conserved currents
$j_\Delta^\mu$.
An example of a deformation of this type is given by
the Freedman-Townsend vertex in three dimensions
\cite{Freedman:1981us,Anco:1995wt}.  In four dimensions,
however, the results of \cite{Torre:1995kb} indicate
that there is no (non trivial) candidate for $j_\Delta^\mu$.
There is an infinite number of conservation laws because
the theory is free, but these do not involve gauge invariant
Lorentz
vectors.
Thus, there is no Poincar\'e invariant deformation
of the type (\ref{FTvert}) in four dimensions.
This strengthens the above result on the uniqueness of the
Yang-Mills cubic vertex, which is the only vertex deforming
the gauge transformations in four dimensions.
Accordingly, in four dimensions, 
the most general deformation of the action for
a set of free abelian gauge fields is given, at first order,
by the Yang-Mills cubic vertex and by strictly gauge invariant
deformations.  We do not know whether the results of 
\cite{Torre:1995kb} generalize to higher dimensions, leaving the
(unlikely in our opinion) possibility of the existence of interactions
of the type (\ref{FTvert}) in $n>4$ dimensions,
which would deform the gauge
transformations without modifying the gauge algebra at first
order in the deformation parameter.

\newpage

\mysection{Three-dimensional Chern-Simons theory}
\label{CS}

\subsection{Introduction -- $H(s)$}

We shall now describe the local BRST cohomology in
3-dimensional pure Chern-Simons theory with general gauge group $G$,
i.e., $G$ may be abelian, semisimple, or
the direct product of
an abelian and a semisimple part. 
Since pure Chern-Simons theory 
is of the Yang-Mills type, theorem \ref{main}
applies to it when the gauge group is semisimple. 
So, the Chern-Simons case is not really special from this
point of view.  However, the results are particularly
simple in this case because the theory is topological. 
As we shall make it explicit below, there
is no non-trivial local, gauge-invariant function and
the BRST cohomology reduces to the Lie algebra cohomology
with coefficients in the trivial representation.

It is because of this, and because of the physical interest
of the Chern-Simons theory, that we devote a special section to it.

The Chern-Simons action is \cite{Deser:1982vy}
\beq
S_\mathrm{CS} = \int 
g_{IJ}[\sfrac 12\, A^I d A^J+\sfrac 16\, e\, \f KLI A^JA^KA^L] .
\label{CS0}
\eeq
where $g_{IJ}=\delta_{IJ}$ for the abelian part of $G$
and $g_{IJ}=\f IKL \f JLK$ for the nonabelian part.
This yields explicitly
\bea
s\, A^{\mu *}_I&=&\sfrac 12\,g_{IJ}\epsilon^{\mu\nu\rho}F_{\nu\rho}^J
+e\, C^J\f JIK A^{\mu *}_K\ ,
\nonumber\\
s\, C^*_I&=&-D_\mu A^{\mu *}_I+e\, C^J\f JIK C^*_K\ .
\label{CS*}
\eea
Again we shall determine
$H(s,\Omega)$ both in the space of all local forms (case I)
and in the space of Poincar\'e-invariant local forms (case II).

We first specify 
$H(s,\Omega)$ in these cases, using the results of Section 
\ref{LieAlgebraCoho}.
Since in pure Chern-Simons theory
the field strengths vanish weakly and do not
contribute to $H(s,\Omega)$ at all,
we find from this section that
in case I  $H(s,\Omega)$ is represented
by polynomials in the $\theta_r(C)$ which can also depend
explicitly on the spacetime coordinates $x^\mu$ and the differentials
$dx^\mu$,
\beq
\mbox{Case I:}\quad
s\omega=0\ \LRA\ \omega=P(\theta(C),x,dx)+s\eta\ .
\label{CS1}
\eeq
In case II, a similar result holds, but now 
no $x^\mu$ can occur and Lorentz invariance enforces
that the differentials can contribute nontrivially only
via the volume form $d^3 x$,
\beq
\mbox{Case II:}\quad
s\omega=0\ \LRA\ \omega=Q(\theta(C))+d^3x P(\theta(C))+s\eta\ .
\label{CS2}
\eeq
(\ref{CS1}) and (\ref{CS2}) provide the solutions
of the consistency condition with a trivial descent.
They also yield the bottom forms which can appear in
nontrivial descents. 
To find all solutions with a nontrivial descent, we
investigate how far these bottom forms can be lifted to
solutions with higher form-degree.
\smallskip

\subsection{BRST cohomology in the case of $x$-dependent forms}

In order to lift a
bottom form $P(\theta(C),x,dx)$
once, it is necessary and sufficient that
$d_x P(\theta(C),x,dx)=0$ where $d_x=dx^\mu \6/\6x^\mu$
(this is nothing but Eq.\ (\ref{rough4}), specified
to $P(\theta(C),x,dx)=I^\tin(x,dx)\Theta_\tin$).
$d_x P(\theta(C),x,dx)=0$ implies $P(\theta(C),x,dx)
=Q(\theta(C))+d_x P'(\theta(C),x,dx)$ 
by the ordinary Poincar\'e lemma, and thus
$P(\theta(C),x,dx)=Q(\theta(C))+d P'(\theta(C),x,dx)
+s [A^I\6_IP'(\theta(C),x,dx)]$ where we used once again
Eq.\ (\ref{rough0}) and the notation 
\[
\6_I=\frac{\6}{\6C^I}\ .
\]
The pieces $d P'(\dots)+s[A^I\6_IP'(\dots)]$ are trivial
and can thus be neglected without loss of generality.
Hence, all bottom forms which can be lifted once
can be assumed to be of the form $Q(\theta(C))$.
Furthermore there are no obstructions to lift
these bottom forms to higher form-degrees.
An elegant way to see this and to construct the 
corresponding solutions at higher form-degree is the following.
We introduce
\beq
\cC^I=C^I+A^I+\star A^{*I}+\star C^{*I}\ ,
\label{CS3}
\eeq
where
\[
\star A^{*I}=\frac 12\, dx^\mu dx^\nu 
\epsilon_{\mu\nu\rho} g^{IJ} A^{*\rho}_J\quad,\quad
\star C^{*I}= d^3x g^{IJ} C^*_J\ ,
\]
(see also for instance \cite{Carvalho:1996uu}).
Using (\ref{CS*}), one verifies
\beq
(s+d)\, \cC^I=\frac 12\, e\, \f JKI\cC^K\cC^J.
\label{CS4}
\eeq
Hence, $(s+d)$ acts on the $\cC^I$ exactly as $s$ acts 
on the $C^I$. This implies $(s+d)\theta_r(\cC)=0$
(which is analogous to $s\theta_r(C)=0$) and thus
\beq
(s+d)\,Q(\theta(\cC))=0.
\label{CS5}
\eeq
This equation decomposes into
the descent equations $s[Q]^3+d[Q]^2=0$, $s[Q]^2+d[Q]^1=0$,
$s[Q]^1+d[Q]^0=0$, $s[Q]^0=0$ where $[Q]^p$ is the
$p$-form contained in $Q(\theta(\cC))$,
\bea
Q(\theta(\cC))&=&\sum_{p=0}^3 [Q]^p
\nonumber\\
{}[Q]^0&=&Q(\theta(C))
\nonumber\\
{}[Q]^1&=&A^I\6_I\,Q(\theta(C))
\nonumber\\
{}[Q]^2&=&
   [\sfrac 12\, A^I A^J\6_J\6_I+\star A^{*I}\6_I]\,\,Q(\theta(C))
\nonumber\\
{}[Q]^3&=&
   [\sfrac 16\, A^I A^J A^K\6_K\6_J\6_I
   +A^I\star A^{*J}\6_J\6_I
   +\star C^{*I}\6_I]\,Q(\theta(C)).
\label{CS6}
\eea
We conclude that the general solution of the consistency
condition is at the various form-degrees given by
\bea
\mbox{Case I:}\quad
H^{*,0}(s|d,\Omega):&&
\omega^0\sim P(\theta(C),x)
\nonumber\\
H^{*,1}(s|d,\Omega):&&
\omega^1\sim [Q]^1+dx^\mu P_\mu(\theta(C),x)
\nonumber\\
H^{*,2}(s|d,\Omega):&&
\omega^2\sim [Q]^2+dx^\mu dx^\nu P_{\mu\nu}(\theta(C),x)
\nonumber\\
H^{*,3}(s|d,\Omega):&&
\omega^3\sim [Q]^3
\label{CS7}
\eea
where one can assume that $dx^\mu P_\mu(\theta(C),x)$ and 
$dx^\mu dx^\nu P_{\mu\nu}(\theta(C),x)$ are not
$d_x$-closed because otherwise they are trivial by the arguments
given above. For the same reason every 
contribution $d^3x P(\theta(C),x)$ to $\omega^3$ is trivial
(it is a volume form and thus
automatically $d_x$-closed) and has therefore not been written
in (\ref{CS7}).
\smallskip

\subsection{BRST cohomology in the case of Poincar\'e invariant forms}
This case is easy. By Eq.\ (\ref{CS2}), all nontrivial
bottom forms that can appear in case II 
are either 0-forms $Q(\theta(C))$ or volume-forms $d^3x P(\theta(C))$.
We know already that 
the former can be lifted to the above
Poincar\'e-invariant solutions $[Q]^1$, $[Q]^2$ and $[Q]^3$.
Hence, we only need to discuss
the volume-forms $d^3x P(\theta(C))$.
They are nontrivial in case II, in contrast to case I, except
for the banal case that $P(\theta)$ vanishes identically.
Indeed, assume that $d^3x P(\theta(C))$ is trivial, i.e.,
that $d^3x P(\theta(C))=s\eta_3+d\eta_2$
for some Poincar\'e-invariant local forms $\eta_3$ and $\eta_2$.
The latter equation has to hold identically in all the
fields, antifields and their derivatives. In particular, it must therefore
be fulfilled when we set all fields, antifields
and their derivatives equal to zero except for the undifferentiated
ghosts. This yields an
equation $P(\theta(C))=sh(C)$ since 
$\eta_2$ does not involve $x^\mu$ in case II (in
contrast to case I). By the Lie algebra cohomology,
$P(\theta(C))=sh(C)$ implies $P(\theta(C))=0$, see Section \ref{LieAlgebraCoho}.
Hence, in the space of Poincar\'e-invariant
local forms, no nonvanishing volume-form $d^3x P(\theta(C))$ is trivial
and the general solution of the consistency condition
reads
\bea
\mbox{Case II:}\quad
H^{*,0}(s|d,\Omega):&&
\omega^0\sim [Q]^0
\nonumber\\
H^{*,1}(s|d,\Omega):&&
\omega^1\sim [Q]^1
\nonumber\\
H^{*,2}(s|d,\Omega):&&
\omega^2\sim [Q]^2
\nonumber\\
H^{*,3}(s|d,\Omega):&&
\omega^3\sim [Q]^3+d^3x P(\theta(C)).
\label{CS8}
\eea

\paragraph{Antifield dependence.}

$[Q]^2$ and $[Q]^3$ contain antifields. 
This antifield dependence can actually be removed
by the addition of trivial solutions,
except when $Q(\theta(C))$ contains abelian ghosts.
For instance, consider a $\theta_r(C)=-\frac 13e\mathrm{Tr}(C^3)$ 
with $m(r)=2$.
We know that this $\theta_r(C)$ can be completed to $q_r(\4C,F)
=\mathrm{Tr}[\4C F-\frac 13e\4C^3]$, $\4C=C+A$, which satisfies
$(s+d)q_r(\4C,F)=0$ in 3 dimensions, see subsection \ref{russian}.
The solutions arising from $q_r(\4C,F)$ do not
involve antifields and are indeed equivalent
to those obtained from 
$\theta_r(\cC)=-\frac 13e\mathrm{Tr}(\cC^3)$.
Namely one has 
\[
\mathrm{Tr}(\tilde C F-\sfrac 13\,e\, \tilde C^3)
\nonumber\\
=-\sfrac 13\,e\, \mathrm{Tr}(\cC^3)+
(s+d)
\mathrm{Tr}(\tilde C\star A^*+\tilde C\star C^*)
\]
where $\star A^*=\star A^{*I}T_I$,
$\star C^*=\star C^{*I}T_I$.
Analogous statements apply to all $\theta_r(C)$ with
$m(r)>1$ and thus to all polynomials thereof. 
In particular, when the gauge group is semisimple,
all $[Q]^p$ can be replaced
by antifield independent representatives
arising from polynomials $Q(q(\4C,F))$. It is then
obvious that (\ref{CS7}) and (\ref{CS8}) reproduce theorem \ref{main}
when the gauge group is semisimple:
in the notation of theorem \ref{main}, one gets
solutions $B^1$, $B^2$ and $B^3$ given by the 1-, 2- and
3-forms in $Q(q(\4C,F))$, and solutions $I^\tin(x,dx)\Theta_\tin$
given by $P(\theta(C),x)$, $dx^\mu P_\mu(\theta(C),x)$ and 
$dx^\mu dx^\nu P_{\mu\nu}(\theta(C),x)$ in case I, and by
$[Q]^0$ and $d^3x P(\theta(C))$ in case II respectively.
In particular there are no solutions $V_{\Delta\tin}$
when the gauge group is semisimple case because there are no nontrivial
Noether currents at all in that case. The latter
statement follows directly
from the results, because $H^{-1,3}(s|d,\Omega)$
is empty when the gauge group is semisimple.

In contrast, the antifield dependence cannot be completely removed
when $Q(\theta_r(C))$ contains abelian ghosts (recall that the
abelian ghosts coincide with those $\theta_r(C)$ which
$m(r)=1$). For instance, the abelian $\cC^I$ satisfy $(s+d)\cC^I=0$
and provide thus
solutions to the descent equations by themselves.
The 3-form solution in an abelian $\cC^I$ is
$d^3x\, C^*_I$. This is a nontrivial solution
because it is also a nontrivial representative
of $H^3_2(\delta|d,\Omega)$,
see Section \ref{Koszul2Section}. Since it is
a nontrivial solution with negative ghost number,
it is impossible to make it antifield independent.

\subsection{Examples}

Let us finally spell out the results
for $H^{g,3}(s|d,\Omega)$, $g\leq 1$,
when the gauge group is either simple and compact, or purely abelian.
The generalization to a general gauge group (product of 
abelian factors times a semi-simple group) is straightforward.

\paragraph{Simple compact gauge group.} In this case
$H^{g,3}(s|d,\Omega)$ vanishes for $g<0$ and $g=1$, while
$H^{0,3}(s|d,\Omega)$ is one-dimensional and represented by
the 3-form $\omega^{0,3}$ contained in
$-\frac 13e\mathrm{Tr}(\cC^3)$ (both in case I and II),
\bea
&H^{g,3}(s|d,\Omega)=0\quad\mbox{for $g<0$ and $g=1$},&
\label{CS9}\\
&\omega^{0,3}=-e\mathrm{Tr}[\sfrac 13A^3+(CA+AC)\star A^*+C^2\star C^*]\ .&
\label{CS10}
\eea

\paragraph{Purely abelian gauge group.}
In this case we have $Q(\theta(C))\equiv
f_{I_1\dots I_k}C^{I_1}\cdots C^{I_k}$ where $f_{I_1\dots I_k}$
are arbitrary constant antisymmetric coefficients.
In case I, this yields the following nontrivial representatives of
$H^{g,3}(s|d,\Omega)$ for $g=-2,\dots,1$:
\bea
\omega^{-2,3}&=& f_{I}\star C^{*I}
\nonumber\\
\omega^{-1,3}&=& 2f_{IJ}(A^{I}\star A^{J}
+C^{I}\star C^{*J})
\nonumber\\
\omega^{0,3}&=& f_{IJK}(
A^IA^JA^K+6C^IA^J\star A^{*K}
+3 C^IC^J\star C^{*K})
\nonumber\\
\omega^{1,3}&=& f_{IJKL}C^I(
4A^JA^KA^L+12 C^J A^K\star A^{*L}
+4 C^JC^K\star C^{*L}).
\label{CS11}
\eea
In case II one gets in addition representatives
of $H^{0,3}(s|d,\Omega)$ and $H^{1,3}(s|d,\Omega)$
given by the volume element $d^3x$ and by
$d^3x\, a_IC^I$ respectively (with $a_I$ arbitrary
constant coefficients).

\newpage

\mysection{References for other gauge theories}

In this section, we give references to works where the previous algebraic
techniques have been used to find the general solution of the
consistency condition with antifields included
(and for the BRST differential associated with 
gauge symmetries) in other field theoretical contexts.

Some aspects of local BRST cohomology for the Stueckelberg model are 
investigated in \cite{Dragon:1997tk}. 
The general solution of the Wess-Zumino consistency
condition for gauged non-linear $\sigma$-models is discussed in
\cite{Henneaux:1998hq,Henneaux:2000dc}.

Cohomological techniques (Poincar\'e lemma, BRST cohomology) have been 
recently analyzed on the lattice in 
\cite{luscher,Fujiwara:1999fi,Suzuki:2000ii}.

Algebraic aspects of gravitational anomalies \cite{Alvarez-Gaume:1984ig}
are discussed in \cite{Bardeen:1984pm,Baulieu:1984pf,Langouche:1984gn,%
Bonora:1986ic,Schucker:1987ca,Dragon:1992fn}.  
The general solution of the consistency
condition without antifields is given in \cite{Brandt:1990et}; this work is
extended to include the antifields in spacetime
dimensions strictly greater than two in \cite{Barnich:1995ap}, where again,
the cohomology of the Koszul-Tate differential is found to play a crucial
r\^ole.

In $2$ spacetime dimensions, these groups have been analyzed in 
the context of the (bosonic) string world sheet action coupled to
backgrounds in
\cite{Baulieu:1986hw,Baulieu:1987jz,Becchi:1988as,%
WerneckdeOliveira:1993ig,Bandelloni:1993cq,Bandelloni:1995vy,%
Buchbinder:1995ns,Tataru:1996ru,Blaga:1995rv}. 
Complete results, with the antifields included, are derived in
\cite{Brandt:1996gu,Brandt:1996nn,Brandt:1998cy}.

Algebraic results on the Weyl anomaly
\cite{Capper:1974ic,%
Deser:1976yx,Deser:1993yx} may be found in \cite{Bonora:1983ff,Bonora:1986cq}.

Algebraic aspects of the consistency 
condition for $N=1$ supergravity in $4$ dimensions
have been discussed in 
\cite{Bonora:1986ug,Baulieu:1987dp,Brandt:1994vd,Piguet:1998bj}.  
The complete treatment, with antifields included, is given in
\cite{Brandt:1997au}.

For $p$-form gauge theories, the local BRST cohomology groups without
antifields have been 
investigated in \cite{Baulieu:1984ih} 
and more recently, with antifields,  in 
\cite{Henneaux:1997nz,Henneaux:1997ws,Verbovetsky:1997fi,
Henneaux:1997ha,Garcia:1998kp,%
Henneaux:1999rp,Knaepen:1999zr}.

\newpage

\section*{Acknowledgements}

\addcontentsline{toc}{section}{Acknowledgements}

We acknowledge discussions with many colleagues in various
stages of development of this work, especially 
Jos\'e de Azc\'arraga, Norbert Dragon, Michel Dubois-Violette,
Jean Fisch, Maximilian Kreuzer, Sergei Kuzenko,
Christiane Schomblond, Jim Stasheff, Raymond Stora,
Michel Talon, Claudio Teitelboim, Jan-Willem van Holten,
Claude Viallet and Steven Weinberg.
G.B. is Scientific Research Worker of the ``Fonds National Belge de la 
Recherche Scientific''. He also acknowledges the hospitality of the 
Department of Theoretical Physics of the University of Valencia.
The work of G.B. and M.H. is supported in part by the ``Actions de
Recherche Concert{\'e}es" of the ``Direction de la Recherche
Scientifique - Communaut{\'e} Fran{\c c}aise de Belgique", by
IISN - Belgium (convention 4.4505.86) and by
Proyectos FONDECYT 1970151 and 7960001 (Chile).
The work of F.B. was supported by the 
Deutsche For\-schungs\-ge\-mein\-schaft
through a Habilitation grant.


\newpage

\addcontentsline{toc}{section}{Bibliography}

\begin{thebibliography}{999}
\baselineskip=12truept
\bibitem{Alexandrov:1997kv}
M.~Alexandrov, M.~Kontsevich, A.~Schwartz and O.~Zaboronsky,
``The Geometry of the Master equation and Topological Quantum Field Theory,''
Int.\ J.\ Mod.\ Phys.\ {\bf A12} (1997) 1405
hep-th/9502010.
\bibitem{Alvarez-Gaume:1984ig}
L.~Alvarez-Gaum\'e and E.~Witten, ``Gravitational Anomalies,''
Nucl.\ Phys.\  {\bf B234} (1984) 269.
\bibitem{Anco:1995wt}
S.C.~Anco,
``New spin-one gauge theory in three dimensions,''
J.\ Math.\ Phys.\ {\bf 36} (1995) 6553.
\bibitem{Anco:1997mw}
S.C.~Anco,
``Novel generalization of three-dimensional Yang-Mills theory,''
J.\ Math.\ Phys.\ {\bf 38} (1997) 3399.
\bibitem{anderson1} I.M.~Anderson and T.~Duchamp, ``On the existence
  of global variational principles,'' Amer.\ J.\ Math.\ {\bf 102}
  (1980) 781. 
\bibitem{Anderson} I.M.~Anderson,
``Introduction to the Variational Bicomplex,''
Cont. Math. {\bf 132} (1992) 51. 
\bibitem{Anselmi:1994ry}
D.~Anselmi,
``Removal of divergences with the Batalin-Vilkovisky formalism,''
Class.\ Quant.\ Grav.\ {\bf 11} (1994) 2181.
\bibitem{Anselmi:1995zx}
D.~Anselmi,
``More on the subtraction algorithm,''
Class.\ Quant.\ Grav.\ {\bf 12} (1995) 319
hep-th/9407023.
\bibitem{arno-deser} R.~Arnowitt and S.~Deser, ``Interaction between Gauge
Vector Fields,'' Nucl.\ Phys.\ {\bf 49} (1963) 133.
\bibitem{AzIz} J.A.~de~Azc\'arraga and J.M.~Izquierdo, ``Lie groups,
    Lie algebras, cohomology and applications in physics,'' Cambridge
  Monographs on Mathematical Physics (Cambridge University Press
  1995). 
\bibitem{Bandelloni:1978ke}
G.~Bandelloni, A.~Blasi, C.~Becchi and R.~Collina,
``Nonsemisimple Gauge Models: 1. Classical Theory And The Properties
Of Ghost States,''
Ann.\ Inst.\ Henri Poincar\'e Phys.\ Theor.\ {\bf 28} (1978) 225.
\bibitem{Bandelloni:1978kf}
G.~Bandelloni, A.~Blasi, C.~Becchi and R.~Collina,
``Nonsemisimple Gauge Models: 2. Renormalization,''
Ann.\ Inst.\ Henri Poincar\'e Phys.\ Theor.\ {\bf 28} (1978) 255.
\bibitem{Bandelloni:1986wz}
G.~Bandelloni,
``Yang-Mills Cohomology In Four-Dimensions,''
J.\ Math.\ Phys.\ {\bf 27} (1986) 2551.
\bibitem{Bandelloni:1987kg}
G.~Bandelloni,
``Nonpolynomial Yang-Mills Local Cohomology,''
J.\ Math.\ Phys.\ {\bf 28} (1987) 2775.
\bibitem{Bandelloni:1993cq}
G.~Bandelloni and S.~Lazzarini,
``Diffeomorphism cohomology in Beltrami parametrization,''
J.\ Math.\ Phys.\  {\bf 34} (1993) 5413.
\bibitem{Bandelloni:1995vy}
G.~Bandelloni and S.~Lazzarini,
``Diffeomorphism cohomology in Beltrami parametrization. 2: The 1 forms,''
J.\ Math.\ Phys.\  {\bf 36} (1995) 1
[hep-th/9410190].   
\bibitem{Bardeen:1969md}
W.A.~Bardeen,
``Anomalous Ward Identities In Spinor Field Theories,''
Phys.\ Rev.\  {\bf 184} (1969) 1848.
\bibitem{Bardeen:1984pm}
W.~A.~Bardeen and B.~Zumino,
``Consistent And Covariant Anomalies In Gauge And Gravitational Theories,''
Nucl.\ Phys.\  {\bf B244} (1984) 421. 
\bibitem{Barnich:1993vg}
G.~Barnich and M.~Henneaux,
``Consistent couplings between fields with a gauge freedom and
deformations of the master equation,''
Phys.\ Lett.\ {\bf B311} (1993) 123
[hep-th/9304057].
\bibitem{Barnich:1994pa}
G.~Barnich, M.~Henneaux and R.~Tatar,
``Consistent interactions between gauge fields and local BRST cohomology:
The Example of Yang-Mills models,''
Int.\ J.\ Mod.\ Phys.\ {\bf D3} (1994) 139
[hep-th/9307155].
\bibitem{Barnich:1994ve}
G.~Barnich and M.~Henneaux,
``Renormalization of gauge invariant operators and anomalies in
Yang-Mills theory,''
Phys.\ Rev.\ Lett.\ {\bf 72} (1994) 1588
[hep-th/9312206].
\bibitem{Barnich:1995cq}
G.~Barnich, F.~Brandt and M.~Henneaux,
``Conserved currents and gauge invariance in Yang-Mills theory,''
Phys.\ Lett.\ {\bf B346} (1995) 81
hep-th/9411202.
\bibitem{Barnich:1995db}
G.~Barnich, F.~Brandt and M.~Henneaux,
``Local BRST cohomology in the antifield formalism. I. General theorems,''
Commun.\ Math.\ Phys.\ {\bf 174} (1995) 57
[hep-th/9405109].
\bibitem{Barnich:1995mt}
G.~Barnich, F.~Brandt and M.~Henneaux,
``Local BRST cohomology in the antifield formalism. II. Application to
Yang-Mills theory,''
Commun.\ Math.\ Phys.\ {\bf 174} (1995) 93
[hep-th/9405194].
\bibitem{Barnich:1995ap}
G.~Barnich, F.~Brandt and M.~Henneaux,
``Local BRST cohomology in Einstein Yang-Mills theory,''
Nucl.\ Phys.\ {\bf B455} (1995) 357
[hep-th/9505173].
\bibitem{Barnich:1999cy}
G.~Barnich, M.~Henneaux, T.~Hurth and K.~Skenderis,
``Cohomological analysis of gauged-fixed gauge theories,''
hep-th/9910201, to appear in Phys.\ Lett.\ {\bf B}.
\bibitem{Batalin:1977pb}
I.A.~Batalin and G.A.~Vilkovisky,
``Relativistic S Matrix Of Dynamical Systems With Boson And Fermion
Constraints,''
Phys.\ Lett.\ {\bf 69B} (1977) 309.
\bibitem{Batalin:1981jr}
I.A.~Batalin and G.A.~Vilkovisky,
``Gauge Algebra And Quantization,''
Phys.\ Lett.\ {\bf 102B} (1981) 27.
\bibitem{Batalin:1983wj}
I.A.~Batalin and G.~A.~Vilkovisky,
``Feynman Rules For Reducible Gauge Theories,''
Phys.\ Lett.\  {\bf B120} (1983) 166.
\bibitem{Batalin:1983jr}
I.A.~Batalin and G.A.~Vilkovisky,
``Quantization Of Gauge Theories With Linearly Dependent Generators,''
Phys.\ Rev.\ {\bf D28} (1983) 2567.
\bibitem{Batalin:1983pz}
I.A.~Batalin and E.S.~Fradkin,
``A Generalized Canonical Formalism And Quantization Of Reducible Gauge
Theories,''
Phys.\ Lett.\ {\bf 122B} (1983) 157.
\bibitem{Batalin:1984ss}
I.A.~Batalin and G.A.~Vilkovisky,
``Closure Of The Gauge Algebra, Generalized Lie Equations And Feynman Rules,''
Nucl.\ Phys.\ {\bf B234} (1984) 106.
\bibitem{Batalin:1985qj}
I.A.~Batalin and G.A.~Vilkovisky,
``Existence Theorem For Gauge Algebra,''
J.\ Math.\ Phys.\  {\bf 26} (1985) 172.
\bibitem{Batalin:1991qy}
I.A.~Batalin, P.M.~Lavrov and I.V.~Tyutin,
``Remarks on the Sp(2) covariant Lagrangian quantization of gauge theories,''
J.\ Math.\ Phys.\  {\bf 32} (1991) 2513.
\bibitem{Batalin:1996mp}
I.A.~Batalin, K.~Bering and P.H.~Damgaard,
``Gauge independence of the Lagrangian path integral in a higher-order
formalism,''
Phys.\ Lett.\ {\bf B389} (1996) 673
[hep-th/9609037].
\bibitem{Batalin:1999gf}
I.~Batalin and R.~Marnelius,
``General quantum antibrackets,''
hep-th/9905083.
\bibitem{Batalin:1999fi}
I.~Batalin and R.~Marnelius,
``Open group transformations within the Sp(2)-formalism,''
hep-th/9909223.
\bibitem{Baulieu:1984pf}
L.~Baulieu and J.~Thierry-Mieg,
``Algebraic Structure Of Quantum Gravity And The Classification Of 
The Gravitational Anomalies,''
Phys.\ Lett.\  {\bf B145} (1984) 53.
\bibitem{Baulieu:1984ih}
L.~Baulieu,
``Algebraic Construction Of Gauge Invariant Theories,''
PAR-LPTHE 84/4
{\it Based on lectures given at Carg\`ese Summer School: Particles and Fields, 
Carg\`ese, France, Jul 6-22, 1983}.
\bibitem{Baulieu:1984iw}
L.~Baulieu,
``Anomalies And Gauge Symmetry,''
Nucl.\ Phys.\ {\bf B241} (1984) 557.
\bibitem{Baulieu:1985tg}
L.~Baulieu,
``Perturbative Gauge Theories,''
Phys.\ Rept.\  {\bf 129} (1985) 1.
\bibitem{Baulieu:1986hw}
L.~Baulieu, C.~Becchi and R.~Stora,
``On The Covariant Quantization Of The Free Bosonic String,''
Phys.\ Lett.\  {\bf B180} (1986) 55.
\bibitem{Baulieu:1987dp}
L.~Baulieu, M.~Bellon and R.~Grimm,
``BRS Symmetry Of Supergravity In Superspace And Its Projection To 
Component Formalism,''
Nucl.\ Phys.\  {\bf B294} (1987) 279.
\bibitem{Baulieu:1987jz}
L.~Baulieu and M.~Bellon,
``Beltrami Parametrization And String Theory,''
Phys.\ Lett.\  {\bf B196} (1987) 142.
\bibitem{Becchi:1974xu} C.~Becchi, A.~Rouet and R.~Stora,
``The Abelian Higgs-Kibble Model. Unitarity Of The S Operator,''
Phys.\ Lett.\ {\bf B52} (1974) 344.
\bibitem{Becchi:1974md} 
C.~Becchi, A.~Rouet and R.~Stora,
``Renormalization Of The Abelian Higgs-Kibble Model,''
Commun.\ Math.\ Phys.\ {\bf 42} (1975) 127.
\bibitem{Becchi:1975nq} C.~Becchi, A.~Rouet and R.~Stora,
``Renormalization Of Gauge Theories,''
Annals Phys.\ {\bf 98} (1976) 287.
\bibitem{BecchiTira} C.~Becchi, A.~Rouet and R.~Stora,
``Renormalizable Theories With Symmetry Breaking",
in ``Field Theory Quantization and Statistical Physics",
E. Tirapegui ed., Reidel, Dordrecht 1981 (preprint dated 1975). 
\bibitem{Becchi:1988as}
C.~Becchi,
``On The Covariant Quantization Of The Free String: The Conformal Structure,''
Nucl.\ Phys.\  {\bf B304} (1988) 513.
\bibitem{Berends:1985rq}
F.A.~Berends, G.J.~Burgers and H.~van Dam,
``On The Theoretical Problems In Constructing Interactions Involving
Higher Spin Massless Particles,''
Nucl.\ Phys.\ {\bf B260} (1985) 295.
\bibitem{Bering:1996kw}
K.~Bering, P.~H.~Damgaard and J.~Alfaro,
``Algebra of Higher Antibrackets,''
Nucl.\ Phys.\  {\bf B478} (1996) 459
[hep-th/9604027].
\bibitem{Bertlmann:1996xk}
R.A.~Bertlmann,
``Anomalies in quantum field theory,''
{\it  Oxford, UK: Clarendon (1996) 566 p. (International series of monographs
                  on physics: 91)}.
\bibitem{Blaga:1995rv}
P.~A.~Blaga, L.~T\v ataru and I.~V.~Vancea,
``BRST cohomology for 2D gravity,''
Rom.\ J.\ Phys.\  {\bf 40} (1995) 773
[hep-th/9504037]. 
\bibitem{Bonneau:1990xu}
G.~Bonneau,
``Some Fundamental But Elementary Facts On Renormalization And Regularization:
A Critical Review Of The Eighties,''
Int.\ J.\ Mod.\ Phys.\ {\bf A5} (1990) 3831.
\bibitem{Bonora:1983ve}
L.~Bonora and P.~Cotta-Ramusino,
``Some Remarks On BRS Transformations, Anomalies And The Cohomology Of
The Lie Algebra Of The Group Of Gauge Transformations,''
Commun.\ Math.\ Phys.\ {\bf 87} (1983) 589.
\bibitem{Bonora:1983ff}
L.~Bonora, P.~Cotta-Ramusino and C.~Reina,
``Conformal Anomaly And Cohomology,''
Phys.\ Lett.\  {\bf B126} (1983) 305.
\bibitem{Bonora:1986ic}
L.~Bonora, P.~Pasti and M.~Tonin,
``The Anomaly Structure Of Theories With External Gravity,''
J.\ Math.\ Phys.\  {\bf 27} (1986) 2259. 
\bibitem{Bonora:1986cq}
L.~Bonora, P.~Pasti and M.~Bregola,
``Weyl Cocycles,''
Class.\ Quant.\ Grav.\  {\bf 3} (1986) 635.
\bibitem{Bonora:1986ug}
L.~Bonora, P.~Pasti and M.~Tonin,
``The Chiral Anomaly In Supersymmetric Gauge Theories 
Coupled To Supergravity,''
Phys.\ Lett.\  {\bf B167} (1986) 191.
\bibitem{Bourbaki} N. Bourbaki, ``Groupes et alg\`ebres de Lie I,''
  (Hermann, Paris 1960).
\bibitem{Brandt:1989rd}
F.~Brandt, N.~Dragon and M.~Kreuzer,
``All Consistent Yang-Mills Anomalies,''
Phys.\ Lett.\ {\bf B231} (1989) 263.
\bibitem{Brandt:1990gy}
F.~Brandt, N.~Dragon and M.~Kreuzer,
``Completeness And Nontriviality Of The Solutions Of The Consistency
Conditions,''
Nucl.\ Phys.\ {\bf B332} (1990) 224.
\bibitem{Brandt:1990gv}
F.~Brandt, N.~Dragon and M.~Kreuzer,
``Lie Algebra Cohomology,''
Nucl.\ Phys.\ {\bf B332} (1990) 250.
\bibitem{Brandt:1990et}
F.~Brandt, N.~Dragon and M.~Kreuzer,
``The Gravitational Anomalies,''
Nucl.\ Phys.\  {\bf B340} (1990) 187.
\bibitem{Brandt:1994sn}
F.~Brandt,
``Antifield dependence of anomalies,''
Phys.\ Lett.\ {\bf B320} (1994) 57
[hep-th/9310080].
\bibitem{Brandt:1994vd}
F.~Brandt,
``Anomaly candidates and invariants of D = 4, N=1 supergravity theories,''
Class.\ Quant.\ Grav.\  {\bf 11} (1994) 849
[hep-th/9306054].
\bibitem{Brandt:1996gu}
F.~Brandt, W.~Troost and A.~Van Proeyen,
``The BRST--antibracket cohomology of $2d$ gravity,''
Nucl.\ Phys.\  {\bf B464} (1996) 353
[hep-th/9509035].
\bibitem{Brandt:1996nn}
F.~Brandt, W.~Troost and A.~Van Proeyen,
``Background charges and consistent continuous deformations
of $2d$ gravity theories,''
Phys.\ Lett.\  {\bf B374} (1996) 31
[hep-th/9510195].                                         
\bibitem{Brandt:1997au}
F.~Brandt,
``Local BRST cohomology in minimal D = 4, N = 1 supergravity,''
Annals Phys.\  {\bf 259} (1997) 253
[hep-th/9609192].
\bibitem{Brandt:1998cy}
F.~Brandt, J.~Gomis and J.~Sim\'on,
``Cohomological analysis of bosonic D-strings and 2d sigma models
coupled to abelian gauge fields,''
Nucl.\ Phys.\  {\bf B523} (1998) 623
[hep-th/9712125].
\bibitem{Browning:1987hc}
A.D.~Browning and D.~McMullan,
``The Batalin, Fradkin, And Vilkovisky Formalism For Higher Order Theories,''
J.\ Math.\ Phys.\ {\bf 28} (1987) 438.
\bibitem{Bryant} R.L.~Bryant and P.A.~Griffiths, ``Characteristic Cohomology
Of Differential Systems I: General Theory,'' 
J.\ Am.\ Math.\ Soc.\ {\bf 8} (1995) 507.
\bibitem{Buchbinder:1995ns}
I.~L.~Buchbinder, B.~R.~Mistchuk and V.~D.~Pershin,
``BRST - BFV analysis of anomalies in bosonic string theory
interacting with background gravitational field,''
Phys.\ Lett.\  {\bf B353} (1995) 457
[hep-th/9502087]. 
\bibitem{Capper:1974ic}
D.~M.~Capper and M.~J.~Duff,
``Trace Anomalies In Dimensional Regularization,''
Nuovo Cim.\  {\bf 23A} (1974) 173.
\bibitem{Carvalho:1996uu}
M.~Carvalho, L.~C.~Vilar, C.~A.~Sasaki and S.~P.~Sorella,
``BRS Cohomology of Zero Curvature Systems I. The Complete Ladder
Case,''
J.\ Math.\ Phys.\  {\bf 37} (1996) 5310
[hep-th/9509047].
\bibitem{ChevEil} C. Chevalley and S. Eilenberg, ``Cohomology
    theory of Lie groups and Lie algebras,'' Trans. Amer. Math. Soc.
{\bf 63} (1948) 85. 
\bibitem{Collins:1984xc}
J.C.~Collins,
``Renormalization. An Introduction To Renormalization,
The Renormalization Group, And The Operator Product Expansion,''
{\it  Cambridge, Uk: Univ. Pr. ( 1984) 380p}, chapter 12.6.
\bibitem{Collins:1994ee}
J.C.~Collins and R.J.~Scalise,
``The Renormalization of Composite Operators in Yang-Mills Theories using
General Covariant Gauge,''
Phys.\ Rev.\ {\bf D50} (1994) 4117
[hep-ph/9403231].
\bibitem{Deser:1976yx}
S.~Deser, M.~J.~Duff and C.~J.~Isham,
``Nonlocal Conformal Anomalies,''
Nucl.\ Phys.\  {\bf B111} (1976) 45.
\bibitem{Deser:1982vy}
S.~Deser, R.~Jackiw and S.~Templeton,
``Three-Dimensional Massive Gauge Theories,''
Phys.\ Rev.\ Lett.\ {\bf 48} (1982) 975.
\bibitem{Deser:1993yx}
S.~Deser and A.~Schwimmer,
``Geometric classification of conformal anomalies in arbitrary dimensions,''
Phys.\ Lett.\  {\bf B309} (1993) 279
[hep-th/9302047].
\bibitem{DeWilde} M.~De Wilde, ``On the Local Chevalley Cohomology of
  the Dynamical Lie algebra of a Symplectic Manifold,'' 
Lett.\ Math.\ Phys. {\bf 5}
  (1981) 351. 
\bibitem{deWit:1978cd}
B.~de Wit and J.W.~van Holten,
``Covariant Quantization Of Gauge Theories With Open Gauge Algebra,''
Phys.\ Lett.\ {\bf 79B} (1978) 389.
\bibitem{DeWitt} B.S.~DeWitt, ``Quantum Theory of Gravity. II. The
  Manifestly Covariant Theory,'' Phys.\ 
Rev.\ {\bf 162} (1967) 1195.
\bibitem{Dickey} L.A.~Dickey, ``On Exactness of the Variational
  Bicomplex'', Cont.\ Math.\ {\bf 132} (1992) 307.
\bibitem{DixonTaylor} J.A. Dixon and J.C. Taylor, ``Renormalization
of Wilson Operators in Gauge Theories", Nucl. Phys. {\bf B78}
(1974) 552.
\bibitem{Dixon:1975si}
J.A.~Dixon,
``Field Redefinition And Renormalization In Gauge Theories,''
Nucl.\ Phys.\ {\bf B99} (1975) 420.
\bibitem{Dixon:1976aa}
J.A.~Dixon,
``Cohomology And Renormalization Of Gauge Theories. 1,''
unpublished preprint.
\bibitem{Dixon:1979bs}
J.A.~Dixon,
``Cohomology And Renormalization Of Gauge Theories. 2,''
HUTMP 78/B64.
\bibitem{Deans:1978wn}
W.S.~Deans and J.A.~Dixon,
``Theory Of Gauge Invariant Operators: Their Renormalization And 
S Matrix Elements,''
Phys.\ Rev.\  {\bf D18} (1978) 1113.
\bibitem{Dixon:1980zp}
J.~Dixon and M.~Ramon Medrano,
``Anomalies of higher-dimension composite fields,''
Phys.\ Rev.\ {\bf D22} (1980) 429.
\bibitem{Dixon:1991wi}
J.A.~Dixon,
``Calculation of BRS cohomology with spectral sequences,''
Commun.\ Math.\ Phys.\ {\bf 139} (1991) 495.
\bibitem{Dragon:1992fn}
N.~Dragon,
``Regular gravitational Lagrangians,''
Phys.\ Lett.\  {\bf B276} (1992) 31.
\bibitem{Dragon:1996md}
N.~Dragon,
``BRS Symmetry and Cohomology,'' 
Lectures given at First
National Summer School for Graduate Students, Saalburg 1995,
hep-th/9602163.
\bibitem{Dragon:1997tk}
N.~Dragon, T.~Hurth and P.~van Nieuwenhuizen,
``Polynomial form of the Stueckelberg model,''
Nucl.\ Phys.\ Proc.\ Suppl.\  {\bf 56B} (1997) 318
[hep-th/9703017].
\bibitem{Dubois-Violette:1985hc}
M.~Dubois-Violette, M.~Talon and C.M.~Viallet,
``New Results On BRS Cohomology In Gauge Theory,''
Phys.\ Lett.\ {\bf 158B} (1985) 231.
\bibitem{Dubois-Violette:1985jb}
M.~Dubois-Violette, M.~Talon and C.M.~Viallet,
``BRS Algebras: Analysis Of The Consistency Equations In Gauge Theory,''
Commun.\ Math.\ Phys.\ {\bf 102} (1985) 105.
\bibitem{Dubois-Violette:1986cj}
M.~Dubois-Violette, M.~Talon and C.M.~Viallet,
``Anomalous Terms In Gauge Theory: Relevance Of The Structure Group,''
Ann.\ Inst.\ Henri Poincar\'e Phys.\ Theor.\ {\bf 44} (1986) 103.
\bibitem{Dubois-Violette:1987uu}
M.~Dubois-Violette,
``Syst\`emes dynamiques contraints: l'approche homologique,''
Ann.\ Inst.\ Fourier {\bf 37} (1987) 45.
\bibitem{Dubois-Violette:1991is}
M.~Dubois-Violette, M.~Henneaux, M.~Talon and C.~Viallet,
``Some results on local cohomologies in field theory,''
Phys.\ Lett.\ {\bf B267} (1991) 81.
\bibitem{Dubois-Violette:1992ye}
M.~Dubois-Violette, M.~Henneaux, M.~Talon and C.~Viallet,
``General solution of the consistency equation,''
Phys.\ Lett.\ {\bf B289} (1992) 361
[hep-th/9206106].
\bibitem{Faddeev:1967fc}
L.D.~Faddeev and V.N.~Popov,
``Feynman Diagrams For The Yang-Mills Field,''
Phys.\ Lett.\  {\bf B25} (1967) 29.
\bibitem{Ferrari:1998jy}
R.~Ferrari and P.A.~Grassi,
``Constructive algebraic renormalization of the Abelian Higgs-Kibble  model,''
Phys.\ Rev.\ {\bf D60} (1999) 065010
[hep-th/9807191].
\bibitem{Feynman:1963ax}
R.P.~Feynman,
``Quantum Theory Of Gravitation,''
Acta Phys.\ Polon.\  {\bf 24} (1963) 697.
\bibitem{Fisch:1989dq}
J.~Fisch, M.~Henneaux, J.~Stasheff and C.~Teitelboim,
``Existence, Uniqueness And Cohomology Of The Classical BRST Charge With
Ghosts Of Ghosts,''
Commun.\ Math.\ Phys.\ {\bf 120} (1989) 379.
\bibitem{Fisch:1990rp}
J.M.L.~Fisch and M.~Henneaux,
``Homological Perturbation Theory And The Algebraic Structure Of The
Antifield - Antibracket Formalism For Gauge Theories,''
Commun.\ Math.\ Phys.\ {\bf 128} (1990) 627.
\bibitem{Fradkin:1975cq}
E.S.~Fradkin and G.A.~Vilkovisky,
``Quantization Of Relativistic Systems With Constraints,''
Phys.\ Lett.\ {\bf B55} (1975) 224.
\bibitem{Fradkin:1977wv}
E.~S.~Fradkin and M.~A.~Vasilev,
``Hamiltonian Formalism, Quantization And S Matrix For Supergravity,''
Phys.\ Lett.\  {\bf B72} (1977) 70.
\bibitem{Fradkin:1977hw}
E.S.~Fradkin and G.A.~Vilkovisky,
``Quantization Of Relativistic Systems With Constraints: Equivalence Of
Canonical And Covariant Formalisms In Quantum Theory Of Gravitational Field,''
CERN-TH-2332.
\bibitem{Fradkin:1978xi}
E.S.~Fradkin and T.E.~Fradkina,
``Quantization Of Relativistic Systems With Boson And Fermion First And
Second Class Constraints,''
Phys.\ Lett.\ {\bf 72B} (1978) 343.
\bibitem{Freedman:1981us}
D.Z.~Freedman and P.K.~Townsend,
``Antisymmetric Tensor Gauge Theories And Nonlinear Sigma Models,''
Nucl.\ Phys.\ {\bf B177} (1981) 282.
\bibitem{Fujiwara:1999fi}
T.~Fujiwara, H.~Suzuki and K.~Wu,
``Non-commutative differential calculus and the axial anomaly in Abelian 
lattice gauge theories,'' hep-lat/9906015.
\bibitem{Garcia:1998kp}
J.A.~Garcia and B.~Knaepen,
``Couplings between generalized gauge fields,''
Phys.\ Lett.\  {\bf B441} (1998) 198
[hep-th/9807016].
\bibitem{gelfand} I.M.~Gel'fand and I.Ya.~Dorfman, ``Hamiltonian
  operators and associated algebraic structures,'' 
Funct.\ Anal.\ Appl.\ {\bf 13} (1979) 174.
\bibitem{Georgi:1992xg}
H.~Georgi,
``Thoughts on effective field theory,''
Nucl.\ Phys.\ Proc.\ Suppl.\ {\bf 29BC} (1992) 1.
\bibitem{Geyer:1999qi}
B.~Geyer, P.~M.~Lavrov and D.~M\"ulsch,
``OSp(1,2)-covariant Lagrangian quantization of reducible 
massive gauge  theories,''
J.\ Math.\ Phys.\ {\bf 40} (1999) 6189
[hep-th/9806117].
\bibitem{Gomis:1995he}
J.~Gomis, J.~Par\'{\i}s and S.~Samuel,
``Antibracket, antifields and gauge theory quantization,''
Phys.\ Rept.\ {\bf 259} (1995) 1
[hep-th/9412228].
\bibitem{Gomis:1993ub}
J.~Gomis and J.~Par\'{\i}s,
``Perturbation theory and locality in the field - antifield formalism,''
J.\ Math.\ Phys.\ {\bf 34} (1993) 2132.
\bibitem{Gomis:1996jp}
J.~Gomis and S.~Weinberg,
``Are Nonrenormalizable Gauge Theories Renormalizable?,''
Nucl.\ Phys.\ {\bf B469} (1996) 473
[hep-th/9510087].
\bibitem{Grassi:1999tp}
P.A.~Grassi, T.~Hurth and M.~Steinhauser,
``Practical algebraic renormalization,''
hep-ph/9907426.
\bibitem{GHV} W. Greub, S. Halperin and R. Vanstone, ``Connections,
    Curvature and Cohomology. Volume III: Cohomology of Principal
    Bundles and Homogeneous Spaces'', Pure and Applied Mathematics. A
  Series of Monographs and Textbooks, vol. 47, eds. S. Eilenberg and
  H.~Bass, (Academic Press, New York: 1976).            
\bibitem{Grigorev:1999qz}
M.A.~Grigorev and P.H.~Damgaard,
``Superfield BRST charge and the master action,''
hep-th/9911092.
\bibitem{Grigorev:2000zg}
M.A.~Grigorev, A.~M.~Semikhatov and I.~Y.~Tipunin,
``BRST formalism and zero locus reduction,''
hep-th/0001081.
\bibitem{gugmay} V.K.A.M.~Gugenheim and J.P.~May, ``On the theory and 
applications of 
torsion products,'' Mem.\ AMS {\bf 142} (1974). 
\bibitem{gug} V.K.A.M.~Gugenheim, ``On a perturbation theory for the 
homology of a 
loop space,'' J.\ Pure Appl.\ Alg.\ {\bf 25}
(1982) 197.
\bibitem{gugsta} V.K.A.M.~Gugenheim and J.D.~Stasheff, ``On perturbations and 
$A_\infty$-structures,'' Bull.\ Soc.\ Math.\ Belgique {\bf 38} (1986) 237. 
\bibitem{guglam} V.K.A.M.~Gugenheim and L.~Lambe, ``Applications of 
perturbation theory 
to differential homological algebra,'' Ill.\ J.\ Math.\ {\bf 33} (1989) 556.
\bibitem{guglamsta} V.K.A.M.~Gugenheim, L.~Lambe and J.D.~Stasheff, 
``Applications of perturbation theory 
to differential homological algebra II,'' Ill.\ J.\ Math.\ {\bf 34} (1990) 485.
\bibitem{Harris:1995tp}
B.W.~Harris and J.~Smith,
``Anomalous dimension of the gluon operator in pure Yang-Mills theory,''
Phys.\ Rev.\ {\bf D51} (1995) 4550
[hep-ph/9409405].
\bibitem{Henneaux:1985kr}
M.~Henneaux,
``Hamiltonian Form Of The Path Integral For Theories With A Gauge Freedom,''
Phys.\ Rept.\ {\bf 126} (1985) 1.
\bibitem{Henneaux:1988ej}
M.~Henneaux and C.~Teitelboim,
``BRST Cohomology In Classical Mechanics,''
Commun.\ Math.\ Phys.\ {\bf 115} (1988) 213.
\bibitem{Henneaux:1989jq}
M.~Henneaux,
``Lectures On The Antifield - BRST Formalism For Gauge Theories,''
{\it Lectures given at 20th GIFT Int. Seminar on Theoretical Physics, 
Jaca, Spain, Jun 5-9, 1989, and at CECS, Santiago, Chile, June/July 1989}, 
Nucl.\ Phys.\ B (Proc.\ Suppl.) {\bf A18} (1990) 47.
\bibitem{Henneaux:1991rx}
M.~Henneaux,
``Space-Time Locality of the BRST Formalism,''
Commun.\ Math.\ Phys.\ {\bf 140} (1991) 1.
\bibitem{Henneaux:1992ig}
M.~Henneaux and C.~Teitelboim,
``Quantization of gauge systems,''
{\it  Princeton, USA: Univ. Pr. (1992) 520 p}.
\bibitem{Henneaux:1993jn}
M.~Henneaux,
``Remarks on the renormalization of gauge invariant operators in
Yang-Mills theory,''
Phys.\ Lett.\ {\bf B313} (1993) 35
[hep-th/9306101].
\bibitem{Henneaux:1996ex}
M.~Henneaux,
``On The Gauge-Fixed BRST Cohomology,''
Phys.\ Lett.\ {\bf B367} (1996) 163
[hep-th/9510116].
\bibitem{Henneaux:1997bm}
M.~Henneaux,
``Consistent interactions between gauge fields: The cohomological  approach,''
{\it Proceedings of a Conference 
on Secondary Calculus and Cohomological Physics, Aug 24-31, 1997,
Moscow, Russia,}
Cont.\ Math.\ {\bf 219} (1998) 93 [hep-th/9712226].
\bibitem{Henneaux:1997ws}
M.~Henneaux, B.~Knaepen and C.~Schomblond,
``Characteristic cohomology of p-form gauge theories,''
Commun.\ Math.\ Phys.\  {\bf 186} (1997) 137
[hep-th/9606181].
\bibitem{Henneaux:1997nz}
M.~Henneaux, B.~Knaepen and C.~Schomblond,
``BRST cohomology of the Chapline-Manton model,''
Lett.\ Math.\ Phys.\  {\bf 42} (1997) 337
[hep-th/9702042]. 
\bibitem{Henneaux:1997ha}
M.~Henneaux and B.~Knaepen,
``All consistent interactions for exterior form gauge fields,''
Phys.\ Rev.\  {\bf D56} (1997) 6076
[hep-th/9706119].
\bibitem{Henneaux:1998hq}
M.~Henneaux and A.~Wilch,
``Local BRST cohomology of the gauged principal non-linear sigma model,''
Phys.\ Rev.\  {\bf D58} (1998) 025017
[hep-th/9802118].
\bibitem{Henneaux:1999rp}
M.~Henneaux and B.~Knaepen,
``The Wess-Zumino consistency condition for p-form gauge theories,''
Nucl.\ Phys.\  {\bf B548} (1999) 491
[hep-th/9812140].
\bibitem{Henneaux:2000dc}
M.~Henneaux and A.~Wilch,
``Semi-invariant  terms for gauged non-linear sigma-models,''
Phys.\ Lett.\  {\bf B471} (2000) 373
[hep-th/9906121].
\bibitem{hirsch} G.~Hirsch, ``Sur les groups d'homologie 
des espaces fibr\'es,'' 
Bull.\ Soc.\ Math.\ Belgique {\bf 6} (1953) 79. 
\bibitem{'tHooft:1971fh} G.~'t Hooft,
``Renormalization Of Massless Yang-Mills Fields,''
Nucl.\ Phys.\ {\bf B33} (1971) 173.
\bibitem{'tHooft:1971rn} G.~'t Hooft,
``Renormalizable Lagrangians For Massive Yang-Mills Fields,''
Nucl.\ Phys.\ {\bf B35} (1971) 167.
\bibitem{'tHooft:1972fi} G.~'t Hooft and M.~Veltman,
``Regularization And Renormalization Of Gauge Fields,''
Nucl.\ Phys.\ {\bf B44} (1972) 189.
\bibitem{'tHooft:1972ue} G.~'t Hooft and M.~Veltman,
``Combinatorics Of Gauge Fields,''
Nucl.\ Phys.\ {\bf B50} (1972) 318.
\bibitem{Howe:1990pz}
P.S.~Howe, U.~Lindstr\"om and P.~White,
``Anomalies And Renormalization In The BRST-BV Framework,''
Phys.\ Lett.\  {\bf B246} (1990) 430.
\bibitem{Hurth:1998nq}
T.~Hurth and K.~Skenderis,
``Quantum Noether method,''
Nucl.\ Phys.\ {\bf B541} (1999) 566
hep-th/9803030.
\bibitem{Hurth:1998ib}
T.~Hurth and K.~Skenderis,
``The quantum Noether condition in terms of interacting fields,''
[hep-th/9811231].
\bibitem{Joglekar:1976nu}
S.D.~Joglekar and B.W.~Lee,
``General Theory Of Renormalization Of Gauge Invariant Operators,''
Ann.\ Phys.\ {\bf 97} (1976) 160.
\bibitem{Kallosh:1977ik}
R.E.~Kallosh,
``Gauge Invariance In Supergravitation. (In Russian),''
Pisma Zh.\ Eksp.\ Teor.\ Fiz.\  {\bf 26} (1977) 575.
\bibitem{Kallosh:1978de}
R.E.~Kallosh,
``Modified Feynman Rules In Supergravity,''
Nucl.\ Phys.\ {\bf B141} (1978) 141.
\bibitem{Kaplan:1995uv}
D.B.~Kaplan,
``Effective field theories,'' {\it Lectures at the Seventh Summer School in 
Nuclear Physics: Symmetries,} (Seattle, 1995),
nucl-th/9506035.
\bibitem{Kluberg-Stern:1975rs}
H.~Kluberg-Stern and J.B.~Zuber,
``Ward Identities And Some Clues To The Renormalization Of Gauge Invariant
Operators,''
Phys.\ Rev.\ {\bf D12} (1975) 467.
\bibitem{Kluberg-Stern:1975xv}
H.~Kluberg-Stern and J.B.~Zuber,
``Renormalization Of Nonabelian Gauge Theories In A Background
Field Gauge: 1. Green Functions,''
Phys.\ Rev.\ {\bf D12} (1975) 482.
\bibitem{Kluberg-Stern:1975hc}
H.~Kluberg-Stern and J.B.~Zuber,
``Renormalization Of Nonabelian Gauge Theories In A Background
Field Gauge. 2. Gauge Invariant Operators,''
Phys.\ Rev.\ {\bf D12} (1975) 3159.
\bibitem{Khudaverdian:1993ji} O.M.~Khudaverdian and A.P.~Nersessian,
``On the geometry of the Batalin-Vilkovsky formalism,''
Mod.\ Phys.\ Lett.\  {\bf A8} (1993) 2377
[hep-th/9303136].
\bibitem{Knaepen:1999zr}
B.~Knaepen,
``Local BRST cohomology for p-form gauge theories,''
hep-th/9912021. 
\bibitem{kobayashiIIsec12} S.~Kobayashi and K.~Nomizu, 
``Foundations of Differential Geometry. Volume II,'' John Wiley and Sons 
(1996), section 12.
\bibitem{Koszul} J.-L. Koszul, ``Homologie et cohomologie des
    alg\`ebres de Lie,'' Bull.\ Soc.\ Math.\ France {\bf 78} (1950)
  65-127.
\bibitem{Lam:1972mb}
Y.P.~Lam,
``Perturbation Lagrangian Theory For Scalar Fields:
Ward-Takahasi Identity And Current Algebra,''
Phys.\ Rev.\ {\bf D6} (1972) 2145.
\bibitem{Lam:1973qa}
Y.P.~Lam,
``Equivalence Theorem On Bogoliubov-Parasiuk-Hepp-Zimmermann
Renormalized Lagrangian Field Theories,''
Phys.\ Rev.\ {\bf D7} (1973) 2943.
\bibitem{lamsta} L.~Lambe and J.D.~Stasheff, ``Applications of perturbation 
theory to iterated fibrations,'' Manus.\ Math.\ {\bf 58} (1987)
363. 
\bibitem{Langouche:1984gn}
F.~Langouche, T.~Sch\"ucker and R.~Stora,
``Gravitational Anomalies Of The Adler-Bardeen Type,''
Phys.\ Lett.\  {\bf B145} (1984) 342.
\bibitem{Lavrov:1985hr}
P.M.~Lavrov and I.~V.~Tyutin,
``Effective Action In General Gauge Theories,''
Sov. J. Nucl. Phys.  {\bf 41} (1985) 1049.
\bibitem{Lee:1972fj}
B.W.~Lee and J.~Zinn-Justin,
``Spontaneously Broken Gauge Symmetries. I. Preliminaries,''
Phys.\ Rev.\ {\bf D5} (1972) 3121.
\bibitem{Lee:1972fk}
B.W.~Lee and J.~Zinn-Justin,
``Spontaneously Broken Gauge Symmetries. II. Perturbation Theory And
Renormalization,''
Phys.\ Rev.\ {\bf D5} (1972) 3137.
\bibitem{Lee:1972fm}
B.W.~Lee and J.~Zinn-Justin,
``Spontaneously Broken Gauge Symmetries. III. Equivalence,''
Phys.\ Rev.\ {\bf D5} (1972) 3155.
\bibitem{Lee:1973fn}
B.W.~Lee and J.~Zinn-Justin,
``Spontaneously Broken Gauge Symmetries. IV. General Gauge Formulation,''
Phys.\ Rev.\ {\bf D7} (1973) 1049.
\bibitem{Lowenstein:1971jk}
J.H.~Lowenstein,
``Differential Vertex Operations In Lagrangian Field Theory,''
Commun.\ Math.\ Phys.\ {\bf 24} (1971) 1.
\bibitem{luscher}
M.~L\"uscher, ``Topology and the axial anomaly in abelian lattice gauge 
theories,'' Nucl.\ Phys.\ {\bf B538} (1999) 515.
\bibitem{Manes:1985df}
J.~Ma\~nes, R.~Stora and B.~Zumino,
``Algebraic Study Of Chiral Anomalies,''
Commun.\ Math.\ Phys.\ {\bf 102} (1985) 157.
\bibitem{Manohar:1996cq}
A.V.~Manohar,
``Effective field theories,'' 
{\it Lectures at the 1996 Schladming Winter School,}
hep-ph/9606222.
\bibitem{Mcmullan:1987hb}
D.~McMullan,
``Yang-Mills Theory And The Batalin-Fradkin-Vilkovisky Formalism,''
J.\ Math.\ Phys.\ {\bf 28} (1987) 428.
\bibitem{misner-wheeler} C.W.~Misner and J.A.~Wheeler,
``Classical physics as geometry: gravitation, electromagnetism, 
unquantized charge and mass as properties of curved empty space,''
Ann.\ Phys.\ (N.Y.) {\bf 2} (1957) 525.
\bibitem{olver} P.J.~Olver, Applications of Lie Groups
to Differential Equations", Springer, New York: 1986. 
\bibitem{Raif} L. O'Raifeartaigh, ``Group structure of gauge
    theories,'' (Cambridge University Press 1986). 
\bibitem{Pich:1998xt}
A.~Pich,
``Effective field theory: Course,'' {\it Talk given at Les Houches
  Summer School in Theoretical Physics, Session 68: Probing the
  Standard Model of Particle Interactions, Les Houches, France, 28 Jul - 5
Sep 1997}, 
hep-ph/9806303.
\bibitem{Piguet:1981nr}
O.~Piguet and A.~Rouet,
``Symmetries In Perturbative Quantum Field Theory,''
Phys.\ Rept.\ {\bf 76} (1981) 1.
\bibitem{Piguet:1995er}
O.~Piguet and S.P.~Sorella,
``Algebraic renormalization: Perturbative renormalization, symmetries 
and anomalies,''
{\it  Berlin, Germany: Springer (1995) 134 p. (Lecture notes in physics:
                  m28)}.
\bibitem{Piguet:1998bj}
O.~Piguet and S.~Wolf,
``The supercurrent trace identities of the N = 1, D = 4 super-Yang-Mills 
theory in the Wess-Zumino gauge,''
JHEP {\bf 9804} (1998) 001
[hep-th/9802027].
\bibitem{Polchinski:1992ed}
J.~Polchinski,
``Effective field theory and the Fermi surface,'' {\it 
Lectures presented at TASI 92, Boulder, CO, Jun 3-28, 1992.} 
Published in Boulder TASI 92:0235-276
[hep-th/9210046].
\bibitem{Postnikov} M. Postnikov, ``Le\c cons de g\'eom\'etrie:
    Groupes et alg\`ebres de Lie'', (traduction
  fran\c caise: Editions Mir, 1985).    
\bibitem{Saunders} D.J.~Saunders, ``The Geometry of Jet bundles",
Cambridge University Press (Cambridge, 1989).
\bibitem{Schucker:1987ca}
T.~Sch\"ucker,
``The Cohomological Proof Of Stora's Solutions,''
Commun.\ Math.\ Phys.\  {\bf 109} (1987) 167.   
\bibitem{Slavnov:1972fg}
A.A.~Slavnov,
``Ward Identities In Gauge Theories,''
Theor.\ Math.\ Phys.\ {\bf 10} (1972) 99.
\bibitem{Stasheff:1991eb}
J.~Stasheff,
``Homological (ghost) approach to constrained Hamiltonian systems,''
hep-th/9112002.
\bibitem{Stasheff:1997fe}
J.~Stasheff,
``Deformation theory and the Batalin-Vilkovisky master equation,''
q-alg/9702012.
\bibitem{Sterman:1978ds}
G.~Sterman, P.K.~Townsend and P.~van Nieuwenhuizen,
``Unitarity, Ward Identities, And New Quantization Rules Of Supergravity,''
Phys.\ Rev.\  {\bf D17} (1978) 1501.
\bibitem{Stora:1976kd}
R.~Stora,
``Continuum Gauge Theories,''
{\it Lectures given at Summer Inst. for Theoretical Physics, Carg\`ese, France,
                  Jul 12-31, 1976}, published in 
{\it New Developments in Quantum 
                  Field Theory and Statistical Physics,} eds. M.~L\'evy 
                and P.~Mitter NATO ASI Series B26 (Plenum, 1977).
\bibitem{Stora:1983ct}
R.~Stora,
``Algebraic Structure And Topological Origin Of Anomalies,''
{\it Seminar given at Carg\`ese Summer Inst. Sep 1-15, 1983},
published in {\it Progress in Gauge Field Theory,} eds. 't Hooft et
al. (Plenum Press, 1984) 
\bibitem{sullivan} D.~Sullivan, ``Infinitesimal computations in
  topology,''
Pub.\ Math.\ IHES {\bf 47} (1977) 269.
\bibitem{takens} F.~Takens, ``A global version of the inverse problem
  to the calculus of variations,'' J.\ Diff.\ Geom.\ {\bf 14} (1979)
  543. 
\bibitem{Suzuki:2000ii}
H.~Suzuki,
``Anomaly cancellation condition in lattice gauge theory,''
hep-lat/0002009.
\bibitem{Talon:1985dz}
M.~Talon,
``Algebra Of Anomalies,''
PAR-LPTHE-85/37,
{\it Presented at Carg\`ese Summer School, Carg\`ese, France, Jul 15-31, 1985}.
\bibitem{Tataru:1996ru}
L.~T\v ataru and I.~V.~Vancea,
``BRST cohomology in Beltrami parametrization,''
Int.\ J.\ Mod.\ Phys.\  {\bf A11} (1996) 375
[hep-th/9504036].
\bibitem{Tataru:1998pn}
L.~T\v ataru and R.~T\v atar,
``Koszul-Tate cohomology for an Sp(2)-covariant quantization of gauge  
theories with linearly dependent generators,''
Int.\ J.\ Mod.\ Phys.\  {\bf A13} (1998) 1981
[hep-th/9708159].
\bibitem{Taylor:1971ff}
J.C.~Taylor,
``Ward Identities And Charge Renormalization Of The Yang-Mills Field,''
Nucl.\ Phys.\ {\bf B33} (1971) 436.
\bibitem{Thierry-Mieg:1984vx}
J.~Thierry-Mieg,
``Classification Of The Yang-Mills Anomalies In Even And Odd Dimension,''
Phys.\ Lett.\ {\bf B147} (1984) 430.
\bibitem{Tonin:1992wf}
M.~Tonin,
``Dimensional regularization and anomalies in chiral gauge theories,''
Nucl.\ Phys.\ Proc.\ Suppl.\  {\bf 29BC} (1992) 137.
\bibitem{Torre:1995kb}
C.G.~Torre,
``Natural symmetries of the Yang-Mills equations,''
J.\ Math.\ Phys.\  {\bf 36} (1995) 2113
[hep-th/9407129].  
\bibitem{Troost:1990} W.~Troost, P.~van Nieuwenhuizen and A.~Van Proeyen,
``Anomalies And The Batalin-Vilkovisky Lagrangian Formalism,''
Nucl.\ Phys.\  {\bf B333} (1990) 727. 
\bibitem{Troost:1994xw}
W.~Troost and A.~Van Proeyen,
``Regularization, the BV method, and the antibracket cohomology,''
hep-th/9410162.
\bibitem{Troost:1993mr}
W.~Troost and A.~Van Proeyen,
``Regularization and the BV formalism,''
hep-th/9307126.
\bibitem{tsujishita} T.\ Tsujishita, ``On variational bicomplexes
  associated to differential equations,'' Osaka J.\ Math.\ {\bf 19}
  (1982) 311. 
\bibitem{tulczyjew} W.M.~Tulczyjew, ``The Euler-Lagrange resolution,''
in Lecture Notes in Mathematics, No.\ 836 (1980) 22.
\bibitem{Tyutin:1975qk} I.V.~Tyutin,
``Gauge Invariance In Field Theory And 
Statistical Physics In Operator Formalism,''
LEBEDEV-75-39.
\bibitem{Tyutin:1981wa}
I.V.~Tyutin and B.L.~Voronov,
``Renormalization Of General Gauge Theories,''
{\it  In *Moscow 1981, Proceedings, Quantum Gravity*, 481-501}.
\bibitem{Unruh} W.G.~Unruh, ``Excluded Possibilities of Geometrodynamical 
Analog to Electric Charge,'' Gen.\ Relativ.\ Gravit.\ {\bf 2}
(1971) 27. 
\bibitem{vanholten} J.W.~van Holten, ``On the construction of supergravity
theories,'' Ph.D.\ thesis, University of Leiden (1980).
\bibitem{Verbovetsky:1997fi}
A.~Verbovetsky,
``Notes on the horizontal cohomology,'' in {\it Proceedings of a Conference 
on Secondary Calculus and Cohomological Physics, Aug 24-31, 1997,
Moscow, Russia,}
Cont.\ Math.\ {\bf 219} (1998) 211 [math.dg/9803115].
\bibitem{vinogradov1} A.M.~Vinogradov, ``On the algebra-geometric foundations
of Lagrangian field theory,'' Sov.\ Math.\ Dokl.\ 
{\bf 18} (1977) 1200.
\bibitem{vinogradov2} A.M.~Vinogradov, ``A spectral sequence associated with a
nonlinear differential equation and algebra-geometric foundations
of Lagrangian field theory with constraints,'' Sov.\ Math.\ Dokl.\
{\bf 19} (1978) 144.
\bibitem{vinogradov3} A.M.~Vinogradov, 
``The ${\cal C}$-spectral sequence,
 Lagrangian formalism, and conservation laws. I. The linear theory. 
II. The non linear theory,'' J.\ Math.\ Anal.\ Appl.\ {\bf 100} (1984) 1.
\bibitem{Voronov:1982cp}
B.L.~Voronov and I.V.~Tyutin,
``Formulation Of Gauge Theories Of General Form. I,''
Theor.\ Math.\ Phys.\ {\bf 50} (1982) 218.
\bibitem{Voronov:1982ur}
B.l.~Voronov and I.V.~Tyutin,
``Formulation Of Gauge Theories Of General Form. II. 
Gauge Invariant Renormalizability And Renormalization Structure,''
Theor.\ Math.\ Phys.\ {\bf 52} (1982) 628.
\bibitem{Voronov:1982ph}
B.~L.~Voronov, P.~M.~Lavrov and I.~V.~Tyutin,
``Canonical Transformations And The Gauge Dependence In 
General Gauge Theories ,''
Sov. J. Nucl. Phys.  {\bf 36} (1982) 292.
\bibitem{wald1} R.M.~Wald, ``Spin-two Fields And General Covariance,''
Phys.\ Rev.\  {\bf D33} (1986) 3613.
\bibitem{Wald} R.M.~Wald, ``On identically closed forms locally constructed 
from a field,'' J.\ Math.\ Phys.\ {\bf 31}
  (1990) 2378.
\bibitem{Weinberg:1979kz}
S.~Weinberg,
``Phenomenological Lagrangians,''
Physica {\bf 96A} (1979) 327.
\bibitem{Weinberg:1995mt}
S.~Weinberg,
``The Quantum theory of fields. Vol. 1: Foundations,''
{\it  Cambridge, UK: Univ. Pr. (1995) 609 p}.
\bibitem{Weinberg:1996kr}
S.~Weinberg,
``The quantum theory of fields. Vol. 2: Modern applications,''
{\it  Cambridge, UK: Univ. Pr. (1996) 489 p}.
\bibitem{Weinberg:1996kw}
S.~Weinberg,
``What is quantum field theory, and what did we think it was?,''
hep-th/9702027.
\bibitem{WerneckdeOliveira:1993ig}
M.~Werneck de Oliveira, M.~Schweda and S.~P.~Sorella,
``Supersymmetric structure of the bosonic string theory in the
Beltrami parametrization,''
Phys.\ Lett.\  {\bf B315} (1993) 93
[hep-th/9305148]. 
\bibitem{Wess:1971cm}
J.~Wess and B.~Zumino,
``Consequences Of Anomalous Ward Identities,''
Phys.\ Lett.\ {\bf B37} (1971) 95.
\bibitem{Witten:1990wb}
E.~Witten,
``A Note On The Antibracket Formalism,''
Mod.\ Phys.\ Lett.\ {\bf A5} (1990) 487.
\bibitem{Yang:1954ek}
C.N.~Yang and R.L.~Mills,
``Conservation of isotopic spin and isotopic gauge invariance,''
Phys.\ Rev.\ {\bf 96} (1954) 191.
\bibitem{zelobenko} D.P.~\v{Z}elobenko, 
``Compact Lie Groups and their Representations,''
(Translations of Mathematical Monographs vol. 40)
American Mathematical Society (Providence, Rhode Island 1973).
\bibitem{Zinn-Justin:1974mc}
J.~Zinn-Justin,
``Renormalization Of Gauge Theories,''
{\it Lectures given at Int. Summer Inst. for Theoretical Physics, Jul 29 -
                  Aug 9, 1974, Bonn, West Germany} published 
in {\it Trends in Elementary Particle Physics,} 
Lectures Notes in Physics 37, H. Rollnik and K. Dietz eds.
(Springer Verlag, Berlin 1975).
\bibitem{Zinn-Justin:1989mi}
J.~Zinn-Justin,
``Quantum Field Theory And Critical Phenomena,''
{\it  Oxford, UK: Clarendon (1989),
914 p.
(International series of monographs
on physics, 77)}. Third ed. 1996, chapter 21.
\bibitem{Zinn-Justin:1999ze}
J.~Zinn-Justin,
``Renormalization of gauge theories and master equation,''
Mod.\ Phys.\ Lett.\ {\bf A14} (1999) 1227
hep-th/9906115.
\bibitem{Zumino:1983ew}
B.~Zumino,
``Chiral Anomalies And Differential Geometry''
{\it Lectures given at Les Houches Summer School on Theoretical Physics, Les
                  Houches, France, Aug 8 - Sep 2, 1983}, published in
                {\it Relativity, Groups and Topology II,} eds. B.S.~
                DeWitt and R.~Stora (North Holland, Amsterdam 1984). 
\bibitem{Zumino:1984rz}
B.~Zumino, Y.~Wu and A.~Zee,
``Chiral Anomalies, Higher Dimensions, And Differential Geometry,''
Nucl.\ Phys.\ {\bf B239} (1984) 477.
\end{thebibliography}
\end{document}